%% file: Testing.tex
\documentclass[12pt,a4paper]{book}
\pdfoutput=1
\usepackage[utf8]{inputenc}

\usepackage{stmaryrd,upgreek,mathrsfs}
\usepackage{amstext,amsmath,amssymb,amsfonts,amsthm, bbm}
\usepackage{hyperref}
\usepackage{authblk}
\usepackage{caption} 
\usepackage{epsfig} 
\usepackage{color} 
\usepackage{graphicx} 
\usepackage{braket} 
\usepackage{slashed} 
\usepackage{dsfont} 
\usepackage{physics}
\usepackage{tensor}
\usepackage{bm}
\usepackage{bbold} 

\usepackage{pdfpages}

\usepackage{cite}
\usepackage{import}

\usepackage{appendix} 

\allowdisplaybreaks[4]

\usepackage{psfrag}
\usepackage{tikz} 
\usetikzlibrary{arrows}
\usetikzlibrary{plotmarks}
\usetikzlibrary{patterns}
\usetikzlibrary{external}
\usepackage{pgfplots}
\pgfplotsset{compat=1.9}
\usetikzlibrary{patterns}
\usetikzlibrary{calc}
\usepackage[compat=1.1.0]{tikz-feynman}
\usetikzlibrary{automata,positioning}

\usepackage{float}  
\usepackage{subfig}
\usepackage[lmargin=50pt,rmargin=60pt,tmargin=60pt,bmargin=65pt]{geometry}

\newenvironment{dedication}
  {
   \thispagestyle{empty}
   \vspace*{\stretch{1}}
   \itshape             
   \raggedleft          
  }
  {\par 
   \vspace{\stretch{3}} 
   \clearpage           
  }

\newcommand{\be}{\begin{equation}}
\newcommand{\ee}{\end{equation}} 
\newcommand{\bga}{\begin{gather}}
\newcommand{\ega}{\end{gather}} 
\newcommand{\nn}{\nonumber}
\newcommand{\f}{\frac}
\newcommand{\p}{\partial}

\newcommand{\la}{\langle}
\newcommand{\ra}{\rangle}

\let\a=\alpha \let\b=\beta  \let\g=\gamma  \let\d=\delta
\let\z=\zeta       \let\k=\kappa \let\l=\lambda
\let\m=\mu    \let\n=\nu    \let\x=\xi      \let\r=\rho \let\om=\omega
\let\s=\sigma \let\t=\tau   \let\vph=\varphi  

\let\G=\Gamma \let\D=\Delta   \let\L=\Lambda \let\X=F
         
\let\Om=\Omega  \let\eps=\epsilon

\newcommand{\xb}{\bar{x}}
\newcommand{\alphab}{\bar{\alpha}}
\newcommand{\betab}{\bar{\beta}}
\newcommand{\gammab}{\bar{\gamma}}
\newcommand{\deltab}{\bar{\delta}}
\newcommand{\phib}{\bar{\phi}}
\newcommand{\psib}{\bar{\psi}}
\newcommand{\vphb}{\bar{\varphi}}

\newcommand{\lb}{\bar{\lambda}}

\newcommand{\cB}{\mathcal{B}}
\newcommand{\cC}{\mathcal{C}}
\newcommand{\cD}{\mathcal{D}}
\newcommand{\cE}{\mathcal{E}}
\newcommand{\cF}{\mathcal{F}}
\newcommand{\cG}{\mathcal{G}}

\newcommand{\cJ}{\mathcal{J}}
\newcommand{\cK}{\mathcal{K}}
\newcommand{\cL}{\mathcal{L}}
\newcommand{\cM}{\mathcal{M}}
\newcommand{\cN}{\mathcal{N}}
\newcommand{\cO}{\mathcal{O}}

\newcommand{\cR}{\mathcal{R}}
\newcommand{\cS}{\mathcal{S}}
\newcommand{\cT}{\mathcal{T}}
\newcommand{\cU}{\mathcal{U}}

\newcommand{\cZ}{\mathcal{Z}}

\newcommand{\tD}{\tilde{\Delta}}

\newcommand{\lt}{\tilde{\lambda}}

\newcommand{\Mt}{\tilde{M}}
\newcommand{\tO}{\tilde{O}}

\newcommand{\overbar}[1]{\mkern 1.5mu\overline{\mkern-1.5mu#1\mkern-1.5mu}\mkern 1.5mu}

\DeclareMathOperator{\im}{\mathrm{i}} 

\newcommand{\sgn}{{\rm sgn}}
\newcommand{\mba}{\mathbf{a}}
\newcommand{\mbb}{\mathbf{b}}
\newcommand{\mbc}{\mathbf{c}}
\newcommand{\mbd}{\mathbf{d}}
\newcommand{\mbe}{\mathbf{e}}
\newcommand{\mbf}{\mathbf{f}}
\newcommand{\mbg}{\mathbf{g}}
\newcommand{\mbh}{\mathbf{h}}
\newcommand{\mbj}{\mathbf{j}}
\newcommand{\mbk}{\mathbf{k}}
\newcommand{\mbm}{\mathbf{m}}
\newcommand{\mbn}{\mathbf{n}}

\newcommand{\mbG}{\pmb{G}}

\allowdisplaybreaks[4]

\theoremstyle{plain}
\theoremstyle{definition}

\newtheorem{proposition}{Proposition}[section]
\newtheorem{definition}{Definition}
\newtheorem{lemma}{Lemma}[section]
\newtheorem{theorem}{Theorem}[section]
\newtheorem{cor}{Corollary}[theorem]

\theoremstyle{remark}

\definecolor{orange}{rgb}{0.88,0.39,0.12} 
\definecolor{rouge}{rgb}{0.8, 0.0, 0.0}
\definecolor{vert}{rgb}{0.4, 0.69, 0.2}
\definecolor{bleu}{rgb}{0.19, 0.55, 0.91}
\definecolor{lavenderpurple}{rgb}{0.59, 0.48, 0.71}

\begin{document}
\frontmatter
\begin{titlepage}
\includepdf[pages=-,pagecommand={},width=\paperwidth]{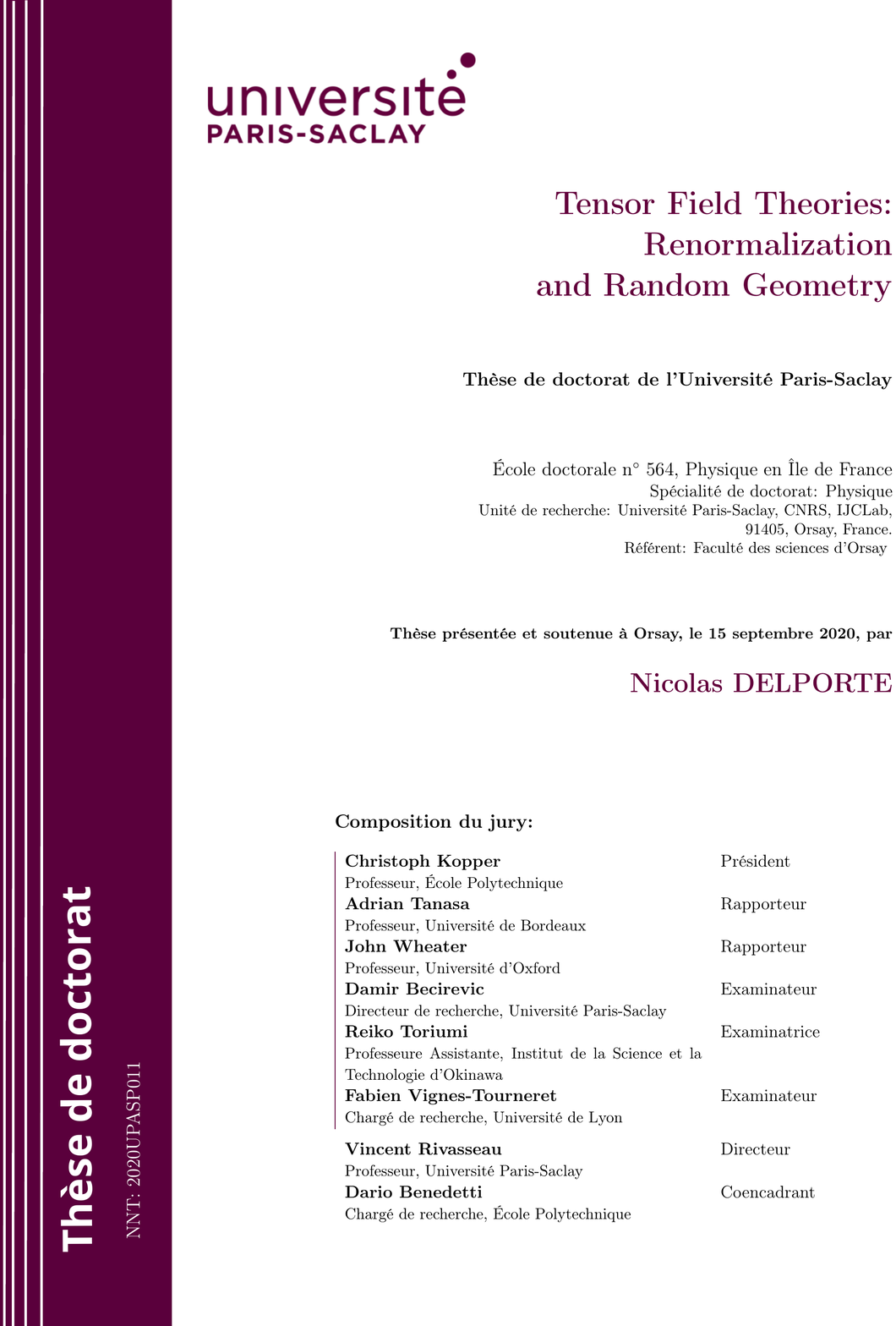}
\end{titlepage}

\begin{dedication}
À ma famille.
\end{dedication}

\chapter*{Acknowledgments}
This section is a personal reflection of the backstage behind those four years in Paris, which never became a routine and resulted in the forthcoming pages. Here I want to express my gratitude to all the people that allowed me to take part and evolve in this adventure, full of wonderfully memorable moments. 

\vspace{5mm}
The first thought turns to Vincent. Intrigued, then charmed by this eloquent and open professor which was transforming mathematics and physics into tales, I did initially an internship that continued into my thesis. Collaboration with Vincent required the ability to follow an endless stream of ideas, almost tumultuous,\footnote{The handwritten notes as well as the blackboards that resulted from those conversations were undecipherable to the non-initiated, hence precious -- even a photographer came once to picture some of them.} that combines a bird’s eye view of numerous fields with a minute care to the details. Tactics elegantly braided with strategy, personified. Besides science, I learned also a whole lot from our discussions on geopolitics, history, literature, music and what not, in any context (zigzaging in an Okinawaian taxi or sharing raviolis in a Viennese coffeeshop). Vincent gave me the example of a creative and complete man, able to fight fiercely for the values he defended. His generous vitality and good humour were contagious. The occasional musical afternoons occurred as a joyful relaxing experience that combined the pleasure of music with the tasteful cakes of Marie-France. 

In parallel, I had the chance to know Dario as second supervisor, which provided a contrasting example on how to work on a project or often allowed me to reassemble my thoughts after experiencing the sheer energy of Vincent. Dario was utterly patient, rightly spotting my countless mistakes, suggesting new directions to explore, helping better manipulate or interpret results. Whereas Vincent was emphasizing ideas and their connections, Dario taught me the importance of small things done carefully. I appreciated a lot our collaborations and the time we have spent around them. 

I thank the members of jury, Damir Becirevic, Christoph Kopper, Adrian Tanasa, Reiko Toriumi, Fabien Vignes-Tourneret and John Wheater, for having accepted to be part of it, in particular the reporters, Adrian and John, whom task I had not made easy. All together, they cautiously read my manuscript (pointing out the many typos I had left), and raised interesting problems to investigate in the future. I esteemed our exchanges and their opinions on those developing subjects.

During my PhD, Gatien and Razvan have been great mentors, supportive and encouraging at all times. The lab directors, Sebastien and later Samuel were both kind and patient until I got used (more or less) to the formalities of the lab, also of good advice in difficult periods. 

It was rewarding to learn from and collaborate with the members of the tensor family: Adrian, Ariane, Astrid, Carlos, Dario, Dine, Fabien, Fidel, Frank, Guillaume, Harold, Johannes, Joseph, Kenta, Luca, Mohamed, Oleg, Razvan, Reiko, Ritam, Romain, Sabine, Stéphane, Sylvain, Taoufik, Thomas, Valentin, Vasily, Victor, and the Vincents. \footnote{The list is non-exhaustive and encompasses the persons I interacted with the most.} Their curiosity, knowledge, honesty and perseverance have always been inspiring. Particularly, thank you Sabine for your patience and rigor, and our myriad of cups of tea; thank you Romain for our beautiful times spent across the world; thank you Vasily for your invitations and our lively discussions; thank you Jo for your trust, your courage and for so much more. Besides the meetings in Paris, the schools, workshops and conferences organised in Bordeaux, Lyon and Okinawa were enthralling, so thank you Adrian, Fabien, Reiko and all others that set them up. Further, I am grateful to Mikhail, Oleg and Ivan for guiding me through Moscow at the same time as in the theory of higher-spins, as well as to Irina and Micha for their warm welcome at the Stekhlov Institute. Dio, Carlo, Junggi, Per and Vaio, thank you for our dense discussions, that I hope we will be able to take over!

Senior and younger members at IJCLab (Asmaa, Bartjan, Benoît, Christos, Damir, Evgeny, Gatien, Henk, Renaud, Robin, Ulrich, Yann...), and CPhT (Andrea, Balt, Balthazar, Blaise, Emilian, Eric, Guillaume, Marios, Thibaut...) have always been gratifying to exchange with, during lunch or over a cup of coffee. Also, thank you Charles for your clear explanations! But without the administrative and IT staff of the labs and of Université Paris-Saclay, I would have been lost long ago in the maze of the bureaucratic formalities and resigned to the daunting technicalities to connect to LAN, printers, etc. From the early days of my internship in LPT to my last days at the lab after the defense, they had always been gladly prompt to aid in any matter. Thank you Danh, Florence, Jean-Pierre, Jocelyne, Malika, Marie, Mireille, Philippe Boucaud, Philippe Molle, Olivier, Sabine, Sarah and Yvette.\footnote{Some are actually researchers.} Claude Pasquier and Véronique Terras, representatives of EDPIF for Paris-Saclay, have been of immense relief for registering each year or preparing the defense. 

During my education, a few teachers and professors shone particularly to me and influenced my worldline and thoughts. In primary and secondary school, Thierry Martin, Libert Legrand, Pierre Bolly, Daniel Caspar taught me to pay attention to others, to watch nature, basics of mathematics and to search for depths and heights. At university, Frank Ferrari advised, supported and pushed me all along until today, helping me to turn slowly from a student into a researcher. Thank you deeply Frank. Outstanding mentors and tutors, Blaja, Céline, Mitia and Victor, have shaped my arsenal during my studies. From the other side of the bench, I thank my fellow teachers and assistants, Axel, Christophe, Michel, Cyril, Gatien, Evangelos, and the others, as well as the students that patiently attended my classes, for making teaching a pleasing counterpoint to research. 

The four years of Parisian life were rich in friendships with whom the many gatherings and outings made concrete the words ``having good time”. I thank the ENS friends that introduced me to the French mountains and the French gastronomy (Adrian, Adrien, Alexandre, Alice, Assaf, Bill, Elie, the Guillaumes and the Nicolases, Manuel, Ségolène), the LPTeamFloFlo (Amaury, Elie, Martin, Giulia, Florentin, Florian, Lydia, Natalie, Simone, Tim, Thomas, Victor), that enriched the life in and around the lab, provided debates of all kinds, good food, tea, poetry or laughs when needed (i.e. always). Especially strong bonds were formed at ENS and later with George, Melih, Sebastian and Tuan, and I hope to continue to climb cliffs and to explore the world (abstract or tangible) with you, guys! I was also pleased to continue growing with friends from high-school, ULB or elsewhere, among many others, Adrien, Amin, Antoine, Camille, Cédric, Clément, Diego, Frédéric, Marius, Marine, Nhan, Noé, Romain, Samuel, Simon, Thierry, Valdo, and Andrei which I was so happy to rejoin in Paris. For being a partner and sensei in go and karate and for our numerous discussions on about just everything, thank you Timothé. Mengkoing, thanks for your thrusting cheerfulness! Ahmed, Alice, Boris, Laurent, Philippe, Renand, Vincent and the others, thank you for those martial evenings! Debtosh, for leaving me a space near the Coulée Verte where the thesis has been largely written. I couldn't also thank enough Vincent that literally shortened the distance between Paris and Brussels since I got to know him. 

Finally, I am particularly glad to count my family at my side (physically or virtually): my parents, siblings, Martin, Damien, Nathalie, Christine and Angelos, grand-parents, aunts, uncles, cousins, nephews and nieces. They all have contributed, in one way or another, a part to the big puzzle I am, pushing me up with faith and nurturing me in all fashions, or simply walking along, across bushes and rivers. I hope we will continue to grow strong and united. 

\vspace{5mm}
For all you have been to me, thank you!

\newpage
\tableofcontents
 
\mainmatter
\chapter*{Introduction}
\addcontentsline{toc}{chapter}{Introduction}
\markboth{INTRODUCTION}{INTRODUCTION}

\input{intro.tex}

\chapter{Tensor Field Theory: Background and Motivations}
\label{ch:TFTintro}
\input{tft.tex}

\chapter{Tensorial Gross-Neveu in $d=3$}
\label{ch:TGN}
\import{}{TGN3D.tex}

\chapter{Renormalization of sextic tensor fields}
\label{ch:sextic}
\import{sexticTM/}{sexticTM-these.tex}

\chapter{Renormalization of a scalar field on Galton-Watson trees}
\label{ch:GW}

\import{}{RenGW.tex}

\chapter{Conclusions and Future Prospectives}
\label{ch:futur}
\import{}{conclusion.tex}

\appendix
\chapter{Résumé en français}

\import{}{Resume.tex}

\providecommand{\href}[2]{#2}\begingroup\raggedright\endgroup

\includepdf[noautoscale]{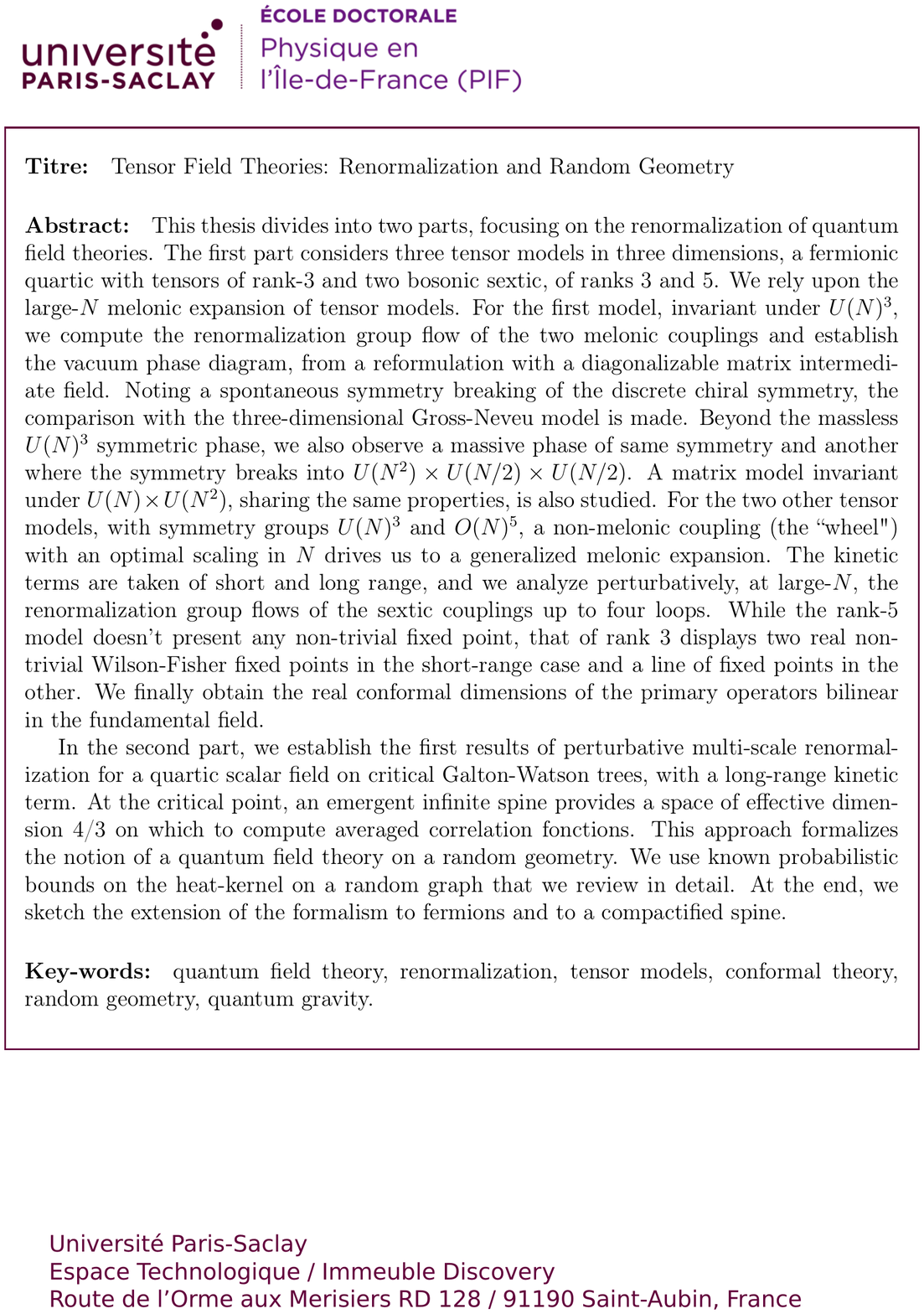}
\includepdf[noautoscale]{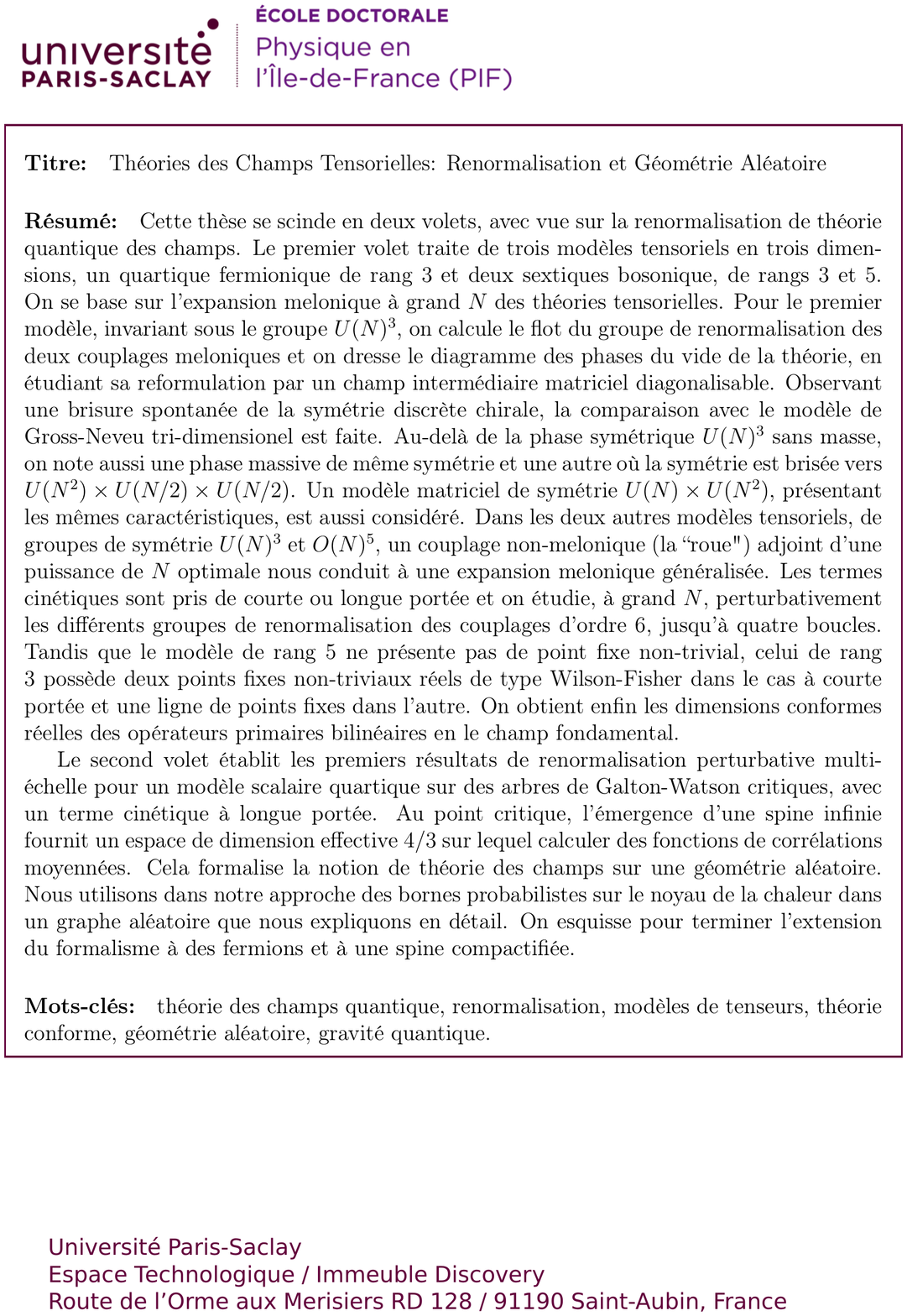}

\end{document}

%% file: intro.tex
\begin{flushright}
\hfill\begin{minipage}{8cm}
{\footnotesize {\it God made the integers, all else is the work of man.}
\vspace{0.2cm}
\newline 
{\footnotesize \textbf{Leopold Kronecker}}}
\end{minipage}
\end{flushright}

\vspace{5mm}
\noindent
Theoretical physicists endeavor to explain mathematically the observed phenomena. In order to formulate the questions, experimental data is paramount and most often, answers are obtained by contemplating and incorporating knowledge from seemingly apart fields. Historical examples are the roots that general relativity bears in the geometry of Riemann or the statistical physics insight on the renormalization group (RG) that helped interpreting divergent scattering amplitudes of particle physics. In the first case, there was intuition but without an adequate formalism, it is faltering; in the second, formalism leads to seemingly nonsensical results, and without proper interpretation, it fails to give any understanding. Hence a multiplicity of points of view can only strengthen and enrich the accumulated knowledge.
 
Today, we sit on stunningly accurate representations of our environment. The microscopic part of the spectrum is captured, within the framework of quantum field theory, by the Standard Model describing elementary matter particles (the Fermionic quarks, leptons and their anti-particles), their interactions (mediated by $SU(3)\times SU(2)\times U(1)$ gauge Bosons), plus the celebrated scalar Higgs Boson. The model arranges inside a single language three of the four fundamental forces: electromagnetic, weak and strong. It is tested with incredible success in particle colliders such as the Large Hadron Collider up to (center-of-mass) energies of 13TeV.\footnote{Recall, $1eV = 1.602176634~10^{-19} J$. In this chapter, we will keep to the metric system for writing our units. Later on, we will switch to the more convenient natural ones, setting $G=c=\hbar=1$, unless for some explicitly stated comments where we return to the first system.}
On the other part of the spectrum, gravity, the weakest forces of all,\footnote{Its coupling strength is about $10^{29}$ times weaker than the weak interaction at scales of order $10^{-15}$m.} rules at large scales. Our sharpest understanding of it comes from Einstein's equations of general relativity
\be 
R_{\mu\nu} - \f{1}{2}(R-2\L)g_{\mu\nu}= \f{8\pi G}{c^4}T_{\mu\nu}\;.
\ee
The left-hand side comprises geometric information in terms of the metric $g_{\mu\nu}$, the Ricci curvature $R_{\mu\nu}$ and scalar $R$ (we have also included a cosmological constant $\L$), and the right-hand side codes for the matter energy content through the stress-energy tensor $T_{\mu\nu}$. 
As summarized by J. A. Wheeler's words: ``Space tells matter how to move, matter tells space how to curve". Again, the equations hold support from a profusion of experiments, such as gravitational lensing and more recently gravitational waves. 

However neither picture is complete. Both theories raise puzzles within themselves, such as the swarm of 19 parameters of the Standard Model or the arising of spacetime singularities in typical cosmological solutions. It is not clear how one should approach those issues and ultimately, if the idea, how so pleasing, of a single and consistent framework unifying all interactions is possible. The problem is that the energy scales at which the gravitational interaction would start compete with other forces, around the Planck energy $\sqrt{c^5\hbar/G}\approx 10^{15}TeV$, lie far beyond what experiments can reach today, but would be relevant near the curvature singularity of black holes or of the Big Bang. So not only we don't know under what laws we could describe the universe at those energies but we don't know what data to fit to. Still, keeping this formidable goal of a quantum theory of gravity in mind, we can only aim at mathematical consistency.

Later in the text, we will revisit a sample of approaches to quantum gravity. For the moment, we will assume that matter and gravitational, in the form of the underlying geometry, degrees of freedom can be considered separately, and will look for a coherent quantization of the latter. The path integral is a preeminent tool inside the arsenal of quantum theory, that in essence sums over all configurations $\Psi$ available to the system, with a particular weight determined by the action $S[\Psi]$, itself motivated by symmetry postulates. In Euclidean signature, which unless stated, we will exclusively be concerned with, it is the partition function
\be 
Z= \int D\Psi e^{-S[\Psi]}
\ee 
that holds in principle all the information about the system. 
In gravitational systems, the configuration space contains all metric structures on manifolds $\cM$ quotiented by diffeomorphism, including different topologies.
A procedure for making sense of this expression is first to discretize the configurations and then to take a continuum limit, hoping to recover in a low-energy limit the Einstein-Hilbert action
\be
S_{EH}[g_{\mu\nu},\cM]=\f{c^4}{16\pi G} \int_{\cM} d^Dx \sqrt{\abs{g}}\left(2\L - R\right)\;.
\ee 
Hence one has to sum over all piecewise linear $D$ dimensional manifolds up to diffeomorphism.\footnote{One assumes at the same time that such discretizations contain enough of gravitational configurations to include in the ``complete" partition function. In other words, what categories of manifold should one sum over?} Those can be obtained from gluing $D$-simplices and it occurs that such geometries appear as the Feynman diagrams of ``colored" tensors of rank $D$. The action of the model is taken as invariant under a symmetry group of size $N$. 
Using tensors of rank 2, i.e. matrices, this point of view lead to much progress regarding two-dimensional quantum gravity, making also contact with string theory. 
The crux is to use $N$ as a perturbative parameter relating it to Newton's constant as $\ln N \sim 1/G$, and to tune the coupling constants appropriately such that the large $N$ limit is non-trivial. Surprisingly, in contrast with the matrix models which require the sum of all planar diagrams at large-$N$, models of rank $D\geq 3$ enjoy a solvable large-$N$ limit, that encompasses a restriction of planar diagrams, coined the \emph{melonic} family.\footnote{This family was also noticed earlier in condensed matter, as a useful, although uncontrolled, simplification of Green's functions for certain phase transitions in models of Bose liquids \cite{Patashinskii}.} Being close relatives of branched polymers, a tree-like phase of Hausdorff dimension $2$ and spectral dimension $4/3$, melons were insufficient to resolve higher dimensional geometries from tensors and different approaches had to be looked for.

A few years after the large-$N$ limit of tensors was established, the same class of diagrams arose in a one-dimensional quantum mechanical model of $N$ interacting Majorana Fermions with quenched disorder, the Sachdev-Ye-Kitaev (SYK) model. Melons came by because a tensor served as the coupling of interaction. More interestingly, in addition to its large $N$ solvability, this Fermionic model showed features akin to the near-horizon limit of near-extremal black holes. Though, set in the framework of holography, which stipulates that a theory of quantum gravity is equivalently described by a gauge theory living on the asymptotic boundary of the first, the presence of the disorder was unsatisfying. Lifting the zero-dimensional tensor models into one dimension rendered properly quantum mechanical theories (i.e. without disorder) with similar features at large $N$ as the SYK model, and all together they were identified as melonic theories. Whereas there has been tremendous progress on working out the gravitational dual of the SYK model, for tensor models it remains challenging. 
Indeed, as opposed to vectors or matrices, tensors don't have a canonical algebra and usual techniques to handle the former have to be revisited. Their contraction patterns become quickly intractable as the number of contracted tensors grows. Until 2015, research constituted a vast supply of techniques to treat their special diagrammatic, or considered models where the covariance broke explicitly the tensor symmetry, allowing the notion of a renormalization group on their indices. 
Yet, defined on a higher dimensional flat background, tensor models define genuine quantum field theories, \emph{tensor field theories}. 
Their symmetry group, smaller than that of vectors or matrices, yields a much richer structure of possible interactions, hence of theories.
This provides a motivation for studying them independently of their origins in quantum gravity.

Within the sea of all quantum field theories, fixed points of the renormalization group are like harbours that control the surrounding flows and are specified by a few features, among others the symmetries of the theory. Very often, there exists a tangible physical system with the corresponding symmetries and its properties at a critical point of its phase diagram will be dictated by the appropriate fixed point.~\footnote{To add to the well-known $\mathbb{Z}_2$ symmetry of the Ising model relevant e.g. for the liquid-vapor transition, $O(m)\times O(n)$ models are relevant for describing noncollinear frustrated spin systems, see \cite{Pelissetto:2000ek} for many more examples.} Thus, a classification of possible fixed points in different dimensions according to their symmetries is a commendable goal. 

Lagrangians of tensor field theories are taken typically invariant under $O(N)^q$ or $U(N)^q$ with the integer $q>2$ related to the rank of the tensor. 
How do they differ with respect to traditional vector and matrix theories? What kind of phase transitions do they support and how do they break their symmetries? Are there properly tensorial interacting fixed points of the renormalization group? If any, what are the associated conformal field theories? Relying on the solvability of the melonic large $N$ limit, we tackle those questions for specific tensor models.

\vspace{5mm}
Accordingly, our first concern during the thesis was to investigate a few properties of melonic tensor field theories. More specifically, we study the renormalization group flows of a Fermionic and a Bosonic model in (or close to) three dimensions, look for their fixed points and respectively understand the structure of the vacuum around the fixed points or the nature of the spectrum of conformal operators at the fixed points. That is exposed in Chapters~\ref{ch:TGN} and~\ref{ch:sextic}.

Our second point of interest, contemplating quantum gravity from afar, was to develop the subject of quantum field theory on a random geometry. Important in the framework of constructive physics, a rigorous analysis of its renormalization group is set up. For ultimate simplicity we consider a scalar $\phi^4$ model on Galton-Watson trees, but the techniques apply to more general random objects, only sufficiently good control over the heat-kernel is necessary. 
This is the content of Chapter \ref{ch:GW}.

\vspace{5mm}
In order to make the next chapters more digest and to provide them with a wider perspective, we continue in Chapter~\ref{ch:TFTintro} with a more extensive overview of the developments we have just sketched. 
Since it is the core of our work, we open in Section~\ref{sec:renormalization} with a review of the renormalization group applied to quantum field theory, explaining how it is implemented through standard regularizations or the rigorous multi-scale analysis. 
Where we are aware of results on constructive models of quantum field theory, we also add some discussion on them.
We end this part with a few examples that we think emblematic of RG or relevant to the cases we study later. 
Going along or against an RG flow we may end up on fixed points, that characterize the low or high energy limit of the theory. In many cases, the scale invariant fixed points enjoy a larger symmetry group forming a conformal field theory (CFT). We will detail some recently developed techniques helpful regarding the melonic CFTs in Section~\ref{sec:CFT}. 
In Section~\ref{sec:QG}, after a brief summary of what we should expect quantum gravity to answer and an account on different approaches to it, we sketch our understanding of studies of random geometry, with an emphasis on the more familiar two-dimensional case connected to random matrices.
The two next Sections~\ref{sec:SYK} and~\ref{section:largeNTM} detail the techniques and results attached respectively to the SYK and tensor models. For the latter, we tried to portray the historical evolution of questions, and how they each build up on previous tools.
In retrospect, although this account seems pretty diverse, we hope that it will also be enjoyable. 
We reserved the last Section~\ref{sec:overview} for a more extended description of Chapters~\ref{ch:TGN} to~\ref{ch:GW}.

\vspace{5mm}
To close, our concluding Chapter \ref{ch:futur} takes on many questions that we left open regarding the two considered directions, sprinkled with some speculative comments.

\newpage
My thesis has lead to the following publications: 
\begin{itemize}
\item two proceedings: 
\begin{itemize}
\item[-] \textit{The Tensor Track V: Holographic Tensors}, with V. Rivasseau, \\ 
Published in: PoS CORFU2017 (2018) [arxiv:1804.11101 [hep-th]],\\
that we extended largely to form the Chapter 2,
\item[-] \textit{The Tensor Track VI: Field Theory on Random Trees and SYK on Random Unicyclic Graphs}, with V. Rivasseau, \\ 
To be published in: PoS CORFU2019 [arxiv:2004.13744 [hep-th]], \\
that serves in Chapter 5,
\end{itemize}
\item two published papers: 
\begin{itemize}
\item[-] \textit{Phase diagram and fixed points of tensorial Gross-Neveu models in three dimensions}, with D. Benedetti, \\
Published in: JHEP 01 (2019) 218, [arxiv:1810.04583 [hep-th]],\\
that is the basis of Chapter 3,
\item[-] \textit{Sextic tensor field theories in rank 3 and 5}, with D. Benedetti, S. Harribey, R.~Sinha, \\
Published in: JHEP 06 (2020) 065, [arxiv:1912.06641 [hep-th]],\\
essentially the Chapter 4,
\end{itemize}
\item one paper accepted in Communications of Mathematical Physics:
\begin{itemize}
\item[-] \textit{Perturbative Quantum Field Theory on Random Trees}, with V. Rivasseau,\\~
[arxiv:1905.12783 [hep-th]], \\
used in the Chapter 5.
\end{itemize}
\end{itemize}

%% file: tft.tex
\section{Renormalization}
\label{sec:renormalization}

Quantum field theory (QFT) is a very powerful framework that describes a landscape of natural phenomena ranging from cosmology to elementary particles. 
It has started as an extension of quantum mechanics to quantize multiple (an infinite amount of!) degrees of freedom in a Lorentz invariant manner. Given quantum fields $\phi(x)$ at position $x$, corresponding to the fundamental degrees of freedom of the theory, the most important object in QFTs is its set of correlation functions:
\be
\expval{O_1(x_1)\dots O_n(x_n)},
\ee
where $O_i$'s are operators depending on $\phi$.\footnote{Nevertheless, naturally plagued with divergences, such operators should be viewed as normal ordered operator-valued distributions, but we will be pedestrian in this overview and always assume that such regularization has been taken care of.} 
The expectation value may be taken on the vacuum, under assumptions of its uniqueness and it generating any other state by acting on it with a finite number of operators. They are the analogs of the moments of a probability distribution. 

In order to calculate correlation functions, our approach will be perturbative. More precisely, we will be concerned with Lagrangian theories with a free and interacting part, for example, in $d$ dimensions:
\be 
S[\phi]=\int \dd[d]x \left[ \dd[d]y \f{1}{2}\phi(x) C^{-1}(x,y) \phi(y) + \l \phi^q(x)\right].
\ee 
Depending on the goal, different representations of the propagator are more adequate. Typically, one deals with the following three:
\begin{itemize}
    \item direct-space: \be C^{-1}(x,y) = [-\partial_x^2  + m^2]\d(x-y)\;,\ee 
    \item momentum: \be \tilde{C}(p) = \int \dd[d]x e^{i p\cdot (x-y)}C(x,y) = \f{1}{(2\pi)^d}\f{1}{p^2+m^2}\;, \ee
    \item parametric (Schwinger proper time):\be C(x,y) = \int_0^\infty  \f{\dd\a}{(4\pi\a)^{d/2}}e^{-\f{\abs{x-y}^2}{4\a} - m^2 \a}\;,\ee
\end{itemize}
the last of which as the rather explicit interpretation of integrating over all times $\a$ the probability that a random walker from $x$ takes to reach $y$, given by the heat-kernel.
The perturbative series comes from Taylor expanding the interaction part and, at fixed order $p$ in the interaction couplings, Wick contracting the fields with respect to the Gaussian term. Then, a correlation function writes as an expansion in Feynman diagrams $G$:
\be
A^{(p)}(x_1,\dots,x_n) = \f{(-\l)^p}{p! sym(G)}\sum_{G\in \cG}\int\prod_{1\leq k\leq p} \dd[d]y_k \prod_{(y_i,y_j) \in E(G)} C(y_i,y_j).
\label{wick-contraction}
\ee
The sum is done over all graphs $\cG$ with $V(G)$ vertices: $p$ $q$-valent vertices (with $q$ half-edges), and $n$ other ``external" single-valent vertices (one half-edge), identified with $(x_i)_{1\leq i\leq n}$. All half-edges are then contracted pairwise with propagators forming the edges $E(G)$ of the graph. 
The factor $sym(G)$ denotes the permutations of the half-edges that preserve the Feynman diagram. 
The integration is done over all internal vertices. Expressed in momentum space, momenta are conserved at the interaction vertices, due to the local and Lorentz scalar interaction term, and to every loop is associated a running momentum.
We will call an amputated graph one from which the propagators attached to the external vertices have been removed.

Perturbative approaches immediately stumble on a number of problems through divergences. First, because of an infinite number of degrees of freedom interacting at arbitrary distances (infrared or IR divergences), that can fluctuate at arbitrary high energy scales (ultraviolet or UV). Second, because at a fixed order $p$ of the expansion, certain correlation functions behave as $p!$ (the renormalons\footnote{As well as instantons, but they will not be discussed here. I thank F. Vignes-Tourneret and V. Rivasseau for the clarification.}) or because of the factorially growing number of Wick contractions to sum over. The first problem is tackled by working in a finite volume. Amazingly, a proper renormalization procedure sweeps over all last three problems.

\subsection{The procedure}

Let us treat the UV divergences that haunted theoretical physics for the first half of the last century. Diverging amplitudes are identified by power-counting on the amputated amplitude. 
To each diagram $G$ is associated a $UV$ degree of divergence $\omega(G)$: 
\be
\omega(G) = dL- 2I\Delta_\phi+ p\Delta_i, 
\ee
with $I$ the number of internal propagators, $p$ the number of interaction vertices and $\Delta_\phi, \Delta_i$ the mass dimensions of the fundamental field and of the interaction. For some number $\l>1$, it gives the scaling dimension one finds after rescaling all positions by $1/\l$. 
The graphs with $\omega(G)\geq 0$ will be called superficially divergent and superficially convergent otherwise. 

After the introduction of regulators rendering the theory finite, the goal is to relate bare (indiced by $0$) to new renormalized variables (indiced by $r$) with an expansion: 
\be 
\l_0 = \l_r + \sum_{n\geq 2} \alpha_n\l_r^n, 
\ee 
and similar expressions holding for all the couplings of the theory, 
through a set of renormalization conditions on a finite minimal ensemble of diverging amplitudes (without regulators), such that the theory expressed in the renormalized variables is finite at any given order in its couplings when the regulators are removed. Those renormalization conditions will fix the dependence of the $\alpha_n$ coefficients on the regulators and on a probing scale $\mu$ at given order. With the standard approach, one employs all means to compute the regulated diverging integral. In the multiscale point of view, preliminary to a constructive analysis, we decompose a propagator in scales and working scale by scale, we need to control convergence properties of the full correlation function as the sum over all scale assignments on the involved propagators is performed.

At a given order in the perturbative parameter, when only a finite number of coupling redefinitions are enough to remove all divergences, the theory is said \emph{renormalizable} and well-defined at all scales. If otherwise, it is said \emph{non-renormalizable} and one must precise its range of validity.

\subsubsection{Stepping back}
Perturbative renormalization proved itself useful after the works of Feynman, Schwinger, Tomonaga and culminating with Dyson \cite{Dyson:1949bp} showing that the theory of quantum electrodynamics was renormalizable at all orders. The calculation of the anomalous magnetic dipole moment of the electron obtained by Schwinger \cite{Schwinger:1948iu} at second order and today up to 10th order \cite{Aoyama:2014sxa}, agreeing with its experimental value up to 10 significant digits, constitute one of the most precise fits between theory and experiment. 

However, renormalization flourished after the brilliant insight of Wilson, connecting the divergences in particle physics with the block spin transformations of Kadanoff and the effective theory point of view of Gell-Mann and Low (see \cite{WilsonNobel} for a historical perspective). The idea is to cutoff the theory at a UV scale $\L$ and an IR scale $\k$, regulating all divergences, and to construct the theory at an intermediate scale $\k<k<\L$. Schematically, after decomposing our field in components associated to different scales (for instance by restricting its Fourier components) $\phi = \phi_{k^\prime<k}+\phi_{k^\prime>k}$ an effective action at scale $k$ is found from integrating out all fluctuations above $k$: 
\be 
e^{-S_k[\phi_{<k}]} = \int_{k<k^\prime<\L}D\phi_{k^\prime}e^{-S\left[\phi_{k^\prime<k}+\phi_{k^\prime>k}\right]}, 
\ee 
generating new effective interactions between the remaining components. In order to form the \emph{renormalization group}, one has to rescale the momenta as well as the field
\begin{gather}
K= l k^\prime\;, \quad \Phi_{K}= Z(l)^{-1/2}\phi_{k^\prime}\;,
\end{gather}
with $l=\L/k$ to recover the original range of momenta and kinetic term with a new action $S^\prime[\Phi]$. This way, the renormalization group produces a trajectory from the UV towards the IR within the space of all possible interactions
\be 
S^\prime[\Phi]= \cR(t)S[\phi]\;,
\ee
writing $t=\log l$.
After a small renormalization step, one can linearize the flow, 
\be
S^\prime[\Phi] = S[\phi] + \f{d}{dt}\left(\cR(t)S[\phi]\right)\bigr\rvert_{t=0}dt + \cO(dt^2)\;,
\ee
which, assembling all couplings under the notation $\{\l_i\}_{i\in \mathbb{N}}$, leads to an infinite set of coupled non-linear differential equations tracing the evolution of the couplings through the beta functions
\be 
\f{d\l_i}{dt} = \b_i(\{\l_j\}).
\ee
Within this view, one relies on techniques of dynamical systems to study generic properties of QFTs, as for example stability analysis or merging of fixed points. This idea is also supporting a notion of universality: the fact that flows are governed by their fixed points \footnote{But it may also contain limit cycles.}, characterized by a few features. Those are the symmetry group of the theory (at the fixed point, the initial theory may have its symmetry spontaneously broken during the flow), the space-time dimension, the number of degrees of freedom and a few extra parameters corresponding to relevant perturbations emanating from the fixed points (FP).

By construction, a fixed point will be a scale-invariant theory. But there are situations where the scale symmetry is enhanced to the full conformal group $SO(d+1,1)$ putting strong constrains on the structure of the correlation functions. In the next section, we will spell out more details on aspects of conformal field theories relevant to our later chapters. 

Generically, in the Lagrangian description of the fixed point, we can understand the RG flow as coming from adding operator perturbations in the action. The effect of the pertubation depends on its dimension $\Delta$, obtained either perturbatively around an RG fixed point or non-perturbatively (for instance by bootstrap constraints), such that for:
\begin{itemize}
    \item $\Delta<d$: the operator is called relevant and it drives away from the FP\;,
    \item $\Delta>d$: the operator is called irrelevant and it drives towards the FP\;,
    \item $\Delta=d$: the operator is called marginal and one needs further quantum corrections to understand its effect. 
\end{itemize}
Note that irrelevant operators are the ones that would lead to non-renormalizable theories and this means that to study the RG flow around some FP, we are only required to consider renormalizable (and marginal) interactions.

\subsubsection{Some regularization schemes}

In order to establish notations and conventions, let us recall how to apply in practice this procedure to quantum fields.
We start with a bare action, defined at the UV scale, around $d=4$ and including only the renormalizable and marginal interactions: 
\be 
S = \int \dd^d x \left(\f{1}{2}(\partial \phi_0)^2 + \f{1}{2}m_0^2\phi_0^2 + \f{1}{4!}\l_0 \phi_0^4 \right).
\ee 
For the purpose of renormalization, we rewrite the action with renormalized variables, that will contain our physical parameters ($m$, $\l$) as well as the counterterms (introduced through $Z$, $Z_m$ and $Z_\l$) and some regulator (implicit below): 
\be 
S = \int \dd^d x \left(\f{Z}{2}(\partial \phi)^2 + \f{Z_m}{2}m^2\phi^2 + \f{m^{4-d}Z_\l}{4!}\l \phi^4 \right).
\ee
Note that we made the renormalized couplings explicitely dimensionless, as it will be easier to track down how they will then flow with the energy scale. 
In a renormalizable theory we will be able to relate bare to renormalized variables such that, when the regulator is removed, all $\phi$ correlation functions are finite at any given order in the renormalized couplings.
For both actions to match, bare and renormalized fields and couplings are connected by
\begin{gather}
\label{eq:bareandren}
\phi_0 = \sqrt{Z}\phi\;,\\
m_0^2 = m^2 Z_m/Z\;,\\
\l_0 = \l m^{4-d}Z_\l/Z^2\;,
\end{gather}
We define those parameters through renormalization group conditions on the superficially diverging one particle irreducible (1PI) diagrams, at some scale $\mu$, for instance:
\be 
\label{eq:renconditions}
\left.\expval{\phi(p)\phi(-p)}\right|_{p^2 = \mu^2} = \f{1}{\mu^2+m^2}, \quad \left.\expval{\phi(p_1)\phi(p_2)\phi(p_3)\phi(-p_1-p_2-p_3)}\right|_{p_i^2 = \mu^2}= -\l \;.
\ee
We emphasize that the first condition fixes the position of the pole of the propagator at $p_1^2 = - m^2$ and its residue to be one.
In case we want to compute correlation functions with insertions of more general operators ($\phi^n$ and derivatives), we should fix similar renormalization conditions on the superficially diverging amplitudes that contain them. 
Enforcing those conditions determines our counterterms at any given order of the coupling expansion, and their divergent part expresses as a series in the regulator. Within dimensional regularization, setting $\eps = 4-d$, the counterterms take all this form: 
\be 
Z_i = 1 + \sum_{n\geq 1} \f{a^{(i)}_n(m,\l,\mu)}{\eps^{n}} + (\text{regular})\;,
\ee  
with terms containing poles in $\eps$ and others regular as $\eps$ vanishes.
Unitarity constraints on the K\"{a}llen-Lehmann representation of the two-point function impose the bounds $0\leq Z<1$ for an interacting theory (see for a standard exposition \cite{Itzykson:1980rh}). 
Eventually, we have the following invertible relations: 
\be 
\l_0 = \l_0(m,Z,\l, \eps, \mu).
\ee 
A convenient regularization procedure is the minimal subtraction, where counterterms are tailored to cancel only the divergent parts of the amplitudes. In this case, only the residue contributes in the $\beta$ function. Naturally, the second condition in eq.~\eqref{eq:renconditions} will generically lead to a polynomial function in the coupling $\l$.

Another renormalization scheme is the BPHZ subtraction formula that expresses directly the renormalized amplitude $R(G)$ of an amplitude $G$ as
\be 
R(G) = \sum_{\G}\prod_{\g_i\in \G}\D_{\g_i}*G_{/\G}. 
\ee 
We have to consider all forests $\G$ of disjoint (no internal line and no vertex in common) or included one in the other UV diverging 1PI subdiagrams $\g_i$ of $G$, including the empty set. $\D_{\g_i}$ is the counterterm associated to the diverging subgraph $\g_i$, $G_{/\G}$ is the (amplitude of the) graph obtained by shrinking the subgraphs in $\G$ to local vertices and the $*$-operation concatenates the amplitude of $G_{/\G}$ with all the corresponding counterterms. Zimmerman \cite{Zimmermann:1969jj} proved that the renormalized amplitude integrated over its external momenta was absolutely convergent when, given a diverging subgraph of degree of divergence $\D$, the counterterms were chosen to subtract the Taylor expansion around zero external momenta of the amplitude up to order $\D$. However, the procedure needs modification to be applied to massless theories \cite{Lowenstein:1975rg}.

Afterwards we can study the effect of changing the renormalization scale $\mu$ on correlation functions with $n$ insertions of operators: 
\be 
\G^{(n)}_0(x_1, \dots,x_n)= Z(\mu,\l)^{n/2} \G^{(n)}_r(\mu;x_1, \dots,x_n)
\ee 
using the bare and the renormalized action and fields, on the left- and right-hand side respectively. Since the left-hand side doesn't depend on the scale $\mu$, differentiating both sides by $\mu\partial_{\mu}$ leads to 
\begin{gather}
\left[\mu \partial_\mu + \beta(\lambda)\partial_\lambda - n\gamma(\lambda)\right]\G^{(n)}_r(\mu;x_1, \dots,x_n)=0, \\ \beta(\lambda) = \f{\partial\l}{\partial\log \mu},\gamma(\lambda) = \f{\partial \log Z}{\partial\log \mu}\;,
\end{gather}
namely the Callan-Symanzik equations, telling that the flowing couplings must be compensated by the wave-function renormalization $Z$ making up the anomalous dimension $\g$ of the field. It naturally generalizes to include a diversity of couplings. If we were to use a UV-cutoff regulator $\L$, since the renormalized couplings are made dimensionless, the dependence in the scale $\mu$ must be replaced by the dimensionless combination $\mu/\L$ and differentiation with respect to $\mu$ can be traded with (minus) one with respect to $\L$.

By definition, $\beta(\lambda)$ gives the variation of $\lambda$ with the probing scale, hence, we can obtain it by doing a perturbative expansion of the renormalization conditions with respect to the bare variables and differentiate with respect to the probing scale, step after which we can express the bare coupling in terms of the renormalized one.
This procedure generalizes to theories with several couplings, but extracting the $\b$ functions is more tedious since one has to keep track of how the different operators mix. Combinatorial factors are crucial since they affect the sign and zeroes of the $\b$ function. In the following chapters, we will work out the details of different such examples. Let us also remark that when considering correlation functions of some composite operators in the fundamental field of canonical dimension $\Delta_O$, their renormalization happens in two steps. First by Wick ordering the operator, second by introducing additional counterterms associated to all operators of dimension lower than $\Delta_O$ (see \cite{zinnjustin}). 

Fixed points $\l_*$ are given by solutions to the equation $\b(\l_*) = 0$ and critical exponents correspond to eigenvalues of the derivatives $\partial \b(\l_*)/\partial\l$ at the fixed points. They give the scaling dimension $\Delta= d-\nu$ of the operator associated to the eigenvector. With several couplings, we will define the scaling operators as the right eigenvectors of the stability matrix. It may happen that the stability matrix is non-diagonalizable, in which case we can extract the conformal data from its Jordan normal form. This signals the presence of a logarithmic CFT, with several conformal primaries of the same dimension (see the next section).

Under an invertible reparameterization of the couplings $\tilde{g} =g+ \alpha g^2 + \cO(g^3)$, coming from a different renormalization scheme, the $\b$ function changes as
\be 
\b(\tilde{g}) = \b(g)\left(\f{\partial \tilde{g}}{ \partial g}\right).
\ee 
Expanding perturbatively the right-hand side, we see that beyond order 2, the coefficients of the $\b$ functions depend on the renormalization scheme. However the fixed points and the eigenvalues of the stability matrix remain invariant. 

\subsection{The constructive point of view}
Getting over the remaining two problems requires to set the theory in a proper mathematical frame. This means obeying a set of axioms (Wightman axioms \cite{Wightman:1956zz,Streater:1989vi}): it must be a Lorentz-invariant formulation of quantum mechanics with operators transforming under given unitary representations, it assumes the existence and uniqueness of Lorentz invariant vacuum state, plus conditions on the domain of definition of operators, it must obey causality and linear combinations of finite number of operators acting on the vacuum must form a dense set. 
Under analytic continuation to Euclidean signature, the Euclidean field theory will then obey the Osterwalder-Schrader axioms \cite{Osterwalder:1973dx,Osterwalder:1974tc} (analyticity, regularity, invariance under Euclidean symmetries, ergodicity and reflection positivity). The theorem implies that one could work equivalently in one or the other signature. Up to now, verifying those axioms on realistic theories as the Standard Model is very hard and was achieved only for a handful of cases, some of which will be mentioned later.
This enters the realm of constructive quantum field theory, recently reviewed in \cite{Summers:2012vr}. In order to control the perturbative expansion, we need to prove its Borel summability through precise bounds on the correlation functions at any order, which also provides its non-perturbative meaning. Non-perturbative techniques are essential in order to explain among others spontaneous breaking of symmetry, phase transitions or contributions of instantons.

Prevailing tools of constructivists are: random walk or current expansions, cluster and multiscale expansions, the lace expansion.
\footnote{Up to now, the main progress has been done for scalar fields in various dimensions. For Fermionic fields, controlling the perturbative expansion is easier as it truncates by Pauli's principle. However, the difficulty lies at low energies, in approaching the extended singularity of the Fermi surface \cite{Salmhofer:2018sgo,Polchinski:1992ed}.} 
For example, a recent notable result is the proof of the triviality of scalar $\phi^4_4$ through its formulation as a spin model and precise bounds of the intersection probability of currents defining correlation functions \cite{Aizenman:2019yuo}. Motivated by many earlier works (see the references in \cite{Aizenman:2019yuo}) and by Monte Carlo simulations \cite{Wolff:2009ke}, it sets rigorously the logarithmic decay of the continuum limit of quartic coupling with respect to the probing scale.\footnote{All the crux is to go beyond Wilsonian perturbation theory and obtain the behaviour of the beta function at large coupling.}
Far ahead for the mathematical physicist lie the gauge theories, as reminds the pending Millenium Prize of constructive four dimensional Yang-Mills theory (which means to prove existence of a mass-gap, quark confinement and chiral symmetry breaking) \cite{Millenium}.

\subsubsection{Multiscale analysis}
Very close to Wilson's original idea of fluctuations of higher energy scales to contribute in effective actions of lower energy scales, the multiscale analysis was developed to set on a firmer ground the earlier analysis and to establish constructive results, that will ensure a proper non-perturbative definition of the theory by tackling the last two problems of the perturbative expansion of amplitudes. Our presentation follows \cite{Rivabook} (see also \cite{VignesTourneret:2006xa}).

\paragraph{Multiscale decomposition.} Given an arbitrary constant $M>1$, the first step is to slice the free propagator into scales:
\begin{gather}
C = \sum_{i=0}^{\rho} C^i, \\
C^i(x,y) = \int_{M^{-2i}}^{M^{-2(i-1)}}\f{\dd\a}{(4\pi\a)^{d/2}}e^{-\f{\abs{x-y}^2}{4\a} - m^2 \a}\;,\\
C^0(x,y) = \int_{1}^{\infty}\f{\dd\a}{(4\pi\a)^{d/2}}e^{-\f{\abs{x-y}^2}{4\a} - m^2 \a}\;.
\end{gather}

In a scalar theory in $d$ dimensions, it is easy to show that for all $0\leq i\leq \rho$, there are constants $\d<1$ and $K, \{K_k\}_{k \in \mathbb{N}} >1$, such that: 
\begin{gather} 
\label{eq:boundonpropa}
\abs{C^i(x,y)} < KM^{(d-2)i}e^{-\d M^i\abs{x-y}}\;,\\
\abs{\partial_{\mu_1}\dots\partial_{\mu_k}C^i(x,y)} < K_k M^{[(d-2)+k]i}e^{-\d M^i\abs{x-y}}.
\end{gather} 

A few remarks:
\begin{itemize}
    \item In renormalizing UV divergences, we set the maximal scale $\rho$ to be the UV cutoff.
    \item There are several ways to define slices, each suitable for different constructive purposes.
    \item There are also different conventions for selecting the scale of the external propagators, that we will denote by $C_\rho$.
\end{itemize}

Then any amplitude of a graph $G$ with internal and external legs $l(G)$ and $\mathbf{l}(G)$ decomposes as a sum over all scales for each propagator in eq.~\eqref{wick-contraction}
\begin{gather}
A(G) = \sum_{\mu \in \mathbb{N}^l}A_{G,\mu} ,\\
A_{G,\mu} = \int\prod_{1\leq k\leq p} \dd[d]x_k \prod_{(x_a,x_b) \in l(G)} C^{\mu(ab)}(x_a,x_b)\prod_{(x_c,x_d) \in \mathbf{l}(G)} C_\rho(x_c,x_d), 
\end{gather}
the assignement $\mu$ attributing the scale $\mu(ab)$ to the internal leg attached to the vertices $x_a$ and $x_b$.

For renormalizable theories, it will be possible to absorb all divergences coming from such sum over all graphs appearing in a given correlation function (up to a given order in the perturbative parameter) into a finite number of redefinitions of parameters of the theory. This is the BPHZ theorem:

\begin{theorem}[BPHZ] 
There is a series for the renormalized constants in terms of the bare ones
such that amplitudes, expanded with respect to the renormalized constants, have a finite limit at each order when the UV regulator is removed. 
\end{theorem}

Instead of proving the theorem, we will give the gist of the ideas necessary to prove it and say how they fit into each other.

\paragraph{High subgraphs.} Given a subgraph $g\subset G$ and a scale assignement $\mu$, we define the scales \be i_g(\mu) = \inf_{l~\text{internal edge of~} g} \mu(l), \quad e_g(\mu)=\sup_{l~\text{external edge of~} g} \mu(l). \ee 
High subgraphs satisfy the condition \be i_g(\mu)>e_g(\mu). \ee In other words, they are ``quasi-local", since from the point of view of the external legs of the subgraph, the inner legs stay at higher or more local scale. Now given some scale $\s$, we write $G^\s$ the subgraph obtained from $G$ by keeping all propagators $l$ with $\mu(l)>\s$, and $\{G^\s_k\}_{1\leq k\leq \k(\s)}$ the corresponding high subgraphs, forming $\k(\s)$ disjoint components (no vertex or edge in common). With respect to the scale $\s$, the collection of high subgraphs \be \bigcup_{\substack{0\leq \s\leq \rho\\1\leq k\leq \k(\s)}}G^\s_k\ee forms an inclusion forest, that is being either included into one another or non-intersecting. The important point is that \emph{only them} appear as an upper bound to the amplitude, giving them their superficial degree of divergence:
\be
A_{G,\mu} \leq K^{V(G)}\prod_{\s}\prod_{k=1}^{\k(\s)}  M^{\omega\left(G^\s_k\right)}.
\ee
Weinberg's theorem follows, stating that a graph without any diverging subgraph is superficially convergent, with a bound exponential in the number of vertices. Let us take the example of a superficially convergent graph with $n\geq 2$ external legs in $\phi^4$ theory, whose divergence degree obeys: 
\be \omega(G) = 4 - n \leq - n/3,\ee and introduce for a vertex $v$ and assignment $\mu$ the scales
\be
i_v(\mu) = \inf_{l \text{~attached at~} v} \mu(l), \quad e_v(\mu)= \sup_{l \text{~attached at~} v} \mu(l).
\ee
We have then the inequalities: 
\begin{align} 
\prod_{\s}\prod_{k=1}^{\k(\s)}  M^{\omega\left(G^\s_k\right)}&\leq \prod_{\s}\prod_{k=1}^{\k(\s)}  M^{-n(G^\s_k)/3}\\
&\leq \prod_{v\in V(G)} M^{-\abs{e_v(\mu) - i_v(\mu)}/3},
\end{align}
since for each scale $\s$, a vertex $v$ belongs to a unique high subgraph $G^\s_k$ only if $\s\leq e_v(\mu)$ and this $G^\s_k$ has external legs attached to $v$ if $\s<i_v(\mu)$, hence the last absolute value. In our case, since each vertex has at most 4 half-edges attached, from which we can form at most 6 pairs of half-edges $(l,l^\prime)$ which naturally obey 
$\abs{\mu(l) - \mu(l^\prime)}<\abs{e_v(\mu) - i_v(\mu)}$, we deduce the crude bound: 
\be 
\prod_{v\in V(G)} M^{-\abs{e_v(\mu) - i_v(\mu)}/3} \leq \prod_{v\in V(G)}\prod_{(l,l^\prime)\text{~attached at~} v} M^{-\abs{\mu(l) - \mu(l^\prime)}/18}.
\ee 
Next, one needs to pick an order for the internal edges $\{l_1, \dots, l_{I(G)}\}$ such that $l_1$ starts at $x_n$ and, for $m\leq \abs{I(G)}$, all subsets $\{l_1,\dots,l_m\}$ are connected. Further, at any edge $l_j$ we associate an edge $l_{p(j)}$ of lower scale with $p(j)<j$ and sharing a vertex, bringing the bound
\be
\prod_{v\in V(G)}\prod_{(l,l^\prime)\text{~attached at~} v} M^{-\abs{\mu(l) - \mu(l^\prime)}/18}\leq \prod_{j=1}^{\abs{I(G)}} M^{-\abs{\mu(l_j) -\mu(l_{p(j)})}/18} 
\ee
Given this order that amounts to organise all edges of the graph in a tree and the above exponential bound, a sum over all scale assignments is bounded by a constant for each internal line. Ultimately, their total $I(G)$ depending linearly on the number of vertices $V(G)$, we recover Weinberg's bound. 

At the same time, only superficially divergent high subgraphs need to be renormalized and bring the only necessary counterterms in diverging amplitudes. With a single strike, this reorganisation solves the overlapping divergence problem (since diverging subgraphs don't overlap anymore) and that of the renormalons. 

To see this last point, we need to explain how the renormalization is carried out. It applies the BPHZ formula through a localization operator $\t$, whose application on a diverging amplitude $g$ of divergent degree $D$ reads in Fourier space.

\be
\t g(k_1,\dots, k_n)= \d\left(\sum_i k_i\right) \sum_{j=0}^D\f{1}{j!} \f{d^j}{dt^j}g(tk_1,\dots, tk_{n})\rvert_{t=0}.
\ee
In direct space, if $a(x_1, \dots, x_n)$ is a test function on which the amplitude is integrated on\footnote{Typically, the external legs attached to the external vertices of $g$.}, this operation amounts to translate the external vertices to a single vertex, here $x_n$
\begin{gather}
\int \prod \dd x_i \tau g(x_1,\dots, x_n) a(x_1, \dots, x_n) = \int \prod \dd x_i g(x_1,\dots, x_n) \t^*a(x_1, \dots, x_n), \\
\t^*a(x_1, \dots, x_n)= \sum_{j=0}^D\f{1}{j!} \f{d^j}{dt^j} a(x_1(t), \dots, x_n(t))\rvert_{t=0},\quad x_i(t) = t(x_i - x_n) + x_n. 
\end{gather}
For an amplitude $A_{G,\mu}$, we define the diverging forest $\mathbf{D}(G,\mu)$ containing the superficially divergent high subgraphs. On each subgraph $g\in\mathbf{D}(G,\mu)$, one recursively selects the external vertex $v_g$ where the localization operator $\t_g$ will attach the other vertices, such that if $g\subset h$ and $v_{h}$ is also external to $g$, one takes $v_g = v_h$. Otherwise and if $g$ and $h$ are disjoint, $v_g$ and $v_h$ are chosen arbitrarily (in accordance with the other inclusion relations). In this way, $\prod_{g\in\mathbf{D}(G,\mu)} \t_g$ is acting commutatively on $\mathbf{D}(G,\mu)$. 

All in all, the renormalized amplitude will be obtained from 
\be 
A_{G,\mu} = \t A_{G,\mu}+ (1-\t)A_{G,\mu}
\ee 
The first term, fully local, repackages all the superficial divergences of the original amplitude. We need to make sure that the second term, rest of the Taylor expansion, can be summed over all scale assignments, leading to the renormalized amplitude. 

The essence of the argument can be gained from the displacement of a single external propagator at scale $j$:
\begin{align}
C^j(x,z) = & C^j(x_e,z) + \int_0^1 \dd t \f{d}{d t} C^j(x_e + t(x-x_e),z)\\
= & C^j(x_e,z) + \int_0^1 \dd t (x-x_e)^\mu \partial_\mu C^j(x_e + t(x-x_e),z). 
\end{align}
Because of the internal propagators $C^i$ in $G$ that must be at higher scales for the divergence to occur, we can use a part of the stronger associated exponential decay \eqref{eq:boundonpropa} to bound the difference $\abs{x-x_e}$
\be
\abs{x-x_e}e^{-\d M^{i}\abs{x-x_e}}\leq \f{2}{\d M^i}e^{-\d M^{i}\abs{x-x_e}/2},
\ee
and another part to change the decaying exponential $e^{-\d M^j(x_e + t(x-x_e) - z)}$ to $e^{-\d M^j(x_e - z)}$, such that in total our amplitude is bounded by a factor $M^{-\abs{i-j}}$, summable over all scale assignments of internal propagators baring scales $i>j$ as we saw for Weinberg's theorem.
At this point appears that the only essential counterterms are coming from internal scales higher than the external ones. The effective expansion of correlation functions takes precisely this consideration seriously to define effective couplings at each scale $j$ such that the above counterterms are introduced to renormalize the diverging high subgraphs over higher scales that $j$. In short, this leads to a series for the coupling $\l_j$ in terms of the couplings $\l_{i>j}$ (and of the other parameters of the theory at the higher scales). The resulting bounds are free of any factorial in the number of internal vertices, in other words of renormalons.
The drawback is that one has to keep track of the external scale, and in this sense the resulting effective counterterms are not strictly local in the ``space of scales". By contrast, the initial renormalized series in the BPHZ works applied the ``localization" operator on all subgraphs including those that were not high. The resulting subtractions are then local but all unnecessary counterterms spawn the renormalons.

\paragraph{Forest formulas.}
In order to deal with the second essential problem of factorially growing number of diagrams to sum over, one has to resort to trees. Indeed, a famous result of Cayley says that there are $n^{n-2}\approx n!$ different trees between $n$ vertices, the largest factor one can afford in the diagrammatic expansion, cancelling the factorial of the denominator from developing the exponential. Here is where the forest formulas come into play. Briefly, they correspond to Taylor expansions of the amplitudes with integral rest. One could say much more about them, see e.g. \cite{Gurau:2014vwa}, but since we didn't use them in later chapters, we will keep the reader on her/his hunger.

\subsection{Key examples and a few remarks}

\paragraph{Massless case.} As we raised the point earlier, for massless theories, one cannot subtract amplitudes at zero momentum, as they diverge, neither at exceptional momenta (for which at least one partial sum vanishes). In that case, one has to introduce a non-zero IR cutoff by hand, for instance with a mass or defining the renormalization conditions with momenta at some non-exceptional point. For example in four dimensions, one can resort to the symmetric point for the four external momenta $k_i$, such that
\be 
k_ik_j = \mu^2(\d_{ij}-1/4)\;.
\ee See \cite{zinnjustin} for more discussion on the simplifications that occur in the Callan-Symanzik equations.

\paragraph{Perturbative parameters.} Rare are the exactly solvable models. And when we hold one, the most courageous hope is that pertubation around it is legit and will lead to non-trivial information. But we can be audacious with the parameter that we choose to expand in. Three are commonly used in field theory: the coupling of the interaction, the space-time dimension and the number of components of the field. Specifically about this last item, commonly called a large-$N$ limit, finding the appropriate scaling of the interaction such that still an infinite set of diagrams contributes in correlation functions and that their maximal power in $N$ is bounded from above, is a non-trivial problem. All the more, when we have several interactions to rescale such that more than one contributes in the limit. We will also be searching for optimal scalings, that is the minimal power beyond which any large-$N$ limit would be doomed. The choice of optimal scalings for several couplings is a hard problem as will be highlighted in Section~\ref{section:largeNTM}.

\vspace{5mm}
Let us recapitulate the basic features of RG through simple examples, that we will rely upon later in the text and incidentally have become cornerstones of QFT. 

\begin{figure}[htbp]
\centering
\tikzsetnextfilename{one_loop_WF2}
\input{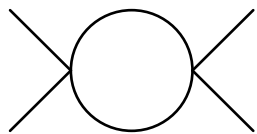}
\caption{\label{fig:one_loop_WF} One loop contribution to the running of the quartic coupling.}
\end{figure}

\paragraph{Wilson-Fisher.}
The textbook example of renormalization is the quartic scalar field in $d$ dimensions, of which the interacting fixed point describes the critical point of models in the Ising universality class. Its action writes
\be 
S = \int \dd^d x \left(\f{1}{2}(\partial \phi)^2 + \f{1}{2}m^2\phi^2 + \f{1}{4!}\l \phi^4 \right), \quad d=4-\eps. 
\ee 
Renormalizing the four-point coupling $\l$ leads to the one-loop beta function (from the diagram~\ref{fig:one_loop_WF})
\be 
\b(\l) = -(4-d)\l + \alpha \l^2, 
\ee 
which, in addition to the Gaussian FP ($\l = 0$), presents in $d<4$ the famous Wilson-Fisher FP \cite{Wilson:1971dc,Wilson:1973jj}: $\l = \eps/\alpha$, where the quartic operator becomes irrelevant with critical exponent $-\eps$. This procedure, applied to $O(N)$ models, allowed to determine perturbatively in $\eps$ critical exponents of many physical systems in three dimensions setting $\eps=1$ using Borel resummation \cite{Guida:1998bx}.

\begin{figure}[htbp]
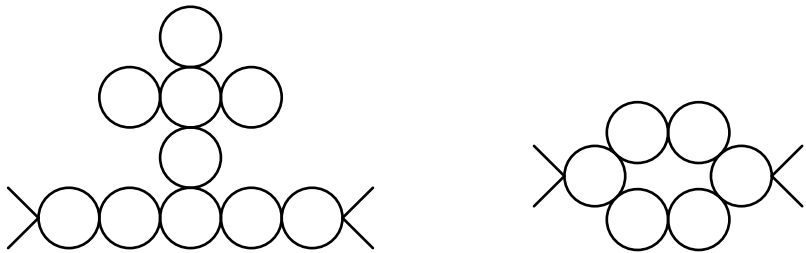

\centering
\captionsetup[subfigure]{labelformat=empty}
\subfloat[]{\tikzsetnextfilename{tree_bubble2}
\input{tree_bubble.tex}}
\hspace{1cm}
\subfloat[]{\tikzsetnextfilename{loop_bubble2}}
\input{loop_bubble.tex}
\caption{\label{fig:bubblechain} Left: A tree of bubbles. Right: A loop of bubbles contributing to $1/N$ correction.}
\end{figure}

\paragraph{Gross-Neveu.}
A second example is a quartic Fermionic theory, with $N$ fields $\psi_i$ \cite{Gross:1974jv} \footnote{From here on, repeated indices imply a summation, unless specified.}

\be 
\label{eq:GN}
S = \int \dd^2 x \left(\bar{\psi}_i\slashed{\partial}\psi_i +\f{\l}{2N}(\bar{\psi}_i\psi_i)^2 \right), 
\ee 
which forbids a mass-term from its chiral symmetry $\psi\rightarrow \g_5\psi$.
The theory is renormalizable in two dimensions and $\l$ is asymptotically free. At leading order in $1/N$, its beta function reads
\be 
\b_\l = -\f{\l^2}{\pi},
\ee 
obtained by summing all diagrams at leading order in $1/N$, trees of bubbles (see Figure~\ref{fig:bubblechain}). Corrections for each order in $1/N$ are given by adding a loop of bubbles. 

%
This can be seen from the intermediate field formalism
\be 
S = \int \dd^2 x \left(\bar{\psi}_i\slashed{\partial}\psi_i - \f{N}{2\l}\sigma^2 + \sigma \bar{\psi}_i\psi_i\right),
\ee
such that integrating out $\sigma$ reproduces \eqref{eq:GN} and on-shell $\sigma = \bar{\psi}_i\psi_i$.
The effective potential for constant $\sigma$, integrating out the Fermions and introducing a UV-cutoff $\L$ is:
\be
V_{eff}(\sigma) =  \f{N}{2\l}\sigma^2 + \f{N}{4\pi}\sigma^2 \left(\log\f{\sigma^2}{\L}-1\right)\;.
\ee
We see that $N$, being a global factor, is the counterpart of $\hbar^{-1}$ and from power counting, each loop of the $\s$ field brings an extra $1/N$ factor. At large-$N$, the saddle-point is given by
\be
\sigma =\L \exp\left(-\pi/\l\right), 
\ee
that is providing a non-perturbative effective mass to the Fermions and spontaneously breaking the chiral symmetry. The saddle is a minimum only when $\l$ stays positive, otherwise the massless Fermions remain the stable vacuum.

In three-dimensions, the interaction is non-renormalizable, however a large-$N$ expansion eliminates all diverging graphs but the bubbles and as will be detailed in the next chapter, leads to a non-trivial UV fixed-point \cite{ROSENSTEINreview}, in fact existing in the range of dimensions $2<d<4$.\footnote{For a discussion in the functional renormalization group framework, see \cite{Gehring:2015vja}.}

Constructive analyses for Fermionic models were performed using phase space expansions.  First, \cite{Feldman:1986ax, Gawedzki:1985ez} established renormalizability and Borel summability of the two-dimensional case by brute force, formalism later facilitated by much simpler inequalities to show analyticity of the partition function \cite{Disertori:1998qe}. Spontaneous chiral symmetry breaking was also considered \cite{Kopper:1993mj,Rivasseau:1995mm}. Finally, the three-dimensional case at large-$N$ was renormalized in \cite{deCalan:1991km,Rivasseau:1995mm}.

\paragraph{Bardeen-Moshe-Bander.}
Three-dimensional scalar fields with sextic interactions have been a model for tricritical points, appearing for instance in mixtures of He3-He4 \cite{Amit:1984ri}. The action writes
\be 
S = \int \dd^d x \left(\f{1}{2}(\partial \phi)^2 + \f{1}{2}m^2\phi^2 + \f{1}{4!N}\l_4 \phi^4 + \f{1}{6!N^2}\l_6 \phi^6 \right), 
\ee 
with, in $d=3$, two relevant directions ($\phi^2$ and $\phi^4$) and one marginal ($\phi^6$).
At large N the beta function of the sextic coupling vanishes, resulting in a line of fixed points at zero renormalized $m_r$ and $\l_{4,r}$, terminating at a UV FP $\l_{6,r}= (4\pi)^2$, above which the non-perturbative effective potential for $\phi^2$ (obtained for instance through a variational principle) becomes unstable at $\phi=0$ (the Bardeen-Moshe-Bander phenomenon \cite{Bardeen:1983rv})
\be V(\phi)= (16\pi^2-\l_{6,r})\abs{\phi}^3,\ee
hence displaying another example of spontaneous symmetry breaking of conformal invariance, and another of an asymptotically safe scalar theory \cite{Litim:2018pxe}.

Let us note that this emergent segment of fixed points is a large $N$ artefact and only the endpoints remain at finite $N$. 
We point to \cite{Yabunaka:2017uox, Fleming:2020qqx} for a recent discussion that explains the nature of the BMB fixed point as the intersection between a line of regular and another of singular fixed points using the functional RG formalism. They demonstrate the presence of non-perturbative fixed points with which the BMB FP can collide in the ($N,d$) phase space, going from infinite to finite $N$. It sheds light on the dependence on $N$ and on the space-time dimension in order to preserve this particular critical behaviour.

\paragraph{Non-abelian gauge theory.} Finally, we had to mention the most notorious example of asymptotically free theory, that is QCD. For $SU(N)$ Yang-Mills coupled to $n_f$ flavors of Fermions
\begin{gather}
S =  \f{1}{2g^2}\int \dd[4]x F^{\mu\nu}F_{\mu\nu} + \sum_{j=1}^{n_f}i \bar{\psi}_j\slashed{D}\psi_j,\\
F^a_{\mu\nu} = \partial_\mu A^a_\nu - \partial_\nu A^a_\mu + f^{abc}A^b_\mu A^c_\nu,\\
D_\mu = \partial_\mu - i t^a A^a_\mu,
\end{gather}
the $\beta$ function is \footnote{A tour de force giving the 2004 Nobel Prize to D. Gross, F. Wilczek and H. D. Politzer.}
\be 
\b_g =-\f{g^3}{4\pi^2}\left(\f{11}{12}N - \f{1}{6}n_f\right),
\ee 
indeed negative for $N=3$ and $n_f=6$. Beyond providing trust to perturbation theory near the ultraviolet, this result also asks for a phase transition at some scale of order $\L_{QCD}=\L \exp(\pi^2/g^2)$ where the coupling grows breaking all convergence of the perturbative series and would lead to hadronization of the quarks. This is the confinement problem. 

Two-dimensional Yang-Mills is quite well-understood (see for instance Witten's preface in \cite{Rivasseau:1995mm}). In contrast, divergences in three and four dimensions are more difficult to control and among approaches to tame them there were lattice discretization (e.g. \cite{Balaban:1985yy, Federbush:1984ee} that begin the long program), continuous with introduction of a mass-term \cite{Magnen:1992wv} or flow equations for an effective action \cite{Efremov:2017sqi}. The interested reader will find a concise review of the current status of rigorous results on this subject in \cite{Chatterjee:2018sgh}.

\section{Aspects of Conformal Field Theories}
\label{sec:CFT}
Fixed points of the RG flow are scale invariant theories. Assuming unitarity and Lorentz invariance, a stronger conformal symmetry typically comes along forming the conformal group $SO(d+1,1)$. 
This was proven in two dimensions with Zamolodchikov’s $c$-theorem that identifies a $c$-function, monotonically decreasing with the RG flow and matching the central charge at the fixed point. There are strong indications that it holds in four, looking at the flow of an anomaly coefficient (Cardy's $a$-theorem). A standard review on this topic is \cite{Nakayama:2013is}.

In two dimensions, proving that statistical models on the lattice preserved conformal invariance was achieved through the introduction of Schramm-Loewner Evolution (SLE), intuitively a continuous curve towards which the interface between domains of different phases converges in the thermodynamic limit at the critical point. It sets a rigorous framework for conformal invariance of the scaling limit of statistical systems (see \cite{DumSmi:2012} for a self-contained introduction). 

The conformal group contains $d$ translations, $d(d-1)/2$ rotations, $d$ special conformal transformations and one dilation.\footnote{Among the countless lecture notes on the topic, we referred to \cite{Simmons-Duffin:2016gjk}.}
The set of eigenvectors of the dilation operator form the primary fields $O_{\D,J}$ characterized by spin $J$ and dimension $\D$. Differentiating the primaries, we get the descendants $O^{\bar{\mu}}_{\D,J}$, \footnote{The multi-index $\bar{\mu} = \mu_1\dots \mu_r$, standing for the derivatives $\partial_{\mu_1}\dots \partial_{\mu_r}$ that act upon the primary.} which form a conformal multiplet, an irreducible representation of the conformal group. Conformal symmetry constrains completely the form of the two- and three-point functions (modulo a normalization factor of the two-point function). For the simplest, scalar fields $\phi_{\D_i}$ of dimension $\D_i$ ($i=1,2,3$), we have
\begin{gather}
\label{eq:2-3ptfunctions}
\expval{\phi_{\D_i}(x_1)\phi_{\D_j}(x_2)} = \f{C\d_{ij}}{\abs{x_1 - x_2}^{2\D_i}},\\
\expval{\phi_{\D_1}(x_1)\phi_{\D_2}(x_2)\phi_{\D_3}(x_3)} = \f{C^{\D_1,\D_2}_{\D_3}}{\abs{x_1 - x_2}^{\D_1 +\D_2 - \D_3}\abs{x_2 - x_3}^{\D_2 +\D_3 - \D_1}\abs{x_3 - x_1}^{\D_3 +\D_1 - \D_2}}, 
\end{gather}
where $C$ and $C^{\D_1,\D_2}_{\D_3}$ are numbers. The coefficients of the three-point functions appear in the operator product expansion (OPE), characterizing the singularity of product of fields at nearby points $x_1\sim x_2$:
\be 
\phi_{\D_i}(x_1)\phi_{\D_j}(x_2) = \sum_{k} C^{\D_1,\D_2}_{\D_k}P_{12k}(x_1-x_2,\partial_{x_2})O_{\D_k}(x_2),
\ee 
the sum being done only on primaries and $P_{12k}(x_1-x_2,\partial_{x_2})$ is a differential operator fully constrained by the conformal invariance of the three-point function. This operator valued equality holds under expectation values.\footnote{Let us note that around a fixed point of which we know the CFT data, one can alternatively find the coefficients in the bare expansion by computing the OPE generating the considered operator (see \cite{Cardy:1996xt} for a pedagogical exposition of this technique).}

All higher-point correlation functions are computed from OPE expansions of the considered operators. 
Hence, in addition to the central charge, \footnote{In $d\geq 2$, it is defined as the coefficient $C$ of the two-point function of the stress-energy tensor and restricting to the contribution of a given field, we define its proper central charge.} the conformal dimensions of the fields and their OPE coefficients fully characterize the CFT. 

A Lorentzian CFT is unitary if all states of the theory have positive norm. In a (Wick-rotated) Euclidean CFT, the equivalent statement by the Osterwalder-Schrader reconstruction theorem is that of reflection positivity of correlators
\be 
\expval{\Theta[O_1(x_1)]\Theta[O_2(x_2)]\dots O_2(x_2)O_1(x_1)\dots} \geq 0,
\ee 
$\Theta$ conjugating and reflecting the operators $O(x)$ with respect to a chosen codimension 1 plane.
This condition imposes that the central charge $c$ is positive and that the conformal dimensions $\D$ of spin $J$ primaries obey
\be
\D \geq 
\begin{cases} \f{d-2}{2}& \text{if } J=0\;, \\
J+d-2 &  \text{if } J>0\;.\end{cases}
\ee

Of course, effective descriptions of many statistical systems are not bound to be unitary. Non-unitary CFTs occur for example: 
\begin{itemize}
    \item in non-integer dimension: it was shown \cite{Hogervorst:2014rta,Hogervorst:2015akt} studying $\phi^4$ in $d=4-\eps$ that the theory at the fixed point disposed of a extra number of evanescent operators that would vanish in integer dimension but not otherwise. Certain of them have negative norm and imaginary conformal dimension. 
    \item from Lagrangians with imaginary couplings: as the Lee-Yang minimal model with a cubic potential $V(\phi)  =ih\phi+i\l\phi^3$, which serves to describe, above the critical temperature, the critical behaviour of the (analytically continued) magnetization of a $2\leq d<6$ dimensionial Ising model with respect to the magnetic field close to a branch point singularity \cite{Fisher:1978pf}. In two dimensions, it has a single primary $\phi$, with negative dimension $h=-1/5$ and a negative central charge $c=-22/5$ \cite{Cardy:1985yy}.
    \item as logarithmic CFTs (for a review \cite{Hogervorst:2016itc}), for which the dilation operator is non-diagonalizable. This induces the presence of extra logarithmic divergences in the above correlations \eqref{eq:2-3ptfunctions}. 
    The fishnet theories, deformations of $\cN = 4$ SYM, belong to this class (cf. for example \cite{Gurdogan:2015csr,Gromov:2018hut}), 
    or special singular limits of statistical systems, as the $n\to 0$ limit of the two-dimensional $O(n)$ model \cite{Cardy:2013rqg}.
\end{itemize}
Later, we will encounter other examples of non-unitary CFTs with particular tensor field theories that develop an imaginary spectrum of conformal dimensions for some bilinear operators in the fundamental fields.\footnote{For a deeper connection between non-unitary CFTs and complex CFTs we refer to \cite{Gorbenko:2018ncu}.}

\subsubsection{Conformal partial wave decomposition}

Let us focus now on the four-point function of scalar fields $\phi_i$ of dimension $\D_i$ ($i=1,\dots,4$). Using two OPEs in the channel $(12\rightarrow 34)$, it also writes as:
\be 
\expval{\phi_1(x_1) \phi_2(x_2) \phi_3(x_3) \phi_4(x_4)} = \sum_{\D,J} C^{\D_1,\D_2}_{\D,J}C^{\D_3,\D_4}_{\D,J} G^{\Delta_i}_{\D, J}(x_i), 
\ee 
with the conformal blocks $G^{\Delta_i}_{\D, J}$, which can be seen as transmitting an operator of dimension $\D$ and spin $J$. We used a condensed notation $\D_i$ and $x_i$ to stand for a dependence on the four conformal dimensions or positions.

An important formalism in recent studies of $d$ dimensional CFTs (e.g. \cite{Simmons-Duffin:2017nub,Liu:2018jhs}) introduces the (unphysical) shadow operators $\tO_{\tD,J}$ of dimension $\tD = d-\D$, dual to operators $\cO_{\D,J}$, and the conformal partial waves \be \Psi^{\D_i}_{\D,J}(x_i) = \int \dd x_0\expval{\phi_1(x_1)\phi_2(x_2) \cO(x_0)}^{cs}\expval{\tO(x_0)\phi_3(x_3)\phi_4(x_4)}^{cs}.
\ee
We denoted by $\expval{\cdot}^{cs}$ the conformal structure of correlators with OPE coefficients set to one. Conformal partial waves form an orthogonal set when integrated over the spatial positions they depend on. It is a complete set when including all integer $J$ and $\D = d/2 + ir$, $r\geq 0$ (called the principal series) in $d>1$ and supplemented by the discrete values $\D = 2n$, $n\geq 1$ for $d=1$.
Contracting a shadow operator with its dual in a three-point function defines the shadow coefficients $S_{O}^{O_2O_3}$:
\be
\int \dd[d]y \expval{\tO_{\tD,J}(x_1)\tO_{\tD,J}(y)}^{cs}\expval{O_{\D,J}(y)O_2(x_2)O_3(x_3)}^{cs} = S_{O}^{O_2O_3}\expval{\tO_{\tD,J}(x_1)O_2(x_2)O_3(x_3)}^{cs},
\ee
or more explicitely 
\be
S^{\D_1,\D_2}_{\D,J} = \frac{\pi^{d/2}\G(\D-d/2)\G(\D+J-1)\G(\f{\D+\D_1-\D_2 + J}{2})\G(\f{\tD+\D_2-\D_1+J}{2})}{\G(\D-1)\G(d-\D + J)\G(\f{\D+\D_1-\D_2+J}{2})\G(\f{D+\D_2-\D_1+J}{2})}.
\ee
Conformal partial waves can be written as linear combination of the conformal blocks: 
\be
\label{eq:CPW}
\Psi_{\D,J}^{\D_i}(x_i)  =  \left(-\f{1}{2}\right)^J S_{\D,J}^{\D_1,\D_2}G^{\D_i}_{\tD,J}(x_i) + \left(-\f{1}{2}\right)^J S_{\tD,J}^{\D_3,\D_4}G^{\D_i}_{\D,J}(x_i) \;.
\ee
Note the exchange of $\D$ with $\tD$ between the two terms.

We will make use later of the irreducible four-point kernel of scalar fields $\phi$ with the same conformal dimension $\D_\phi$, in the channel $(12\rightarrow 34)$, defined as: 
\be 
K(x_1,x_2;x_3,x_4)= \int \dd x_a \dd x_b G_{1a}G_{2b} \f{\d\Sigma_{34}}{\d G_{ab}},
\ee 
with $\Sigma$ and $G$ corresponding to the self-energy (1PI amputated two-point function) and the two-point function of the field $\phi$. Their indices are here a shorthand for the position of the field ($1$ for $x_1$, etc.). Notice that under conformal transformations, $K$ transforms as two $\d$ functions since $\Sigma$ transforms as an inverse two-point function. 
In section~\ref{section:largeNTM} with the 2PI language, we will show that the four-point function in this same channel decomposes as a series in $K$: 
\be 
\label{eq:seriesK}
\expval{\phi_1\phi_2\phi_3\phi_4}_{(12\rightarrow 34)} = \int \dd x_a \dd x_b \f{1}{1 - K}(x_1,x_2;x_a,x_b)\left(G_{a3}G_{b4} + G_{a4}G_{b3}\right).
\ee 
Representation theory also tells us that \footnote{To obtain the full four-point function, one has also to take into account non-normalisable $\Psi$'s in order to subtract unphysical poles from closing the contour of the principal series.}
\be 
\expval{\phi_1\phi_2\phi_3\phi_4} = \f{1}{\abs{x_{12}}^{2\D_\phi}\abs{x_{34}}^{2\D_\phi}} + \sum_J\int_{d/2}^{d/2 + i\infty}\f{\dd \D}{2\pi i}\rho(\D,J)\Psi_{\D,J}^{\D_\phi}(x_i), 
\ee 
the first term coming from the insertion of the identity operator and the second term corresponding to the expansion of the propagation through the channel $(12\rightarrow 34)$. One can show that the first two terms in this channel are given by
\begin{gather}
\expval{\phi_1\phi_3} \expval{\phi_2\phi_4} + (1\leftrightarrow 2) = \sum_j \int_{d/2}^{d/2+i\infty} \f{d\D}{2\pi i}\rho^0(\D,J)\Psi^{\D_\phi}_{\D,J}(x_i),\\
\rho^0(\D,J) = \f{1 + (-1)^J}{n_{\D,J}}t_0 S_{\tD_\phi}^{\tD_\phi,(\D,J)}S_{\tD_\phi}^{\D_\phi,(\D,J)},
\end{gather}
where $n_{\D,J}$ and $t_0$ are known constants depending on the dimensions $d,\D$ and spin $J$.
Conformal invariance implies that
\be
\int \dd[d] x_3\dd[d] x_4 K(x_1,x_2;x_3,x_4)\expval{\phi_3\phi_4O_{\D,J}(x)} = k(\D,J)\expval{\phi_1\phi_2O_{\D,J}(x)}.
\ee
From eq.~\eqref{eq:seriesK} follows 
\be
\rho(\D,J) = \f{1}{1 - k(\D,J)}\rho^0(\D,J). 
\ee
The symmetry of the density $\rho(\D,J) = \rho(\tD,J)$ and the decomposition~\eqref{eq:CPW} allows to extend the integration contour on the whole imaginary axis:
\be 
\expval{\phi_1\phi_2\phi_3\phi_4}_{(12\rightarrow 34)} =\sum_J\int_{d/2-i\infty}^{d/2 + i\infty}\f{\dd \D}{2\pi i}\f{1}{1 - k(\D,J)} \rho^0(\D,J)\left(-\f{1}{2}\right)^JS^{\D_\phi\D_\phi}_{\tD,J}G^{\D_\phi}_{\D,J}(x_i)\;.
\ee 
Closing the contour on the right takes up poles coming from $k(\D,J)$, the $\G$ functions hiding in $\rho^0$, the shadow coefficients and the conformal blocks, but the only ones that don't cancel are those such that $k(\D_n,J) = 1$, leading to 
\be 
\expval{\phi_1\phi_2\phi_3\phi_4}_{(12\rightarrow 34)} = \sum_{J,n}\left(C^{\D_\phi\D_\phi}_{\D_n,J}\right)^2G_{\D_n,J}^{\D_\phi}(x_i),
\ee 
with
\be 
C^{\D_\phi\D_\phi}_{\D_n,J} = Res\left(\f{1}{1-k(\D,J)};\D_n\right)\rho^0\left(-\f{1}{2}\right)^JS^{\D_\phi\D_\phi}_{\tD,J}.
\ee 
Actually, the question of what poles contribute is more subtle \cite{Simmons-Duffin:2017nub,Caron-Huot:2017vep}. Since this formula was proved when the dimensions of the external operators lie on the principal series ($\D_\phi= d/2+ir$, $r\in \mathbb{R}_{\geq 0}$), and closing the contour on the right, we keep the poles on the right of $d/2$ (the situation for poles on $d/2$ is more tricky). When we consider more generic external dimensions, the contour has to be deformed to pass through those dimensions all the while leaving only the poles taken up earlier on its right. This peculiarity will be met in Chapter~\ref{ch:sextic}.

\subsubsection{Remarks}
\paragraph{Long-range models.}
We will later study a model with fractional power of the inverse Laplacian as free propagator
\be 
S = \int \dd[d]x \phi (-\partial^2)^\z\phi + S_{int}[\phi].
\ee 
Such rescaling corresponds physically to long-range models, for instance spins correlated at arbitrary distance. Without interactions they are known as generalized free fields, that are trivially conformal.
With interaction, models have also been proved to possess conformal fixed points, by embedding them in a space of larger dimension $D = d+ 2-2\z$, localizing the kinetic term while the interaction appears as a defect, and by considering the Ward identities under scaling and special conformal transformations, show that they are proportional to the $\beta$ functions of the system. This implies that conformal invariance is restored at the fixed point. 
Concerning Bosonic models, a large work focused on $\phi^4$ models around dimension 3 \cite{Paulos:2015jfa}. Depending of the value of $\z$, different critical properties are laid out: Gaussian ($\z<d/4$), short-range ($\z>\z^*$) or long-range ($d/4<\z<\z^*$), for some $\z^*$. Perturbative and rigorous estimates of critical exponents have been obtained as well as a perturbative proof of conformal invariance.\footnote{See also \cite{Paulos:2015jfa} for progress on non-perturbative results.}

With Fermions, \cite{Gawedzki:1985ed} used such rescaling on a two-dimensional quartic model, such that the marginal interaction becomes non-renormalizable, but leading to a UV fixed point that could be reached perturbatively in $\epsilon$. Finally for models that are close to conformal (such as the SYK model that we will discuss later), it is worthwhile to tune their dimension to match the nearly-conformal one, in order to study the conformal sector of the theory \cite{Gross:2017vhb}. 

One important point is that the free propagator is not renormalized, being non-local, while radiative corrections contribute only local divergences. A second remark is that by manipulating the field dimension, we can tune the critical dimension until the interaction becomes marginal, while the space-time dimension remains fixed. This provides another way to study perturbatively the RG flow without generating any anomalous dimension of the bare fields.

\section{Quantum Gravity and Random Geometry}
\label{sec:QG}
\subsection{Panorama of Quantum Gravity} 
The two most powerful and predictive theories at hand for now are general relativity and quantum mechanics, which unfolded into quantum field theory, or quantum mechanics at each point of space. The first one, deterministic, flourishes at large distances, it gives dynamics to space-time and predicted among others deflection of light around massive objects, gravitational waves and black holes. Its natural mathematical language is that of geometry. 
The second one excels at microscopic scales, with the LHC reminding continuously the precision of the Standard Model, lasers displaying the emergence of collective phenomena or the development of quantum computers that may revolutionize the way we deal with information or simulate systems. We are more familiar seeing it expressed with algebra or analysis. But its intrinsically random nature, the status of the observer in measurement, disconcert and are subject to unceasing philosophical discussions. 

Naive attempts to reconcile both in order to write a theory of gravity valid at all scales describing all the while its interactions with matter seemed flawed since general relativity is non-renormalizable as a four dimensional field theory. To proceed, we would need to abandon at least one of the axioms that seed each theory: locality, general covariance, unitarity or something else. Among the questions that divide are: the treatment of the geometry as a background on which matter distributes or as emerging with the matter from a single entity, the existence of a minimal length scale, the validity of quantum field theoretic language at the Planck scale, etc. Also, the extent to which a simple Wick-rotation from Euclidean signature is enough to learn about the four dimensional space-time remains blurry.

But what is sure, is that a quantum theory of gravity should provide a resolution of a pressing problem deeply rooted in Einstein's equations: cosmological singularities, such as the Big Bang and black holes. We will not discuss the first, nevertheless the second already highlights many curious aspects of the quantum nature of gravity. 

Until recently, black holes remained quite hypothetical predictions of Einstein's theory of general relativity, with only indirect evidence of their existence such as observation of their accretion disks and X-ray emissions, or of stars orbiting around them. Today, gravitational waves detectors allow to decipher much finer signals, such as that emitted from the merging and ringdown of two stellar mass black holes \cite{Abbott:2016blz} matching precisely with numerical simulations. More recently, the EHT collaboration is taking a closer look at the event horizon of very compact objects and made available the first ``picture" of a black hole in April 2019 \cite{Akiyama:2019cqa}. 

From a theoretical standpoint, black holes are filled with puzzles. To start, according to the ``no-hair theorem" obtained in the early 1970s, a generic stationary black hole solution to Einstein's equations is given by the Kerr–Newman metric, characterized by only three parameters: mass, angular momentum, and electric charge. At about the same time, a formula for its entropy was derived \cite{Bekenstein:1973ur}, proportional to the area of its horizon, $S_{bh} = c^3 k_BA/4G\hbar$, followed further by the four laws of black hole thermodynamics \cite{Bardeen:1973gs}, assimilating a black hole with a thermodynamical system at equilibrium. At what temperature? In 1975, Hawking quantized a scalar field on the background of a Schwarzschild black hole of mass $M$, concluding that the ''vacuum state" was filled with thermal radiation at temperature $T_H= \hbar c^3/8\pi G  k_B M$ \cite{Hawking:1974sw}. Hence the black hole was evaporating and an initial pure state associated to the matter before the collapse would evolved into a final thermal state.\footnote{Let us remark that a quick calculation tells us that in the 2.7K CMB, black holes of mass smaller than $10^{22}$ kg would start evaporate and that those of mass smaller than $10^{11}$ kg would do so within the age of the universe. Contrariwise, astronomical black holes have a mass typically larger that $10^{30}$ kg (1\(\textup{M}_\odot\)).}
Where did the initial information contained in the matter that formed the black hole go to?
In other words, how could this evaporation correspond to a unitary process in a quantum theory? This is the information paradox. In order to preserve unitarity, Page \cite{Page:1993wv}, considering a finite quantum mechanical system, has argued that the Von Neumann entropy of the radiation should, starting from zero, first obey a linear increase in time as it is emitted by the black hole, and after reaching a maximum (at the Page time), should return to zero (in a power law, non-universal manner). As we will point later, recovering this curve for a realistic system in a fully controlled way, may not be completely out of sight today. Nevertheless, understanding how evaporation happens dynamically or what happens to an observer traversing an event horizon would require a clearer picture of the involved fine-grained gravitational degrees of freedom. At the same time, this problem questions basic assumptions in the physicists' toolbox such as the use of effective field theory far from the horizon, the decay of exponential corrections approaching the horizon, the validity of quantum mechanics, etc. For a short survey of the information paradox, some of the tentatives to tackle it and relation of astronomical data, see \cite{Compere:2019ssx}.

The preceding elaboration seems to hint that gravity is an emergent phenomenon, much as hydrodynamics is a effective description at large scale of numerous microscopic interacting degrees of freedom.\footnote{Incidentally, relations between Navier-Stokes and Einstein's equations have been established \cite{Hubeny:2011hd}.} Except that in view of the dependence of the entropy on the area of the horizon, the fundamental degrees of freedom seem to hide in a ``spacetime" of one dimension less. This argument, developed in the 1990s, lead to the ``holographic principle" \cite{tHooft:1993dmi, Susskind:1994vu}.
Recently, considerable emphasis has been put on establishing connections between gravity and quantum information theory, in particular entanglement properties of quantum field theories (e.g. \cite{Witten:2018zxz}). For example, qubit circuits have been designed to model black hole evaporation \cite{Osuga:2016htn}. 

\subsubsection{At the crossroad}

This said, different approaches have emerged to tackle the more general problem of quantizing gravity with very different conceptual foundations. They essentially divide into continuum and discrete approaches, the second akin to the lattice version of QFT, except that the lattice must be dynamical. 

\paragraph{String Theory.} In the same way that quantum field theory considers elementary particles as point-like objects propagating in (a fixed) space-time, string theory quantizes string degrees of freedom (of string length $l_s$) that propagate in space-time (the target space), sweeping a two-dimensional surface, the worldsheet \cite{Polchinski:1994mb}. Two formulations of the theory are employed, as a sigma model of fields (forming ultimately the target space) on the worldsheet, from which, when the fields are quantized, the spectrum is extracted, scattering amplitudes computed, etc. or as an effective low energy classical field theory background above which fields are quantized, reproducing the preceding spectrum.\footnote{However, for many non-trivial backgrounds, it is difficult to find the equivalent worldsheet path-integral formulation, hence saying that string theory has no complete non-perturbative description at the moment.} Among its achievements, we count the existence of a massless spin two particle in its spectrum \cite{Scherk:1974ca}, the UV-finiteness of correlation functions up to two loops \cite{Morozov:2008wz} due to the non-locality of the string interactions and a profusion of dualities \cite{Polchinski:2014mva}. Mathematical coherence has also contributed many new mathematical results and interpretations of them in number theory, algebraic geometry, etc. However, for the theory to be consistent, supersymmetry and a ten-dimensional space-time are required and getting away of these two conundrums leads to an extraordinary number of models and vacua.
In addition to strings, second fundamental dynamical objects in the theory are Dp-branes. Of $p$-dimensional spatial extension, they may serve perturbatively as endpoints of open strings but arise also as non-perturbative solutions in the string coupling. Given $N$ coincident branes, the two ends of the string may be described by an effective hermitian $N\times N$ matrix. More precisely, massless excitations of open strings transverse to the brane correspond to $SU(N)$ gauge fields on the brane. When a large number of branes is stacked together, they form black holes. The entropy of extremal supersymmetric black holes has been computed \cite{Strominger:1996sh} and found to match the 1/4 factor of Bekenstein-Hawking. 

\paragraph{Gauge/gravity duality.} The idea of gravitational degrees of freedom being encoded in a lower dimensional spacetime, as a hologram, was made concrete through a duality arising in string theory, between a gauge theory and a gravitational one living in one more dimension \cite{AdS}. It gives a dictionary that relates symmetries, states, operators and correlation functions of both sides (see e.g. the review \cite{Hubeny:2014bla}). The duality originally grew and is much better understood in asymptotically AdS spacetimes, where the quantum theory is viewed as living on the timelike asymptotic boundary which presents conformal symmetry.\footnote{Today, flat and dS variants are being developed in parallel (e.g. \cite{Laddha:2020kvp,Cotler:2019dcj}).}
It is a strong/weak duality in the following sense. If the gauge coupling of the boundary $SU(N)$ gauge theory is given by $g_{YM}$, with 't Hooft coupling $\l = g_{YM}^2 N$, those parameters are related to the string coupling $g_s$ and bulk curvature $l/l_s$ by 
\be
4\pi g_s = g_{YM}^2 = \f{\l}{N}\quad \l^{1/4}\sim (4\pi g_s N)^{1/4}= \f{l}{l_s}, 
\ee
such that the bulk approximates to classical gravity in the limits $N\gg \l \gg 1$, that is computations in a strongly coupled field theory are reformulated in terms of a semi-classical process on a gravitational background. Incidentally, it brings an interesting geometrization of the renormalization group: boundary correlation functions at different energy scales are computed at different radial positions in AdS, UV degrees of freedom close to the boundary and moving deep in the bulk as the RG flows. 
Quite curiously this conjecture has brought contributions in many different fields: string theory, condensed matter, quantum information theory, pure mathematics, etc.

\paragraph{Higher-spins.} Higher-spin gauge theories generalize the structure equations of the spin-connection and the vielbein to involve in non-linear equations an infinite tower of massless higher spins gauge fields (reviews include \cite{Bekaert:2005vh,Didenko:2014dwa}). In particular, they always contain a spin two gauge field, the graviton. They use techniques of twistor and non-commutative theories. They are consistently written on background with non vanishing cosmological constant, in any dimension, and have obtained support from the holographic conjecture. In particular, gauged vector models are believed to be the boundary duals of higher-spin theories, the gauging condition being necessary for matching the spectra of primary operators on both sides. Further substance to the correspondence is given by precise equivalence of three-point functions, non-trivial to compute in the bulk.\footnote{However, the theory incorporating fields with an infinite number of derivatives is questioning the usual notion of locality while even Riemanian geometry has to be reconsidered.} Duality between correlation functions of Fermions, Bosons, free and critical models has been brought to light, with similar relations in the higher-spin duals going from the free to the interacting model by changing the boundary conditions of the bulk fields \cite{Giombi:2016ejx}. Extensions of this line of work go towards getting the spectrum of operators when the higher-spin symmetry is (weakly) broken. Recently, tensor models have been conjectured to be dual to multi-particle versions of higher-spin theories \cite{Vasiliev:2018zer}. 

\paragraph{Asymptotic safefy.} The program of asymptotic safety postulates that gravity may be non-perturbatively renormalizable, or said otherwise, present a non-Gaussian UV fixed point of the renormalization group with a finite number of relevant operators, an idea vented by Weinberg \cite{Weinberg:1980gg}. One way to test such idea relies on the non-perturbative functional RG to generate flow equations of a truncated gravitational effective action including couplings for the (dimensionless) Newton's and cosmological constants. Notably such a non-trivial fixed point has been found \cite{Reuter:2001ag} and resists in larger truncations. The late \cite{Bonanno:2020bil} contains a critical overview of the program. Surprising at first sight, a reduction of the effective spectral dimension (cf. Subsection~\ref{subsec:RGBU}) of the spacetime was observed \cite{Lauscher:2005qz}, changing from four at large scales to two at UV scales, dimension at which general relativity becomes renormalizable. This property was also encountered in another approach (see below).

\paragraph{Non-commutative geometry.} Another point of view assumes that a minimal lengthscale renders non-trivial commutation relations between space (or spacetime) coordinates of space. Such a scale arises naturally if one attempts to probe a small enough region of spacetime, which according to Heisenberg's principle, would necessitate enough energy such as to create a black hole. A characteristic problem in this framework is the appearance of mixing of UV and IR divergences when attempting to renormalize the theory. The addition of a harmonic potential for the non-commutative parameter cured the problem and interacting quantum field theories were shown asymptotically safe, with restoration of translation and Lorentz invariance in the planar sector. One can find excellent reviews in \cite{Wulkenhaar2019} and for a broader perspective on non-commutative geometry in \cite{Connes:2019vcx,Poulain:2018mtr}.

\paragraph{Loop Quantum Gravity.} Loop Quantum Gravity goes by quantizing general relativity as a constrained system under diffeomorphism invariance. Focusing on the background independent side of the story, it has a covariant formulation in terms of the partition function for spin-foams, or 2-complexes, whose graph structure codes for an $SU(2)$ connection at each edge, leading to Lorentz invariance. Group Field Theories are a Lagrangian formulation that generate those spin foam amplitudes from their Feynman diagrams. They are close relatives to the tensor models we will discuss in details later, in that the field components take their values on copies of a group. Interestingly many of those models are asymptotically free \cite{Carrozza:2016vsq}.

\paragraph{Dynamical triangulation.} Others assume that spacetime could emerge as a continuum limit of a discrete lattice built by gluing simplices together, tuning the partition function of the system towards a UV critical point; crudely it is statistical physics for random geometry. The Regge action of a discrete configuration mimics the Einstein-Hilbert action, depending on the number of simplices of different dimensions (giving a phase space parametrized by a coupling to the volume and another to a discrete notion of curvature). Mostly relying on Monte-Carlo simulations, they construct diffeomorphism invariant observables (such as mean volume of subsimplices, curvature, etc.) to characterize phases and phase transitions.
Causal dynamical triangulation, where a preferred direction is imposed such that gluing the simplices foliate the spacetime, features notable emergent properties: dimensional reduction \cite{Ambjorn:2005db, Reuter:2011ah} and a second order phase transition from which a continuous limit can be taken. 
For an overview of recent developments in the discrete approach to quantum gravity, we refer to the special issue \cite{Carrozza:2020akv}.

\subsection{2d Quantum Gravity}
\label{subsec:2dQG}

Since in four dimensions, gravity is non-renormalizable and manifolds are difficult to classify, it is worthwhile to start quantizing in lower dimensions, more specifically two. 
Dividing the partition function into a gravitational and a matter part, we have
\begin{gather}
Z = \sum_{\text{topologies}} \int DX Dg \exp\left(-S_{EH}[g_{ab}] - S_{m}[g_{ab},X]\right),\\
S_{EH}[g_{ab}] = \f{1}{4\pi}\int d^2\xi \sqrt{\abs{g}}( R -2 \L),\\
S_m[g_{ab},X] = \f{1}{8\pi}\int \dd^2\xi \sqrt{\abs{g}} g^{ab}\partial_aX^\mu\partial_bX_\mu\;, 
\end{gather}
where we are supposed to integrate over all distinct differentiable structures and topologies are classified by the genus of the surface. The integral of the Ricci curvature $R$ gives $4\pi (2 - 2h)$, from Gauss-Bonnet theorem, with $h$ the genus of the surface, hence is topological. Assigning a coupling $\log g_s$ in front of the gravitational action, one finds the string expansion in the genus $g_s^{2(h-1)}$ describing the different topologies of the worldsheet parametrized by the coordinates $\xi$. 
The matter action $S_m$ could contain some extra potential $V(X)$. Different procedures were adopted to work out this partition function and matter correlation functions, splitting in a continuum or a discrete perspective. 

\subsubsection{From the continuum}

In the path-integral approach, by the Riemann uniformization theorem that allows to gauge transform any two-dimensional metric $g$ to a reference metric $\hat{g}$
\be
g= e^\phi \hat{g} \;, 
\ee up to a dilaton $\phi$, leaving a residual two-dimensional conformal group, the partition function rewrites as Liouville conformal field theory
\begin{gather}
Z = \int D\phi DXD(gh) e^{-S_m [X,\hat{g}] - S_{gh}[b,c,\hat{g}] + S_L[\phi,\hat{g}]}\;,  \\
S_L[\phi,\hat{g}] =\f{1}{4\pi} \int\dd^2\xi \sqrt{\hat{g}} \left(\hat{g}_{ab}\partial_a\phi\partial_b\phi + Q\hat{R}\phi + 4\pi\L e^{\g\phi}\right)\;, 
\end{gather}
where $\L$ is a cosmological constant, $\hat{R}$ is the curvature and the $bc$ ghosts have central charge $c_g = -26$. The Liouville action $S_L$ is conformally invariant if we take
\begin{gather}
Q = \sqrt{\f{25-c_m}{6}}= \f{2}{\g}+\f{\g}{2}\;,\\
\hat{g} \rightarrow e^\sigma \hat{g} \quad \phi \rightarrow \phi - \sigma/\g\;
\end{gather}
and the central charge is $c_L = 1 + 6Q^2$. 
Conformal field theory was born and techniques were developed to extract correlation functions and associated critical exponents. 

In particular \cite{Knizhnik:1988ak} studied how the critical exponents of minimal CFTs ($0<c_m<1$) coupled to 2d gravity (in the light-cone gauge) changed with respect to pure gravity.
When the theory was defined on a surface of fixed area, they found that the string susceptibility $\g_{str}$, controlling the scaling of the partition function with respect to the area, was given by
\be 
\g_{str} = \f{1}{12}\left(c_m-1 - \sqrt{(1-c_m)(25-c_m)}\right),
\ee 
which remains real only for $c_m\leq 1$ and $c_m\geq 25$. This was the reason of the $c_m=1$ barrier in non-critical strings. Further the conformal dimensions $\D$ primary fields involving matter are modified with respect to those on fixed geometry $\D_0$, in a typical relation encoded by the KPZ relation:  
\be 
\D_0= \D+ \f{\g^2}{4} \D(\D-1), \quad \g = \sqrt{\f{25-c_m}{6}} - \sqrt{\f{1-c_m}{6}}.
\ee 
These relations were extended to higher genera in \cite{David:1988hj,Distler:1988jt}.
For a more comprehensive account, we refer to \cite{DiFrancesco:1993cyw,Anninos:2020ccj}.

\vspace{5mm}
However a proper definition of the dilaton measure was missing, as well as a proper treatment of the exponential term. The upgrade came from probability theory and was phrased in the terms of ``Gaussian multiplicative chaos" (a distributional limit), from which the measure was rigorously defined. Among other results, the three-point function (DOZZ formula) was recovered \cite{Vargas:2017swx} and correlation functions on the sphere and surfaces of higher genus were computed in \cite{David et al.(2016)}.

\subsubsection{From Feynman Graphs}
Feynman diagrams are used as a perturbative tool for evaluating amplitudes in a QFT Taylor expanding the interaction term and Wick-contracting the fields. From another point of view, they generate particular (piecewise-linear) discretized geometries to which the associated amplitude is a natural measure, vertices corresponding to interactions and edges to the propagators. This observation was particularly fruitful for realizing two dimensional quantum gravity from a perturbative expansion of matrix models. By analogy, it will be tempting to increase the number of indices for discretizing higher-dimensional geometries, hence tensor models.

\begin{figure}[htbp]
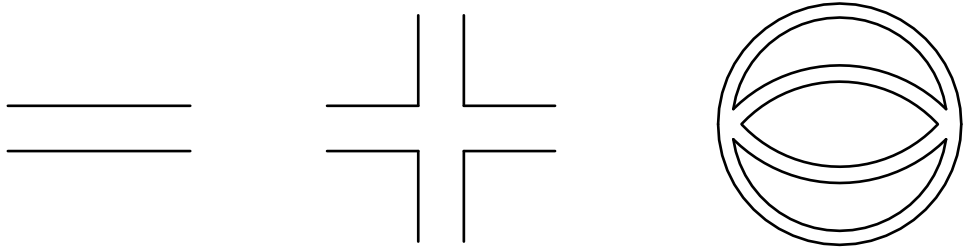

\centering
\captionsetup[subfigure]{labelformat=empty}
\subfloat[]{\tikzsetnextfilename{propa-inter-matrix2}
\input{propa-inter-matrix.tex}}
\hspace{1cm}
\subfloat[]{\tikzsetnextfilename{2vertex-matrix3}
\input{2vertex-matrix.tex}}
\caption{The propagator, the interaction and a two-vertex vacuum graph, in ribbons.}
\label{fig:propa-inter-matrix}
\end{figure}

Let us consider a zero-dimensional real matrix field $M$ of size $N\times N$ and quartic potential 
\be 
Z(g) = \int dM \exp\left(-N \Tr S(M)\right)\quad S(M) = \f{1}{2}M^2 + \f{g}{4}M^4, 
\ee 
where the integration measure is taken as follows: 
\be 
dM = \prod_{i=1}^N dM_{ii}\prod_{1\leq i<j\leq N}dM_{ij} = \prod_{1\leq i<j\leq N}\abs{\l_i-\l_j}^2 \prod_{i=1}^Nd\l_i d\Om_N\;,
\ee 
with the first factor being the Vandermonde determinant and $\Omega_N$ the Haar measure on the sphere $S^N$.
The partition function $Z$ can be computed using Wick's theorem with the propagator
\be 
\expval{M_{ij}M_{kl}} = \f{1}{N}\d_{ik}\d_{jl},
\ee 
depicted as two parallel strands, while interaction vertices are drawn as crossroads (Fig.~\ref{fig:propa-inter-matrix}).
This leads to an expansion in terms of connected graphs $F_n$ with $n$ interaction vertices:
\be 
Z(g) = e^{F(g)}\quad F(g) = \sum_{n\geq 0} (-1)^n g^n N^{n-E+L}F_n(g).
\ee 
We made explicit the factors of $N$ coming for the insertions of interactions ($n$), the propagators from the Wick contractions (edges $E$) and the loops $L$ coming from contracting the matrix indices, since propagators and interaction bring each a factor of $N$, and each loop another factor of $N$.

From the Euler formula $n-E+L = 2 - 2h$ between the number of faces $L$, edges $E$ and vertices $n$ of a graph and its genus $h$, the partition function sums as: 
\be 
Z = N^2\sum_G N^{2 - 2h(G)}\;.
\ee 
Hence, matrix models feature a topological expansion, entirely analogous to that encountered earlier with $g_s=1/N$, with planar diagrams dominating the series at large-$N$. This simplification was first noted by 't Hooft for treating QCD \cite{tHooft:1973alw}.  \footnote{$1/3$ is sufficiently small to serve as a perturbative parameter.}

\vspace{5mm}
To determine the partition function, one has now to enumerate planar graphs or more precisely maps. The Cori-Vauquelin-Schaeffer bijection allows to identify those maps, picked at random, with particular correlated random walks. The original graph supports a metric structure (e.g. the graph distance) and the problem amounts to establish convergence in the Gromov-Hausdorff limit of this random space to a continuous limit.

Two schools have grown, working either in the direction just outlined (see e.g. \cite{Le Gall(2014),Miermont:2014}) or making sense of the Liouville path-integral measure straight in the continuum, studying two-dimensional conformally invariant growth processes (the earlier mentioned SLE curves), depending on the above $\gamma$ parameter (see \cite{Miller:2018ICM,Gwynne:2019hwi} for recent reviews). 
At the special value $\g = \sqrt{8/3}$, physically corresponding to pure gravity, both points of view were shown to match \cite{MillerSheffield}, identifying the canonical metric structure of the limiting two-dimensional surface. Other values of $\g\in (0,2)$ describe statistical models coupled to the planar maps, corresponding to models with central charge $c<1$, and can be studied in the matrix point of view by decorating the planar map with loops. 
More precisely, through polynomial potentials at least of order $m$ and by an appropriate tuning of the couplings, one can reach a multi-critical point of order $m$ and change the string susceptibility to $\g_{str} = -1/m$. This family is described in the continuum limit by the non-unitary $(2m-1,2)$ minimal CFT. The generic $(p,q)$ minimal models whose central charge is $c_{p,q} = 1 - 6(p-q)^2/pq$, are obtained from multi-matrix models \cite{Anninos:2020ccj}.
More recently, a generalization for $c\in (1,25)$ has been devised \cite{Gwynne et al.(2020)}. 
Its physical interpretation is that the vacuum on which correlation functions are computed is unstable \cite{David:1996vp}. The KPZ relations have been also confirmed in the planar map context \cite{Garban(2012),Gwynne and Pfeffer(2019)}.

Looking for analogs of the KPZ relations between critical behaviour in flat and dynamical quantum geometry in higher dimensions remains an important challenge.

\vspace{5mm}
Random matrix models (seen as formal or convergent integrals) dispose of a wealth of techniques to be confronted with \cite{Eynard:2015aea,Eynard:2016yaa}. For instance, with normal matrices in a potential $V(M)$, it can be helpful to diagonalize the matrix and study the effective action for the eigenvalues
\be 
S[\l_1,\dots,\l_N] = \f{1}{N}\sum_{i=1}^NV(\l_i) -\f{1}{N^2} \sum_{1\leq i<j\leq N}\log (\l_i - \l_j)^2,
\ee 
the logarithmic repulsion coming from the Vandermonde in the earlier Jacobian.\footnote{Let us note that for one-dimensional matrix models, a hamiltonian formalism translates the dynamics of the eigenvalues into a decoupled system of $N$ Fermions in a non-trivial potential \cite{Brezin:1977sv}.} When we will consider the intermediate formalism in tensor models, the logarithmic contribution of the matrix intermediate field will be subleading with respect to the potential.
In the large-$N$ limit, one can solve for the continuous eigenvalue density (a kind of emergent geometry). One can also write (loop) equations for correlation functions, all encoded in the solution to an algebraic equation, the spectral curve; they were formalized as topological recursion and had a large impact for proving Witten's conjecture on intersection numbers of moduli spaces of Riemann surfaces. The reasoning was also generalized to two-matrix models \cite{Bergere:2011pp}. 

\subsection{Random Geometry from the bottom up}
\label{subsec:RGBU}
Our approach to random geometry will also start from discrete graphs with a finite number of vertices connected between each other by edges. The graph is naturally equipped by the symmetric graph distance. 
A scaling limit is taken by sending the number of vertices to infinity and the graph distance appropriately to zero, such that the limiting geometry is continuous. One should provide an appropriate measure to the initial graphs and study various properties of the continuum limit (topology, metric, Hausdorff or spectral dimensions, etc.). If one eyes the quantization of gravity, the measure should be related to the Einstein-Hilbert action and the resulting continuous geometry should be a non-trivial higher than three-dimensional manifold, which we are, to some extent, supposed to experience. 

Recurring themes once we are looking for the scaling limit of random ensembles of objects are the fractal nature of the continuous object and the emergence of universality, i.e. that the limit does not depend on details of the initial set. 

Correspondingly, on fractal spaces, depending on the information we are concerned with, two definitions of dimension are especially useful: 
\begin{itemize}
    \item Hausdorff dimension: it gives a global picture, close to the topological dimension. More precisely, given a set $A$, its Hausdorff dimension is: \be d_H(A) = \inf\left\{d\geq 0: \lim_{r\to 0}\inf \left(\sum_i r_i^d\right)=0, \text{such that balls $S$ of radii $r_i\in (0,r)$ cover $A$}\right\},\ee 
    \item spectral dimension: Formally, given a set $A$ and a random walk $\{X_t\}_{t\geq 0}$ starting at $x\in A$, noting the probability that the random walker is at $y\in A$ at time $t$ by $q_t(y,x)$, the spectral dimension of $A$ is 
    \be d_s(A) = -2 \lim_{t\to \infty}\f{\log q_t(x,x)}{\log t}. \ee 
    The spectral dimension provides a more local picture of the landscape.
\end{itemize}

\vspace{5mm}
Two categories of random geometries are very well-understood: trees and surfaces. Their associated field theoretic models are respectively vector and matrix models in their large-$N$ limit for planar surfaces. 

In the same way that the scaling limit of discrete random walks is the Brownian motion, trees and surfaces both have their continuous representative. For the first it is the continuous random tree (CRT) of Aldous, a random metric space, scaling limit of various critical models. For example, the branching process corresponding to Galton-Watson trees (which we will encounter later), at criticality (such that the growth continues indefinitely with an average of a child per generation), converges towards the CRT.
Galton-Watson trees appear in many physical models related to growth.
The Hausdorff dimension of the CRT has been shown to be 2 \cite{Duquesne and Le Gall(2005),Duquesne and Le Gall(2005)2}, same as for the Brownian motion, and its spectral dimension to be $4/3$ \cite{DJW, BarlowKumagai}.
Subcritical trees present a different scaling limit with a single vertex of infinite degree to which are attached Galton-Watson trees.
Branching processes with infinite variance ($\alpha$-stable laws, $\a\in (1,2]$) have also been considered and shown to lead to still different scaling limits, of Hausdorff dimension $d_H=\a/(\a-1)$ \cite{Duquesne and Le Gall(2005), Duquesne and Le Gall(2005)2} and spectral dimension $d_s=2\a/(2\a - 1)$ \cite{CroydonKumbeta, Croydon:2010}. 

The universal planar random surface, bearing the name Brownian sphere (or Brownian map), has been constructed in the two ways we mentioned earlier. 
The Brownian map is homeomorphic to the 2-sphere (has topological dimension 2) \cite{Le Gall and Paulin(2006), Miermont(2007)} and is reached independently of the polygon used to pave the sphere \cite{Le Gall(2011), Bettinelli et al.(2013), Addario-Berry and Albenque(2019)}.
In \cite{Le Gall(2007)}, it was proved that the Brownian map had Hausdorff dimension 4, the intuitive reason being that it could be constructed by gluing two Brownian random walks motions on each other.
It was later shown in \cite{Ambjorn et al.(1995), Gwynne and Miller(2017)} that its spectral dimension was two for the whole range $\g\in (0,2)$. 
The dependence of the Hausdorff dimension with the central charge is on the other hand less clear, with the propositions of Watabiki \cite{Watabiki:1993fk}, and Ding and Goswami \cite{Ding:2019} closest to numeral simulations \cite{Barkley:2019kvp}. 
Other topologies than the sphere were also worked out, such as the disk \cite{Bettinelli and Miermont(2015)} or the half-plane \cite{Baur et al.(2016)}.

\vspace{5mm}

Essential tools for analytically establishing the above properties are bijections with trees, the Brownian motion, SLE curves or with pairs of them \cite{Gwynne:2019hwi}. 
In particular, to determine the spectral dimension, one has different tools available. Within the discrete approaches to quantum gravity, Monte-Carlo simulations of diffusion process on the considered geometry allow a direct estimate of $d_s$, e.g. \cite{Benedetti:2009ge}. 
Otherwise, estimates have been obtained either from the generating function of random walks or from a direct control of the heat-kernel, as we shall develop in Chapter~\ref{ch:GW}. In the Appendix~\ref{app:proba}, we explain in more details how the spectral dimension can be obtained from bounds on the heat kernel.

\vspace{5mm}
Sorting out ensembles of random geometries of dimensions higher than two is much harder, since the genus itself is not enough any more to distinguish the non-equivalent differentiable structures (the non-orientable case can be amended with some surgery, that is opening a hole and gluing a Moebius strip). 

Inspiration may come from numerical simulations gluing a large number of simplices with a Regge-like action as done in dynamical triangulation, or without any a priori, to determine average observables of a ($r+1$)-regular colored graphs uniformly picked at random \cite{carra2019}.
Both approaches lead however to a ``crumpled" phase, that is a highly connected object, with average distance between two points uniformly bounded. With dynamical triangulations, one also observes a first-order phase transition towards a branched polymer phase.
In 4d causal dynamical triangulation, a Hausdorff four-dimensional phase (with properties analogous to a de Sitter geometry) is present, separated from three unphysical phases, relatives of the preceding crumpled and branched polymer phases, by first and at least one second order phase transitions \cite{Ambjorn:2019lrm,Ambjorn:2020rcn}. 

A novel idea consists in iterating the mating procedure of two trees that allowed to construct the Brownian map. Namely, Lionni and Marckert proposed the first random geometric object that could generalize the Brownian sphere in higher dimensions \cite{Lionni:2019tdb}. The first iteration reproduces the CRT ($d_H=2$), the second gives the Brownian map ($d_H=4$) and the following $D$ ones are conjectured to have $d_H=2^D$. Spectral dimension, topology, etc. still need to be analyzed.

\section{The SYK Model and Black Holes}
\label{sec:SYK}

Devised initially as a model of strange metal \cite{Sachdev:1992fk}, the Sachdev-Ye-Kitaev model \cite{Kitaev2015} is a disordered quantum mechanical system of $N$ strongly interacting Majorana Fermions, with Hamiltonian: 
\begin{equation*}
    H = i^{q/2}\sum_{1\leq i_1<\dots<i_q\leq N} J_{i_1\dots i_q}\psi_{i_1}\dots \psi_{i_q}\;,\quad \expval{J^2_{i_1\dots i_q}} = \f{J^2(q-1)!}{N^{q-1}}\;.
\end{equation*}
It is parametrized by the inverse temperature $\b$, the order of interaction $q$ and the number of Fermions $N$.
Within the holographic duality, it describes at low temperature the near-horizon geometry of near-extremal black holes, since at large $N$ it saturates the chaos bound and displays the same effective action breaking reparameterization invariance through the Schwarzian action \cite{Kitaev2015,Polchinski:2016xgd,Maldacena:2016hyu,Maldacena:2016upp}. 
Let us explain a bit deeper those two features.

A notion of a quantum chaos has been developed for quantum systems, by computation of a specific correlator, evaluated at inverse temperature $\b$: 
\be 
C(t) = \expval{[V(t),W(0)]^2}_{\b}. 
\ee 
Seen as the quantization of Poisson brackets between the classical observables associated to $V$ and $W$, this correlator would probe the sensitiviy to initial conditions. If $V$ was the position and $W$ the conjugate momentum, classically, it would grow exponentially in a chaotic system. Such a growth however can only happen at short times, and $C(t)$ reaches a plateau at longer times, as the system equilibrates. The approach to equilibrium in a quantum system is dictated by Pollicott–Ruelle resonances, to which the black hole analogs are the quasi-normal modes (e.g. \cite{Zwo:2017}).
The early Lyapunov exponential growth is related to the exponential redshift observed by an observer near the horizon, for a wave-packet escaping to infinity after bouncing on the event horizon \cite{Polchinski:2015cea}. 
The most relevant term that shapes its evolution is the out-of-time-order correlator (OTOC): 
\be 
OTOC(t) = \expval{V(t)W(0)V(t)W(0)}_\b. 
\ee 
For quantum systems with a large number $N$ of degrees of freedom, this correlator follows a typical structure \cite{Maldacena:2015waa}: 
\be 
\f{OTOC(t)}{\expval{VV}_\b\expval{WW}_\b} = 1 - \f{A}{N}f(t) + \cO\left(\f{1}{N^2}\right), 
\ee 
where $f(t)\leq e^{\k t}$ and $\k \leq 2\pi/\b$. The bound was argued to be saturated when probing systems with a dual black hole and this was verified in multiple settings, notably from scattering on a BTZ black hole \cite{Shenker:2013pqa,Shenker:2013yza}. In case of gravitational scattering with massive particles of spin $J>2$, the bound generalizes to $\k \leq 2\pi(J-1)/\b$ ($J$ being the highest spin present) \cite{Perlmutter:2016pkf}.

\subsubsection{Two-Point Function}
We follow from here the analysis of \cite{Maldacena:2016hyu}.
\begin{figure}
\centerline{\includegraphics[width=12cm]{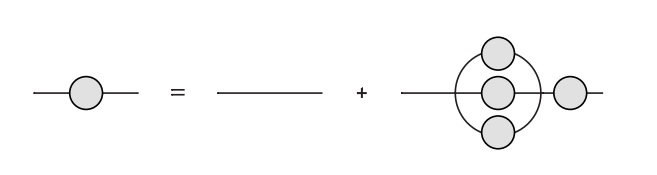}}
\caption{Two-Point Melonic Function}
\label{fig : 2 point melonic}
\end{figure}
In order to build a dimensionless parameter to describe different regimes of the theory, one can use $\beta J$, the UV-dimension of $J$ being $1$. Then a low-temperature (or infrared) regime, at fixed coupling, can be as well described by a strong coupling regime at fixed temperature, or $\beta J \gg1$ (finite temperature or not will be distinguished by the domain of 
the Euclidean time $\tau$). 

To start, we need to obtain the Euclidean two-point function 
\be
G(\tau,\t^\prime) = \frac{1}{N}\sum_i\expval{\psi_i(\tau)\psi_i(\t^\prime)}
\ee
or from time translation invariance, its Fourier transform, \footnote{Abusing notation twice.}
\be
G(\omega) = \int_{-\infty}^\infty \dd{\tau} e^{i\omega \tau}G(\tau).
\ee
At finite temperature, the (Matsubara) frequencies are quantized: $\omega_n= \frac{2\pi }{\beta} (n + 1/2)$ and the Euclidean time is bounded $0\leq \tau\leq \beta$. It is also convenient to define the quantities $\Delta = 1/q$, $\cJ^2 = \frac{qJ^2}{2^{q-1}}$.

At leading order in $1/N$, with the free propagator $G_0^{-1} (\omega) = i \omega$ and the self-energy $\Sigma(\omega)$, the Schwinger-Dyson (SD) equations in the large $N$ limit read
\be G^{-1} (\omega) = G_0^{-1} (\omega) - \Sigma (\omega),\quad  \Sigma (\tau) = J^2 G(\tau )^{q-1} ,
\ee
that is depicted in Fig. \ref{fig : 2 point melonic}. \footnote{A diagrammatic proof identifying the leading order graphs was provided later in \cite{Bonzom:2018jfo}.} Notice that they have this simple form written in terms of Fourier and direct space respectively. Given the resemblance of the self-energy with melons, the limiting equations bear the name ``melonic". The ``blob" indicates a full propagator and a quenched average was taken.
The first equation is the usual one linking the complete two-point function to the self-energy.
Then, taking advantage of the form of the free propagator, the IR limit simplifies the above equation to the simpler convolution
\be 
\label{eq:convolution}
\int \dd \tau^\prime J^2 G(\tau-\tau^\prime) G(\tau^\prime - \tau^{\prime \prime} )^{q-1} = -\delta (\tau - \tau^{\prime\prime}).
\ee
Reparametrization invariance of eq. \eqref{eq:convolution} under any differentiable function $f$:
\be G (\tau, \tau') \rightarrow [ f'(\tau ) f'(\tau ') ]^\Delta  G(f(\tau), f(\tau ')), \quad  \Sigma (\tau, \tau') \rightarrow [ f'(\tau ) f'(\tau ') ]^{\Delta (q-1)}  \Sigma(f(\tau), f(\tau ')),
\ee
suggests to search for a particular solution of type
\be \label{2point}
G_c (\tau)  = b \vert \tau \vert^{-2 \Delta}  {\rm sign} \ \tau, \quad
J^2 b^q \pi = \left(\frac{1}{2} - \Delta\right) \tan (\pi \Delta ) .
\ee 
 The equation for $b$ comes from the formula
\be \int_{- \infty}^{+ \infty} d \tau e^{i \omega \tau}  {\rm sign}  \tau \vert \tau \vert^{-2 \Delta} = 2^{1 - 2 \Delta} i \sqrt{\pi} \frac{\Gamma (1 - \Delta )}{\Gamma (\frac{1}{2} + \Delta)} \vert \omega \vert^{2 \Delta -1}  {\rm sign} \ \omega .
\ee
Applying reparametrization $f_\beta ( \tau) = \tan (\pi\tau/\beta )$ leads to 
the finite temperature two-point function
\be G_c ( \tau) = \left[\frac{\pi }{\beta \sin (\pi \tau / \beta)} \right]^{2 \Delta} b \ {\rm sign } \;\tau.
\ee
Recalling that $\Delta = 1/q$,  the anomalous dimension for $ G_c (\tau)  \propto  \tau^{-2/q}$ at small $\omega$
corresponds to the theory being just renormalizable. 

It is important to notice that the reparametrization symmetry has been spontaneously broken by the solution \eqref{2point} to an $SL(2,\mathbb{R})$: 
$\tau \rightarrow f(\tau) = \frac{a \tau + b}{c\tau + d}$.

\subsubsection{Four-Point Function}
The equation for the four-point function in the large $N$ limit is
 \begin{equation}
 \frac{1}{N^2}\sum_{1\leq i,j\leq N}\expval{T\left(\psi_i(\tau_1)\psi_i(\tau_2)\psi_j( \tau_3)\psi_j(\tau_4)\right)} =  G(\tau_{12})G(\tau_{34}) + \frac{1}{N} 
 \mathcal{F}(\tau_1,\tau_2, \tau_3, \tau_4) + \cdots
\end{equation}
where we write $\tau_{12} = \tau_1 - \tau_2$. 

\begin{figure}
\centering
\captionsetup[subfigure]{labelformat=empty}
\subfloat[]{\tikzsetnextfilename{four-point-ladder-intro2}
\input{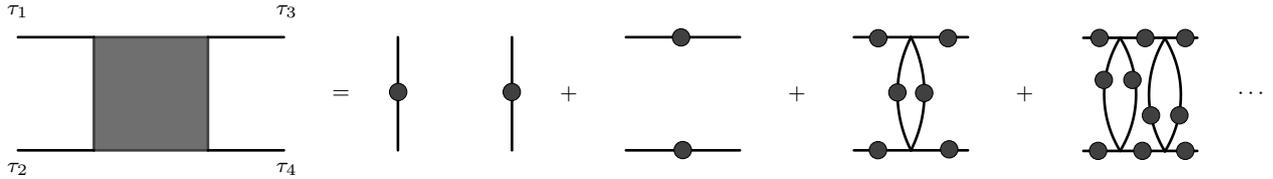}}
\caption{Four-Point Melonic Function (first contribution of order $\cO(1)$ and remaining of order $\cO(1/N)$).}
\label{fig:four-point-ladder-intro}
\end{figure}

\begin{figure}
\centering
\captionsetup[subfigure]{labelformat=empty}
\subfloat[]{\tikzsetnextfilename{rung-op-intro2}
\input{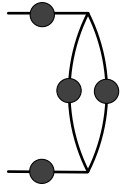}}
\caption{The Rung Operator.}
\label{fig:rung-intro}
\end{figure}

The function $ \mathcal{F}$
then develops as a geometric series in \emph{rungs}. Calling $\mathcal{F}_n$ the ladder with $n$ ``rungs" we have
$ \mathcal{F} = \sum_{n \ge 0}  \mathcal{F}_n$, with $\mathcal{F}_0 = -G(\tau_{13})G(\tau_{24}) + G(\tau_{14})G(\tau_{23})$. The induction rule is
\begin{align}
\nonumber
\mathcal{F}_{n+1} (\tau_1,\tau_2, \tau_3, \tau_4) =  
\int d \tau d \tau^\prime K(\tau_1,\tau_2, \tau, \tau')  \mathcal{F}_{n} (\tau,\tau', \tau_3, \tau_4) \;.
\end{align}
The rung operator $K$, cf. Figure~\ref{fig:rung-intro}, adds one rung to the ladder. It acts on the space of ``bilocal" functions
with kernel
\be K(\tau_1,\tau_2, \tau_3, \tau_4) = -J^2(q-1)G(\tau_{13})G(\tau_{24})G(\tau_{34})^{q-2}\;.
\ee

The geometric series gives:
\be   \mathcal{F} = \sum_{n \ge 0}  \mathcal{F}_n = 
\sum_{n \ge 0} K^n \mathcal{F}_0 = \frac{1}{1-K} \mathcal{F}_0 
\ee
or, more explicitly 
\be
\mathcal{F}(\tau_1,\tau_2, \tau_3, \tau_4) = \int d \tau d \tau^\prime 
\frac{1}{1-K} (\tau_1,\tau_2, \tau, \tau^\prime)\mathcal{F}_0(\tau, \tau^\prime, \tau_3, \tau_4) .
\ee

A way to proceed is to diagonalize the rung operator $K$.
However if 1 is an eigenvalue of $K$, we will face a divergence and need to return to the full SD equations. 

Recalling the formula for the two point function in the approximate conformal (infrared) limit at zero temperature
$G_c ( \tau) = \frac{b }{\vert \tau \vert^{2 \Delta}} \sgn \tau $
with $ b^q J^2 \pi = \left(\frac{1}{2}  - \Delta\right) \tan (\pi \Delta)$
we find that in this limit the kernel $K$ becomes (after symmetrizing with respect to $(\tau_1,\tau_2) \leftrightarrow (\tau_3, \tau_4))$
\begin{gather}
K_c (\tau_1,\tau_2, \tau_3, \tau_4) = - \frac{1}{\alpha_0} 
\frac{\sgn (\tau_{13}) \sgn (\tau_{24}) }{\vert \tau_{13}\vert^{2 \Delta}
\vert \tau_{24}\vert^{2 \Delta} \vert \tau_{34}\vert^{2-4 \Delta}  }\;,\\
\alpha_0 = \frac{2 \pi q}{(q-1)(q-2) \tan (\pi / q)}\;.
\end{gather}

Conformal invariance allows us to simplify the problem by reexpressing $K$ as a function of the cross ratio
$\chi = \frac{\tau_{12} \tau_{34} }{\tau_{13}\tau_{24}}$
acting on single variable rung functions 
\be \mathcal{F}_{n+1} (\chi) = \int \frac{d \tilde \chi }{\tilde \chi ^2} K_c (\chi, \tilde \chi)  \mathcal{F}_{n} (\tilde \chi) .
\ee
To further simplify the diagonalization, it is important to find out operators commuting with $K_c$. 
Recalling the $SL(2,\mathbb{R})$ invariance of the two-point function, the associated Casimir operator $C  = \chi^2 (1 - \chi) \partial^2_\chi  - \chi^2 \partial_\chi$ is such an operator, with a known 
complete set of eigenvectors $\Psi_h (\chi)$ with eigenvalues $h(h-1)$. They are therefore 
also the eigenvectors of $K_c (\chi, \tilde \chi) $. The strategy to compute $\mathcal{F} $
can then be roughly summarized as

\begin{figure}
\centering
\includegraphics[width=10cm]{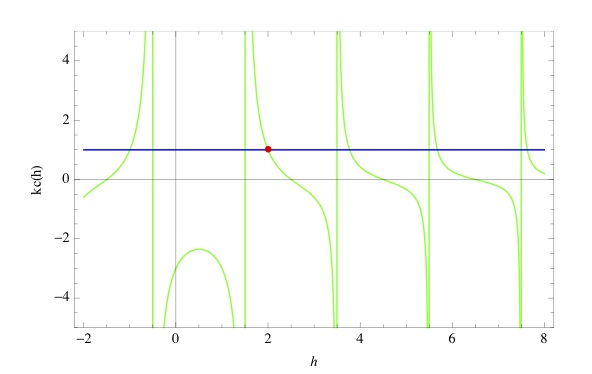}
\caption{\label{fig:spectrumSYK}The spectrum of bilinears of the SYK model and the $h=2$ mode, in red.}
\end{figure}

\begin{itemize}
\medskip\item

Find properties of $\mathcal{F}_n (\tilde \chi)$ and	
the eigenvectors $\Psi_h (\chi)$ of the Casimir operator $C$
with these properties.
\item
Deduce conditions on $h$. One finds two families,
$h=2n$ with $n \in {\mathbb N}^\star$ and
$h = \frac{1}{2} + is$, $s \in {\mathbb R}$
\item
Compute the eigenvalues $k_c (h)$ of the kernel $K_c$
and the inner  products $\expval{\Psi_h , \mathcal{F}_{0} }$ and 
$\expval{\Psi_h , \Psi_h }$.
\item
Conclude that the 4 point  function is
\be \mathcal{F} = \frac{1}{1-K} \mathcal{F}_0 = \sum_h \Psi_h (\chi )
\frac{1}{1 - k_c (h)} \frac{\expval{\Psi_h , \mathcal{F}_{0} }}{\expval{\Psi_h , \Psi_h }}\;.
\ee
\item
But... \emph{one finds a single $h=2$ mode with $k_c (h) =1$}, which requires a special desingularization.
\end{itemize}
For $h = \frac{1}{2} + is$ or $h = 2n$ one can compute, for $\chi<1$,
\be
\Psi_h = A \frac{\Gamma(h)^2}{\Gamma(2h)}\chi^h \tensor[_2]{F}{_1}(h,h,2h,\chi) + B \frac{\Gamma(1-h)^2}{\Gamma(2 - 2h)}\chi^h 
\tensor[_2]{F}{_1}(1-h,1-h,2-2h,\chi)\;,
\ee
with 
$A = \frac{1}{\tan\frac{\pi h}{2}}\frac{\tan \pi h}{2} $ and $B = -\tan \frac{\pi h}{2} \frac{\tan \pi h}{2}$. For $ \chi >1$
\begin{align}
\Psi_h =& \frac{\Gamma(\frac{1-h}{2})\Gamma(\frac{h}{2})}{\sqrt \pi}\tensor[_2]{F}{_1}\left(\frac{h}{2},\frac{h}{2},\frac{2 - 2h}{2},\left(\frac{2 - 2\chi}{\chi}\right)^2\right)\;,\\
\Psi_h =& \frac{1}{2}\int_{-\infty}^\infty d y \frac{\abs{\chi}^h}{\abs{y}^{h}\abs{\chi - y}^{h}\abs{1 - y}^{1- h}}\;.
\end{align}
The conformal spectrum of $K$ follows as:
\be
k_c(h)= -(q-1) \frac{\Gamma(\frac{3}{2} - \Delta) \Gamma(1 - \Delta)}{\Gamma(\frac{1}{2} + \Delta)\Gamma(\Delta)}\frac{\Gamma(\frac{h}{2}+ \Delta)}{ \Gamma(\frac{3 - h}{2} - \Delta)}\frac{\Gamma(\frac{1-h}{2} + \Delta)}{\Gamma(1 + \frac{h}{2} - \Delta)}\;,
\ee
and is depicted in Figure~\ref{fig:spectrumSYK}.

The conformal part of the four-point function
\begin{align}
\frac{\mathcal{F}_{h\neq 2}(\chi)}{\alpha_0} 
=& \int_0^\infty \frac{\dd s}{2\pi}\frac{2h - 1}{\pi \tan(\pi h)} \frac{k_c(h)}{1 - k_c(h)} \Psi_h(\chi) + \sum_{n\geq 2} \left[\frac{2h-1}{\pi^2}\frac{k_c(h)}{1 - k_c(h)}\Psi_h(\chi)\right]_{h = 2n}\nonumber\\ 
=& -\sum_{m\geq 0} Res \left[\frac{h-1/2}{\pi\tan(\pi h/2)}\frac{k_c(h)}{1 - k_c(h)}\Psi_h(\chi)\right]_{h = h_m}\nonumber\\
=& \sum_{m\geq 0} c_m^2 \left(\frac{N}{\alpha_0}\right)\left[\chi^{h_m} \tensor[_2]{F}{_1}(h_m,h_m,2h_m, \chi)\right],\\ 
c_m^2 =& \frac{- \alpha_0}{N}\frac{h_m - 1/2}{\pi \tan(\pi h_m/2)}\frac{\Gamma(h_m)^2}{\Gamma(2 h_m)}\frac{1}{-k'(h_m)},
\end{align}
is written as a sum over conformal blocks, indicating the spectrum of operators of the model. 
However this regular contribution is subdominant in the IR regime. 
A careful treatment of the dominant, although conformally 
divergent mode $h=2$ will show that the MSS bound is indeed saturated. 

\subsubsection{The $h=2$ Divergent Mode}

To treat this divergent mode,
we need to compute the deviations to conformal
invariance at least to first order in $1/\beta J$. Corrections to the rung operator eigenvalues can be obtained as in time-independent perturbation theory in quantum mechanics. Anticipating the analytic continuation, it is convenient to work on the thermal circle: $\theta = 2\pi \tau/\beta$. Then, varying the conformal SD equations, one finds that reparametrizations of the two-point function are $K_c$-eigenfunctions with eigenvalue $k(h)= 1$ and of proper conformal weight $h=2$. For linearized reparametrizations $\theta \rightarrow \theta + \epsilon(\theta)$, written as $\epsilon_n = e^{-in\theta}$, the reparametrization modes $\Psi_{h=2}$ mode break into an infinite family
\begin{gather}
\Psi_{2,n} = \gamma_n \frac{e^{-iny}}{2\sin \frac{x}{2}}f_n(x), 
\quad f_n(x) = \frac{\sin \left(\frac{nx}{2}\right)}{\tan \frac{x}{2}} - n \cos \left(\frac{nx}{2}\right), \\
x = \theta_{12},  \quad y = \frac{\theta_1 + \theta_2}{2},  \quad \gamma_n^2 = \frac{3}{\pi^2 \abs{n}(n^2 - 1)}.
\end{gather}
At large $q$, an analytic expression for the $K$-eigenfunctions and their eigenvalues can be found for all couplings. Selecting among the first those that lead to the above functions in the IR, the associated eigenvalues are then used to get the first IR-corrections to the $K$-eigenvalues 
\begin{equation}
k(2,n) =  1 - \frac{3\abs{n}}{\beta \mathcal{J}} + \frac{7n^2}{(\beta \mathcal{J})^2} + \mathcal{O}\left((\beta \mathcal{J})^{-2}\right).
\label{eq:correction_k}
\end{equation}
Supported by numerical solutions, \cite{Maldacena:2016hyu} extrapolated large $q$ results to estimate for finite $q$
\begin{equation}
k(2,n) = 1 - \frac{\alpha_K}{\beta \mathcal{J}}\abs{n} + \mathcal{O}\left((\beta \mathcal{J})^{-2}\right),  \qquad \alpha_K \equiv -q k^\prime(2) \alpha_G. 
\end{equation}
Plugging into ${\mathcal F}$ and taking down the extra two-point functions of the reparametrizations we get at leading order
\begin{align}
\frac{\mathcal{F}_{h=2} (\theta_1, \theta_2, \theta_3, \theta_4)}{G(\theta_{12} )G(\theta_{34}) } = \frac{6\alpha_0}{\pi^2 \alpha_K}\beta \mathcal{J} \sum_{\abs{n}\geq 2}
\frac{e^{in(y^\prime - y)}}{n^2(n^2 - 1)}&\left[ \frac{\sin \frac{nx}{2}}{\tan \frac{x}{2}} - n \cos \frac{nx}{2}\right]
\left[ \frac{\sin \frac{nx^\prime}{2}}{\tan \frac{x^\prime}{2}} - n \cos \frac{nx^\prime}{2}\right].
\end{align}

For $\theta_1 > \theta_3 >  \theta_2> \theta_4$ :
\be
\frac{\mathcal{F}_{h=2} (\theta_1, \theta_2, \theta_3, \theta_4)}{G(\theta_{12} )G(\theta_{34}) } =  \frac{6\alpha_0}{\pi^2 \alpha_K}\beta \mathcal{J} \left(\frac{\theta_{12}}{2\tan \frac{\theta_{12}}{2}} - 1 - \pi \frac{\sin \frac{\theta_1}{2}\sin \frac{\theta_2}{2}}{\abs{{\sin \frac{\theta_{12}}{2}}}}\right) ,
\ee
with $\theta_3 =\pi,\theta_4 =0$. For the ``regularized" OTOC \cite{Maldacena:2015waa}, $\theta_2 = \pi/2 - 2\pi i t/\beta = \theta_1 + \pi$
\be
\frac{\mathcal{F}_{h=2} (\theta_1, \theta_2, \theta_3, \theta_4)}{G(\theta_{12} )G(\theta_{34}) } 
= \frac{6\alpha_0}{\pi^2 \alpha_K}\beta \mathcal{J} 
\left(1 - \frac{\pi}{2}
\cosh \frac{2\pi t}{\beta}
\right),
\ee
hence $\lambda_L = 2 \pi / \b $. In units where $\hbar =1$ this is the \emph{MSS bound}.

Other approaches leading to the chaos exponent are worth mentioning. For instance, correlation functions of the Fermions can be obtained from the (non-local) effective action of the two-point function and self-energy. A simpler way, if one is only interested in the chaotic regime, is to look for eigenfunctions of the ladder kernel (using analytically continued propagators) with eigenvalue $1$ and exponential in time (see \cite{Maldacena:2016hyu}, or \cite{Murugan:2017eto} when applied to the two-dimensional, Bosonic and supersymmetric variant). 

This large $N$ computation allows to see the initial growth in time of the correlation function. However one needs to sum the full perturbative series in order to see it reaching the equilibrium values. This is done in \cite{Maldacena:2016upp} from the bulk point of view (where the $1/N$ expansion becomes a $G$ expansion).

At leading order, we can take the annealed partition function \cite{Gurau:2017xhf}, and integrating out the Fermions, we are left with an effective action for the two-point function $G$ and its Lagrange multiplier $\Sigma$: 
\be 
\f{I_{eff}[G,\Sigma]}{N} = -\f{1}{2}\log\det\left(\d(\t-\t^\prime) \partial_\t - \Sigma(\t,\t^\prime)\right) + \f{1}{2}\int_{\t,\t^\prime} \left(\Sigma(\t,\t^\prime) G(\t,\t^\prime) - \f{J^2}{4}G(\t,\t^\prime)^4\right). 
\ee
Note the bilocality of the right-hand side. Varying the action with respect to $G$ recovers the SD equation. Besides, looking fluctuations around the saddle-point $G_c$, $\Sigma_c$, leads to 
\begin{gather}
I_{eff} = I_{CFT} + I_S\;,\\
\f{I_{CFT}}{N} = -\f{1}{2}\log\det\left( - \Sigma(\t,\t^\prime)\right) + \f{1}{2}\int_{\t,\t^\prime} \left(\Sigma(\t,\t^\prime) G(\t,\t^\prime) - \f{J^2}{4}G(\t,\t^\prime)^4\right)\;,\\
\f{I_S}{N} = -\f{C}{J}\int^{\infty}_{-\infty}\dd \t Sch\left[f(\t),\t \right]\;, \quad Sch[f(\t),\t]=\f{f^{\prime\prime\prime}(\t)}{f^\prime(\t)}-\f{3}{2}\left(\f{f^{\prime\prime}(\t)}{f^\prime(\t)}\right)^2\;,
\end{gather}
where the constant $C$ in front is obtained from a numerical match \cite{Maldacena:2016hyu,Sarosi:2017ykf}. The key to understand the low-energy dynamics of near-extremal black holes is precisely the Schwarzian action of this reparameterization symmetry breaking $h=2$ mode, since in their near-horizon limit, they have an $AdS_2$ factor with a dilaton, the boundary action of which, under a suitable regularization, is precisely described by such Schwarzian (e.g. \cite{Maldacena:2016upp,Yang:2018gdb, Suh:2019uec} for a selection of works).

The leading order diagrammatic structure of higher $n$-point correlation functions has been worked through \cite{Gross:2017aos} and comes out as typical of the melonic theories we will detail below.

\subsubsection{More recent progress}

JT gravity was simple enough that it allowed to apply explicitly the prescriptions to compute the entanglement entropy of boundary subregions and extract the Page curve of two copies of SYK models coupled to a bath serving to absorb the emitted radiation \cite{Almheiri:2019yqk,Penington:2019kki}, with arguments that the lessons would apply also in higher dimensions \cite{Almheiri:2019psy}. 

Non-perturbative contributions to the bilocal effective action given by replica-symmetry breaking terms have been also considered \cite{Arefeva:2018vfp}. A very subtle point concerns the applicability of those lessons to quantum systems without disorder averaging, that the replica wormholes seemed to heavily rely upon, see especially the discussions in \cite{Penington:2019kki,Marolf:2020xie}.

In short, this toy model has galvanized intense activity in numerous directions, that either stiffen the holographic dictionary, study quantum chaos \cite{Kobrin:2020xms}, connect with random matrix integrals \cite{Stanford:2019vob}, quantum groups \cite{Berkooz:2018jqr}, or still return to its condensed matter roots as a simple model of strange metals \cite{Facoetti:2019rab} -- and often several at the same time. It has also seen proposals for experimental realizations \cite{Pikulin:2017mhj}. 

\section{Large $N$ limits of Tensor Models}
\label{section:largeNTM}

The idea to use tensors $T_{i_1\dots i_r}$ of rank $r\geq 3$, with invariance group of size $N$, as higher dimensional analogs of matrices was around since the 90's, coding for simplices that glued together would form $r$-dimensional manifolds \cite{Ambjorn:1990ge, Sasakura:1990fs, Boulatov:1992vp}. However too much advanced in their time, models with indistinguishable indices were considered, such that the geometries they were summing over were singular \cite{Gurau:2010nd}, allowing multiple and self gluings between simplices, spoiling a $1/N$ expansion akin to the topological expansion of matrix models. This technicality was overcome with the introduction of \emph{colored} models, where the indices stood distinguishable in contrast to their predecessors. They were first formulated in the language of Group Field Theory \cite{Gurau:2009tw} and subsequently, preserving the essential structure of the interactions, rewritten as \emph{tensor models}.

To start, their Feynman diagrammatic expansion provides a discretization of piecewise-linear manifolds of topological dimension $d\geq 3$, obtained by gluing simplices, \footnote{Hereby they are also linked to crystallization theory \cite{Casali:2017tfh}.} weighted by an action analogous to that of Regge that discretized the Einstein-Hilbert action \cite{Regge:1961px}. The recurrent themes are an expansion in integer powers of $N$ characterized by a half-integer, the Gurau degree, and their large-$N$ solvability due to the iterative structure of the leading diagrams based on ``melons". Beyond their roots in quantum gravity, their relevance to holography and black holes through the melons, tensor models present an unusual large-$N$ field theoretic flavor in view of the peculiar CFTs they possess as RG fixed points, with some remaining tough challenges, such as the nature of their subleading corrections, the difficulty to get around a branched polymer phase or the rapid growth of group invariants. We begin in Subsection~\ref{subsec:combi} by recalling the definitions of the main models, their different scalings, their combinatorial amplitudes and the revisited 2PI formalism. Subsection~\ref{subsec:1D} will adress the one dimensional tensor versions of the SYK model and Subsection~\ref{subsec:melonicCFTs} will discuss their higher dimensional cousins in search of properly tensorial CFTs. References for the first part are \cite{Gurau-book, Gurau:2019qag, Lionni:2017yvi,Valette:2019nzp}, whereas the last two can be supplemented with the set of lectures notes \cite{Klebanov:2018fzb,Gurau:2019qag,Benedetti:2020seh}. 

\subsection{Combinatorics and other tools}
\label{subsec:combi}

Many variations of the models have been studied, playing with their symmetries, their rank or their dimension and before surveying the main results and concerns, we will set the frame by detailing two different classes of tensor models with which the $1/N$ expansion started: colored and uncolored. 

\paragraph{Colored.} Colored models consist of $r+1$ tensors $T^{(j)}_{a^j_1\dots a^j_r}$ ($0\leq j\leq r$, $1\leq a^j_i\leq N$, $1\leq i\leq r$) of rank $r$, and their conjugates $\bar{T}^{(j)}_{a^j_1\dots a^j_r}$, indices of different position being assigned a different color. Given a pair of indices $(ij)$ ($i\neq j$), they transform respectively as
\begin{gather}
T^{(i)}_{a^i_1\dots a^i_j\dots a^i_r}\rightarrow T^{(i)}_{a^i_1\dots b^i_j\dots a^i_r} = O^{(ij)}_{b^i_ja^i_j}T^{(i)}_{a^i_1\dots a^i_j\dots a^i_r}\;,\\
T^{(j)}_{a^j_1\dots a^j_i\dots a^j_r}\rightarrow T^{(j)}_{a^j_1\dots b^j_i\dots a^j_r} = O^{(ij)}_{b^j_ia^j_i}T^{(j)}_{a^j_1\dots a^j_i\dots a^j_r}\;, 
\label{eq:transfo_tensor}
\end{gather}
with $O^{(ij)}\in U(N)$ and similarly for $\bar{T}$ with $O^\dagger$ instead of $O$.

The action is chosen as
\begin{gather}
\label{eq:colored}
S = N^{r/2}\left(\sum_{j=1}^{r+1}\sum_{A^j}T^{(j)}_{A^j}C^{-1}\bar{T}^{(j)}_{A^j} + \l I(T)+ \bar{\l}  I(\bar{T})\right),\\
I(T) = \prod_{i<j} \d_{a^i_ja^j_i}\prod_{j=1}^{r+1}T^{(j)}_{A^j}, 
\end{gather}
with invariance group $U(N)^{r(r+1)/2}$. We used the collective notation $A^j=(a^j_1\dots a^j_r)$. In the original zero-dimensional tensor models, the covariance was taken trivial $C=1$. 

The associated Feynman graphs are usually depicted in either the stranded or the colored representation. The first has a line propagator labeled by a pair ($ij$) denoting the color $i$ of the tensor and the position $j$ of the contracted index. It is a natural extension of the matrix ribbon graphs. As in lower ranks, the number of faces is identified by counting closed loops. However, diagrams become quickly cluttered.
In the colored representation, we draw a dot for each tensor (filled or hollow for $T^{(j)}$ or $\bar{T}^{(j)}$ interactions respectively), with $r+1$ colored half-legs attached for each tensor. Then we join the half legs that are Wick-contracted. Hence the interaction vertex is a canonically colored $r$-simplex, coloring the subsimplices according to the colors of their boundaries. Further, the Feynman diagram, an $(r+1)$-colored graph, becomes a gluing of simplices that respect the matching of colors. Note that the $U(N)$ symmetry constrains the Feynman diagrams to be bipartite, i.e. white vertices Wick-contracted only to black vertices and reciprocally. 

\begin{figure}[htbp]
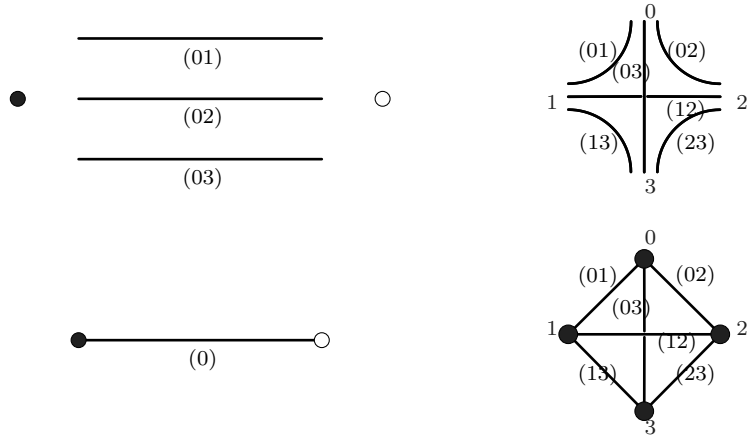

\centering
\captionsetup[subfigure]{labelformat=empty}
\subfloat[]{\tikzsetnextfilename{propa-intro2}
\input{propa-intro.tex}}
\hspace{1cm}
\subfloat[]{\tikzsetnextfilename{inter-intro2}
\input{inter-intro.tex}}
\caption{The stranded and the colored representations of the propagator $\expval{T^{(0)}\bar{T}^{(0)}}$ and of the tetrahedral interaction vertex.}
\label{fig:propa_inter}
\end{figure}

\begin{figure}[htbp]
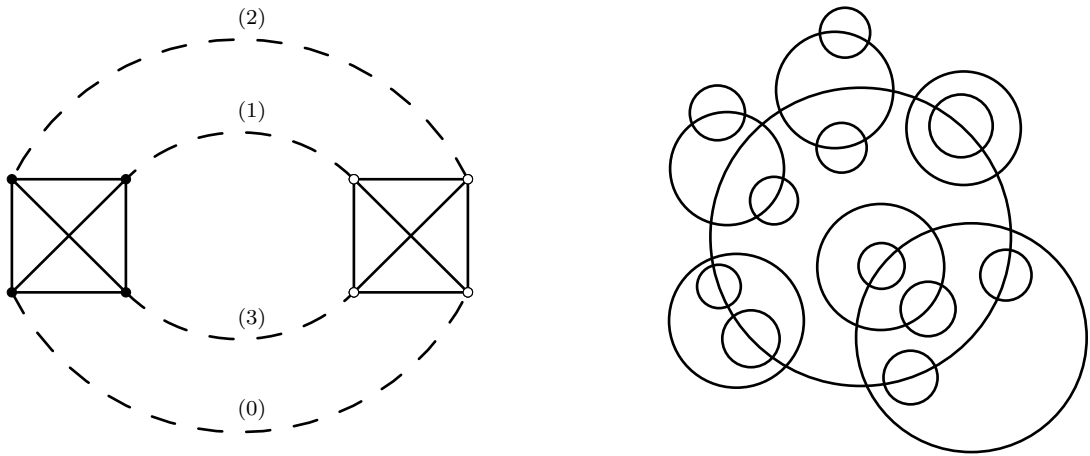

\centering
\captionsetup[subfigure]{labelformat=empty}
\subfloat[]{\tikzsetnextfilename{fundmelon2}
\input{fundmelon.tex}}
\hspace{1cm}
\subfloat[]{\tikzsetnextfilename{melons2}
\input{melons-intro.tex}}
\caption{Feynman diagrams of a vacuum correlation function (the fundamental melon on the left and a member of the family on the right).}
\label{fig:melons-intro}
\end{figure}

\vspace{5mm}
Let us evaluate the power of $N$ associated to a vacuum graph. Each propagator brings a factor $N^{-r/2}$ and each interaction vertex a factor $N^{r/2}$. Additionally, one needs to count the total number of faces $\Phi(G)$, that is closed bicolored cycles for all pairs of colors $(ij)$ labelling the strands, which bring each a factor of $N$. 
In total, an amplitude has a power: 
\be 
A(\cG) \sim N^{-\f{r}{2}(E(\cG) - V(\cG)) + \Phi(\cG)}. 
\ee 
It is interesting to see how from this exponent we come to the $1/N$ expansion. 
\begin{proposition} The number of faces can be written as:
\be 
\Phi(\cG) =r C(\cG) + \f{r(r-1)}{4}V(\cG)- \bar{\omega}(\cG),\quad \bar{\omega}(\cG)= \f{1}{2(r-1)!} \sum_{\cJ}k(\cJ)\;.
\label{eq:faces-colored}
\ee 
\end{proposition} 
Here $C(\cG)$ count the disconnected components of $\cG$ (in the sense of $(r+1)$-colored graphs) and $\cJ$ are the jackets, i.e. the set of permutations (modulo orientation and cyclicity) of $r+1$ colors attached at each vertex. Once a permutation is fixed, it allows an embedding of the Feynman graph on a surface to which a (non-orientable) genus $k(\cJ)$ can be now assigned. For any $(r+1)$-colored graph, there are $r!$ such jackets. $\bar{\omega}(\cG)$ is the (Gurau) degree of the graph and is positive being a sum of genera. Let us show how the relation between faces and degree~\eqref{eq:faces-colored} comes about. 

\proof Let us focus on a single connected component of $\cG$. Given a jacket $\cJ_\pi$ associated to a permutation $\pi$ of the colors $(0,\dots, r)$, Euler's formula tells that the genus of the jacket $k(\cJ_\pi)$ obeys
\be
V(\cG) - \f{r+1}{2} V(\cG) + \sum_{p=0}^r \Phi^{(\pi^p(0)\pi^{p+1}(0))}(\cG) = 2 - k(\cJ_\pi)\;, 
\ee
using for the second term the graph relation $2E(\cG) = (r+1) V(\cG)$ on the vacuum graph, $\Phi^{(ij)}$ standing for the faces of alternating edges $(ij)$ (and we sum over the ones selected by the jacket). However, each face of type $(ij)$ will appear in $2(r-1)!$ cycles: $(ij\dots)$ and $(ji\dots)$. Hence, summing over all the jackets both members of the equality, we obtain eq.~\eqref{eq:faces-colored}.\qed

\vspace{5mm}
Again from the relation $(r+1)V(\cG) = 2E(\cG)$, we find the large-$N$ expansion indexed by the degree
\be
A (\cG) \sim N^{r-\bar{\omega}(\cG)}.
\ee

It was shown \cite{Gurau:2010ba, Gurau:2011aq, Gurau:2011xq} that the leading order graphs, with $\bar{\omega} = 0$ have a recursive structure starting from a ring graph and iteratively inserting, on every edge, two vertices connected by $r$ propagators. The first insertion with two vertices, the fundamental melon, stands on the left of Fig.~\ref{fig:melons-intro} -- looking like a Cantaloup melon, or a bunch of ``bananas'' according to the amplitude community \cite{Broedel:2019kmn}). All together they form the melonic family, a member of which is shown on the right of Fig.~\ref{fig:melons-intro}. Notice that the interaction vertices have been reduced to a point and that the edges\footnote{Corresponding to the dashed edges of the left drawing.} are assumed to respect the color labeling.
The melons are planar for every jacket, hence can be embedded on the sphere $S^r$.

Correlation functions of fundamental fields are obtained from successively opening edges. For instance, leading order two-point functions come by from cutting a propagator of a vacuum graph. The full two-point function is determined by the usual Schwinger-Dyson equation in terms of  the self-energy $\Sigma$, at leading order taking a simple form
\be 
\label{eq:melonic_colored}
G^{-1} = C^{-1} - \Sigma\;, \quad \Sigma = \l \bar{\l}G^{r-1} + \cO(1/N). 
\ee 
We will see below how solving this equation provides information on the geometry lurking behind.

\begin{figure}
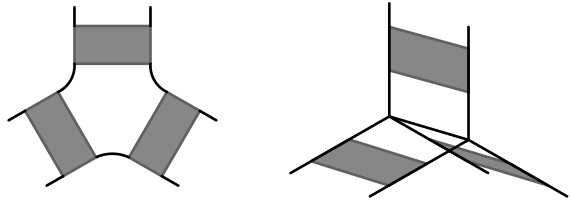

\centering
\captionsetup[subfigure]{labelformat=empty}
\subfloat[]{\tikzsetnextfilename{6ptladderintro13}
\input{6ptladderintro1.tex}}
\subfloat[]{\tikzsetnextfilename{6ptladderintro12}
\input{6ptladderintro2.tex}}
\caption{Leading order six-point functions ($r=3$).}
\label{fig:6-pt-ladder-intro}
\end{figure}

Regarding higher-point correlation functions, since we will not be concerned with colored models in later chapters, let us point that dominating four-point functions have a more varied structure, function of the color pattern of external legs \cite{Gurau:2016lzk} (see also \cite{Bonzom:2017pqs} for explicit drawings and an extensive analysis). But their essential ingredient is the earlier met ladder, cf. Fig.~\ref{fig:four-point-ladder-intro}, a melonic graph with two opened propagators. To keep the discussion simple, we assume that the external fields have all the same color. 
Here, an important result is in order. If one takes out all melons and reduces all insertions of ladders to single vertices, It was shown in \cite{GurauSchaeffer} that for each order in $N$, there is only a finite number of possible graphs, the schemes, connecting those vertices together that contribute.
For instance, with six-point functions, there are two ways to glue chains of ladders at leading order, drawn in Fig.~\ref{fig:6-pt-ladder-intro}.

\begin{figure}[htbp]
\centering
\captionsetup[subfigure]{labelformat=empty}
\subfloat[]{\tikzsetnextfilename{quarticbubble-intro2}
\input{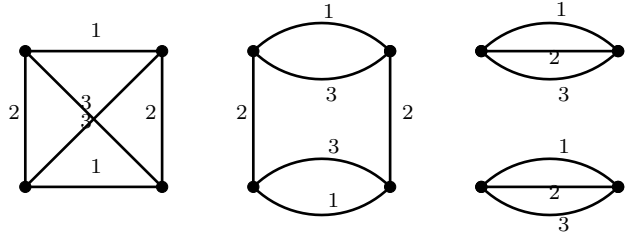}}
\caption{All quartic bubbles of the rank 3 $O(N)^3$ tensor model. From left to right: tetrahedron, pillow (up to color permutation), double-trace.}
\label{fig:quarticbubbles}
\end{figure}

\paragraph{Uncolored.} Uncolored models have gathered much attraction lately since, compared to the colored ones, they cope with less constraints. They are invariant under the tensor product representation of a group for each index of a single tensor. A restriction to $U(N)^r$ symmetry was first obtained in \cite{Bonzom:2012hw}, taking a rank $r$ tensor $T$ and its conjugate $\bar{T}$, that transform with a distinct group $U(N)$ per index: 
\be 
T_{a_1\dots a_i\dots a_r}\rightarrow O^{(i)}_{a_ia_j}T_{a_1\dots a_j\dots a_r} \quad \bar{T}_{a_1\dots a_i\dots a_r}\rightarrow O^{(i)*}_{a_ja_i}\bar{T}_{a_1\dots a_j\dots a_r}\;.
\ee 
Generic observables of the theory are tensor invariants or \emph{bubbles}, $\cB(T,\bar{T})$, where lower indices of the tensors are contracted pairwise forming $r$-regular graphs, not necessarily connected as show the real quartic examples of Fig.~\ref{fig:quarticbubbles}. They can be given different roles inside correlation functions. Either, written as interactions, they come down in the perturbative expansion, or they can be taken as an external operator, or still, they can appear as the boundary graph of a specific Wick-contraction, once one keeps only the external tensors and contracts their indices according to the Wick contraction pattern (see Fig.~\ref{fig:4colored} for an example at rank 3 that combines two sextic bubbles to form a different sextic operator). 

We will write their action as
\be 
\label{eq:uncol-action}
S = T \bar{T} + N^{s(\cB)} t_\cB \cB(T,\bar{T}), \quad s(\cB) = -\f{2}{(r-2)!}\omega(\cB),
\ee 
$\omega(\cB)$ denoting again the degree of the bubble. As usual, correlation functions arise from Taylor expanding the interactions and Wick contracting (the propagator adding a color $0$ to each node), in effect vacuum graphs drawn as $(r+1)$-regular graphs.\footnote{Computing the generating function of 1PI graphs, one needs to sum over connected graphs when the interaction vertices are constitute the vertices of the graph and the propagators its edges. But seen as a colored graph, expanding the vertices back, it may look disconnected if the interaction bubble is disconnected.}
The scaling with $N$ of the amplitude of a diagram $\cG$ is given by 
\be
A(\cG) \sim N^{\sum_{\cB\in\cG} s(\cB) + \sum_i\Phi^{(0i)}(\cG)}\;.
\ee 
Note that, compared to the colored models, only the bicolored cycles $(0i)$ for $i = 1,\dots,r$ bring factors of $N$ by closed loops. However we can view the graph as colored and compute the total number of faces 
\be \Phi(\cG) = \sum_{1\leq i\leq r} \Phi^{(0i)}(\cG) + \sum_{1\leq i<j\leq r}\Phi^{(ij)}(\cG)\;,
\ee
where the last bicolored faces occur only in the interaction bubbles. Then, using the degree relation~\eqref{eq:faces-colored} for the left-hand side ($r+1$ colors) and the last term of the right-hand side ($r$ colors), we can rewrite the amplitude as
\be
 A(\cG) \sim N^{\sum_{\cB\in\cG} s(\cB) - r \left[1 + \sum_{\cB\in \cG}(C(\cB) - 1)- C(\cG)\right] -\left[\bar{\omega}(\cG)- \f{r}{r-1}\bar{\omega}(\cG^0)\right]}\;.
\ee
Denoting $\cG^0$ the graph where the propagators (the edges of color 0) have been removed, we have that the existence of a large-$N$ expansion relies on the two inequalities:\footnote{We note that the degree or genus of disconnected components is the sum of that of the different components.}
\begin{gather}
\bar{\omega}(\cG)\geq \f{r}{r-1}\bar{\omega}(\cG^0),\\
1 + \sum_{\cB\in \cG^{0}}(C(\cB) - 1)\geq C(\cG)\;.
\end{gather}
With the jackets of $\cG^0$ associated to permutations $\pi^0$, the first is equivalent to 
\be 
\sum_{\pi}k(\cJ_\pi)\geq r \sum_{\pi^0}k(\cJ_{\pi^0})\;. 
\ee
Indeed, we obtain a jacket of $\cG$ from a jacket of $\cG^0$ by adding the edges of color $0$ and there are $r$ possible ways to insert them around the vertices. Furthermore, the genus can only increase by adding an edge\footnote{When removing an edge, either the graph stays connected and the number of faces can only decrease, or the graph disconnects and the total number of faces is unchanged.} such that $k(\cJ_\pi)\geq k(\cJ_{\pi^0})$ and one can conclude. 

The second inequality holds for graphs $\cG$ that are connected when considering the interaction bubbles as reduced to a vertex (i-connected in the language of \cite{Benedetti:2020yvb}), which occur in the logarithm of the partition function. Then the left-hand side counts the edges of a tree that joins the interaction bubbles (forming a minimal i-connected graph), this way linking some of the connected components of each interaction, plus the left alone disconnected components of each interaction bubble. This gives indeed the maximal number of possible disconnected components of an i-connected graph $\cG$. 

\vspace{5mm}
Once again, the leading order graphs appearing in the partition function displayed the same melonic structure as above, \footnote{Remember that in the colored representation of the colored model, the vertices of the melonic diagrams corresponded to interaction vertices, whereas now they are associated to the individual tensors.} 
and melonic bubbles dominated in all correlation functions: 
\be 
\label{eq:factorization}
\lim_{N\to \infty}\f{1}{N}\expval{\cB(T,\bar{T})} =\begin{cases} G^{V(\cB)} & \text{(if }\cB \text{ is melonic)}\\ 0 & \text{(else),}\end{cases}
\ee 
Wick contractions preserving the melonic structure of the Feynman diagrams, $G$ solving the same equation~\eqref{eq:melonic_colored}.
The fact that at leading order expectation values of several melonic bubbles disconnect indicates a large-$N$ factorization property, providing the tensor models with a universality class of Gaussian character \cite{Gurau:2011kk}. 
This contrasts strikingly with matrix models, for which the large-$N$ planar diagrams entail many more contraction patterns.

\subsubsection{The melonic universality class}
From another perspective, an analysis of the Schwinger-Dyson equation of the two-point function brings about the distinctive continuum limit of the melonic universality class \cite{Bonzom:2012cu}.\footnote{A similar discussion can be done with the free-energy, since the two-point function can be obtained from selecting a particular edge of a vacuum graph.} 
In the 0-dimensional case, the melonic equation \eqref{eq:melonic_colored} can actually be solved exactly by Fuss-Catalan numbers: 
\be
G = \sum_{n\geq 0} \f{1}{rn+1}{rn+1 \choose n}\left(\f{\l\bar{\l}}{(r-1)!}C^r\right)^n C\;.
\ee
One can show by recursion that edge-colored rooted $r$-ary trees are counted by the same numbers, establishing in this way a bijection between the latter and melonic diagrams \cite{Bonzom:2011zz,Bonzom:2012zf}. A detailed computation of the Hausdorff and spectral dimension of the geometry associated with melonic diagrams, 2 and $4/3$ respectively \cite{Gurau:2013cbh}, cemented their connection to branched polymers, recurrent in euclidean and causal dynamical triangulations as seen in Section~\ref{sec:QG}.

Given the series expansion of the leading order two-point in terms of the coupling $\l\lb=z$, we can look for their radius of convergence $\abs{z_c} = \f{r^r}{(r+ 1)^{r+1}}$ and the way they approach the singularity
\be 
G (z) = \sum_{n\geq 0} c_n z^n \sim (z - z_c)^{1-\g}.
\ee 
As realized in the matrix models, Subsection~\ref{subsec:2dQG}, the (string) susceptibility $\g$ controlling the non-analytic behaviour
is a critical quantity that gives information about the continuum limit of the theory. In this limit, the coupling is tuned to approach the large $N$ singularity in a way that an infinite number of interaction vertices contribute and at the same time that the graph dual of the Feynman diagrams have a finite volume. The exponent can then be extracted from the asymptotics of the series expansion of $G$ (see for example \cite{Lionni:2017xvn})
\be 
c_n \sim \alpha z_{c}^{-n}n^{\g-2}.
\ee 
While planar matrix models have $\g = -\f{1}{2}$, associated to the Brownian sphere, the melonic limit of tensor models possess $\g = \f{1}{2}$ , characteristic of branched polymers or the continuum random tree. 
As for vector and matrix models, multicritical points were also obtained \cite{Bonzom:2012hw} with susceptibilities $\g = m/(m+1)$ matching those of branched polymers.
They were given a physical interpretation in terms of hard dimers on random lattices, tuning activities appropriately \cite{Bonzom:2012np}.
In the quest to generate higher-dimensional manifolds from a large-$N$ limit, it is thus imperative to escape the melonic limit. 

\vspace{5mm}
Stepping in the footprints of the rank two predecessor \cite{DiFrancesco:1993cyw}, an idea was to consider a \emph{double scaling limit}, where additional terms from all subleading orders are taken into account by tuning the approach to the critical value of the coupling with a dependence in $N$. Applied to matrix models, this boosts contributions of all genera and although it can be analyzed using orthogonal polynomials \cite{bleher2002double}, the factorial growth of the relevant graphs makes the partition function non-summable.
In tensor models of rank strictly less than 6, with melonic interactions, a double-scaling limit was seen to retain the same non-analytic behaviour corresponding to branched polymers \cite{Bonzom:2014oua}. The graphs contributing in the limit were mapped to trees with the end leaves decorated by loops, an exponentially bounded family. To the contrary, in higher rank than 6, all 3-valent graphs were contributing at the critical point, hence non-summable. 

\vspace{5mm}
A more drastic option is to search for \emph{optimal scalings} of the interactions, or enhancements with respect to the scaling~\eqref{eq:uncol-action}, so that a large-$N$ limit still exists, while making the dominant class of graphs larger than the melonic family \cite{Bonzom:2016dwy}. If further boosted, the associated interaction would contribute in amplitudes with arbitrary high powers of $N$. A necessary criterion for optimality is that the interaction appears infinitely often at any order in $N$. 
Against all odds, the melonic universality class is a pretty strong attractor.

Since melonic bubbles already appear at leading order, Gurau's degree is the \emph{only} scaling that allows for a non-trivial large-$N$ limit as far as they are concerned. If other interactions are also rescaled by their Gurau degree, we had that the large-$N$ limit remains Gaussian since melonic bubbles dominate the Feynman diagrams. 

\begin{figure}[htbp]
\centering
\tikzsetnextfilename{bidipolesintro2}
\input{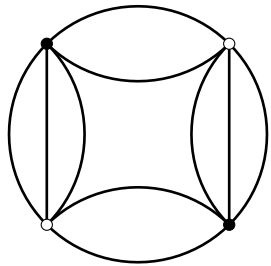}
\caption{A quartic bubble of a rank $5$ tensor with two 2-bidipole insertions.}
\label{fig:bidipole}
\end{figure}

This uniqueness has been extended to \emph{generalized melonic} interactions \cite{Bonzom:2019kxi}, built from insertions of $C$-bidipoles, with $C\leq r/2$. An example of $2$-bidipole for a rank 5 tensor is given in Fig.~\ref{fig:bidipole}. The unique choice of enhancement is:
\be
s(\cB)=\sum_C\abs{C}b_C - \f{r(V(\cB) - 2)}{2}\;,
\ee
where $b_C$ counts the number of $C$-bidipole insertions to obtain the bubble $\cB$.

If the action doesn't include bubbles with the symmetric $(r/2)$-dipole insertions, then the leading order theory is Gaussian. Otherwise, planar diagrams are generated and the critical exponent of the Brownian map can be obtained. An intermediate critical phase of ``baby universes" with $\g = 1/3$, also present in matrix models e.g. \cite{Das:1990, AlvarezGaume:1992np}, could be obtained with particular adjustements of the couplings, the melonic interactions balancing the planar bubbles (see also \cite{Bonzom:2015axa, Lionni:2017xvn} for the first tensor models to observe it).

Melonic and generalized melonic interactions have the convenient feature to be equivalent to (multi-)matrix models through an intermediate field reformulation, which will serve in the next chapter. This formalism is at the heart of a beautiful bijection between edge colored graphs and particular combinatorial (stuffed Walsh) maps, simplifying the enumeration of graphs at any order \cite{Bonzom:2015kzh, Bonzom:2019kxi}.

\vspace{5mm}
A different extension of the melonic family to encompass \emph{generalized melonic} Feynman diagrams has been obtained as follows \cite{Carrozza:2015adg,Ferrari:2017jgw}, see also \cite{Benedetti:2020yvb}. For rank $r$ tensors, the action takes the form:
\begin{gather}
S_N(T) = N^{r/2}\left(TC^{-1}T + \sum_\cB t_\cB N^{-\rho(\cB)}I_\cB\right), \quad \rho(\cB) = \f{F_{\cB}}{r-1} - \f{r}{2},
\end{gather}
$F_b$ counting the number of faces (bicolored cycles) of the bubble $b$.

\begin{figure}
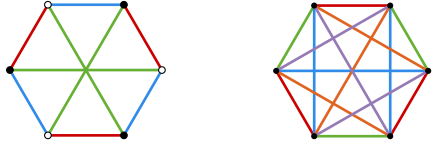

\centering
\captionsetup[subfigure]{labelformat=empty}
\subfloat[]{\tikzsetnextfilename{wheel-intro3}
\input{wheel-intro.tex}}
\hspace{1cm}
\subfloat[]{\tikzsetnextfilename{5simplex-intro2}
\input{5simplex-intro.tex}}
\caption{Two MST interactions of order 6, with complex tensors of rank 3 and real ones of rank 5.}
\label{fig:MSTbubbles}
\end{figure}

If the tensor is of prime rank and the interaction bubble $\cB^*$ is a complete graph, \footnote{Enumerations of possible complete interactions up to rank 13 were studied in \cite{Gubser:2018yec}.} 
it was proven \cite{Ferrari:2017jgw} in the framework of matrix-tensor models that the enhanced scaling $\rho(\cB^*)$ is optimal. In an expansion characterized by an integer generalizing the degree, the \emph{index}, the dominant graphs were melons formed by mirrored pairs of $\cB^*$ and the usual melons constructing from melonic bubbles. Applied for example to $O(N)^3$ quartic models with tetrahedron, pillow and double-trace interactions, with respective scalings $\rho(\cB)=0,1/2,3/2$ that allows the tetrahedron to enter the large-$N$ expansion, always in pairs. Other examples with bubbles of order 6 will be looked at in Chapter~\ref{ch:sextic} with the MST interactions in figure~\ref{fig:MSTbubbles}. 

Regarding tensors of any odd rank, this scaling is optimal for a particular type of invariants, called maximally-single-trace (MST) which similarly form generalized melonic diagrams from pairwise contractions.\footnote{However listing all the leading order graphs in this case is a harder problem.} They possess the minimal number of faces allowed, that is one face for each two colors. Furthermore, their connected correlations also obey the bound \cite{Benedetti:2020yvb}
\be 
\expval{\cB_1\dots \cB_n}_c \leq N^{r-rn/2},
\ee 
which applied on a generic correlation function of MSTs gives at leading order
\be 
\expval{\cB_1\dots \cB_n} =\sum_{P} \prod_{B\in P} \expval{\prod_{j\in B}\cB_j}_c,
\ee 
where we sum over all partitions $P$ into blocks $B$ of the $n$ bubbles, will make dominate correlations of pairs of MSTs if we assume that the one-point functions vanish. This continues the Gaussian large-$N$ factorization we saw earlier~\eqref{eq:factorization}.

\vspace{5mm}
Other symmetry groups were also explored for uncolored models, each allowing for a different class of interactions. For instance, multi-orientable models $U(N)\times O(N)\times U(N)$ \cite{Tanasa:2011ur,Tanasa:2015uhr}, symplectic groups $Sp(N)^3$ \cite{Carrozza:2018psc}, or in the much harder case of reduced symmetry, two coupled symmetric rank 3 tensors \cite{Gurau:2017qya} and a single tensor in the three irreducible representations of $O(N)$ \cite{Benedetti:2017qxl} were all shown to display melonic limits. With motivations that spring from the original string theoretic matrix models and the large dimension limit of general relativity, tensors with symmetry groups of different dimension $O(D)\times U(N)\times U(N)$ (and their real counterpart) enter the melonic universality class, in a large $N$ followed by a large $D$ limit, that extracts the summable class from the planar diagrams \cite{Ferrari:2017ryl}. When discussing SYK-like models, we will touch upon their uncommon low-energy physics. 

\vspace{5mm}
Another attempt to overcome the melonic barrier went by explicitly breaking the color-symmetry of tensors. In a quartic theory with a single color pillow interaction, \cite{Benedetti:2015ara} found, from the matrix-like intermediate field theory, a regime where the $O(N)$ got spontaneously broken. Considering then a double-scaling limit in the $O(N)$ symmetric or broken phases, the former had at leading order the summable 3-valent planar graphs, symptomatic of a Brownian map scaling limit, whereas the later was composed of a tree of planar 3-valent graphs. We shall see in Chapter~\ref{ch:TGN} that the color symmetry may also break spontaneously (breaking the continuous symmetry at the same time).

\subsubsection{The 2PI formalism} 
To deal with effective actions of tensor models, a formalism mostly used in many-body theory has been revisited \cite{Benedetti:2018goh, Gurau:2019qag}. The 1PI effective action Legendre transforms the Lagrangian of the generating function of connected correlations, with respect to a source for one-point function. The 2PI effective action takes a Legendre transform with respect to a bilinear in the fields. 

Given the generating function of connected correlations $W[J,\cJ]$
\be
W[J,\cJ] =\ln \int \mathcal{D}\phi \exp \left(-S[\phi] +J_a \phi_a +  \frac{1}{2}\phi_a \cJ_{ab} \phi_b\right),
\ee
in the compact DeWitt notation where indices stand for spacetime dependence and all other internal degrees of freedom, we have its Legendre transform\footnote{We use the functional derivative with symmetric projector $\f{\d\cJ_{ab}}{\d\cJ_{cd}} =\f{1}{2} \left(\d_{ac}\d_{bd} + \d_{ad}\d_{bc}\right) =: \cS_{ab;cd}$.} $\G[\Phi,G]$
\be
\label{eq:effaction}
\Gamma[\Phi,G]=  - W[J,\cJ] + J_a \Phi_a + \frac{1}{2} \Phi_a \cJ_{ab} \Phi_b+\frac{1}{2}G_{ab}\cJ_{ab}.
\ee
In the middle and right equation, $(J,\cJ)$ are implicit functions of $(\Phi,G)$ obtained by inverting the relations that define the connected one- and two-point functions:
\begin{align}
\label{eq:sources}
\Phi_a[J,\cJ] = \fdv{W}{J_a}[J,\cJ], \quad G_{ab}[J,\cJ] = \frac{\delta^2 W}{\delta J_a \delta J_b}[J,\cJ] = 2\f{\d W}{\d \cJ_{ab}} - \f{\d W}{\d J_a} \f{\d W}{\d J_b}.
\end{align}
and others analogous with $\Gamma$ exchanging sources with fields, assumed invertible for $(J,\cJ)$ and $(\Phi,G)$ respective inverse, allow to go from one set of variables to the other.
Denoting the free propagator $C^{-1} = \fdv[2]{S}{\phi}$, we obtain in a loop expansion (expanding the field $\phi$ around its expectation value $\phi = \Phi + f$)
\begin{align}
\label{eq:2PI}
\Gamma[\Phi,G]= S[\Phi] + \frac{1}{2} \Tr[\ln G^{-1}] + \frac{1}{2} \Tr[C^{-1} G] +  \Gamma^{2PI}[\Phi,G].
\end{align}
The first terms in the last equality come from quadratic terms in $f$. $\Gamma^{2PI}$, containing the higher orders, is (minus) the generating function of the 2PI graphs, i.e. graphs that stay connected cutting 2 lines, drawn with with propagators $G^{-1}$ and interaction vertices of $S_{int}$. The equations of motion for $\Phi$ and $G$ obtained by varying the effective action can be studied in a $1/N$ expansion keeping the leading order graphs in $\Gamma^{2PI}$.  The source $J$ is typically introduced to study symmetry-breaking phases and since we will assume that the symmetry is preserved ($\Phi = 0$), we let $J=0$ in the following\footnote{The classical solution $\Phi$ should obey the equations of motion derived from the action, and the equations derived from the effective action should be consistent. However, non-trivial solutions that preserve a smaller symmetry group ($SO(3)$ invariant instead of $O(N^3)$) also exist \cite{Benedetti:2019sop}, with evidence for their stability. They also possess Feynman diagrams dominated by melonic graphs with insertions of background field and they have a similar 2PI effective action to that of the CTKT model we will encounter shortly.}.

We recover the self-energy $\Sigma$ (amputated 1PI two-point function) by differentiation and analogously introduce the irreducible four-point kernel $K$: 
\be 
\label{eq:irreducible}
\Sigma_{ab} = -2 \f{\d \Gamma^{2PI}}{\d G_{ab}}, \quad K_{ab;cd} = G_{aa^\prime}G_{bb^\prime}\f{\d \Sigma_{cd}}{\d G_{a^\prime b^\prime}}.
\ee 
Its interest lies in that the four-point function is expressed from the kernel $K$. Indeed the four-point function in the channel $(ab\rightarrow cd)$ is 
\be 
\expval{\phi_a\phi_b\phi_c\phi_d}_{ab;cd} := 4\f{\d^2 W}{\d \cJ_{ab}\d \cJ_{cd}} = \expval{\phi_a\phi_b\phi_c\phi_d} - \expval{\phi_a\phi_b}\expval{\phi_c\phi_d} .
\ee  
Besides we have from eq.~\eqref{eq:effaction} and \eqref{eq:sources}: 
\be
\f{\d^2 \G}{\d G_{ab}\d G_{cd}} = \f{1}{2}\f{\d \cJ_{cd}}{\d G_{ab}} = \f{1}{4}\left(\f{\d^2W}{\d \cJ_{ab}\d \cJ_{cd}}\right)^{-1},
\ee
while from \eqref{eq:2PI} and \eqref{eq:irreducible}
\be
\f{\d^2 \G}{\d G_{ab}\d G_{cd}} = \f{1}{2}G^{-1}_{aa^\prime}G^{-1}_{bb^\prime}\left(\cS -K\right)_{a^\prime b^\prime;cd}.
\ee
Combined, the last equations mean, with all the effect of the projector $\cS$ moved onto the numetor: 
\be 
\expval{\phi_a\phi_b\phi_c\phi_d}_{ab;cd}= \left(\frac{1}{1 - K}\right)_{ab;c^\prime d^\prime}\left(G_{c^\prime c}G_{d^\prime d}+G_{c^\prime d}G_{d^\prime c}\right).
\ee
This relation is generic, but the luxury of tensor models, and by extent of melonic theories, stands upon the fact that we control the truncation of the kernel $K$ at large-$N$. 

Taking advantage of the $1/N$ expansion, the restriction to 2PI graphs in tensor models allows to write an effective action. In particular, melonic insertions are not allowed as they would form 2PR graphs.

\paragraph{Renormalizing in tensor theory space.} Before we jump to tensor field theories, let us say a few words on results that have been obtained for RG flows when the tensor indices serve to discretize the space. The locality principle in usual renormalization is replaced by an invariance principle of the interactions under a certain symmetry group and a non-trivial propagator in index space will break this symmetry, such that the IR direction corresponds to smaller and smaller values of indices. This point of view was initiated with matrices \cite{Brezin:1992yc}, and other variations, including the Grosse-Wulkenhaar model, shown asymptotically safe at one loop (e.g. \cite{Wulkenhaar:2014sqh}). Using a multiscale analysis \cite{Geloun:2016bhh} or functional RG methods \cite{Carrozza:2016vsq, Eichhorn:2018phj} a plethora of models have been analysed: with rescaled momenta $p^{2\a}$ ($\a\in (0,1]$), with an extra $U(1)$ or $SU(2)$ gauge invariance, etc.
Constructive results have also been obtained, relying on the (multiscale) Loop-Vertex-Expansion, ending up proving convergence and analyticity of free energy in ``cardioid" domains for theories with melonic interactions (quartic in rank 3 \cite{Delepouve:2014hfa} and 4 \cite{Rivasseau:2017xbk}).

\subsection{SYK-like models}
\label{subsec:1D}
The melonic structure of leading order diagrams in the two- and four-point functions allowed to write quickly colored and uncolored Fermionic tensor models enjoying a low-energy behaviour similar to SYK and saturating the chaos bound. 

In the colored case, with $r+1$ Majorana fields $\psi^{(0)},\dots ,\psi^{(r)}$, transforming as in \eqref{eq:transfo_tensor}, the action of the Gurau-Witten (GW) model, invariant under $O(N)^{r(r+1)/2}$ is taken as \cite{Witten:2016iux}
\be
S = \int \dd t \left(\f{i}{2}\sum_{k=0}^r \psi^{(k)}_{A^k}\partial_t\psi^{(k)}_{A^k} - i^{(r+1)/2}\f{J}{N^{r(r-1)/4}}\psi^{(0)}\dots \psi^{(r)}\right)
\ee
(the $i$ in the interaction makes the term hermitian) where the same notation as in eq.~\eqref{eq:colored}.
Based on the preceding analysis of \cite{GurauSchaeffer}, \cite{Gurau:2016lzk} identified the structure of two- and four-point correlation functions, respectively in terms of melons, and unbroken or broken ladders. 

An uncolored version, nicknamed Carrozza-Tanasa-Klebanov-Tarnopolsky (CTKT), was written in \cite{Klebanov:2016xxf}, using a rank-3 tensor $\psi_{abc}$: 
\be 
S = \int \dd t\left(\f{i}{2}\psi_{abc}\partial_t\psi_{abc}+\f{g}{4}\psi_{a_1b_1c_1}\psi_{a_1b_2c_2}\psi_{a_2b_1c_2}\psi_{a_2b_2c_1}\right),
\ee 
where the invariance subgroup is $O(N)^3$.
Naturally, both colored and uncolored models present the same leading order melonic SD equations that we have just discussed in Subsection~\ref{subsec:combi} and in the SYK context.

The diagonalization of the four-point ladder kernel could be reproduced to determine the conformal spectrum of bilinear operators in the (broken) CFT characterizing the low-energy theory. In particular, the spectrum contained the mode dual to the stress-energy tensor, signalling the dynamical gravitational dual and the saturation of chaos. 

Extending the results of \cite{GurauSchaeffer} on the next-to-leading orders of colored tensor models, \cite{Bonzom:2017pqs} characterized in detail the structure of the ladders interlaced in the four-point function for a colored version of the SYK model and all tensor SYK-like models discussed here.  Although in both cases the diagrams involved are similar, since their faces are counted differently, beyond the leading order similar diagrams contribute at different orders in $1/N$: for tensors all bicolored cycles are counted, whereas the vector case considers only $0i$ faces. In a sense, the tensors were breaking the degeneracy of graphs of SYK. 

As we mentioned earlier, the bilocal effective action was crucial in the SYK model to understand the departure from conformal invariance leading to quantum chaos. Using the formalism of the 2PI effective action, \cite{Benedetti:2018goh} derived a similar bilocal action for the above tensor models. Applying this formalism first to SYK, \cite{Benedetti:2018goh} recovered at leading order in $1/N$ the SD equations of SYK and at next-to-leading order a $\Tr\log$ term that could be interpreted as arising from integrating out a quadratic field with the same covariance as in \cite{Jevicki, Kitaev:2017awl}.\footnote{However, it is inaccurate to extend this analysis at following orders: disorder averaging before or after setting the theory on shell leads to different results, in the same way that at subleading orders, the different replicas interact and a replica-diagonal ansatz is not enough \cite{Arefeva:2018vfp}. }
For the CTKT model, they rewrote the effective action for the two-point function as follows: \footnote{Here the trace $\Tr$ is tracing on, or picturally closing the strand of, every color separately and integrating over the two times.}
\begin{gather}
\G_{CTKT}[G] = -\f{1}{2}\Tr[\ln G^{-1} (t,t^\prime)] -\f{1}{2}\Tr [\partial_t G(t,t^\prime)] -\f{1}{8}\l^2\int_{t,t^\prime} G^4(t,t^\prime),\\
\f{1}{2}\Tr \partial_t G(t,t^\prime) = -\f{\alpha}{2}\int \dd t \partial_tH\partial_tH + \cO(H^3).
\end{gather}
The first and third terms are gauge invariant under $O(N)^3$, \footnote{If the Gaussian term in the action is discarded, the global $O(N)^3$ symmetry becomes trivially local.} and give the melonic SD equations in the IR limit. The second term brings corrections taking the form of a non-linear $\sigma$-model for the generators $H$ of the $O(N)^3$ symmetry locally broken by the second term. $\alpha$ stands for a regulator that comes from evaluating at a gauge-invariant saddle the second order derivative of $\partial_tG$ term with respect to $G$.
This derivation provided support for the conjecture of \cite{Choudhury:2017tax} about the effective action of tensor models. The GW model features a similar leading order 2PI effective action \footnote{The analysis was made easier by \cite{Bonzom:2017pqs} where the graph structure of the following orders in $1/N$ was detailed.}
\begin{equation} \label{eq:GW-freeEn}
 \begin{split}
\G_{GW}[G^{(0)}]  = & N^{q-1} \frac{q}{2} \Tr [\ln(G^{(0)})] - N^{q-1}\frac{q}{2} \Tr[\partial_t G^{(0)}] - N^{q-1} \frac{\lambda^2}{2} \int_{t,t'} G^{(0)}(t,t')^{q}  \\
&   + \bigg[ \frac{N(N-1)}{2} \binom{q}{2} \bigg] \frac{1}{2} \Tr\bigg[ \ln\bigg( 1 - \lambda^4 [ \underline{\hat \cK}^{(0)}]^2 I_+\bigg) \bigg] \\
  &  + \bigg[ \bigg(\frac{N(N-1)}{2} + (N-1)\bigg) \binom{q}{2} \bigg] \frac{1}{2} \Tr \bigg[ \ln \bigg( 1 - \lambda^4 [ \underline{\hat \cK}^{(0)}]^2 I_-\bigg) \bigg]  \\
  &+ (q-1) \frac{1}{2}   \Tr\bigg[ \ln\bigg( 1 + \lambda^2 [ \underline{\hat \cK}^{(0)}] I_- \bigg)   \bigg]  +  \frac{1}{2}  \Tr\bigg[ \ln\bigg( 1 - (q-1)\lambda^2 [\underline{\hat \cK}^{(0)}] I_- \bigg)   \bigg] \;,
 \end{split}
\end{equation}
with the ladder kernel
\be 
 \hat \cK^{(0)} (t_a,t_{a'} ; t_b, t_{b'}) = (-1) G^{(0)}(t_a,t_b) G^{(0)}(t_{a'},t_{b'}) [G^{(0)}(t_b,t_{b'})]^{q-2}
\ee
and the four-point projectors
\begin{align}
I_{\pm} & =\f{I^{=} \pm I^{\times}}{2} \;, \crcr
I^{=}(t_a,t_{a'} ; t_b, t_{b'}) & = \delta(t_a - t_b) \delta(t_{a'} - t_{b'})  \;, \crcr
I^{\times}(t_a,t_{a'} ; t_b , t_{b'}) & = \delta(t_a - t_{b'}) \delta(t_{a'} - t_b) \;.
\end{align}
Notably the subleading orders correspond to those obtained from a postulated effective action \cite{Choudhury:2017tax}
\be \label{eq:Seff-GW}
\begin{split} 
{\bf S}_{\rm eff}[G] = & \f{1}{2} \sum_{c=1}^q \Tr[\ln (G^{(c)} )] - \f{ 1 }{2} \sum_{c=1}^q\Tr[(G_0^{(c)})^{-1}G^{(c)}] \\
& - \f{\l^2}{2 N^{(q-1)(q-2)/2}} \int_{t,t'} \prod_{c=1}^q G^{(c)}_{\bf a_c b_c}(t,t') \prod_{c_1<c_2} \d_{a_{c_1 c_2} a_{c_2 c_1}} \d_{b_{c_1 c_2} b_{c_2 c_1}} \;
\end{split}
\ee
expanded around a symmetric saddle point $G^{(0)}$ 
\be
G^{(c)}_{\bf a_c b_c}(t,t') = G^{(0)}(t,t') \d_{\bf a_c b_c} + g^{(c)}_{\bf a_c b_c}(t,t')\;.
\ee
Here fluctuations $g^{(c)}$ ($1\leq c \leq q$) for each color $c$ are an $N^{q-1}\times N^{q-1}$ matrix. The important observation is that the covariance of fluctuations factorizes, when the latter are decomposed as 
\be \label{eq:g-decomp}
g^{(c)}_{\bf a_c b_c}(t,t') = g^{(c)}(t,t') \prod_{i\neq c} \d_{a_{ci} b_{ci}}+ \sum_{i\neq c} g^{(ci)}_{a_{ci} b_{ci}}(t,t') \prod_{j\neq i,c} \d_{a_{cj} b_{cj}} + \hat{g}^{(c)}_{\bf a_c b_c}(t,t') \;.
\ee
We recognize traces (giving the last line of eq.~\eqref{eq:GW-freeEn}), a 2-color submatrix, splitting into antisymmetric (second line of eq.~\eqref{eq:GW-freeEn})) and symmetric traceless terms (third line of eq.~\eqref{eq:GW-freeEn})), and finally the rest, which doesn't contribute above.\footnote{In fact such clearcut interpretation for each order in $N$ was holding for $q\geq 6$, whereas in $q=4$, $1/N$ corrections to $G^{0}$ have to be included as well as an extra set of 2PI diagrams \cite{Bonzom:2017pqs}.} 

In short, the global symmetry of tensor models is reflected in the low-energy effective action as correcting the Schwarzian term. Further studies \cite{Yoon:2017nig} have shown that only the Schwarzian mode contributes to chaos in the four-point function, the others bringing subleading or subexponential behaviours. Connections to the SYK model with extra global symmetries are worth further exploration, e.g. \cite{Gross:2016kjj,Liu:2019niv,Kapec:2019ecr}.

\subsubsection{Remarks}

\paragraph{Energy Spectrum.} After Wigner's footsteps, a complementary approach to characterize quantum chaos comes from the distribution of the energy spectrum. After some relaxation time, it follows the distribution of an ensemble of random matrices, with symmetries specific to the model. This was explored in depth for the SYK model \cite{Cotler:2016fpe}, mainly relying on numerical diagonalization of the Hamiltonian.\footnote{A peculiar part of the approach towards equilibrium could be recovered from a bulk analysis \cite{Saad:2018bqo}.} It was alluring to attempt a similar analysis for tensors. The prevailing challenge is to cope with the size of the Hamiltonian $2^{N^r}$. Latest progress \cite{Pakrouski:2018jcc} (restricting to $N=4$ and heavily relying on the discrete symmetries that hide in the Hamiltonian) seems to advocate that the random matrix ensemble followed by tensors is different than that of SYK (see also \cite{Gaitan:2020zbm}). Simulations with a more consequent size of the tensors are tantamount to outline finer details of their holographic dual and to get a handle on the effect of subleading corrections on the dominant melons. 

\paragraph{Invariants.} A major difference between the vector, matrix and the tensor models (hence between SYK and its tensor analogues) is the number of connected group invariants at fixed number of fields. For vectors, there is only one: $\sum_i\phi_i^2$. For matrices, there is one single-trace invariant per number of matrices: $\Tr M^n$. In constrast, without a canonical algebra, the indices of tensors can be contracted in multiple ways. At quartic order for an $O(N)^3$ tensor, there are five different possible contractions, cf. Fig.~\ref{fig:quarticbubbles}. 
At order six there are already 18.

First enumerations of contractions depending on the number of vertices or fields, identified a factorial growth, \footnote{As emphasized for instance in \cite{Bulycheva:2017ilt}, the number of operators involving $2k$ tensors grows like $2^kk!$.} using decomposition formulas of characters of the permutation group $S_r$ (the latest review is \cite{BenGeloun:2020lfe}). Subsequently, using characters of $O(N)^3$ decomposition of the partition function, \cite{Bulycheva:2017ilt, Beccaria:2017aqc} interpreted this growth as a vanishing Hagedorn temperature $T_H = 1/\log N$, that is a temperature above which the partition function doesn't converge anymore.\footnote{Let us note that those formulas are using however the UV dimensions of the operators which aren't necessarily appropriate to study the theory near its IR fixed point.} 
Obviously, this calls for further investigation.

\paragraph{Towards black holes.}
Two melonic models $U(N)^2\times O(D)$ and $U(N)\times O(D)$ (with quartic, sextic or both interactions, and included a mass $m$) were studied in imaginary and real time, presenting quite remarkable features \cite{Ferrari:2019ogc}. The first presented a line of first-order transition between an SYK-like phase and a Gaussian one, terminating at a critical point and was saturating the quantum chaos bound. The second showed a crossover between those two phases, but in the zero-temperature limit, broke spontaneously the $SL(2,\mathbb{R})$ symmetry with a two-point function possessing two different power law decays in the limits of $t\rightarrow\pm\infty$. While it showed the same residual zero-temperature entropy as the first model, it was not maximally chaotic. From a string theoretic perspective, those models\footnote{Although less clear for the $U(N)\times O(D)$ case.} could describe the formation of a black hole, as we go from the Gaussian to the SYK-like phase. Understanding how those models behave when coupled to additional matter fields might provide a quantum mechanical description of matter crossing an event horizon.

\subsection{Melonic CFTs}
\label{subsec:melonicCFTs}
SYK-like models were one-dimensional and not only it was natural to examine higher-dimensional versions of the non-disordered models in the quantum chaos perspective, but a better understanding of the field theoretic content of higher-dimensional tensor models was pending. As we will review, Bosonic and Fermionic versions were explored, \footnote{And even supersymmetric ones \cite{Popov:2019nja}, although we will not discuss them.} featuring those characteristic \emph{melonic} graphs in leading order Schwinger-Dyson equations. What relevant perturbations would generate an RG flow with a non-Gaussian IR fixed point? Is the fixed point conformal, unitary? 

A quick power counting analysis determines the critical dimension in the different models. With an interaction of order $q$ and a Gaussian term of dimension $2\z$ ($\z=1$ for Bosons with standard Laplacian and $0<\z<1$ to preserve reflection positivity), we deduce that the interaction is relevant if 
\be
d\left(1-\f{q}{2}\right) + q\z>0.
\ee
Thus, quartic Fermions with standard propagator ($\z=1/2$) have critical dimension 2. The critical dimension of quartic Bosons is 4, but one can reduce it to $d=4\z$. 

\begin{figure}[htbp]
\centering
\captionsetup[subfigure]{labelformat=empty}
\subfloat[]{\tikzsetnextfilename{3pt-ladder-intro2}
\input{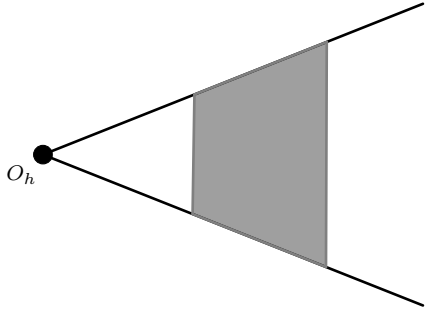}}
\caption{Leading order three-point function between a primary operator $O_h$ and two fundamental fields.}
\label{fig:3pt-ladder-intro}
\end{figure}

The treatment of melonic CFTs follows the following schematic procedure. Having worked out that the leading two-point function obeys a melonic structure 
\be 
G^{-1} = C^{-1} - \l^2 G^{q-1},
\ee we have, when the free covariance $C$ is neglegible, \footnote{In the case of long-range models, one selects the free covariance to scale as the full two-point function.} a power-law solution:
\be 
G(x) = \f{c}{\abs{x}^{2\D_\phi}}\;, \quad \D_\phi = d/q\;,
\ee
and $c$ is a normalization factor. 
Given that the four-point function in the channel $(12\rightarrow 34)$ involves an infinite sum of ladders generated by a kernel $K$ itself fully determined by the two-point function: 
\be 
F =\sum_{n\geq 0} K^n= \frac{1}{1 - K}, \quad K(x_1,x_2;x_3,x_4) = \l^2 G(x_{13})G(x_{24})G(x_{34})^{q-2}
\ee 
we head for the spectrum of the (to be shown) conformal primaries. The three-point function between $\cO_h$, a bilinear in the fields of dimension $h$, and two fundamental fields also contains a sum of ladders at leading order in $N$, cf. Fig~\ref{fig:3pt-ladder-intro}, hence it must stay invariant under the addition of a rung:
\be 
\expval{\cO_h(x_1)\phi(x_2)\phi(x_3)} = \int \dd x_a\dd x_b K(x_1,x_2;x_a,x_b) \expval{\cO_h(x_1)\phi(x_a\phi(x_b)} = k(h)\expval{\cO_h\phi\phi}.
\ee 
The eigenvalues $k(h)$ can be computed analytically inserting the dependence of $K$ on the two-point function $G$ and the assumed conformally invariant form for the three-point correlator. Solving for $k(h)=1$ gives the spectrum of bilinear conformal operators. 
This equation can also be generalized to operators involving spin (descendants) and informs us about the content in higher-spin (in a local theory, the stress-energy tensor of dimension $h=d$ and spin $s=2$ would be one of them). 

\subsubsection{Fermions}
Inspired by SYK, Fermionic quartic tensor theories were first analyzed \cite{Prakash:2017hwq} in generic $d$, confirming the existence of a non-trivial IR fixed point in $d<2$, with real conformal dimensions.\footnote{Writing such tetrahedral $\psi^4$ term, tight to SYK physics, required an $SU(N)\times O(N) \times SU(N)$ symmetry to overcome the anticommutation relations.} Furthermore, in $d>2$, a UV limit was necessary in order to discard the free propagator leading to melonic SDE. Eigenvalues of the four-point kernel presented complex values.  

The critical dimension $d=2$ was more tricky to analyze \cite{Benedetti:2017fmp}. Dirac and Majorana Fermions were considered separately, the last allowing a tetrahedral contraction, the first having only pillow and double-trace interactions, with respective couplings $\l_1$ and $\l_0$. It can be seen as a generalization of the Gross-Neveu model with $N^3$ Fermions breaking the $O(N^3)$ symmetry into $O(N)^3$. A conformal fixed point was searched by renormalizing the model. Regarding first the SDE for the two-point function, assuming a generic ansatz $G^{-1} =i\slash{p}+m$ a mass shouldn't be generated. In the Dirac case, the generated mass could vanish if $\l_0 + 3\l_1 = 0$ and for establishing the stability of this line, an intermediate field formalism was introduced: 
\be 
S[\psi, M] = \int \dd[2] x\left[\bar{\psi}_{abc}\slashed{\partial}\psi_{abc}+\f{1}{2}\left(\Tr[M^2] -\f{\l_0\l_1}{(\l_0+\l_1)N}\Tr[M]^2\right)+\f{\sqrt{2\l_1}}{N}\bar{\psi}_{abi}\psi_{abj}M_{ij}\right].
\ee 
Integrating out the Fermions left an effective potential for the matrix $M$. The vacua and stability analysis justified the presence of a first order phase-transition between the trivial solution $M=0$ and diagonal matrices $M=(\m, \dots, -\m)$. Nevertheless this spontaneous breaking of the continuous symmetry $U(N)$ to $U(N/2)\times U(N/2)$ cannot be effective due to the Coleman-Mermin-Wagner theorem. It was advocated that, as in the $SU(N)$ Thirring model (a version of the GN model with $U(1)$ chiral symmetry), symmetry restoration was at play by non-perturbative contributions to the correlation function of the would-be Goldstone mode, such that its two-point function would behave as $\abs{x}^{-1/N}$ which indeed decays at large distances, but still is non-vanishing if $N$ is taken large first, seemingly signalling long-range order \cite{Witten:1978qu}. 

In the Majorana case, the three types of interactions together didn't allow for a massless solution when $\l_t\neq 0$, returning to the Dirac case. Focusing on $\l_t$ alone, and imposing that neither mass nor wave-function renormalization was generated, and that radiative corrections to the couplings $\l_{d}$, $\l_p$ vanished at leading order, didn't leave any room for a non-trivial $\l_t$. The form of the leading order beta function $\b_t$ was however interesting: 
\be 
\b_t = \l_t^3,
\ee 
with IR attractive Gaussian fixed point, and taking $d = 2- \eps$, generates an IR attractive fixed point $\l_* = \sqrt{\eps}$.

\subsubsection{Bosons}
Bosonic tensor field theories were first written in \cite{Klebanov:2016xxf}. With a quartic potential, they deduced that a real spectrum appeared only for $d>4$, but at the fixed point the values of the melonic couplings turned complex. Formally, it was also possible to analytically continue the ladder kernel to generic $d$ and $q$ \cite{Giombi:2017dtl}, 
\begin{gather}
K = (q-1)\l^2 G(x_{13})G(x_{24})G(x_{34})^{q-2}\\
G(x) =C_\phi\l^{2/q}\f{1}{\abs{x}^{2d/q}},\quad 
C_\phi=\left(-\f{\G(\f{d}{q})\G\left(\f{d(q-1)}{q}\right)}{\pi^d\G\left(\f{d(2-q)}{2q}\right)\G\left(\f{d(q-2)}{2q}\right)}\right)^{1/q},
\end{gather}
and search for the critical values $d_{c}$, that only contained real operator spectrum. It allowed to conjecture the existence of another melonic fixed point for a sextic model in $2.97<d<3$, with again imaginary couplings at the fixed points. \footnote{For the quartic case, they also used a finite $N$ expression of the $\b$ functions to find appropriate scalings that lead to a non-trivial fixed point, whereas the finite $N$ $\b$ functions of the sextic theory were written down in \cite{Giombi:2018qgp}. } 

A way to circumvent the above inconveniences proceeded from the introduction of a non-local propagator in the bare action while at the same time allowing the tetrahedral coupling to take complex values \cite{Benedetti:2019eyl}. From a controlled renormalization group analysis, the quartic model: 
\be 
S =\int\dd[d]x \phi_{abc}(-\partial_\mu\partial^\mu)^{\z}\phi_{abc} + S_{int}, 
\ee 
presented four lines of fixed points, parametrized by the value of the quartic coupling $\l_t$ up to a critical value. Besides, for an imaginary $\l_t=ig$, the spectrum of bilinear operators $O_{h_n,J}$ was found real: 
\begin{gather}
h_0 = \f{d}{2}\pm 2\f{\G(\f{d}{4})^2}{\G(\f{d}{2})}\sqrt{-3g^2}+\cO(g^3)
\\ h_{n,J} = \f{d}{2}+J+ 2n+ 2\f{3g^2\G(\f{d}{4})^4\G(n+J)\G(n+ 1-\f{d}{2}) \sin(-\pi\f{d}{2})}{\G(\f{d}{2}+J+n)\G(n+ 1)\pi} + \cO(g^4), 
\end{gather}
above the unitarity bound and asymptoting to their free value $h_{n,J}=d/2 +J+ 2n$ at large $n$.
In a subsequent study \cite{Benedetti:2019ikb}, the OPE coefficients from three-point functions of bilinear primaries with two fundamental fields were also shown to be real. 

Note that at any $d$ there is no $h=2$ solution, associated to the stress-tensor, and that the imaginary coupling $\l_t$ comes in pairs in the leading order correlation functions in order to form melons, hence only its absolute value appears above.

Yet, further work was required in order to ensure conformal invariance and unitarity of those long-range fixed points. A more recent work \cite{Benedetti:2020yvb}, inspired by the techniques of \cite{Paulos:2015jfa} showed that the vanishing of the beta functions at the fixed points implies the validity of the conformal Ward identities at all orders in perturbation theory, by an embedding in a higher dimensional space that localized the kinetic term. Computation of two- and three-point functions of (renormalized) operators corroborated this fact, suggesting that at leading order in $1/N$ the fixed point is a unitary CFT.

A different non-melonic large-$N$ limit was devised in \cite{Giombi:2018qgp}, by starting from a two-tensor model, interacting through a tetrahedron: 
\be 
S[\phi, \chi] = \int \dd[d]x \left(\f{1}{2}\left(\partial_\mu\phi_{abc}\partial^\mu\phi_{abc}\right)+\f{\l_t}{3!N^{3/2}}\phi_{a_1b_1c_1}\phi_{a_1b_2c_2}\phi_{a_2b_1c_2}\chi_{a_2b_2c_1}+\f{1}{2}\chi_{abc}\chi_{abc}\right),
\ee 
which, when integrating out the field $\chi$ lead to a prism-type interaction: 
\be 
S[\phi]= \int \dd[d]x \left(\f{1}{2}\left(\partial_\mu\phi_{abc}\partial^\mu\phi_{abc}\right)+ \f{\l_t^2}{(3!)^2N^3}\phi_{a_1b_1c_1}\phi_{a_1b_2c_2}\phi_{a_2b_1c_2}\phi_{a_3b_3c_1}\phi_{a_3b_2c_3}\phi_{a_2b_3c_3}\right).
\ee 
Three bilinears could be written $\psi\psi$, $\psi\chi$ and $\chi \chi$, resulting in a 3-by-3 ladder kernel to diagonalize. They numerically studied the spectrum of conformal dimensions of bilinears and windows of reality opened for $d<1.68$ and $2.81<d<3$. An RG analysis was also made for $d=3-\eps$, including all eight $O(N)^3$ invariant sextic couplings, using finite-$N$ $\beta$ function results of multi-scalar interactions. A unique real fixed point was found, that included all eight couplings.

\paragraph{On the importance of being gauged.} From the holographic standpoint, it is difficult to interpret a number of bulk fields, growing with a power of $N$, dual to the fundamental tensors. Hence, the global (here $O(N)$ or $U(N)$) symmetry is typically gauged.\footnote{Nevertheless, \cite{Maldacena:2018vsr, Berkowitz:2018qhn} considered the difference between gauged and ungauged boundary $U(N)$ global symmetry of a $D0$-brane matrix model and argued from the bulk and the boundary (in the latter case, with support of Monte-Carlo simulations), that their free energies near zero temperature differed by non-perturbative corrections in the temperature.}
In the one-dimensional case, since the gauge field is non-dynamical, this constraint leaves as only observables the gauge invariant operators.
In higher-dimensions, the dynamics of the gauge field has to be taken into account. Particularly in three dimensions, one can couple to Chern-Simons theory which has the particularity that the gauge coupling takes only integer values, thus it cannot flow under the renormalization group. This simplification, as well as the already well studied 3d gauged vector models, motivated us to look at a similar question for tensors. Much the same conclusions were reached in \cite{Popov:2019nja}, in the meantime discussing supersymmetric generalizations. 

\section{Overview and Summary}
\label{sec:overview}
\subsection{Melonic field theories}

When discussing the renormalization group, we have emphasized that symmetries are decisive criteria to distinguish universality classes determining critical properties of quantum field theories. Their breaking, spontaneous or explicit, is as important. Matrix and tensor models can be seen as breaking the continuous symmetry of a vector into a smaller group. For instance, a vector $\phi_I$ ($1\leq I\leq \cN$) transforming under $O(\cN)$, can also be reassembled as a matrix $\Phi_{ab}$ ($1\leq a,b\leq \sqrt{\cN}$) and ask for invariance under $O( \sqrt{\cN})\times O( \sqrt{\cN}) $ or as a tensor $\varphi_{abc}$ ($1\leq a,b,c\leq \cN^{1/3}$), with invariance $O(\cN^{1/3})\times O(\cN^{1/3}) \times O(\cN^{1/3})$ and so on. The allowed interactions are different, so are their respective large $N$ limits and this can lead to theories with very different content. 

Over the next two chapters, we will explore the tensorial counterpart of two well-known three-dimensional vector theories: the quartic Fermionic and the sextic Bosonic. The underlying thread is to pinpoint how their field theoretic structures differ, in terms of symmetry patterns of the vacuum as well as their renormalization group flows and fixed points. 

\vspace{5mm}
First, we will be concerned with the tensorial analog of the Gross-Neveu model. The large-$N$ diagrammatics of the self-energy will in a certain regime of couplings uphold spontaneous chiral symmetry breaking through mass-generation. We extract the $\beta$ functions and in addition to the trivial IR fixed point and to the standard UV fixed point of the Gross-Neveu model (where the tensorial interactions that break the $U(N^3)$ symmetry are absent), we find two new fixed points: one with two relevant directions, and another with one relevant and one irrelevant direction. Since the properly color-invariant $U(N)^3$ tensor model presents spontaneous breaking of color-symmetry, we first analyze the resulting $U(N)\times U(N^2)$ effective theory, which essentially is a rectangular matrix model. Eventually, we identify three phases of the vacuum by analyzing the effective potential of the intermediate field: a massless phase preserving all the symmetries of the model; a phase with dynamically generated mass, breaking only the chiral symmetry; and a phase breaking the chiral symmetry, as well as one of the $U(N)$ subgroups of the symmetry group (and as a consequence breaking also the color symmetry, when present in the original model). We find that the conformal dimension of the unconstrained component of the intermediate field (and thus of the corresponding composite operator $\psib \psi$) is the same at all the nontrivial fixed points, $\D_{\psib \psi}=1$ (to be compared to the Gaussian fixed point value, $\D_{\psib \psi}=2$), thus matching that of the critical vector GN model.
The dimensions of the (integrated) quartic operators are instead different, as they change from relevant to irrelevant from one fixed point to another, but are the same in modulus, which is always equal to one.

A natural extension of this model is to gauge the $SU(N)$ symmetry with a Chern-Simons term, as had been previously done with vector models (which lead to a wonderful set of dualities between Fermionic and Bosonic models, free and interacting, and the unifying picture of a dual higher-spin theory \cite{Giombi:2016ejx}). In our case, we argue that at large-$N$, the gauge field doesn’t affect the tensor field equations of motion, while restricting observables to $SU(N)$ singlets.

\vspace{5mm}
The second chapter focuses on sextic Bosonic models with bipartite (rank 3) and non-bipartite interactions (rank 5).
In both cases, we allow long-range propagators, tuning the dimension of the field in order to keep the sextic interactions marginal for $d<3$.
Short-range sextic tensor models have been considered before, but either without actually studying the existence of fixed points \cite{Giombi:2017dtl} (and only for rank 5), or for a different scaling in $N$ of the couplings than the optimal one \cite{Giombi:2018qgp}.
Here, we compute the beta functions for our models, at leading order in the $1/N$ expansion, and at four-loop order, expanding on top with respect to a small parameter, such as $\epsilon=3-d$ in the short-range case, or an exactly marginal coupling in the long-range case.

Our main results are: in rank 3, we find two non-trivial infrared fixed points for the short-range model in $d<3$, and for the long-range model a line of infrared fixed points parametrized by one of the sextic couplings. In both cases, the couplings are real and we find a window with real spectrum of bilinear operators. However, the short-range model has a non-diagonalizable stability matrix, sign of a non-unitary, logarithmic CFT. Surprisingly, in rank 5, the only fixed point is non-interacting.

We will compare to two other series of works. Firstly, as we had mentioned in Subsection~\ref{subsec:melonicCFTs}, to the studies of quartic long-range models~\cite{Benedetti:2019eyl}, where in order to find a \emph{real} line of fixed points parametrized by the non-renormalized tetrahedral coupling, the analog of our sextic wheel coupling, this quartic coupling is taken \emph{imaginary}. Real OPE coefficients~\cite{Benedetti:2019ikb} and correlation functions between primaries displaying a conform structure~\cite{Benedetti:2020yvb} strongly support that the fixed points form a line of unitary CFTs. 
Secondly to those of~\cite{Giombi:2017dtl}, as we give solid ground to the conjectured melonic fixed point for sextic model, that we disprove in rank-5 but show to exist in a rank-3 model.
Common to both these lines, reality of the spectrum at the fixed points will again constrain the expansion parameter (be it $\eps$ or the marginal coupling) in a very small window.

\subsection{Perturbative quantum field theory on random trees}

In the last chapter of our work, we turn to a quite different topic, with different formalism, tools and goals. Our idea was that since the critical dimension of the interacting quartic Fermionic theory discussed earlier was two, and the theories in one and two-dimensions have been extensively discussed, we wondered if we could reach interesting conclusions about theories in intermediate dimensions. In other words, fractals. The object we had in mind was the Aldous continuum random tree, known to have Hausdorff dimension $d_H = 2$ and spectral dimension $d_s = 4/3$, and closely connected to the large $N$ limit of tensors. We mentioned earlier some results on theories in fractional dimensions stating for instance their non-unitarity, but they were relying on analytic continuation of formulas obtained in integer dimensions. We wanted more constructive results. Hence, in order to define a quantum theory on such a beast, we needed a propagator. Once we rewrite the latter as a sum over random walks and look for the probability to reach one point from another, we open the door to a vast topic in probability theory dealing with random walks on random environments. Concretely, we were looking for precise estimates of the heat-kernel on the tree. 

Actually, the use of random walks in QFT was introduced, after Symanzik, as a constructive tool to prove that correlators obeyed OS axioms. The idea is to associate the renormalization group scale with the proper time of the random walker, which then limits the distance it may travel on average. In the perturbative expansion, we need to evaluate the probability of intersection of multiple random walks. This point of view provided a constructive proof of the triviality of quartic scalar field theory in $d\geq 4$ and non-triviality in $d=2,3$ \cite{Fernandez:1992jh}. 

We will limit ourselves to perturbative results on Feynman amplitudes
for a self-interacting scalar theory. We take as propagator a fractional rescaled Laplacian as in \cite{Abd,Gross:2017vhb} to put ourselves in the interesting just renormalizable case. Our basic tool is
the multiscale analysis of Feynman amplitudes \cite{FMRS}, which remains available on random trees since it simply slices the proper time --  i.e. Feynman's parameter in high energy physics language
-- of the random path
representation of the inverse of the Laplacian. 

Combining this slicing with the probabilistic estimates of Barlow and Kumagai \cite{BarlowKumagai,Kumagai}
we establish basic theorems on power counting, convergence and renormalization
of Feynman amplitudes. Our main results, Theorems \ref{theoconv} and \ref{theoconv1} below, use the Barlow-Kumagai technique of ``$\lambda$-good balls" to \emph{prove} that the averaged amplitude of \emph{any} graph without superficially divergent subgraphs is finite and that logarithmically divergent graphs and subgraphs can be renormalized via local counterterms.

The conclusion is that the superficial degree of divergence of amplitudes is expressed in terms of the spectral dimension of the geometry ($4/3$ in our case), instead of the usual spacetime dimension. This approach provides field theory amplitudes as a different tool to probe the spectral dimension of a geometry, a local property. An analogous result was phrased by Eyink \cite{Eyink} in an earlier study of quantum field theory on a fixed fractal geometry, and although the estimates on heat-kernels were much less understood, still allowed a rigorous Wilson block renormalization construction for hierarchical models. 

Finally, let us remark that Barlow and Kumagai obtained heat-kernel bounds \cite{BarlowKumagai} for the Incipient Infinite Cluster (IIC) on Cayley trees (regular and rooted). This graph contains as subgraphs all clusters connected to the root of given size $n$, for all $n\in \mathbb{N}$, when considering the critical percolation on Cayley trees \cite{Kesten}. In the continuum limit it is the Aldous Continuum Random Tree (CRT) \cite{aldous}. \cite{Croydon1} showed that the scaling limit of random walks on Galton-Watson trees is the Brownian motion on the CRT and \cite{Croydon2} obtained quenched bounds on heat-kernel on the CRT compatible with the ones of \cite{BarlowKumagai}, although, in our understanding, with less explicit ranges of validity. Their techniques generalize more easily to the random graphs known as Random Conductance Models (see for instance Ch.~8 of \cite{Kumagai} for an overview of results and their proofs).

%% file: TGN3D.tex
In this chapter, Section~\ref{sec:vectorGN} begins by shortly reviewing the large-$N$ study of the three-dimensional Gross-Neveu model. Going from vectors to rectangular matrices, Section~\ref{sec:S2} undertakes the analysis the RG flow and effective potential of the matrix intermediate field, where equations of motion can be explicitly solved. Appendix~\ref{app:rectangular} discusses a different intermediate field decomposition of the rectangular matrix case. Follows in Section~\ref{S:2} the analysis of the color-symmetric tensor model, where a similar phase diagram is found, while the details of our analysis of the effective potential are assembled in Appendix~\ref{app:traceless_matrices}. Finally in Section~\ref{sec:CS}, we explain what are the consequences of gauging the global symmetries and compare with the vector model analysis. The Appendix~\ref{sec:appendix gamma} details our conventions on the used $\g$ matrices.

\section{A brief reminder of the vectorial Gross-Neveu model}
\label{sec:vectorGN}
The Gross-Neveu model \cite{Gross:1974jv} has been extensively studied, in particular in two dimensions, where it provides a model of asymptotic freedom and dynamical mass generation, which is also integrable. Here, we are rather interested in its three-dimensional version, which despite being perturbatively non-renormalizable, is renormalizable in the $1/N$ expansion \cite{Parisi:1975im} and admits an ultraviolet fixed point at large $N$ \cite{Rosenstein-PRL,deCalan:1991km} which renders the model meaningful at arbitrarily high energies (see \cite{ROSENSTEINreview} for a review). The nontrivial fixed point theory has been conjectured to be dual to a particular version of higher spin theory in $AdS_4$ \cite{Klebanov:2002ja,Sezgin:2003pt}, a conjecture which has passed several tests (see \cite{Giombi:2016ejx} and references therein).

In view of the upcoming generalizations, we define here the model for the case of $N^3$ Dirac Fermions in Euclidean signature (see Appendix \ref{sec:appendix gamma} for conventions on $\g$ matrices). The action is
\be
S_{\rm GN}[\psi,\bar{\psi}] = \int d^3 x \; \left( \psib_i  \slashed{\p} \psi_i -\f{\l}{N^3} (\psib_i \psi_i)^2 \right) \;.
\ee
Expressing the four-Fermion interaction in terms of an intermediate field $\sigma$, the action writes
\begin{equation} \label{eq:GNY}
S[\psi,\bar{\psi},\sigma]=\int d^3 x \;\left(\bar{\psi}_i\slashed{\partial}\psi_i + \sigma\bar{\psi}_i\psi_i + \frac{N^3}{4\l}\sigma^2\right).
\end{equation}
Besides the $U(N^3)$ invariance (with the Fermions transforming in the fundamental representation), the model has also a discrete chiral symmetry, which acts as
\begin{equation}
\psi\rightarrow \gamma^5\psi \quad \bar{\psi} \rightarrow -\bar{\psi}\gamma^5 \quad \sigma \rightarrow - \sigma.
\end{equation}

In the large-$N$ limit one can write a closed Schwinger-Dyson equation for the Fermion 2-point function, which reduces to a gap equation for the Fermion mass $m=\expval{\sigma}$:
\begin{equation} \label{eq:GN-gap}
\frac{m}{\l} = 8 m \int_\Lambda \frac{\dd[3]p}{(2\pi)^3}\frac{1}{p^2+m^2}\;,
\end{equation}
where the divergent integral is regulated by a UV cutoff $\L$. The integral on the right-hand side of the gap equation is a monotonically decreasing function of $m$, hence it has a maximum at $m=0$, which defines a critical coupling
\begin{equation}
\frac{1}{\l_c} \equiv 8\int_\Lambda \frac{\dd[3]p}{(2\pi)^3}\frac{1}{p^2} \;,
\end{equation}
above which the gap equation \eqref{eq:GN-gap} admits a real solution $m\neq 0$, besides the trivial one.
Using the intermediate field formulation \eqref{eq:GNY}, and integrating out the Fermions, one finds that for $\l>\l_c$ the stable solution of the effective potential is the non-zero solution. Therefore, the theory has a dynamically generated mass for $\l>\l_c$, and this in turn means that the chiral symmetry is spontaneously broken. 

Using the gap equation for $\l>\l_c$, the effective potential writes
\begin{equation} \label{eq:V_GN}
V_{\rm eff}(\sigma) = \frac{1}{\pi}\left(\frac{1}{3}\abs{\sigma}^3 - \frac{m}{2}\sigma^2\right) \;,
\end{equation}
with an evident minimum at $\s=m$.

For $0\leq \l\leq \l_c$ the symmetry is instead preserved, as $m=0$ is stable. The phase transition at $\l=\l_c$ is second order.

The $\beta$-function of the adimensional coupling $\lt \equiv \L \l$ is obtained from eq.~\eqref{eq:GN-gap} derivating both sides with respect to $\L$, leading to 
\begin{equation}
\b = \L\p_\L \lt = \lt - \frac{4}{\pi^2}\lt^2.
\end{equation}

\section{$U(N)\times U(N^2)$-symmetric model}
\label{sec:S2}

The $U(N)\times U(N^2)$-symmetric model is obtained by first rearranging the label $i=1,\ldots,N^3$ as a set of two labels $a$ and $A$, so that we rewrite $\psi_i \to \psi_{aA}$, with $a=1,\ldots,N$ and $A=1,\ldots,N^2$, and $\psi_{aA}$ transforming in the fundamental of the product group.\footnote{There is a slight redundancy in denoting the symmetry group as $U(N)\times U(N^2)$: its action on $\psi_{aA}$ is not faithful, because the action of the two $U(1)$ subgroups of $U(N)$ and $U(N^2)$ are indistinguishable. Therefore, a faithfully acting symmetry group of the theory would be $U(1)\times (SU(N)/\mathbb{Z}_N\times SU(N^2)/\mathbb{Z}_{N^2})$, where we have quotiented also by the residual centers of the special unitary groups. A similar caveat applies of course also to the symmetry group of section \ref{S:2}. In the rest of the chapter, for compactness of notation we will stick to the non-faithful denotation of the symmetry group.} 
In order to explicitly break the symmetry from $U(N^3)$ to $U(N)\times U(N^2)$, while preserving the discrete chiral invariance, we add the following interaction to the GN model:
\be
\frac{\l_p}{N^2} \bar{\psi}_{a A}\psi_{a^\prime A}\bar{\psi}_{a^\prime A'}\psi_{a A'} \;.
\ee

\begin{figure}
\centering
\begin{minipage}{0.4\textwidth}
           \centering 
            \includegraphics[width=0.5\textwidth]{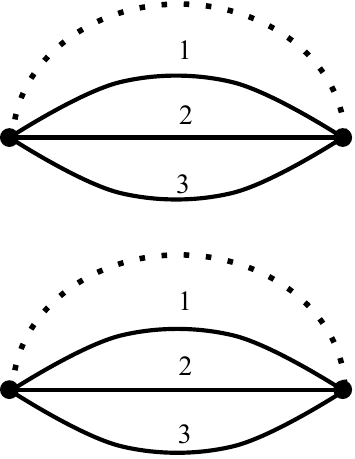}
        \end{minipage}
        \hspace{0.01\textwidth}
\begin{minipage}{0.4\textwidth}
            \centering
            \includegraphics[width=0.5\textwidth]{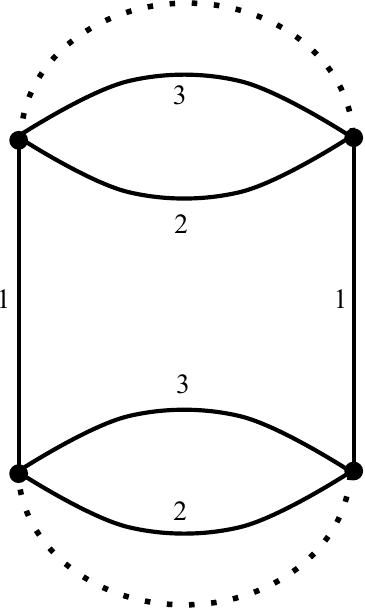}
        \end{minipage}
\caption{\label{fig:interactions}Graphical representation of the interaction vertices (GN and pillow).}
\end{figure}

In view of the next generalization, we will actually replace also the index $A$ by a pair of indices, each taking values from 1 to $N$, i.e.\ we write $\psi_i \to \psi_{abc}$. The total action then reads
\be \label{eq:action}
S[\psi,\psib] = S_{\rm free}[\psi,\psib] + S_{\rm int}[\psi,\psib] \;,
\ee
with
\be
S_{\rm free}[\psi,\psib]= \int d^3 x \;  \bar{\psi}_{abc}\slashed{\partial}\psi_{abc} \;,
\ee
\be \label{eq:S_int1}
S_{\rm int}[\psi,\psib]= -\frac{\l}{N^3}  \int d^3 x \; (\bar{\psi}_{abc}\psi_{abc})^2 - \frac{\l_p}{N^2} \int d^3 x \; \bar{\psi}_{abc}\psi_{a^\prime bc}\bar{\psi}_{a^\prime b^\prime c^\prime}\psi_{ab^\prime c^\prime} \;.
\ee
Having written the rectangular matrix as a cubic tensor, we can depict the interactions as in Fig.~\ref{fig:interactions}, where each vertex represents a tensor, and the solid lines with label $n=1,2,3$ represent the contraction of two indices in the $n$-th position. The dotted lines represent instead the spin contraction (as in \cite{Benedetti:2017fmp}, we could consider also other interactions in which such contraction is mediated by a $\g_5$ or $\g_\m$ matrix). The solid-line graph on the right of Fig.~\ref{fig:interactions} is commonly called the \emph{pillow} graph, hence the subscript $p$ for its coupling $\l_p$.
Notice that it comes with a different power of $N$ in \eqref{eq:S_int1}, as required for a non-trivial large-$N$ limit (see for example \cite{Bonzom:2016dwy}).

\subsection{\texorpdfstring{$\beta$}{b}-functions and flow diagram}
\label{sec:SD1}

\begin{figure}
\centering 
\includegraphics[width=0.6\textwidth]{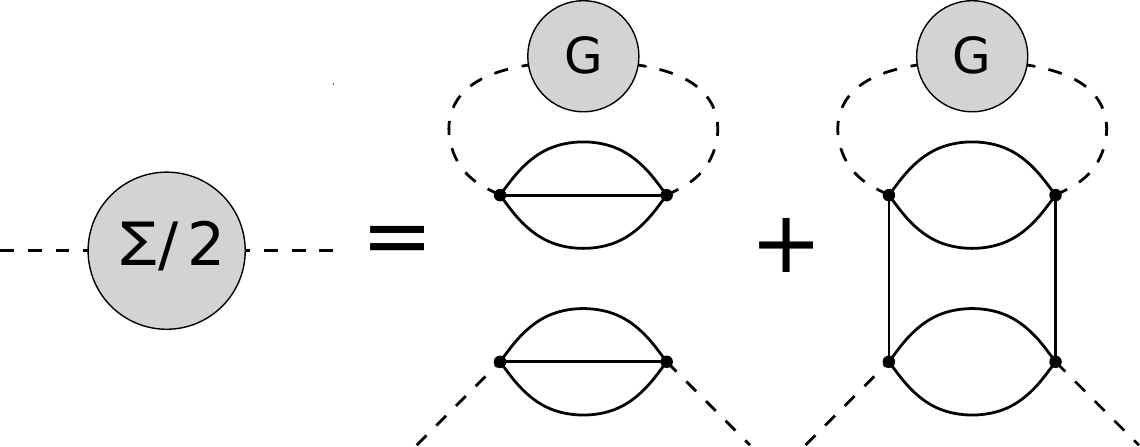}
\caption{\label{fig:SDE}Schwinger-Dyson equation at large $N$ for the self-energy.}
\end{figure}

Assuming that the $U(N)\times U(N^2)$ invariance is unbroken, we write the two-point function as
\be \label{eq:2pt-sym-ansatz}
\langle \psi_{a_1 a_2 a_3}(x) \psib_{b_1 b_2 b_3}(x') \rangle = G(x,x')\, \d_{a_1 b_1} \d_{a_2 b_2} \d_{a_3 b_3} \;.
\ee
In the large-$N$ limit, the self-energy $\Sigma$ is expressible in terms of tadpole diagrams with full two-point function on the internal propagator, as depicted on Fig.~\ref{fig:SDE}.
As a consequence, the large-$N$ Schwinger-Dyson equation is a closed equation for $G(x,x')$, which for its Fourier transform $\hat{G}(p)$ reads
\begin{equation}
\hat{G}(p)^{-1} 	= i\slashed{p} -\Sigma(p) = i\slashed{p} + 2(\l + \l_p)\int\frac{\dd[3]q}{(2\pi)^3}\tr[\hat{G}(q)] \mathbb{1}\;,
\end{equation}
the trace and the identity being defined in spinor space (in the following we will generally omit the identity matrix, unless we want to emphasize its presence).
Since the tadpole integral is momentum-independent, we can write $\Sigma = -m\mathbb{1}$, resulting in the gap equation
\begin{equation}
m = 2(\l + \l_p)\int\frac{\dd[3]q}{(2\pi)^3} \frac{\tr(-i\slashed{q} + m)}{q^2 + m^2}, 
\end{equation}
or
\begin{equation}
\frac{1}{\l+\l_p}=8 \int_{\abs{p}<\Lambda}\frac{\dd[3]p}{(2\pi)^3}\frac{1}{p^2+m^2} = \frac{4}{\pi^2}\left(\Lambda - m \arctan\left(\frac{\Lambda}{m}\right)\right)\;,
\label{eq:gap-1c}
\end{equation}
that is the analog of \eqref{eq:GN-gap}.

\begin{figure}[!h]
\begin{minipage}{0.5\textwidth}
  	\centering 
	\includegraphics[width=0.8\textwidth]{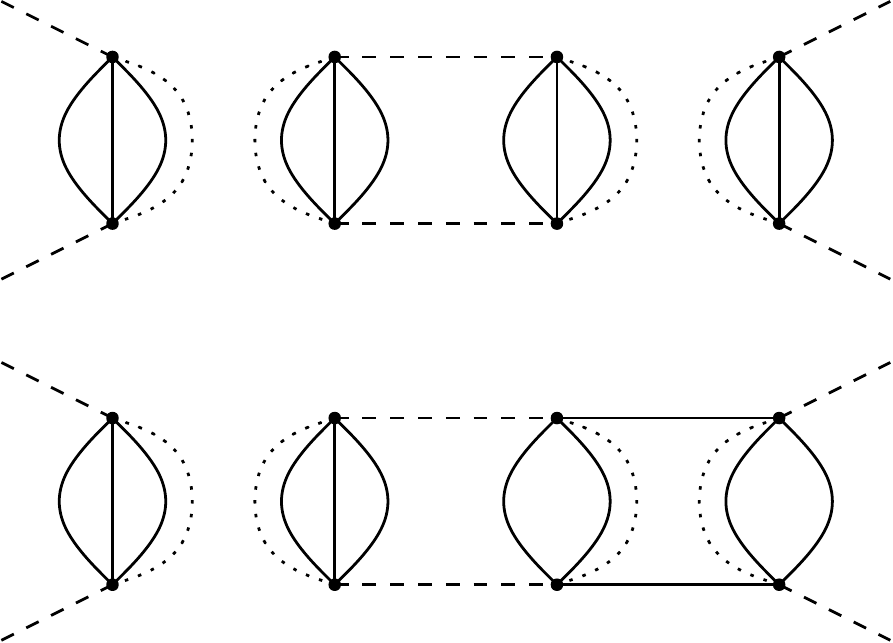}
        \end{minipage}
\begin{minipage}{0.5\textwidth}
      	\centering
	\includegraphics[width=0.8\textwidth]{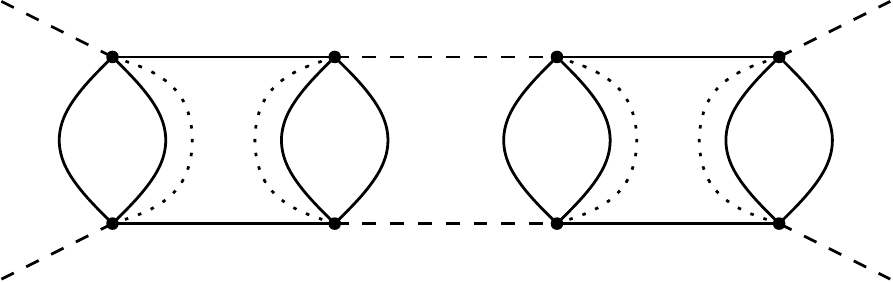}
        \end{minipage}
\caption{\label{fig:4-point graphs}Leading order graphs renormalizing the $\lt$ and $\lt_p$ couplings respectively.}
\end{figure}

\begin{figure}[!h]
\centering
\includegraphics[width=0.6\textwidth]{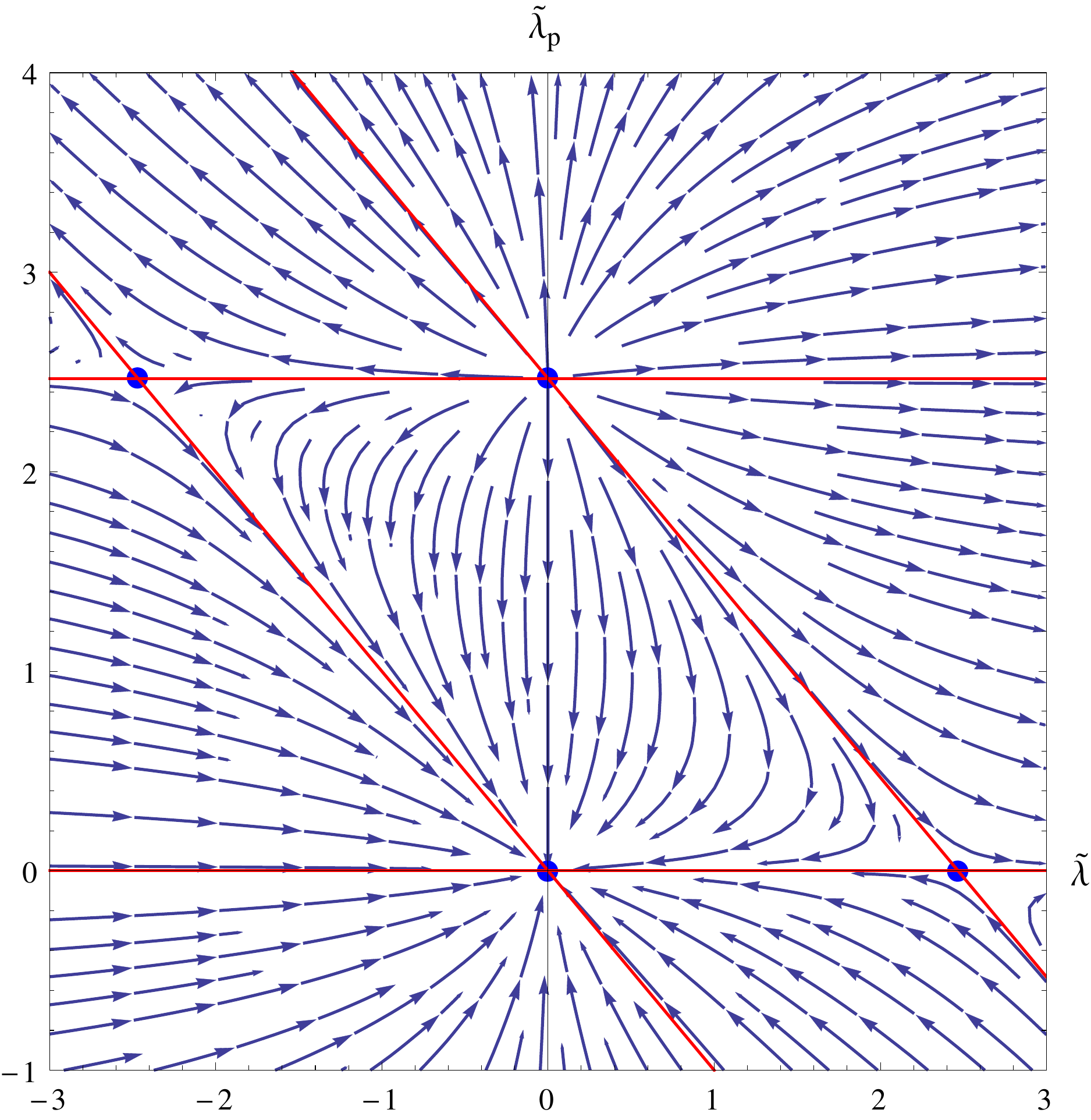}
\caption{Renormalization group flow in the $\left(\lt,\lt_p\right)$-plane. Arrows point towards the $IR$, blue dots denote the fixed points, and red lines mark the zeros of either $\beta_p$ or $\beta+\beta_p$. }
\label{fig:RG-flow-Fermions}
\end{figure}

After a rewriting in terms of the dimensionless couplings ($\lt_p \equiv \Lambda\l_p$ and $\lt \equiv \Lambda\l$), and derivating both sides with respect to $\Lambda$, we get 
\begin{equation*}
\frac{1}{\lt+\lt_p} - \frac{\Lambda}{\left(\lt+\lt_p\right)^2}\partial_\Lambda\left(\lt+\lt_p\right) = \frac{4}{\pi^2}\left(1 - \left(\frac{m}{\Lambda}\right)^2\right)\;.
\end{equation*} 
Defining $\kappa \equiv 4/\pi^2$, and taking $\L\gg m$, we find the following combination of beta functions:
\be
\beta+\beta_p  \equiv \Lambda\partial_\Lambda \lt + \Lambda\partial_\Lambda\lt_p= \left(\lt + \lt_p\right) - \kappa\left(\lt + \lt_p\right)^2 \;.
\ee
Taking into account the different structure of diagrams that contribute to the flow of the couplings (see Fig.~\ref{fig:4-point graphs}), we can disentangle the beta functions and obtain
\begin{align}
\beta &= \lt - \kappa\left(\lt^2 + 2\lt\lt_p\right)\;,\\
\beta_p &= \lt_p - \kappa\lt_p^2 \;.
\label{eq:beta-1c}
\end{align}
This leads to the flow diagram of Fig.~\ref{fig:RG-flow-Fermions}.

One sees, in addition to the trivial IR-fixed point $\left(\lt,\lt_p\right) = (0,0)$, three new non-trivial fixed points at $\left(\frac{1}{\kappa},0\right)\equiv {\rm FP}_1$, $\left(-\frac{1}{\kappa},\frac{1}{\kappa}\right)\equiv {\rm FP}_2$, and $\left(0,\frac{1}{\kappa}\right)\equiv {\rm FP}_3$. The first of them is the usual interacting CFT of the vector GN model, while the other two are new interacting CFTs. The fixed point at the origin is IR-stable, while ${\rm FP}_2$ is UV-stable, and the other two are saddles. The irrelevant directions at ${\rm FP}_1$ and ${\rm FP}_3$ can be understood as a statement of the fact that their universality classes are stable against symmetry breaking perturbations (for ${\rm FP}_1$, which being the usual GN fixed point has a larger symmetry, namely $U(N^3)$) and trace perturbations (for ${\rm FP}_3$, which lies on the tracelessness constraint subspace of the diagram).
The critical exponents are all $\pm1$ with the signs determined by the corresponding eigenperturbation being relevant or irrelevant.\footnote{In our convention, the critical exponent corresponds to the mass-dimension of the integrated operator (opposite to that of the relative coupling), hence relevant ones have negative critical exponents.}

\subsection{Effective potential and phase diagram}
\label{subsection: effective potential}
The next question to raise concerns the nature of the phase diagram of the stable vacuum.

Following \cite{Benedetti:2017fmp}, we introduce a Hermitian matrix $M_{ij}$ as intermediate field, such that the interaction terms rewrite as\footnote{In App.~\ref{app:rectangular}, we will explain what happens if instead we choose to introduce an intermediate field by cutting the pillow interaction along the index of size $N^2$ (or double index in the tensor notation).}
\begin{equation}
\label{eq: sint with psi}
S_{\rm int}[\psi,\psib,M]=\frac{1}{2}\left[\Tr(M^2) + \frac{b}{(1-b)N}(\Tr M)^2\right] + \frac{\sqrt{2\l_p}}{N}\bar{\psi}_{ibc}\psi_{jbc}M_{ij},
\end{equation}
$b \equiv \frac{-\l}{\l_p}$, allowing to integrate out the Fermions and obtain
\begin{equation}
\label{eq:S_M}
S[M]=\int \frac{1}{2}\left[\Tr(M^2) + \frac{b}{(1-b)N}(\Tr M)^2\right] - N^2 \tr \Tr\left[\ln\left(\slashed{\partial} + \frac{\sqrt{2\l_p}}{N} M\right)\right].
\end{equation}
To remove the $N$ factor from inside of the log-term, we rescale $M \rightarrow NM$. We are interested in the effective potential, which in the large-$N$ limit is simply given by \eqref{eq:S_M} evaluated at constant $M$.\footnote{Remember that the effective potential is defined as the effective action $\G[M]$ (the one-particle-irreducible generating functional) at constant field, and that $\G[M]$ is the Legendre transform of $W[J]$, the generating functional of connected $n$-point functions. Since the latter is given in the large-$N$ limit simply by the Legendre transform of $S[M]$, and since the Legendre transform is an involutive transformation, we conclude that $\G[M]=S[M]$.} Then the last term, up to a constant independent of $M$, gives
\be
\begin{split}
\tr \Tr \int \frac{\dd[3]k}{(2\pi)^3} \ln\left(\slashed{\partial} + \sqrt{2\l_p} M\right) &= -4 \Tr \int \frac{\dd[3]k}{(2\pi)^3}\sum_{n>0}\left(\frac{ik}{k^2}\sqrt{2\l_p} M\right)^{2n}\frac{1}{2n} \\
&= \frac{1}{\pi^2}\int_0^\Lambda \dd k k^2 \Tr\log\left(1 + \frac{2\l_p M^2}{k^2}\right) \\
&= \frac{\Tr}{3\pi^2} \left[4\l_p\Lambda M^2 - 2(2\l_p M^2)^\frac{3}{2}\left(\arctan \frac{\Lambda}{\sqrt{2\l_pM^2}}\right) \right. \\
&\quad\qquad \left. + \Lambda^3\log\left(1 + \frac{2\l_pM^2}{\Lambda^2}\right)\right].
\end{split} 
\ee

Switching to dimensionless variables and couplings, we find the following effective potential:
\begin{equation} \label{eq:effective action}
\begin{split}
V_{\rm eff}[\Mt]\equiv \frac{S[\Mt = \text{const.}]}{N^2\Lambda^3 {\rm Vol}}= & \frac{1}{4\lt_p}\left[\Tr(\Mt^2) + \frac{b}{(1-b)N}(\Tr \Mt)^2\right] \\
& - \frac{1}{3\pi^2}\Tr\left[2 \Mt^2 - 2\abs{\Mt}^3\arctan{\frac{1}{\abs{\Mt}}} + \log\left(1 + \Mt^2\right)\right] \;,
\end{split}
\end{equation}
where we defined
\begin{equation}
\Mt \equiv \frac{\sqrt{2\l_p} M }{\Lambda}, \quad \lt \equiv \l \Lambda, \quad \lt_p \equiv \l_p \Lambda, \quad {\rm Vol}=\int d^3 x \;.
\end{equation}
Owing to the $U(N)$-invariant form of the effective potential \eqref{eq:effective action}, we can diagonalize the matrix $\Mt$ and recover an effective potential for its set of eigenvalues\footnote{The Vandermonde determinant originating in the change of variables is subleading in $1/N$ (the action is of order $N^3$ and the logarithm of the Vandermonde determinant is of order $N^2$, see \cite{Nguyen:2014mga}), hence it is not included.} 
$\mu_i$, $1\leq i\leq N$:
\begin{align} \label{eq:effective potential}
V_{\rm eff}[\{\mu_i\}]  &= \sum_i \left[\frac{1}{4\lt_p}\mu_i^2 + \frac{1}{3\pi^2}\kappa(\mu_i)\right] - \frac{\lt}{4\lt_p (\lt+\lt_p) N}\left(\sum_i\mu_i\right)^2 \;,\\
\kappa(\mu) &= 2\mu^3\arctan\frac{1}{\mu} - 2\mu^2 - \log(1+\mu^2) \;.
\end{align}
An important point is that the potential \eqref{eq:effective potential} is unbounded from below in the regions $\lt_p<0$ and $\lt_p+\lt<0$ (this is most easily seen by studying special symmetric configurations such as those we will encounter below for the stationary points), therefore considered unphysical.

The only extremum of the potential which preserves all the symmetries of \eqref{eq:action} is the trivial solution $M=0$, for which our potential is normalized such that $V_{\rm eff}[0] = 0$.
Stationary points with non-zero eigenvalues $\mu_i = \mu ~ (1\leq i\leq N)$, spontaneously break the chiral symmetry of \eqref{eq:action} (reflected in the symmetry $M \rightarrow -M$), whereas if the eigenvalues are not all equal, the original $U(N)$ symmetry of \eqref{eq:action} is spontaneously broken as well.

\begin{figure}
\centering
\includegraphics[width=0.7\textwidth]{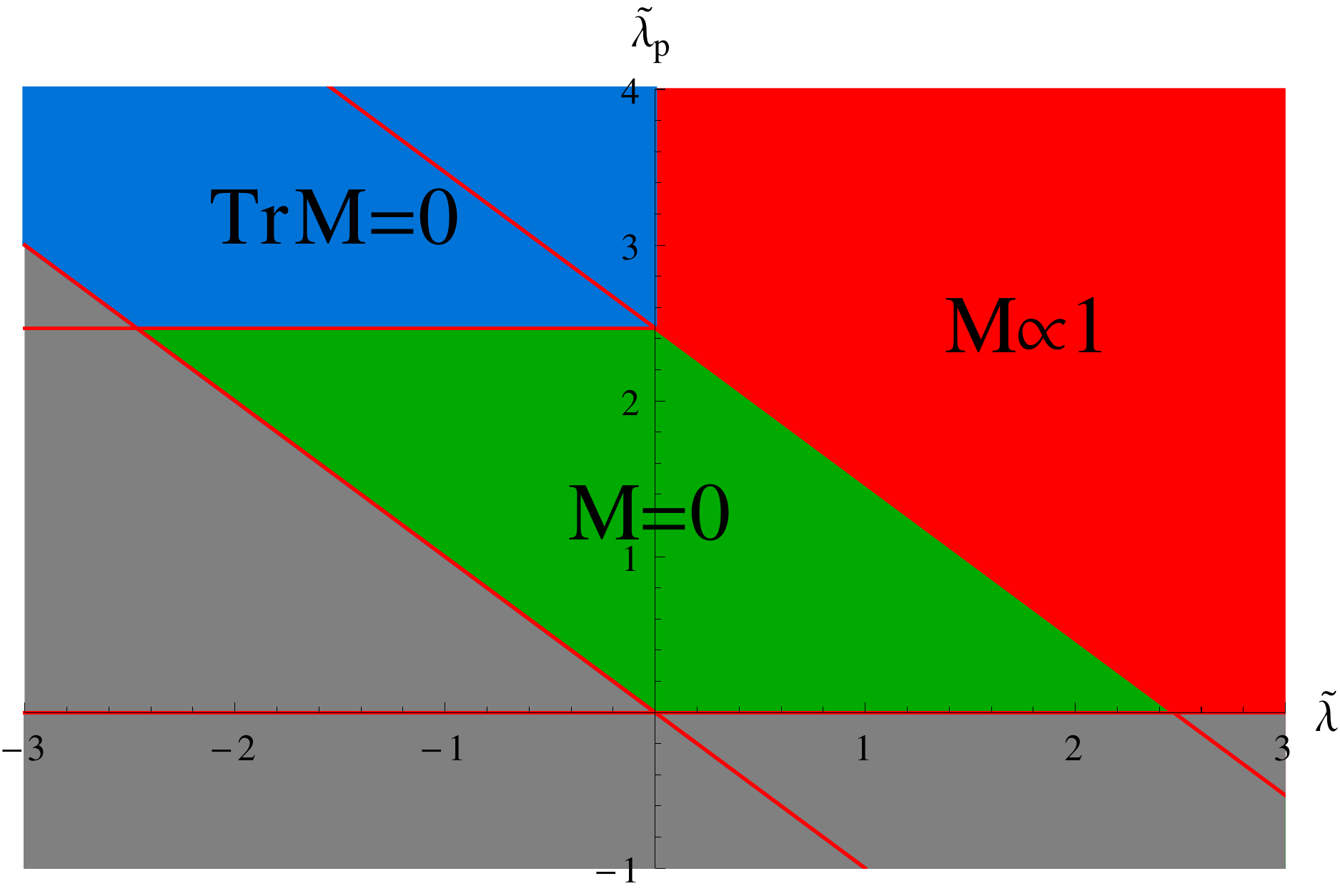}
\caption{\label{fig:phase diagram}Phase diagram of the Fermionic TGN model in 3$d$. The red zone has a vacuum $M\propto \mathbb{1}$, the blue one to a traceless vacuum and the green parallelogram to $M=0$. The grey zone corresponds to an unstable model.}
\end{figure}

In the green parallelogram region of the phase diagram in Fig.~\ref{fig:phase diagram}, we can show that the potential is non-negative, and the solution $\mu = 0$ gives a global vacuum. Indeed, the term in square brackets of eq. \eqref{eq:effective potential} is (for each $i$) non-negative and convex (with a global minimum at the origin) in the range $0<\lt_p<\pi^2/4$. 
Consequently, if $\sum_i\mu_i\neq 0$ and $\lt< 0$ then $V_{\rm eff}[\{\mu_i\}]  > 0 = V_{\rm eff}[0]$. In the case $\lt> 0$, we can use the Cauchy-Schwarz inequality to bound
\begin{equation}
V_{\rm eff}[\{\mu_i\}]  \geq 
\sum_i \underbrace{\left[\frac{1}{4\lt_p}\mu_i^2 + \frac{1}{3\pi^2}\kappa(\mu_i) - \frac{\lt}{4\lt_p (\lt+\lt_p)}\mu_i^2\right]}_{\equiv w(\mu_i)},
\end{equation}
which is convenient, as the eigenvalues decouple. By taking first and second derivatives of each term we can now prove that $\mu_i = 0$ is the unique minimum of $w(\mu_i)$, and hence $\mu_i =  0 ~\forall i$ is the global minimum of $V_{\rm eff}[\{\mu_i\}]$ for $0\leq \lt+\lt_p \leq \pi^2/4$ and $0\leq \lt_p \leq \pi^2/4$:
\begin{align}
w^\prime(x)& = \frac{2x}{\pi^2 \alpha}\left[1 - \alpha + \alpha x \arctan\frac{1}{x}\right] >0 \quad (x>0,\; 0<\a<1),\\
w^{\prime\prime}(x) &= \frac{2}{\pi^2 \alpha}\left[1  -\alpha\f{1+2x^2}{1 + x^2}  + 2\alpha x  \arctan\frac{1}{x}\right] >0 \quad (0<\a<1) \;,
\end{align}
having introduced $\alpha= 4(\lt+\lt_p)/\pi^2$. At $\a>1$ the origin becomes unstable.

The stability of the trivial solution is more properly analyzed by studying the full Hessian.
Coming back to eq.~\eqref{eq:effective action}, we can want to compute the second derivative around the point $\Mt = 0$, for which we can discard the $\arctan$ term, as it is of cubic (and higher) order in the fluctuations. The first derivative gives 
\begin{gather}
\pdv{V}{\Mt_{ij}} = A \Mt_{ji} + B \f{\Tr \Mt}{N} \d_{ij}+ C \Mt_{ji} \; ,\\
A = \f{1}{2\lt_p}\; ,\quad B = \f{b}{2\lt_p(1-b)}\; , \quad C = - \f{2}{\pi^2}\;.
\end{gather}
The second derivative
\begin{equation}
\pdv{V}{\Mt_{ij}}{\Mt_{kl}} = (A + C) \d_{ik}\d_{jl} + B\f{\d_{ij}\d_{kl}}{N}\;,
\end{equation}
can be rewritten as follows
\begin{gather}
H = \a(1 - P)+ \b P \\
\a = A + C\; \quad \b = A + B + C\;,\quad P_{ij,kl} \equiv \frac{\d_{ij}\d_{kl}}{N}\;,
\end{gather}
introducing $P$ that projects on the trace.

Articulated as such, the Hessian $H$ is easy to diagonalize, as the eigenfunctions are easily found to be: \begin{itemize}
\item[-] traceless matrices, with eigenvalue $\a = \f{1}{2}\left(\f{1}{\lt_p} - \f{4}{\pi^2}\right)$,
\item[-] matrices proportional to the identity, with eigenvalue $\b = \f{1}{2}\left(\f{1}{\lt + \lt_p} - \f{4}{\pi^2}\right)$. 
\end{itemize}
This suggests that the trivial solution becomes unstable towards traceless perturbations at $\lt_p~\geq~\pi^2/4$ and towards trace perturbation at $\lt+\lt_p ~\geq~ \pi^2/4$. 
Therefore, the following two particular non-zero solutions are examined:\footnote{Other solutions are possible, as in appendix D of \cite{Benedetti:2017fmp}. In that case it was possible to show that such solutions are never global minima of the potential; here the analysis is more complicated and we limit ourselves to conjecture that the analysis of the following two types of solutions suffices to understand the full phase diagram of the model.}
\begin{itemize}
\item $\mu_i = \mu\neq 0 ~\forall i$: The potential takes the form
\begin{equation}
\label{eq:uniform potential}
\f1N V_{\rm eff}(\mu)=\frac{\mu^2}{4\lt_p}\left(1 + \frac{b}{1-b} - \frac{8\lt_p}{3\pi^2}\right) + \frac{2}{3\pi^2}\abs{\mu}^3\arctan{\frac{1}{\abs{\mu}}} - \frac{1}{3\pi^2}\log(1+\mu^2) \;,
\end{equation}
and the equation of motion that $\mu$ must satisfy is
\begin{equation}
\label{eq: eom uniform}
\mu \arctan{\frac{1}{\mu}} = 1 - \frac{\pi^2}{4(\lt + \lt_p)} \;.
\end{equation}
The range of values of the left-hand side tells us that such a solution exists only for $\lt + \lt_p \geq \pi^2/4$.
\item $\Tr \Mt= 0$: Then the potential reduces to a sum over the eigenvalues. We obtain
\begin{equation}
\label{eq:traceless potential}
V_{\rm eff}[\Mt]=\sum_i v(\mu_i)\;,
\end{equation}
where
\be
v(\mu) = \frac{\mu^2}{4\lt_p}\left(1 - \frac{8\lt_p}{3\pi^2}\right) + \frac{2}{3\pi^2}\abs{\mu}^3\arctan{\frac{1}{\abs{\mu}}} - \frac{1}{3\pi^2}\log(1+\mu^2) \;.
\ee
The equation of motion for $\m_i$ is
\begin{equation}
\label{eq: eom traceless}
\mu_i \arctan{\frac{1}{\mu_i}} = 1 - \frac{\pi^2}{4\lt_p} \;.
\end{equation}
The range of the left-hand side tells us that such a solution exists only for $\lt_p \geq \pi^2/4$.
Furthermore, being an even function, monotonic on each semiaxis, there are only two solutions $\mu_i = \pm \t$. The tracelessness condition finally tells us that the two must come in equal number (for odd $N$ we necessarily have either a zero eigenvalue or a violation of the tracelessness condition, which amounts to a subleading effect in $1/N$).
\end{itemize}

Using the equations of motion, we need to compare the values of the potential at the above critical points:
\begin{equation}
V_{\rm eff}(q) = \frac{\tau(q)^2}{12 q} -\frac{1}{3\pi^2}\log(1 + \tau(q)^2),
\end{equation}
with $q = \lt+\lt_p$ in the uniform case and $q=\lt_p$ in the traceless one, and $\tau(q)$ being the solution of $\t \arctan{(1/\t)} = 1 - \pi^2/(4 q)$. Since $\tau(q)^2$ is a monotonically increasing function, and since as a function of $q$, $V_{\rm eff}(q)$ is decreasing monotonically starting from $0$ (the trivial solution), we conclude that
\begin{equation}
\lt_p > \lt+\lt_p \implies V_{\rm eff}(q_{traceless})<V_{\rm eff}(q_{uniform})
\end{equation}
and reciprocally. In other words, the traceless solution wins over the uniform one for $\lt<0$, while the uniform wins for $\lt>0$.

Such transition can be qualitatively understood in terms of the double-trace term: we see that if $\lt<0$, then the double-trace term comes with a positive sign and has to be minimized, showing why the traceless solution wins (when it exists, i.e.\ for $\lt_p>\pi^2/4$), while if $\lt>0$, then the coefficient of the double-trace term is negative and has to be maximized, leading to the uniform solution.

At last, the phase diagram is as shown in Fig.~\ref{fig:phase diagram}.
Knowing the value of the potential at the different vacua, we see that the transition from red and blue to green are continuous, hence of second order. Indeed, comparing eq. \eqref{eq: eom uniform} with eq. \eqref{eq: eom traceless}, we see that taking the limit $\lt_p+\lt$ or $\lt_p$ to $\pi^2/4$, $\mu$ and $\tau$ decrease monotonically to zero. 
On the other hand, a first order transition separates the two non-trivial phases: at $\lt = 0$ and $\lt_p>\pi^2/4$, they both have same potential energy (but are distinct), and as the coupling $\lt$ grows or decreases, the uniform or traceless solution become global minima, respectively.

\subsection{Schwinger-Dyson equations in the traceless phase}
\label{sec:SD1bis}

In Sec.~\ref{sec:SD1} we derived the gap equation and beta functions from the Schwinger-Dyson equations in the $U(N)\times U(N^2)$-symmetric phase. Although we expect the beta functions to be independent of such choice it is instructive to do an explicit check, now that we discovered a broken phase.

Assuming that the two-point function breaks the $U(N)\times U(N^2)$ symmetry into $U(N/2)^2\times U(N^2)$, with the following ansatz:
\begin{gather}
M_1 = m (\mathbb{1}_N- 2\mathbb{P}_N), 
\quad \mathbb{P}_N = \begin{pmatrix}
\mathbb{0}_{N/2} & \mathbb{0}_{N/2} \\
\mathbb{0}_{N/2} &\mathbb{1}_{N/2}\end{pmatrix},\\ \label{eq:broken-2pt-ansatz}
G^{-1}(p) = i\slashed{p} \,\mathds{1}_1\otimes \mathds{1}_2 \otimes \mathds{1}_3 + M_1\otimes \mathds{1}_2 \otimes \mathds{1}_3 \;,
\end{gather}
the Schwinger-Dyson equation becomes
\begin{equation}
\label{eq:SDE U(N)-broken}
G^{-1}(p) =  i\slashed{p} \mathds{1}_1\otimes \mathds{1}_2 \otimes \mathds{1}_3 + 2  \f{\l_p}{N^2}\int\f{\dd[3]q}{(2\pi)^3}  (\Tr_{\backslash 1} \tr \left[ G(q)  \right])\otimes \mathds{1}_2 \otimes \mathds{1}_3\;,
\end{equation}
where $\Tr_{\backslash c}$ is a trace on all indices except the one of color $c$.
The coupling $\l$ is missing from the equation because it multiplies a full trace of $G(q)$, which is zero for the ansatz above.
Forgetting momentarily the trace over the $\g$-matrices, we have
\begin{equation}
\Tr_{\backslash 1} \left[G(q)\right] = N^2 \f{-i\slashed{p} + M_1}{p^2+m^2}\;,
\end{equation}
such that the Schwinger-Dyson equation reduces to
\begin{equation}
M_1 = 2\l_p \int \f{\dd[3]q}{(2\pi)^3} \f{\tr(-i\slashed{p} + M_1)}{p^2+m^2}\;,
\end{equation}
or the mass gap equation
\begin{equation}
m = 8\l_p\int \f{\dd[3]q}{(2\pi)^3} \f{m}{p^2+m^2}\;,
\end{equation}
that is, nothing more than eq.~\eqref{eq:gap-1c} with $\l = 0$, thus leading directly to \eqref{eq:beta-1c}. 

\subsection{Anomalous dimension}

We have three non-trivial fixed points (out of which one has two relevant directions while the others have one), and for each of them we can compute the conformal dimension of the intermediate field. It happens, that in all cases, the computation is almost unchanged and gives the same result:

\begin{itemize}

\item $(\lt,\lt_p)=(\pi^2/4,0)$

This is the UV fixed point of the usual GN model. The limit $\l_p\to 0$ constrains to zero the traceless part of the intermediate field \cite{Benedetti:2017fmp}, and thus
it is equivalent to starting from the action \eqref{eq:GNY}. Integrating out the Fermions:
\be
S_{\rm int}[\s] = N^3 \int d^3 x \;\left( \frac{1}{4\l}\sigma^2 - \log\left(\slashed{\partial} + \s \right)\right).
\ee
At the fixed point, $\la \s\ra =0$ and the inverse propagator is obtained by the second functional derivative with respect to $\s$, computed at $\s=0$, i.e.:
\begin{equation} \label{eq:effective_prop_case1}
\f{1}{N^3}\la\s(p)\s(-p)\ra^{-1} = \frac{1}{2\l} - \tr \int\frac{\dd[3]q}{(2\pi)^3}\frac{\slashed{q}(\slashed{q}-\slashed{p})}{q^2(q-p)^2}
=\frac{1}{2\l} - 4 \int\frac{\dd[3]q}{(2\pi)^3}\frac{q^2-q\cdot p}{q^2(q-p)^2}.
\end{equation} 

The last integral can be computed as
\be
\begin{split}
4 \int\frac{\dd[3]q}{(2\pi)^3}\frac{q^2-q\cdot p}{q^2(q-p)^2} &= \f{2}{\pi^2} \int_0^\L dq - \f{2p^2}{(2\pi)^2} \int_0^{+\infty} dq \int_{-1}^{+1} d(\cos\theta) \f{1}{q^2+p^2-2 q p \cos\theta}\\
&= \f{2}{\pi^2} \L - \f{p}{4} \;.
\end{split}
\ee
The linear divergence is cancelled by the fixed point condition $\l = \frac{\pi^2}{4\Lambda}$, thus yielding
\be
\la\s(p)\s(-p)\ra =  \f{4}{N^3 p} \;,
\ee
corresponding to a conformal dimension $\D_\s=1$, which is also the dimension of $\psib_{abc}\psi_{abc}$.

\item $(\lt,\lt_p)=(0,\pi^2/4)$

Let us recall the effective action of the intermediate field 
\begin{align}
S_{\rm int}[M] = &\int \frac{1}{2}M^*_{ij}K_{ij;kl}M_{kl} - N^2 \tr \Tr\left[\log\left(\slashed{\partial} + \frac{\sqrt{2\l_p}}{N}M\right)\right],\\
&K_{ij;kl} = \delta_{ik}\delta_{jl} + \frac{b}{(1-b)N}\delta_{ij}\delta_{kl}.
\end{align}
It is convenient to introduce again the rescaled matrix $\Mt = \frac{\sqrt{2\l_p}}{N}M$. Derivating twice with respect to an eigenvalue $m_i$ of $\Mt$, and setting $b=0$, gives, after a Fourier transform
\begin{equation}
\f{1}{N^2} \frac{\delta^2 S_{\rm int}}{\delta m^2}\big|_{m=0} = \frac{1}{2\l_p} -\tr \int\frac{\dd[3]q}{(2\pi)^3}\frac{\slashed{q}(\slashed{q}-\slashed{p})}{q^2(q-p)^2}\;,
\end{equation} 
namely the same expression as \eqref{eq:effective_prop_case1}, with $\l_p$ replacing $\l$.
The linear divergence is cancelled by the fixed point condition, $\frac{1}{2\l_p} = \frac{2\Lambda}{\pi^2}$, and we arrive at the same propagator (hence same conformal dimension) as in the usual three-dimensional Gross-Neveu model.

\item $(\lt,\lt_p)=(-\pi^2/4,\pi^2/4)$

Here, because $b=1$ is a singular point of $K$, we need to take a few steps back \cite{Benedetti:2017fmp}. The Fermionic interaction action was written with a matrix-like field $B_{ij} = \bar{\psi}_{ibc}\psi_{jbc}$ as
\begin{align}
&S_{\rm int} = -\frac{\l_p}{N^2}\int B^*_{ij}C_{ij;kl}B_{kl}\;,\\
C_{ij;kl} = \delta_{ik}\delta_{jl} - \frac{b}{N}\delta_{ij}\delta_{kl} &= (\bm{1} - \bm{P})_{ij;kl} + (1 - b)\bm{P}_{ij;kl}\;, \qquad \bm{P}_{ij;kl} = \frac{\delta_{ij}\delta_{kl}}{N}\;.
\end{align}
Since $\bm{P}$ projects on the trace part of the matrix, it appears clearly that $b=1$ restricts us to work with a traceless $B$. Except for this constraint on the fields (which then follows for the matrix-like intermediate field $M$ \footnote{The trace of $M$ couples to that of $B$ and its effective action will be identical to eq. \eqref{eq:S_M}, except for an absent double-trace term.}), the computation of the effective propagator will be identical to other two cases above.
\end{itemize}

\section{$U(N)\times U(N)\times U(N)$-symmetric model}
\label{S:2}

Adding to the model defined in \eqref{eq:action}-\eqref{eq:S_int1} other pillow interactions which differ by simultaneous permutations of the three tensor indices of all the fields, we break the $U(N)\times U(N^2)$ symmetry down to $U(N)\times U(N)\times U(N)$. In the tensor model literature it is usual to refer to an index location (first, second, or third index, in our case) as a color, and hence such permutations of indices are called color permutations. There are three distinguishable colorings for the pillow interaction, one for each choice of {\emph{transmitted color}}, i.e.\ for each choice of index being associated to the vertical lines of Fig.~\ref{fig:interactions}. Considering that of course there is only one coloring for the double-trace interaction, we have in general four independent couplings. We will restrict the theory space by demanding \emph{color symmetry} of the action, i.e. invariance under permutations of the indices, thus writing for the new interacting part of the action
\be \label{eq:S_int123}
S_{\rm int}[\psi,\psib]= -\frac{\l}{N^3}  \int d^3 x \; (\bar{\psi}_{abc}\psi_{abc})^2 - \frac{\l_p}{N^2} \sum_{\ell =1}^3  \mathcal{P}_\ell[\psi,\psib] \;,
\ee
where $\mathcal{P}_\ell[\psi,\psib]$ is the pillow interaction with transmitted color $\ell$.

\subsection{Schwinger-Dyson equations and $\beta$-functions}
\label{sec:SD123}

\begin{figure}
\centering 
\includegraphics[width=0.7\textwidth]{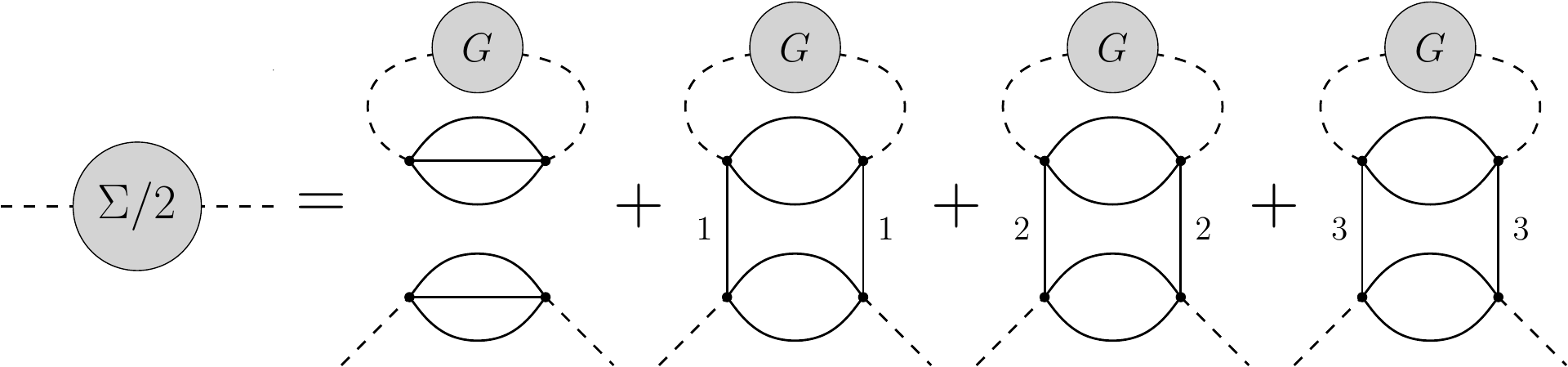}
\caption{\label{fig:SDE-3c}Schwinger-Dyson equation at large $N$ for the self-energy.}
\end{figure}

\begin{figure}
\begin{minipage}{0.5\textwidth}
  	\centering 
	\includegraphics[width=0.8\textwidth]{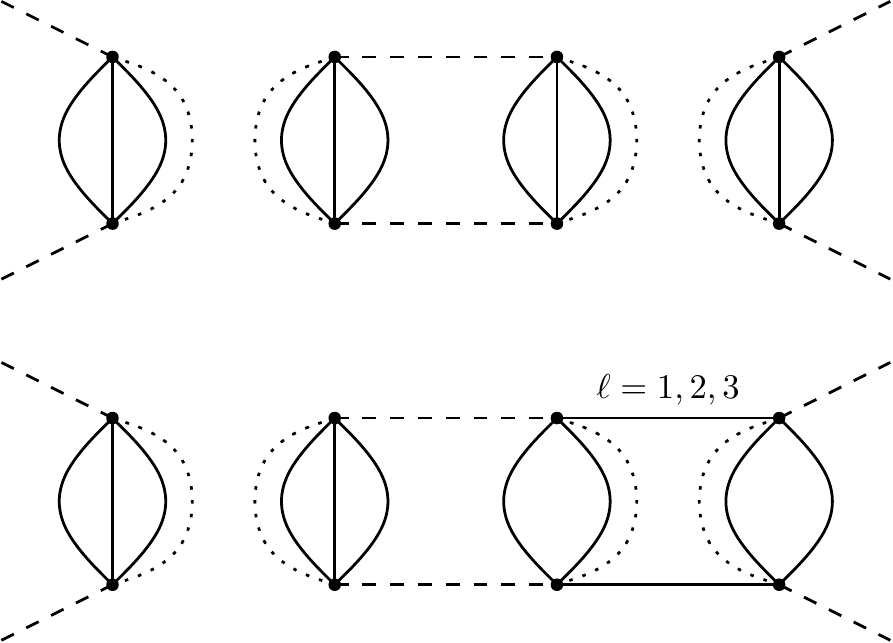}
        \end{minipage}
\begin{minipage}{0.5\textwidth}
      	\centering
	\includegraphics[width=0.8\textwidth]{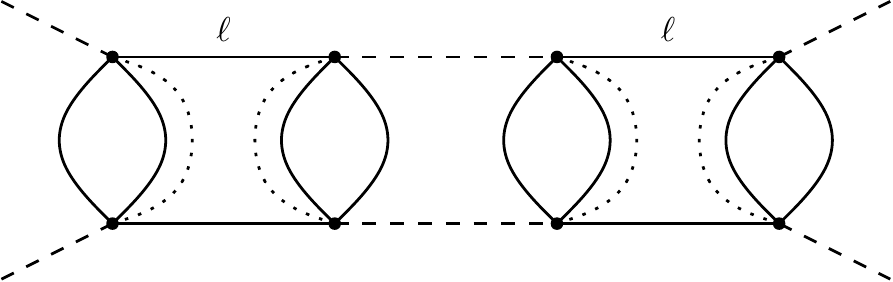}\\	
	\vspace{.6cm}
	\includegraphics[width=0.8\textwidth]{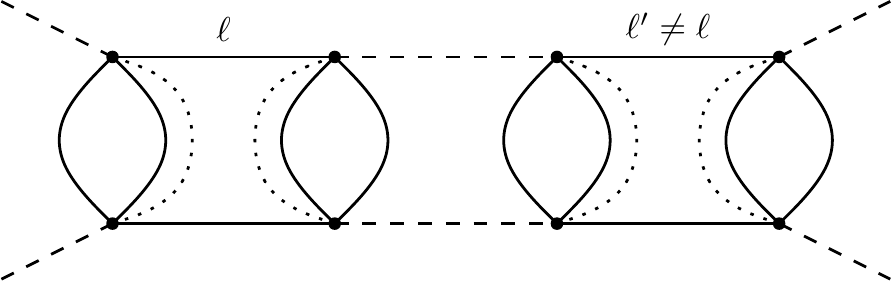}
        \end{minipage}
\caption{\label{fig:4-point graphs-3c}Leading order graphs renormalizing the couplings $\lt_p$ (top right) and $\lt$ (all the others).}
\end{figure}

Following \cite{Benedetti:2017fmp}, the SD equations in momentum space write as \footnote{$\Tr$ is a trace on all the color indices, while with $\Tr_{\backslash c}$ the color $c$ is not traced on. As before, $\tr$ is a trace on the $\g$-matrix space.}
\begin{equation}
\label{eq:SDE-3c}
G^{-1}(p) = G^{-1}_0(p) + 2 \int\f{\dd[3]q}{(2\pi)^3}  \tr \left[\f{\l}{N^3} \Tr G (q) + \f{\l_p}{N^2} \left(\Tr_{\backslash 1} G(q)  + \Tr_{\backslash 2} G(q) +\Tr_{\backslash 3} G(q)\right)\right]\;,
\end{equation}
as also depicted in Fig.~\ref{fig:SDE-3c}.
Assuming that the $U(N)$ symmetry is unbroken by the vacuum, an ansatz for the full propagator is again given by \eqref{eq:2pt-sym-ansatz}, with the diagonal $\hat{G}(p)^{-1}= i\slashed{p} + m $; thus, similarly to subsection \ref{sec:SD1}, the gap equation is
\begin{equation}
m = 2(\l + 3\l_p)\int\frac{\dd[3]q}{(2\pi)^3} \frac{\tr(-i\slashed{q} + m)}{q^2 + m^2}, 
\end{equation}
or
\begin{equation}
\frac{1}{\l+3\l_p}=8 \int_{\abs{p}<\Lambda}\frac{\dd[3]p}{(2\pi)^3}\frac{1}{p^2+m^2} = \frac{4}{\pi^2}\left(\Lambda - m \arctan\left(\frac{\Lambda}{m}\right)\right).
\end{equation}
In terms of the dimensionless couplings ($\lt_p \equiv \Lambda\l_p$ and $\lt \equiv \Lambda\l$) and with $\kappa \equiv 4/\pi^2$, the $\beta$-functions read 
\be
\label{eq:beta-unbroken}
\beta+3\beta_p = \left(\lt + 3 \lt_p\right) - \kappa\left(\lt + 3 \lt_p\right)^2 \;.
\ee
By direct inspection of the one loop diagrams at leading order in $1/N$, depicted in Fig.~\ref{fig:4-point graphs-3c}, we can disentangle the two beta functions, obtaining:
\begin{align} \label{eq:beta-3c}
\beta &=  \lt - \k\left(\lt^2 + 6\lt \lt_p + 6\lt_p^2\right)\;,\\
\beta_p &= \lt_p - \kappa\lt_p^2 \;.
\end{align}

With the experience of the previous section, we can also consider the case of broken $U(N)$ symmetry, and disentangle the two beta functions by combination of the symmetric and broken results. 
Anticipating the results of section \ref{sec:potential-3c}, it turns out that there is a broken phase where we can make precisely the same ansatz as in \eqref{eq:broken-2pt-ansatz}. The calculation then proceeds exactly as in that section, after having noticed that in \eqref{eq:SDE-3c} the $ \Tr_{\backslash 2}$ and $ \Tr_{\backslash 3}$ terms vanish because the trace on color 1 is zero.
Therefore, the beta function for $\l_p$ is unchanged, and combining it with that of the unbroken case (eq. \eqref{eq:beta-unbroken}), we find again \eqref{eq:beta-3c}, as expected.

It is also interesting to point out that $\b + 2\b_p = 0$ along $\l+2\l_p=0$.

\begin{figure}
\centering
\includegraphics[width=0.6\textwidth]{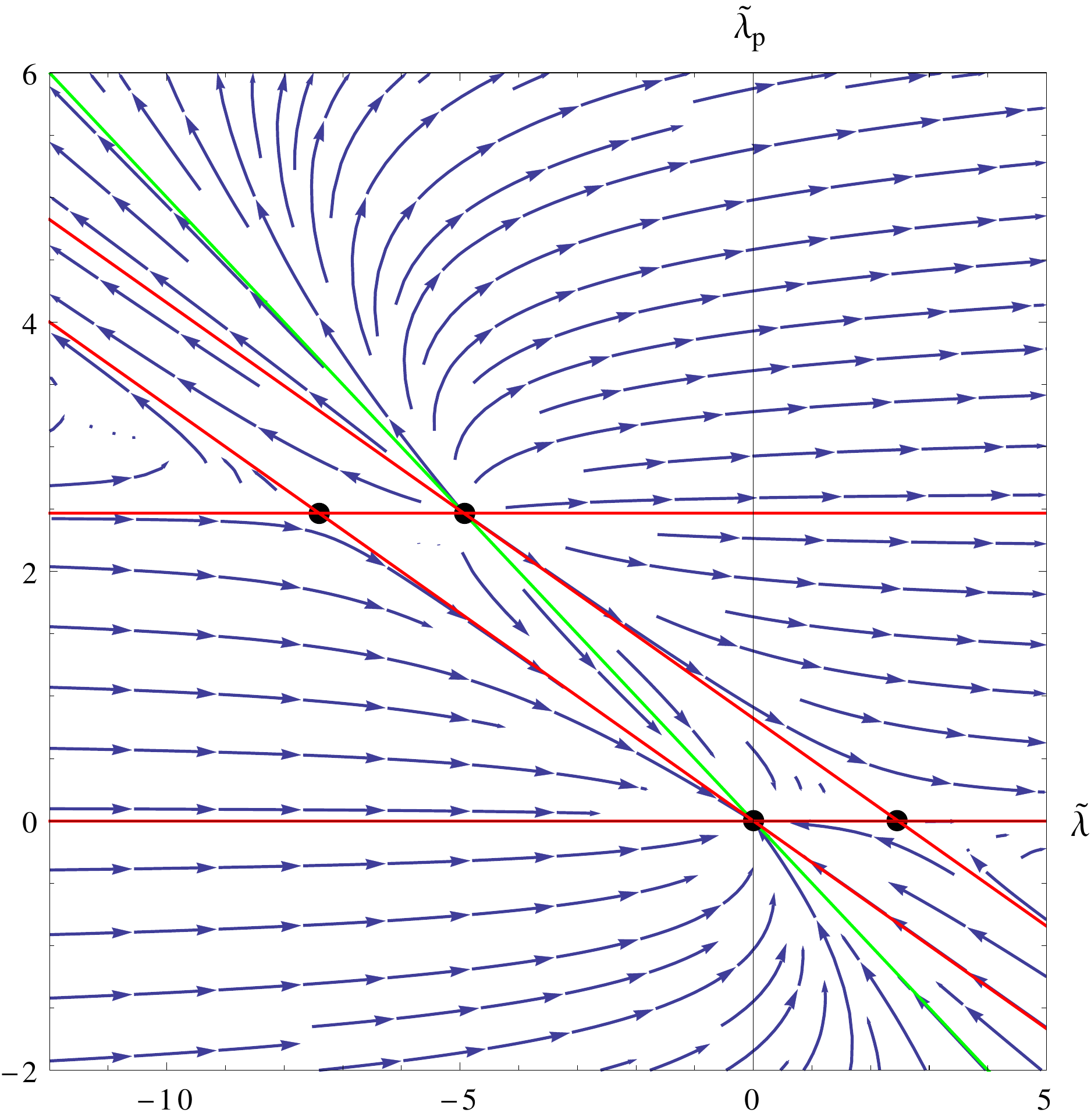}
\caption{Renormalization group flow in the $\left(\lt,\lt_p\right)$-plane. Arrows point towards the $IR$, black dots denote the fixed points, and red lines mark the zeros of either $\beta_p$ or $\beta+3\beta_p$. The green line corresponds instead to $\beta+2\beta_p=0$, which morally replaces the vertical line $\beta=0$ of Fig.~\ref{fig:RG-flow-Fermions}.}
\label{fig:RG-flow-3c}
\end{figure}

We picture the resulting flow on Fig.~\ref{fig:RG-flow-3c}, seemingly a distorted version of Fig.~\ref{fig:RG-flow-Fermions}.
In fact the critical exponents are also the same as in the previous model.

\subsection{Effective potential and phase diagram}
\label{sec:potential-3c}

Rewriting all the quartic interactions in terms of intermediate fields, the action takes the following form
\begin{gather}
\label{eq:action_cIF}
S[M_1,M_2,M_3] = \int \frac{1}{2} \sum_{c=1,2,3}\left[\Tr(M_c^2) + \frac{b}{(1-b)N}(\Tr M_c)^2\right] - \tr \Tr\left[ \ln\left(\slashed{\partial} + \frac{\sqrt{2\l_p}}{N} R\right)\right]\\
R \equiv M_1\otimes \mathds{1}_2 \otimes \mathds{1}_3 + \mathds{1}_1 \otimes M_2 \otimes \mathds{1}_3 + \mathds{1}_1 \otimes \mathds{1}_2 \otimes M_3\;; \quad b = -\f{\l}{3\l_p},
\end{gather}
with the three intermediate fields $M_1, M_2, M_3$ needed for the three pillow interaction terms. The original GN-interaction is split into three identical parts, thus in effect changing $\l \rightarrow \l/3$ in each of the $(\Tr M_c)^2$ terms.
Again, coming to adimensional variables, rescaling such that $\Mt_i \equiv N \L^{3/2}M_i, ~ \forall i$, and using the $U(N)$ symmetry to diagonalize, we arrive at
\be
\begin{split}
V &\left[\{\mu_{1,i}, \mu_{2,j} , \mu_{3,k}\}\right] \equiv \f{S\left[\Mt_1,\Mt_2,\Mt_3\right]_{|_{\Mt_i={\rm const.}}}}{N^3\L^3 {\rm Vol}} \\
&\quad = \f{1}{N}\sum_{c} \frac{1}{4\lt_p} \left[\sum_i\mu_{c,i}^2 - \frac{\lt}{(\lt+3\lt_p) N}\left(\sum_i\mu_{c,i}\right)^2 \right]\\
& \qquad \qquad + \f{1}{N^3}\sum_{1\leq i,j,k\leq N}\frac{1}{3\pi^2}\kappa(\mu_{1,i},\mu_{2,j},\mu_{3,k})\;,
\end{split}
\ee
\begin{gather}
\kappa(\mu_{1,i},\mu_{2,j},\mu_{3,k}) = 2\mu^3\arctan\frac{1}{\mu} - 2\mu^2 - \log(1+\mu^2), \\
\mu \equiv \mu_{1,i}+\mu_{2,j}+\mu_{3,k} \;.
\end{gather}
We are looking for the vacua of this potential. The equations of motion for $\mu_{c,i}$ read 
\begin{equation}
\f{1}{2\lt_p}\left[\m_{c,i} - \f{\lt}{(\lt +3 \lt_p)N}\sum_j\m_{c,j}\right] + \f{2}{\pi^2N^2}\sum_{1\leq j,k\leq N}\left[\mu^2\arctan\f{1}{\mu} - \mu\right] = 0.
\label{eq:eom_3c}
\end{equation}

To begin, a useful remark is that the three different color-intermediate fields must have the same trace at the saddle points. Indeed, this is seen by summing eq. \eqref{eq:eom_3c} over $i$. Because the second term in squared brackets depends only on $\mu$, all colors end up with the same equation (of the type $\Tr[\Mt_c]=F[\{\m\}]$, with the same right-hand side), hence the equality of traces. 

Another straightforward observation is that $M_i = 0 ~ \forall i$ is always a solution. In order to find other solutions, it is helpful to search for unstable directions around this point, i.e. analyse the Hessian of the potential. 

Developing around the point $R = 0$ allows to discard the $\arctan$ term, of higher order. The first derivative gives\footnote{The exponent $(c)$ in $\d_{ij}^{(c)}$ only serves to keep track of what color the indices belong to.}
\begin{gather}
\pdv{V}{M_{c,ij}} = (A +C) M_{c,ji} + B \f{\Tr M_{c}}{N} \d_{ij}^{(c)} + \sum_{c^\prime \neq c} C \f{\Tr M_{c^\prime}}{N} \d_{ij}^{(c)}\; ,\\
A = \f{1}{2\lt_p}\; ,\quad B = \f{b}{2\lt_p(1-b)}\; , \quad C = - \f{2}{\pi^2}\;.
\end{gather}
The second derivative is 
\begin{equation}
 \pdv{V}{M_{c,ij}}{M_{c^\prime,kl}} = \d_{cc^\prime}\left((A + C) \d_{ik}\d_{jl} + B\f{\d_{ij}\d_{kl}}{N}\right) + C\f{\d_{ij}^{(c)}\d_{kl}^{(c^\prime)}}{N},
\end{equation}
and can be rewritten as a matrix in color-space
\begin{gather}
H = 
\begingroup
\renewcommand*{\arraystretch}{1.5}
\begin{pmatrix}
\a(1 - P_1)+ \b P_1 & C \f{\Pi_1\Pi_2}{N} &  C \f{\Pi_1\Pi_3}{N}\\
C \f{\Pi_2\Pi_1}{N} &\a(1 - P_2)+ \b P_2 & C \f{\Pi_2\Pi_3}{N}\\
C \f{\Pi_3\Pi_1}{N} & C \f{\Pi_3\Pi_2}{N} &\a(1 - P_3)+ \b P_3
\end{pmatrix},
\endgroup \\
\a = A + C\; \quad \b = A + B + C.
\end{gather}
We defined the projector $P_c$ on the trace of a matrix of given color $c$
\begin{equation}
P_{c;ij,kl} \equiv \frac{1}{N}\Pi_{c,ij}\Pi_{c,kl}, \quad \Pi_{c,ij}\equiv \d_{ij}^{(c)}.
\end{equation}
Articulated as such, the Hessian $H$ is easy to diagonalize\footnote{Using a simplified notation in which, for instance, an element $a$ on the second line of the column vector is intended to represent the element $\mathbb{1} \otimes a \otimes \mathbb{1}$, and analogously with the other lines.}:
\begin{itemize}
\item we have the set of eigenvectors associated to traceless matrices, of form 
\begin{equation}
E^1 = \begin{pmatrix}Q\\0\\0\end{pmatrix}, \quad E^2 = \begin{pmatrix}0\\Q\\0\end{pmatrix},\quad E^3 = \begin{pmatrix}0\\0\\Q\end{pmatrix}, \quad \Tr Q = 0,
\end{equation}
with eigenvalue $\alpha = \f{1}{2}\left(\f{1}{\lt_p} - \f{4}{\pi^2}\right)$ ; 
\item the eigenvector associated to matrices proportional to the identity, of the form
\begin{equation}
E^s = \g\begin{pmatrix}1\\1\\1\end{pmatrix},
\end{equation}
with eigenvalue $\b + 2C = \f{3}{2}\left(\f{1}{\lt + 3\lt_p} - \f{4}{\pi^2}\right)$ ; 
\item and finally the are eigenvectors
\begin{equation}
e^1 = \g_1 \begin{pmatrix}1\\-1\\0\end{pmatrix}, \quad e^2 = \g_2 \begin{pmatrix}1\\0\\-1\end{pmatrix},
\end{equation}
with eigenvalue $\b - C = \f{3\lt_p}{\lt + 3\lt_p}$.
\end{itemize}
The first two correspond to instabilities that decrease the potential (in the ranges where the eigenvalues become negative), while the last always increases it for the range of couplings allowed by the requirement of boundedness of the potential (and in addition they do not satisfy the equations of motion, because the traces of the three matrices are not equal).

\begin{figure}
\centering
\includegraphics[width=0.7\textwidth,height=0.48\textwidth]{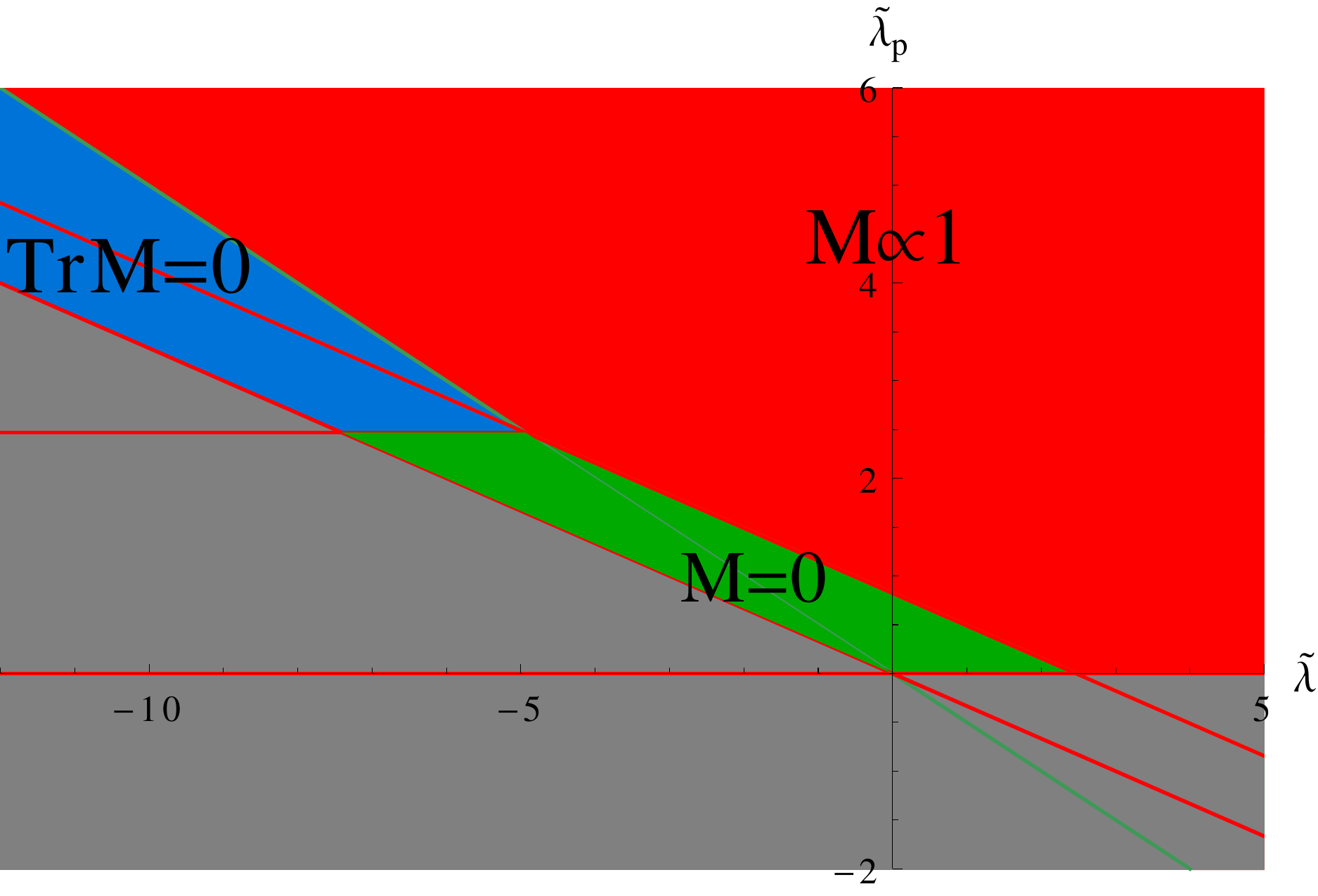}
\caption{Phase diagram of the color-symmetric Fermionic TGN model in 3$d$. The color code is the same as in Fig. \ref{fig:phase diagram}.}
\label{fig:phase-diagram-3c}
\end{figure}

In light of this analysis, we can constrain the form of the intermediate field in our search of new minima of the potential: 
\begin{itemize}
\item Assuming $M_i = m\mathbb{1}_N ~ (i=1,2,3)$, the equations of motion imply for $m$ the equation
\be
1 - 3m\arctan\f{1}{3m} = \f{\pi^2}{4} \f{1}{\lt+3\lt_p},
\label{eq:constraint-m}
\ee
allowing such solutions to exist only for $\lt+3\lt_p>\pi^2/4$. After using eq. \eqref{eq:constraint-m} to remove the $\arctan$ term, their potential energy is given by
\be
V[m] = \f{1}{12 (\lt+3\lt_p)}(3m)^2 - \f{1}{3\pi^2}\log\left(1+(3m)^2\right),
\label{eq:potential-m-3c}
\ee
corresponding to a non-trivial minimum as soon as $\lt + 3\lt_p>\pi^2/4$.
\item Assuming that the intermediate fields are traceless reduces the equations of motion \eqref{eq:eom_3c} to the following:
\begin{equation}
\mu_{1,i} \left(\f{\pi^2}{4\lt_p} - 1\right) + \f{1}{N^2}\sum_{jk}(\mu_{1,i}+\mu_{2,j}+\mu_{3,k})^2\arctan\f{1}{\mu_{1,i}+\mu_{2,j}+\mu_{3,k}}=0\; .
\label{eq:eom_traceless}
\end{equation}
Taking only one non-trivial matrix $M_1 = w(\mathbb{1}_N-2 \mathbb{P}_N)$, and $M_2 = M_3 = \mathbb{0}_N$, $w$ is hold by eq. \eqref{eq:eom_traceless} to 
\be 
1 - w \arctan\f{1}{w} = \f{\pi^2}{4\lt_p}\;,
\label{eq: single-traceless-eom}
\ee
while the potential energy is 
\be
V[w] = \f{1}{12 \lt_p}w^2 - \f{1}{3\pi^2}\log\left(1+w^2\right),
\label{eq:potential-w-3c}
\ee
again corresponding to a non-trivial minimum for $\lt_p>\pi^2/4$.

The cases of two and three traceless matrices are examined in App. \ref{app:traceless_matrices}, where we show in detail that the single-traceless potential dominates double- and triple-traceless solutions above $\lt_p = \pi^2/4$.
\end{itemize}
The potentials of eq. \eqref{eq:potential-m-3c} and \eqref{eq:potential-w-3c} compare easily: $m$ and $w$ are controlled by similar equations, and they grow monotonically with the coupling. Consequently, we have that $V[w]<V[m]$ for $\lt + 3\lt_p>\lt_p$ or $\lt+2\lt_p>0$. In other words, the green line outlined in Fig. \ref{fig:RG-flow-3c} and \ref{fig:phase-diagram-3c} sets apart the traceless vacuum from the traceful one.

\

Let us summarize the results.
Integrating out the Fermions, we obtained the effective potential of the intermediate fields. The associated equations of motion were too difficult to solve in full generality. Nevertheless, after a stability analysis around the trivial vacuum, showing that the unstable perturbations beyond some critical lines are given by traceless and pure trace modes, we explored those two types of solutions (i.e.\ traceless matrices and matrices proportional to the identity) and showed in what range of the couplings they minimized the energy. The conclusion is very similar to that of the previous section. Indeed, three phases appear: the trivial, traceful or traceless solutions dominate in turn, as displayed on the phase diagram of Fig. \ref{fig:phase-diagram-3c}. However, in the last case, a startling new feature is the breaking of color-symmetry. The transitions between trivial and traceful or traceless are continuous, while a discontinuous transition separates the non-trivial phases.

\section{Coupling to Chern-Simons}
\label{sec:CS}
Motivated by the well-entrenched higher-spin/gauged vector model duality \cite{Giombi:2016ejx}, we were curious to look for similar patterns with tensor models. In particular, gauging three dimensional Bosonic or Fermionic vector fields with Chern-Simons has lead to non-trivial solvable flows and phase structures at large-$N$, and dualities between the Bosonic and Fermionic theories. For a sample of works, see \cite{Giombi:2011kc,Aharony:2011jz,Jain:2013py}. The advantage of coupling to Chern-Simons is that its coupling is quantized hence doesn't run under the RG flow \cite{Chern-Simons}. In the footsteps of \cite{Giombi:2011kc}, we introduce the light-cone coordinates
\begin{equation}
p^{\pm} = p_{\mp} = \f{p^1 \pm i p^2}{\sqrt{2}}, 
\end{equation}
and similarly for other letters. The partition function reads:
\begin{equation}
Z = \int \cD A \cD \psi \cD\bar{\psi} e^{-S}\;, 
\end{equation}
where the action is
\begin{align}
S =& -\f{\im k}{2\pi}\sum_{c = 1,2,3}\int \f{\dd[3]p}{(2\pi)^3} \Tr\left(A^{(c)}_3(-p) p_-A^{(c)}_+(p)\right) + \int \f{\dd[3]p}{(2\pi)^3}  \bar{\psi}_{ijk}(-p)\left(\im\slashed{p} + m\right) \psi_{ijk}(p) \\
& -\im \int \f{\dd[3]p}{(2\pi)^3}\int \f{\dd[3]q}{(2\pi)^3} \bar{\psi}_{ijk}(-p)\left[\g^+ A^{(1)ii^\prime}_+(-q) + \g^3A^{(1)ii^\prime}_3(-q)\right] \psi_{i^\prime jk}(p+q) \\
&+ (i\rightarrow j\rightarrow k \rightarrow i ~;~1\rightarrow2\rightarrow3)+ S_{pillow}[\psi]\;,
\end{align}
with $A^{(c)}_\mu$ hermitian matrices generating $U(N)$. Integrating out the gauge field of color $1$ brings an additional, momentum dependent, quartic pillow interaction to the Fermions:
\begin{equation}
\label{eq:effective-pillow}
\f{2\pi i}{k} \int \f{\dd[3]p~\dd[3]q~\dd[3]r}{(2\pi)^9} \f{1}{q^+} \psib_{ijk}(-p)\g^+\psi_{i^\prime jk}(p-q) \psib_{i^\prime j^\prime k^\prime}(-r) \g^3 \psi_{ij^\prime k^\prime}(r+q)
\end{equation}
and analogously for the other colors. Thus, in order to have a proper large $N$ limit, we take $k \equiv \f{N^2}{\lambda}$, with $\lambda$ fixed.

Propagators in the lightcone gauge are given by
\begin{gather}
\expval{\psi_{ijk}(p)\bar{\psi}_{lmn}(-q)}_0 =  \f{-i \slashed{p}+m}{p^2+m^2} \d(p-q) \d^{ijk}_{lmn}=:\cG_0(p)\;, \\
\expval{A^{(c)ij}_\mu(p)A^{(c^\prime)mn}_\nu(-q)}_0 = (2\pi)^3\d(p-q)G_{0,\mu\nu} \d^{cc^\prime} \d^{in}\d^{jm}\;,  \\
G_{0,+3}(p) = -G_{0,3+}(p) = \f{4\pi i}{k p^+}\;.
\end{gather}

\subsection{Free and interacting Fermions}
\begin{figure}
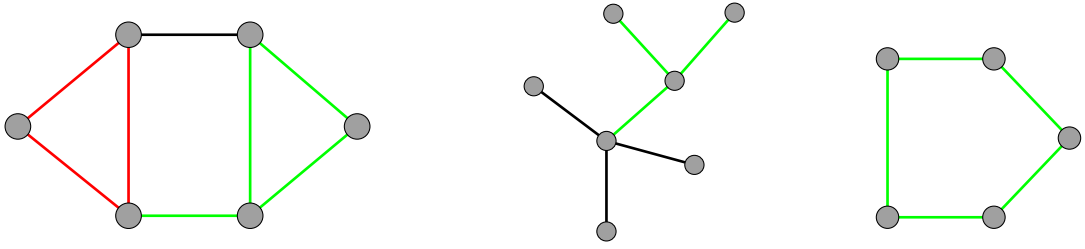

\centering 
\captionsetup[subfigure]{labelformat=empty}
\subfloat[]{\tikzsetnextfilename{Example-Loop-TGN.tex3}
\input{Example-Loop-TGN.tex}}
\hspace{.5cm}
\subfloat[]{\tikzsetnextfilename{LO-tree-TGN3}
\input{LO-tree-TGN.tex}}
\hspace{.5cm}
\subfloat[]{\tikzsetnextfilename{NLO-tree-TGN2}
\input{NLO-tree-TGN.tex}}
\caption{\label{fig:LOandNLOgraphs}Different diagrams, to illustrate the counting of $N$. The grey dots represent the Fermion degrees of freedom, while the colored lines stand for gauge field propagators. On the left, the red and green edges form a loop each, the green ones a branching loop and the black one a string. Isolated vertices are neither connected to a string, nor to a branching loop -- ($l_r = l_v = 1, l_b = 0, v_r = 3, v_g = 2, v_b = 4$). Examples of leading (center) and next-to-leading order graphs (right).}
\end{figure}

First, let us assume that the Fermions don't have any self-interaction in $S_{pillow}$. From the stranded representation, the connected diagrams can be classified with respect to $N$ as follows. Each gauge propagator brings a factor $1/N^2$ and each loop (alternating gauge field and Fermion lines, or purely Fermionic), also called face, gives a factor of $N$. For a given color $i$, in order to count the total factor of $N$, one removes the edges of different color and counts: $l_i$ loops strings ($s$) and $v_i$ isolated vertices, illustrated on diagrams of Fig.~\ref{fig:LOandNLOgraphs}. 
Writing the total number of edges $E = \sum_i E_i$, with $E_i$ edges of color $i$, then the factor of $N$ associated to a given diagram is
\begin{equation}
\label{eq:Nfactor}
N^{-2E + \sum_i (L_i + v_i)}\;,
\end{equation}
introducing $L_i = \sum_{c = 1}^{C_i} (l_{i,c} + 1)$, the sum being made on each connected component (after removal of the other colors). Using the well-known relation $l_i = E_i - V_i + C_i$ where $V_i$ counts all vertices of color $i$ in connected components (differing from a single point) i.e. $v_i = V - V_i$, we have that $L_i = E_i - V_i + 2C_i$ and eq.\eqref{eq:Nfactor} can be reshaped into
\begin{equation}
N^{\sum_i (2E_i - 2V_i +2C_i) + 3(V-E)} = N^{2\sum_i l_i + 3(1 - l)}\;,
\end{equation}
this time $l$ counting all loops in the diagram disregarding the colors associated to each edge. We define then 
\begin{equation}
\omega \equiv  3(l -1) - 2\sum_i l_i = \sum_i l_i + 3\mathcal{L} - 3\;,
\end{equation}
denoting with $\mathcal{L}$ the loops that are made by mixing edges of different colors. We need to minimize $\omega$ in order to find the leading order vacuum graphs. It is then clear that the leading order graphs don't have any loops of color $i$, neither disregarding the colors. In a word, they are trees of the form displayed on the center of Figure~\ref{fig:LOandNLOgraphs}.
However, all those graphs made of tadpoles vanish. In order to find the next-to-leading order graphs, it is useful to introduce $V_n$ counting the vertices of valence $n$ (without consideration for the color of the edges), such that 
\begin{equation}
2E = \sum_n nV_n\;.
\end{equation}
Then on a connected graph (without tadpoles), 
\begin{equation}
l = E-V + 1 = \sum_{n>1} \left(\f{n}{2} - 1\right) V_n + 1\;.
\end{equation}
This relation shows that the diagrams that minimize $l$ have only two-valent vertices, as represented on the right of Figure~\ref{fig:LOandNLOgraphs}, of specific color. The power of $N$ decreases progressively with each additional loop, as in usual vector models, and when different colors mix, it decreases even more. This allows a full classification of the diagrams contributing at given order.

\begin{figure}
\centering 
\includegraphics[height=0.3\textwidth]{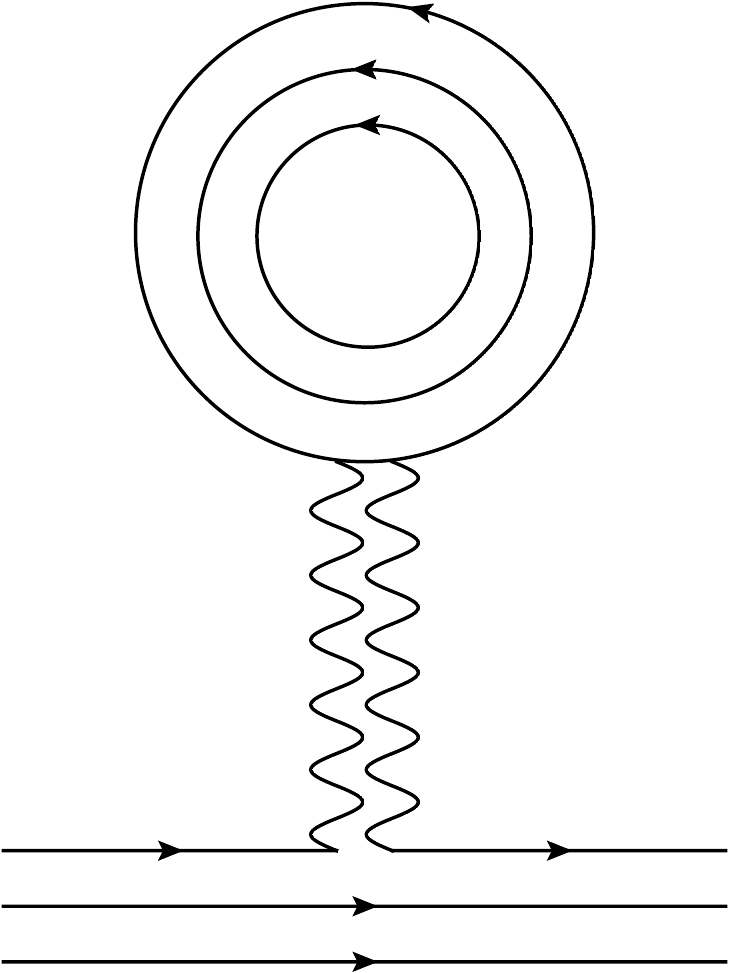} \hspace{0.05\textwidth}
\includegraphics[height=0.2\textwidth]{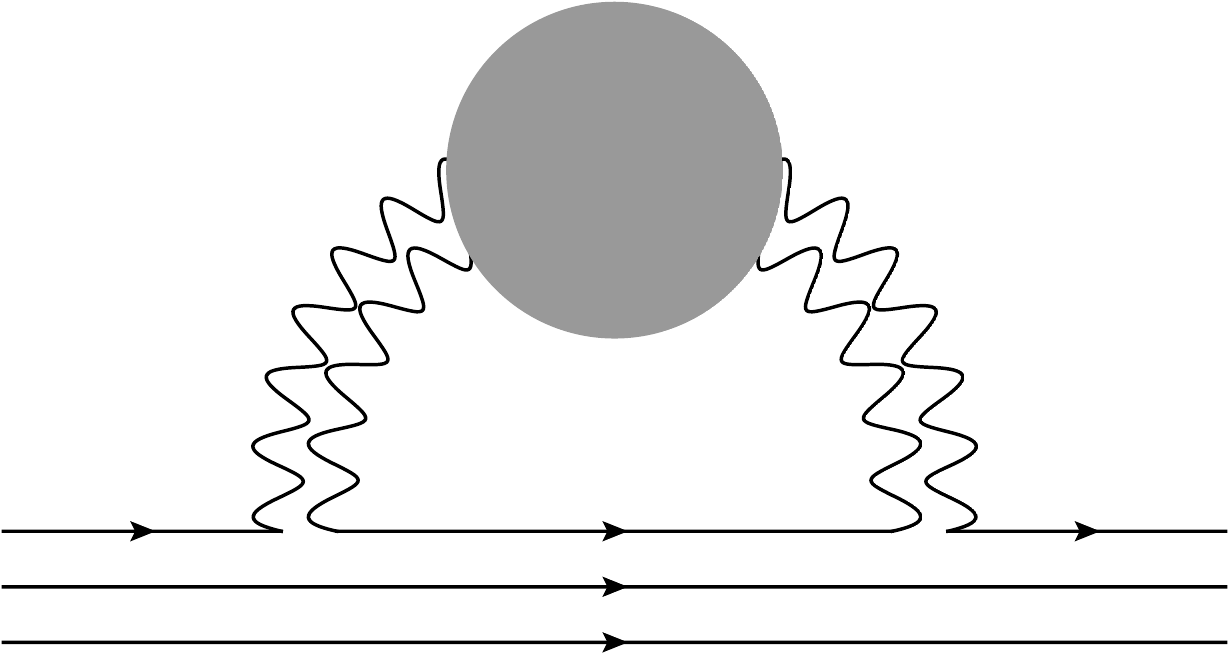}
\caption{\label{fig:tadpole}Diagrams contributing to the full Fermion propagator. Left: Tadpole, $\cO(1)$. Right: $\cO(1/N)$.}
\end{figure}

Then, to go from a vacuum graph to a two-point Fermion graph, we simply cut a Fermion propagator, i.e. divide the scaling by $N^2$. Graphically, this corresponds to open a vertex, as in Fig.\ref{fig:tadpole}. 

\begin{figure}
\centering 
\includegraphics[height=0.15\textwidth]{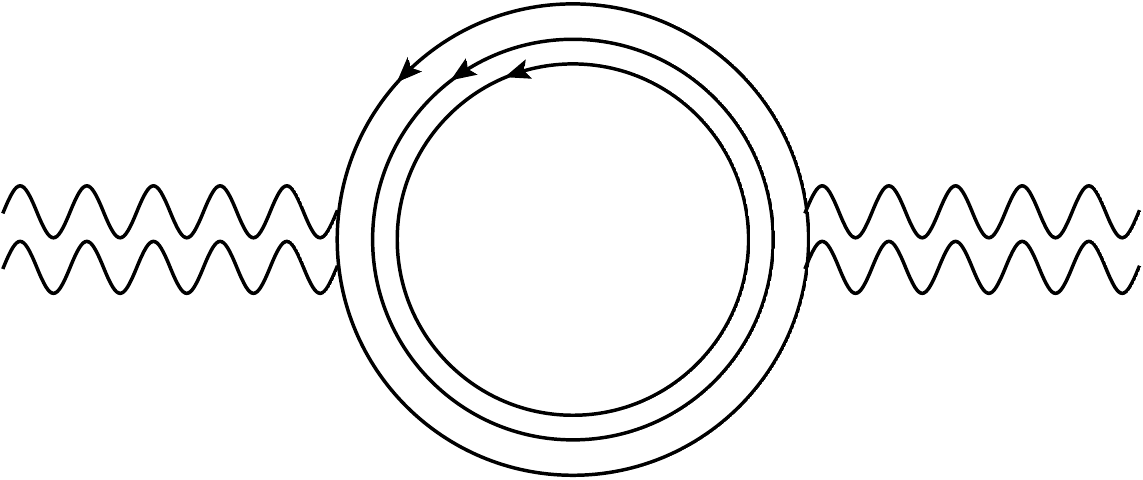}
\caption{\label{fig:selfgauge}Self-energy of the gauge field.}
\end{figure}

At leading order in $1/N$, the Schwinger-Dyson equations for the 2-point functions of the Fermions $\cG(p)$ and of the gauge fields $G^{(i)}(p)$ are
\begin{gather}
\mathcal{G}(p) = \mathcal{G}_0(p) + \f{1}{N}\sum_i \int \f{\dd[3]q}{(2\pi)^3} \mathcal{G}_0(p)\gamma^\alpha \mathcal{G}_0(p-q)\gamma^\beta \mathcal{G}_0(p)G^{(i)}_{\alpha\beta}(q) ,\\
G^{(i)-1} (p)= G^{(i)-1}_{0}(p) - \mathcal{S}(p),
\end{gather}
with the self-energy \footnote{The first line uses $[\g^a,\g^b] = 2i \epsilon^{abc} \g^c$.}
\begin{align}
\mathcal{S}(q) &= \int \f{\dd[3]p}{(2\pi)^3} \Tr \left(\g^+ \f{1}{i\slashed{p} + m} \g^3 \f{1}{i(\slashed{p} + \slashed{q}) + m} \right) - (\g^3 \leftrightarrow \g^+) \\
&=6imq^+\int \f{\dd[3]p}{(2\pi)^3}\f{1}{p^2+m^2}\f{1}{(p+q)^2+m^2}\\
&=\f{3i}{4} q^+\underbrace{\int_0^1\dd x \f{1}{R}}_{\equiv  F(\abs{\a})}\;,\qquad R^2 \equiv 1-\a^2x(1-x)\;,\hspace{5mm}q^2=-\a^2/m^2\;,\\
&F(\a) = \f{1}{\a}\log\left(\f{1+\a/2}{1-\a/2}\right)\;.
\end{align}
This $q^+$ appearance in the self-energy allows to rewrite the full gauge propagator at large-$N$ as
\begin{equation}
\label{eq:full-gauge-propa}
G^{(i)} (p)=\f{4\pi i}{kp^+} \f{1}{1 + \pi F(\abs{p/m})/3k}.
\end{equation}
The first corrections to the Fermion propagator are depicted on Fig.~\ref{fig:tadpole} and the self-energy of the gauge field on Fig.~\ref{fig:selfgauge}. 

\subsubsection{Interacting Fermions}
Bringing back the original pillow interactions in the action, we find that the Fermions don't see the gauge field at large $N$, as the tadpole diagram dominates the two-point function (cf. the right-hand side of Figure~\ref{fig:SDE}), along with the free propagator, thus returning to the preceding sections. At leading order, no other divergent diagram with the gauge field renormalizes the Fermionic interactions. 

The only sign of the presence of the gauge field is an additional non-local effective pillow interaction. Furthermore, because of how the pillows assemble themselves to form chains of bubbles (if one reduces a pillow to a point, one recovers Figure~\ref{fig:bubblechain}), this non-local vertex will not affect the flow of the local vertices constructed earlier, whereas the flow of the former is completely taken into account by the full gauge propagator~\eqref{eq:full-gauge-propa}.

\subsection{Relation to the vector case analysis}
\begin{figure}
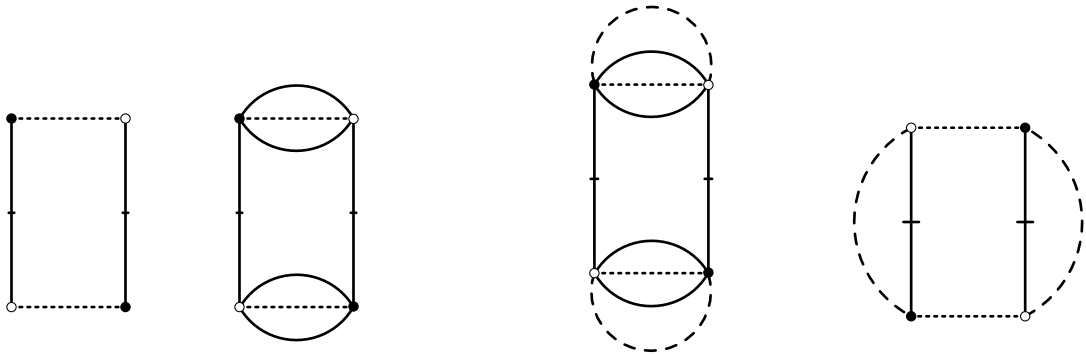

\centering 
\captionsetup[subfigure]{labelformat=empty}
\subfloat[]{\tikzsetnextfilename{effective_pillow2}}
\input{effective_pillow.tex}
\hspace{1cm}
\subfloat[]{\tikzsetnextfilename{2PILO2}}
\input{2PILO.tex}
\subfloat[]{\tikzsetnextfilename{2PILOvector2}}
\input{2PILOvector.tex}
\caption{\label{fig:vecvsteninter} Effective vertices of the vector (first) and the tensor (second) case. The third diagram represents the only leading order two-particle irreducible graph with Fermion propagators dashed. The rightmost diagram is also at leading order in the vector case.}
\end{figure}

Let us explain why our result differs from those of \cite{Giombi:2011kc}, considering a single color for tensors. \footnote{We focus on the case of non-self-interacting Fermions.} For vectors, the interaction between the gauge field and the Fermions takes the form
\begin{equation}
A^{ij}_{\mu} \psib_{i}\g^\mu \psi_{j}\;,
\end{equation}
while for tensors, we only added two extra indices
\begin{equation}
A^{ij}_{\mu} \psib_{iab}\g^\mu \psi_{jab}\;.
\end{equation}
In both cases, integrating out the gauge field leads to an effective quartic interaction for the Fermions, eq.~\eqref{eq:effective-pillow}. We depict both vertices on the left of Fig.~\ref{fig:vecvsteninter}. The full line represents contraction in flavor space, the dotted line stands for contraction in spinor space and the marked line, residue of the gauge field, transports momentum. Then at large-$N$, the Fermion propagator (without the free part) is a tadpole in both cases, but of different structure (see the right diagrams of Fig.~\ref{fig:vecvsteninter}). In particular, the left tadpole of Fig.~\ref{fig:tadpole} vanishes by parity and because of the trace over the $\g$-matrices. 

Similarly to what we did above, we could integrate out the Fermions and study the effective action of the intermediate field $A_\mu$. However, we do not expect any spontaneous symmetry breaking that would imply a breaking of Lorentz invariance. Another approach is instead to integrate over $A_\mu$ and compute the 2PI effective action $\G(\Psi,\cG)$ of the Fermions. In the limit $\l\rightarrow 0$, we expect no breaking of $U(N)$ or chiral symmetries (hence take $\Psi = 0$). The leading order diagram (right of Fig.~\ref{fig:vecvsteninter}) vanishes since it is two Fermionic tadpoles, such that we arrive at an effective action for the Fermions as:
\begin{equation}
\G(0,\cG) = +\Tr\ln \cG^{-1} + \Tr\left(\cG_0^{-1}\cG\right)\;,
\end{equation}
or on-shell up to an irrelevant constant
\begin{equation}
\G(0,\cG_0) = \Tr\ln (i\slashed{p}).
\end{equation}

\begin{figure}[htbp]
\centering
\captionsetup[subfigure]{labelformat=empty}
\subfloat[]{\tikzsetnextfilename{2PILO-rectangular2}
\input{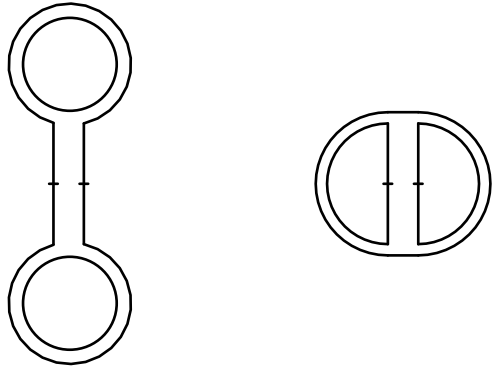}}
\caption{Stranded representation of the leading order 2PI graphs in the rectangular case (keeping as above the marked line standing for the gauge field).}
\label{fig:stranded-rectangular}
\end{figure}

The conclusion is that at large $N$, the gauged tensor forgets about the gauge field (Fermion self-interactions contribute at leading order), whereas the analysis is non-trivial for vectors. 
Indeed, the second tadpole leads to a non-trivial effective action of the form, 
\be
\f{2\pi i}{k}\int \f{\dd[3]q\dd[3]p}{(2\pi)^6}\f{1}{q^+}\Tr\left(\cG(-p)\g^+\cG(p-q)\g^3\right)
\ee
from which all the discussion of \cite{Giombi:2011kc} proceeds.
In fact, using $U(N^2)$ gauge fields in the tensor case (assembling the above $N^2$ indices $(a,b)$ into $N^2$ capital indices $A$, cf. App.~\ref{app:rectangular})
\begin{equation}
A^{AB}_{\mu} \psib_{iA}\g^\mu \psi_{iB}\;,
\end{equation} 
that is as if the color symmetry were broken, \footnote{This is reminiscent of the Chern-Simons coupling to bifundamental matter fields with $U(N)\times U(M)$ treated in the limit $M/N\ll 1$ \cite{Gurucharan:2014cva}.}
one could also make the rightmost diagram contributing at leading order Schwinger-Dyson equations, as is most clearly seen from the stranded diagram Fig.~\ref{fig:stranded-rectangular}. It would be interesting to analyse what the consequences of both ``triviality" in the color-symmetric phase and return to a matrix theory when the color-symmetry is broken, would be in the potential higher-spin duals of tensor models \cite{Vasiliev:2018zer}. 

\newpage

\begin{subappendices}
\section{\texorpdfstring{$\gamma$}{g}-matrices in odd-dimensions}
\label{sec:appendix gamma}
Construction irreducible representations of the Clifford algebra in any dimension $D$
\begin{equation}
\{\g^\mu,\g^\nu\} = 2\eta^{\mu\nu} \;,
\end{equation}
is quite standard (see for instance \cite{Park:2005cyw}). Even dimensions admit a unique irreducible representation, with matrices of dimension $2^{D/2}$, that can be constructed with the following cyclic structure\footnote{All definitions below are correct up to $i$ factors required accord ing to the signature.}:
\begin{align}
\g^1 &\equiv \s_1\otimes \mathbb{1}\otimes \mathbb{1}\dots\\
\g^2 &\equiv \s_2\otimes \mathbb{1}\otimes \mathbb{1}\dots\\
\g^3 &\equiv \s_3\otimes \s_1 \otimes \mathbb{1}\dots\\
\g^4 &\equiv \s_3\otimes \s_2 \otimes \mathbb{1}\dots\\
\g^5 &\equiv \s_3\otimes \s_3 \otimes \s_1\dots, 
\end{align}
and so on, stopping at $D$. 
There, one can also introduce an extra hermitian matrix, squaring to one and anticommuting with $\g^i ~ (1\leq i \leq D)$
\begin{equation}
\g^{(D+1)} \equiv \g^1\cdots\g^D\;.
\end{equation} 
Irreducible representations in odd dimensions $D$ use the same $2^{\f{D-1}{2}}$ dimensional representation as in $D-1$ dimensions taking $\g^{D} = \pm\g^{(D)}$ as the last needed to complete the set. Two inequivalent representations exist, differing by this ``$\pm$'' sign.

In the main text, it was convenient to work with a reducible representation of the Clifford algebra, in order to define a chiral transformation in terms of $\g_5$. We use the following realization of the three dimensional Euclidean Clifford algebra. The $\gamma$-matrices write as
\begin{equation}
\gamma^\mu  = 
\begin{pmatrix}
\tilde{\gamma}^\mu & 0 \\
0 & -\tilde{\gamma}^\mu\end{pmatrix} \quad 
\tilde{\gamma}^1 = \sigma_1 = 
\begin{pmatrix}
0 & 1\\
1& 0 \end{pmatrix} \quad
\tilde{\gamma}^2 = \sigma_2 = 
\begin{pmatrix}
0 & -i\\
i & 0\end{pmatrix} \quad
\tilde{\gamma}^3 = \sigma_3 = 
\begin{pmatrix}
1 & 0\\
0 & -1\end{pmatrix}\;.
\end{equation}
The Fermion fields write as
\begin{equation}
\psi = \begin{pmatrix}\psi_1\\\psi_2\end{pmatrix},
\end{equation}
where $\psi_i$'s are 2-component Dirac spinors. Finally our choice for a ``$\gamma^5$''-matrix, such that $\{\gamma^\mu,\gamma^5\}=0$ and $(\g^5)^2 = \mathbb{1}$, is
\begin{equation}
\gamma^5 = \begin{pmatrix}
& \mathbb{1}\\
\mathbb{1} & \end{pmatrix}.
\end{equation}

\section{Other representations of the $U(N)\times U(N^2)$-symmetric model}
\label{app:rectangular}

Given the rectangular matrix structure of the $U(N)\times U(N^2)$ model of Sec.~\ref{sec:S2}, one might wonder why we perform the intermediate field analysis on the index taking values from one to $N$ and not on the one taking values from one to $N^2$, and whether choosing the second option we would discover also a spontaneous symmetry breaking of the $U(N^2)$ subgroup.
The intuitive reason why such option is not favorable is that we would have an intermediate field which is an $N^2\times N^2$ matrix, thus containing more variables than the original Fermionic field. We would then expect such formulation to contain redundant information.

In order to illustrate this in a simplified context, let us consider the zero dimensional Bosonic analogue of our model, with only one pillow interaction and no double-trace, i.e.\ the model studied in \cite{Benedetti:2015ara}, which however we now rewrite in rectangular matrix notation.\footnote{For tensor models with a rectangular-matrix interpretation see also \cite{doubletens,Bonzom:2013lda}.} The model is defined by the partition function
\be \label{eq:Z-0d}
Z(g) = \int [d\vph d\vphb] e^{- N^2\left( \vphb_{aA}\vph_{aA}-\f{g}{2} \vphb_{aA}\vph_{a'A}\vphb_{a'A'}\vph_{aA'}\right)} \;,
\ee
where the measure is normalized with respect to the free theory, as usual we have summation on the repeated indices, and with respect to \eqref{eq:action} we have rescaled the fields by $N$ to pull out an $N^2$ in front of the action.
Notice that the action can be viewed as a square-matrix action for either
\be
\phi_{ab} \equiv \vph_{aA}\vphb_{bA}
\ee
 or 
\be
\Phi_{AB}\equiv \vphb_{aA}\vph_{aB}\;.
\ee 
Naively, one would expect an action of order $N^3$ in the first case, and an action of order $N^4$ in the second, as the action writes in terms of single traces of square matrices of size $N$ and $N^2$, respectively. However, being one and the same model, the free energy must be the same in the two cases. The reason why the naive reasoning fails is that the two square matrices $\phi_{ab}$ and $\Phi_{AB}$ have the same number of non-zero eigenvalues, which are the squares of the non-zero singular-values of the rectangular matrix $\vph_{aA}$.
The singular-value decomposition was applied to rectangular matrix models in \cite{Anderson:1990nw,DiFrancesco:2002mvz}, and we concisely repeat it here. We write
\be
\vph_{aA} = P_{ab} X_{bB} Q^\dagger_{BA} \;,
\ee
where $P\in U(N)$ is a unitary transformation that diagonalizes $\phi_{ab}$, $Q\in U(N^2)$ is a unitary transformation that diagonalizes $\Phi_{AB}$, and $X_{bB}=\d_{bB} x_b$ is a rectangular matrix whose only non-zero entries are the positive square roots of the eigenvalues of $\phi_{ab}$. Due to the invariance of such decomposition, the matrices $P$ and $Q$ are not unique, hence one should restrict the angular variables to $(P,Q)\in (U(N) \times U(N^2)/(U(N^2-N))\times U(1)^N)$.
Including the Jacobian of the transformation $(\vph,\vphb)\to(X,P,Q)$ \cite{Morris:1990cq,Anderson:1990nw,DiFrancesco:2002mvz} we rewrite the partition function \eqref{eq:Z-0d} as
\be \label{eq:Z-0d-eig}
Z(g) = \f{1}{\cN} \int \left( \prod_{i=1}^N dx_i\, x_i^{2(N^2-N)+1} e^{-N^2( x_i^2 - \f{g}{2} x_i^4)} \right) \D(x^2)^2 \, ,
\ee
where the angular variables have been factored out, and canceled with the normalization. The leftover normalization is in the factor $\cN$, equal to the integral in \eqref{eq:Z-0d-eig} evaluated at $g=0$. Lastly, $\D(x^2)=\prod_{1\leq i<j\leq N}(x_i^2-x_j^2)$ is the standard Vandermonde determinant for the eigenvalues of $\phi_{ab}$. The latter is subdominant in the large-$N$ limit, but the factor $x_i^{2(N^2-N)}$ is not, and must be taken into account. However, unlike the Vandermonde determinant, such term does not couple different eigenvalues, hence in the large-$N$ limit we have a simple saddle point equation in which the eigenvalues are mutually decoupled:
\be
x_i-g x_i^3 - \f{1}{x_i} = 0 \;\;\;\; \text{for}\;\;\; i=1,\ldots, N\,.
\ee
The solutions are
\be
x_\pm^2 = \f{1\pm \sqrt{1-4g}}{2g} \, ,
\ee
in agreement with \cite{Benedetti:2015ara},\footnote{The eigenvalues of the intermediate field $H_{ab}$ in \cite{Benedetti:2015ara} are $a_\pm=\sqrt{g} x_\pm^2$, because $H_{ab}$ is conjugate to $\sqrt{g}\phi_{ab}$.} where they were obtained by an intermediate field representation of the square of $\phi_{ab}$, as we did in Sec.~\ref{sec:S2}. 
For $g\leq 1/4$, the free energy at leading order is obtained by evaluating the effective action for the eigenvalues on the saddle point solution $x_i=x_-$, $\forall i$, which as shown in \cite{Benedetti:2015ara} is the only minimum of the action (and contrary to $x_+$, it is regular at $g=0$):
\be
-\f{1}{N^3} \ln(Z(g)) = x_-^2 - \f{g}{2} x_-^4 - \ln (x_-^2) - 1\;, 
\ee
where the minus one comes from the normalization factor $\cN$.
By explicit check, the result above coincides with the one in \cite{Benedetti:2015ara}.

Now, what happens if we introduce instead an intermediate field representation for the square of $\Phi_{AB}$?
The partition function \eqref{eq:Z-0d} rewrites
\be \label{eq:Z-0d-H}
\begin{split}
Z(g) &= \int [d\vph d\vphb dH] e^{- N^2 \left(\vphb_{aA}\vph_{aA} -\sqrt{g} \vph_{aA}\vphb_{aB} H_{AB} + \f12 H_{AB} H_{BA}  \right)}\\
& = \int [dH] e^{- \f{N^2}{2} \Tr(H^2)-N\Tr\ln(1- \sqrt{g}H)}
 \;.
\end{split}
\ee
Diagonalizing the matrix $H$, and counting the trace as a contribution of order $N^2$ (since the matrix $H$ is of size $N^2$), we conclude that the logarithmic term in the action is subleading, while the logarithm of the Vandermonde determinant originating in the change of variables is of $N^4$ as the Gaussian part of the action. On the other hand, the Gaussian integral on $H$ is normalized to one by construction, hence the free energy at leading order (LO) (order $N^4$) is correctly zero. In order to obtain the first non-trivial contribution to the free energy, we have to go to next-to-leading order (NLO) (order $N^3$).
Let us denote the eigenvalues of $H$ by $y_i$, $i=1\ldots N^2$. Writing explicitly the normalization as before, the partition function \eqref{eq:Z-0d-H} now writes
\be \label{eq:Z-0d-H-eig}
Z(g) = \f{1}{\cN'} \int \left(\prod_{i=1}^{N^2} dy_i\right) e^{- S_{\rm eff}(y)}\;,
\ee
with
\begin{align} \label{eq:Seff}
S_{\rm eff}(y) &\equiv N^4 S_0(y) + N^3 S_1(y)\;,\\
S_0(y)  &= \f{1}{2 N^2} \sum_{i=1}^{N^2}  y_i^2   -\f{2}{N^4} \sum_{1\leq i<j\leq N^2} \ln |y_i-y_j| \;, \\
S_1(y) &=  \f{1}{N^2} \sum_{i=1}^{N^2} \ln(1- \sqrt{g}y_i) \;.
\end{align}
Both $S_0(y)$ and $S_1(y)$ are of order one, and the order in $N$ has been made explicit in \eqref{eq:Seff}. Notice that the subleading term (of order $N^3$) is dominant over the one-loop correction, which is of order $N^2$. Therefore, we will not need to compute any loop corrections, and the NLO free energy is given by simply evaluating the subleading term of the action on the LO saddle point.
In fact, the saddle point $\bar{y}$ has an expansion (omitting the indices)
\be
\bar{y} = y_{{}_{\rm LO}} +\f1N y_{{}_{\rm NLO}} +O\Big(\f{1}{N^2}\Big)\;,
\ee
and the action then expands as
\be
S_{\rm eff}(\bar{y}) =  N^4 S_0(y_{{}_{\rm LO}}) + N^3\left(\f{\p S_0}{\p y}{\big|}_{{y=y_{{}_{\rm LO}}}} \cdot y_{{}_{\rm NLO}}+ S_1(y_{{}_{\rm LO}}) \right) +O(N^2)\;.
\ee
Since $\f{\p S_0}{\p y}{\big|}_{{y=y_{{}_{\rm LO}}}}  =0 $ by definition of $y_{{}_{\rm LO}} $, and since $\ln (\cN') = -N^4 S_0(y_{{}_{\rm LO}})+O(N^2)$, we obtain
\be
-\f{1}{N^3} \ln(Z(g)) =  S_1(y_{{}_{\rm LO}}) +O\Big(\f{1}{N}\Big)\;.
\ee
In order to complete the calculation, we need the explicit solution $y_{{}_{\rm LO}}$, which is given in terms of the famous Wigner semicircle law. 
This is given in terms of the density of eigenvalues $\r(y) = \f{1}{2\pi} \sqrt{4-y^2}$, from which we obtain
\be
S_1(y_{{}_{\rm LO}}) = \int_{-2}^{+2} dy \r(y) \ln(1- \sqrt{g}y) = x_-^2 - \f{g}{2} x_-^4 - \ln (x_-^2) - 1\;,
\ee
precisely the result we obtained before.

In conclusion, we have learned that it is possible to obtain the same results in three different ways: singular-value decomposition, intermediate field on the $U(N)$ sector, and intermediate field on the $U(N^2)$ sector. The first one is maybe the most enlightening for the zero dimensional case, but it becomes not viable in higher dimensions, where the angular variables do not decouple due to the derivative term (the action has a global invariance, not a local one). The last one is instead definitely the most difficult, as one needs to go to NLO and use the matrix model solution. For these reasons, in Sec.~\ref{sec:S2} we have followed the second option, which is of course also the standard one.

There remains to comment on the question of spontaneous breaking of the $U(N^2)$ symmetry. The fact that the saddle point of the intermediate field $H$ is given by the Wigner law would seem to indicate a complete breaking of the symmetry: the eigenvalues are spread along the interval $[-2,+2]$, hence the matrix is far from being proportional to the identity, as demanded by invariance under the symmetry group.
However, it is wrong to identify the saddle point solution of $H$ with its expecation value. In fact, in going from \eqref{eq:Z-0d-H} to \eqref{eq:Z-0d-H-eig} we have canceled the integral over the unitary group with the normalization, but if we have an insertion of a non-invariant observable, such as $\la H_{AB} \ra$, the integral does not drop. Furthermore, since the unitary matrices appearing in the change of variables from $H$ to its eigenvalues have no action term (because the action is invariant), we cannot do anything but perform the full integral.
We have:
\be
\begin{split}
\la H_{AB} \ra &= \f{1}{\cN'} \int \left(\prod_{i=1}^{N^2} dy_i\right) e^{- S_{\rm eff}(y)} Y_{CD} \int [dU] U_{AC}U^\dagger_{DB}\\
&= \f{\d_{AB}}{N^2} \f{1}{\cN'} \int \left(\prod_{i=1}^{N^2} dy_i\right) e^{- S_{\rm eff}(y)} \sum_i y_i\\
&= 0\;,
\end{split}
\ee
where we introduced the diagonal matrix of eigenvalues $Y_{AB}=y_A \d_{AB}$ and used the known formula for integration over the unitary group
\be
 \int [dU] U_{AC}U^\dagger_{DB} = \f{1}{N^2} \d_{AB} \d_{CD} \;.
\ee
Lastly, owing to the fact that the distribution of eigenvalues in the Wigner law is symmetric around the origin, $\Tr[Y]=0$ when evaluated on the saddle point solution. 

To conclude, we should remark that a similar phenomenon of symmetry restoration (\`a la Mermin-Wagner) will happen also in $d>0$. In higher dimensions, due to terms with derivatives in the action, the angular variables do not drop out of the action in general because we only have a global symmetry, not a local one. However, in the intermediate field representation the derivatives only appear in the higher-dimensional generalization of the logarithmic term in the action \eqref{eq:Z-0d-H}, which is subleading in this case. Hence, the same reasoning as in the zero dimensional case above applies also to the higher dimensional case.
In brief, the $U(N^2)$ part of the symmetry group does not undergo spontaneous symmetry breaking.

\section{Details on two and three traceless matrices}
\label{app:traceless_matrices}
\subsection{Two traceless matrices}
Imposing the intermediate field to have the form
\begin{equation}
M_1 = m_1(\mathbb{1}_N-2 \mathbb{P}_N), \quad M_2 = m_2(\mathbb{1}_N-2\mathbb{P}_N) , \quad M_3 = 0 \; ,
\end{equation}
the equations on motion have, for $m_1$, the form
\begin{equation}
\f{N^2}{2}\left(\f{1}{\lt_p} - \f{4}{\pi^2}\right)(\pm m_1) + \f{N^2}{\pi^2}\left[(\pm m_1 + m_2)^2\arctan\f{1}{\pm m_1 + m_2} + (\pm m_1 -m_2)^2\arctan\f{1}{\pm m_1 - m_2} \right] = 0\;,
\end{equation}
which combined with a similar one exchanging $m_1$ and $m_2$, lead to 
\begin{gather}
\Box\left(\f{1}{\lt_p} - \f{4}{\pi^2}\right) +  \f{4}{\pi^2}\Box^2\arctan\f{1}{\Box} = 0\;,\\
\triangle \left(\f{1}{\lt_p} - \f{4}{\pi^2}\right) + \f{4}{\pi^2} \triangle^2\arctan\f{1}{\triangle}= 0\;,\\
\Box \equiv m_1+m_2 \; , \quad \triangle \equiv m_1 - m_2\;.
\end{gather}
This implies either $\Box=0$ or $\triangle=0$, which after renaming gives $m_1 = m_2 = m$, or if neither is zero we have $\Box=\triangle$ forcing $m_1 = m_2 = 0$.
In the non-trivial case, the consequence for the effective potential is that we compare
\begin{gather}
V_1= \f{1}{4\lt_p}m^2 + \f{1}{3\pi^2}\k(m),\\
V_2= \f{1}{2}\left[\f{1}{4\lt_p}\Box^2 + \f{1}{3\pi^2}\k(\Box)\right],\\
\kappa(x) = 2x^3\arctan\frac{1}{x} - 2x^2 - \log(1+x^2), 
\end{gather}
with $m$ and $\Box$, obeying the same algebraic equation linking them to $\lt_p$. Now the dominance of the single over the double traceless matrices appears straightforwardly, as soon as they acquire negative values, that is for $\lt_p>\pi^2/4$.

\subsection{Three traceless matrices}
Imposing the intermediate field to have the form
\begin{equation}
M_1 = m_1(\mathbb{1}_N-2\mathbb{P}_N) , \quad M_2 = m_2(\mathbb{1}_N-2\mathbb{P}_N) , \quad M_3 = m_3(\mathbb{1}_N-2\mathbb{P}_N) \;,
\end{equation}
the equations of motion can be shaped into 
\begin{gather}
\left(\f{1}{\lt_p} - \f{4}{\pi^2}\right)\a + \f{3}{\pi^2}\alphab + \f{2}{\pi^2}\left(\betab + \gammab + \deltab \right) =0\;,\\
\left(\f{1}{\lt_p} - \f{4}{\pi^2}\right)\b + \f{3}{\pi^2}\betab + \f{2}{\pi^2}\left(\alphab + \gammab + \deltab \right) =0\;,\\
\left(\f{1}{\lt_p} - \f{4}{\pi^2}\right)\g + \f{3}{\pi^2}\gammab + \f{2}{\pi^2}\left( \alphab + \betab +\deltab \right) =0\;,\\
\left(\f{1}{\lt_p} - \f{4}{\pi^2}\right)\d + \f{3}{\pi^2}\deltab + \f{2}{\pi^2}\left( \alphab + \betab +\gammab \right) =0\;,\\
\xb \equiv x^2\arctan\f{1}{x}\;,\\
\a \equiv m_1 + m_2 + m_3\;,\quad \b \equiv m_1 - m_2 + m_3\;,\quad \g \equiv m_1 + m_2 - m_3\;,\quad \d \equiv -m_1 + m_2 + m_3\;.
\end{gather}
We then rearrange into
\begin{equation}
\left(\f{1}{\lt_p} - \f{4}{\pi^2}\right)\a + \f{1}{\pi^2}\a^2\arctan\f{1}{\a}=(\a \rightarrow \b) =(\a \rightarrow \g) = (\a \rightarrow \d) \overset{!}{=} K\;,
\end{equation}
for $K$ some constant. 

First, if we assume that $K$ vanishes, then the variables that are non-zero must obey the single-traceless equation of motion \eqref{eq: single-traceless-eom}. And two different options come: $\a=\b$ (or $\a=\g$ or $\a=\d$) letting us with the double-traceless case $(m_1 = m_2 = m ~ ;  m_3 = 0)$ (or permutations), or $\b=\g$ (or $\g=\d$ or $\b=\d$) giving $m_1 = m_2 = m_3 = m$.\footnote{Of course, we also have the combination of the two options which results to $m_1 = m_2 = m_3 = 0$.}  

Secondly, for non-zero $K$ and introducing $\lt_p = (1+\eps)\pi^2/4$, we found that non-trivial solutions (not resulting in $\a=\b=\g=\d$) existed only in the range $\eps \in [0,\eps_*]$ with $\eps_* \approx 0.37$. In this region, again by numerical exploration, we found two non-trivial solutions:
\begin{itemize}
\item $(\a = 3\b,~ \b = \g =\d) \implies m_1 = m_2 = m_3 = m$, with associated equation of motion and potential
\begin{gather}
4 \left(1 - \f{1}{1 + \eps}\right) - m \left(9 \arctan\f{1}{3 m} + \arctan\f{1}{m}\right) = 0 \;,\\
V_3 = \f{3 m^2}{\pi^2 (1 + \eps)} + \f{\k(3 m) + 3 \k(m)}{12 \pi^2}\; ;
\label{eq: triple-traceless1}
\end{gather}
\item $(\a = 0.08\b,~ \d = -1.92 \b,~ \b = \g) \implies -0.46 m_1 = m_2 = m_3$, with associated equation of motion and potential
\begin{gather}
4 \left(1 - \f{1}{1 + \eps}\right)0.92 + m \left(0.08^2 \arctan\f{1}{0.08 m} - \arctan\f{1}{m}\right) = 0\;,\\
V_3^\prime = \left(\f{0.92^2}{2} + 1\right) \f{m^2}{\pi^2 (1 + \eps)} + \f{\k(0.08 m) + 2 \k(m) + \k(-1.92 m)}{12 \pi^2}.
\label{eq: triple-traceless2}
\end{gather}
\end{itemize}

We studied numerically the differences $V_1-V_3$ and $V_1-V_3^\prime$, and found both to be negative, hence we conclude that the single-traceless solution is dominant.

\end{subappendices}

%% file: sexticTM/sexticTM-these.tex
This chapter is organised as follows. In Section~\ref{sec:models}, we start by setting the scene with definitions of our models in rank 3 and 5, long- and short-range, and a description of the leading order diagrams. We continue in Section~\ref{sec:SDeq} and \ref{sec:kernels} by computing the two- and four-point functions. Section~\ref{sec:betas} contains a detailed derivation of the $\beta$-functions of our sextic couplings, as well as their fixed points. Before concluding, we compute in Section~\ref{sec:BSeq} the spectrum of bilinears (including spin dependence) through the now standard eigenvalue equation. In three appendices, we spell out details on our conventions and on the main loop integrals. 

\section{The models}
\label{sec:models}

Both models we are going to consider can be viewed as symmetry-breaking perturbations of a free $O(\cN)$-invariant action for $\cN$ scalar fields $\phi_{\bf a}(x)$, with ${\bf a}= 1,\ldots,\cN$, $x\in \mathbf{R}^d$:\footnote{As usual, a summation is implied for repeated indices.}
\be
S_{\rm free}[\phi,\phib] = \int d^d x  \, \phib_{\bf a}(x) (   - \p_\m\p^\m)^{\zeta} \phi_{\bf a}(x) \, .
\ee
The scalar fields will be either complex or real (in the latter case $\phib_{\bf a}=\phi_{\bf a}$ and we multiply the action by a factor $1/2$).
 $\zeta$ is a free parameter, which must be positive in order to have a well-defined thermodynamic limit, and it must be bounded above by one in order to satisfy reflection positivity.
We will later fix it to be either $\zeta=1$, as in~\cite{Klebanov:2016xxf,Giombi:2017dtl,Giombi:2018qgp}, or $\zeta=d/3$, as in~\cite{Benedetti:2019eyl,Benedetti:2019ikb}.\footnote{For $\zeta<1$, the fractional Laplacian can be defined in several ways \cite{Kwasnicki_2017}. In Fourier space, with the convention that $f(x) = \int\,\f{d^d p }{(2\pi)^d} e^{-i p\cdot x}\,f(p)$, we simply have:
\begin{equation*}
S_{\rm free}[\phi,\phib] = \int \f{d^d p}{(2\pi)^d}  \, \phib_{\bf a}(p) (p^2)^{\zeta} \phi_{\bf a}(p) \, .
\end{equation*}
In direct space we can instead write it as a kernel:
\begin{equation*}
S_{\rm free}[\phi,\phib] =  c(d,\zeta) \int d^d x\, d^d y  \, \f{\phib_{\bf a}(x)  \phi_{\bf a}(y)}{|x-y|^{d+2\z}} \, ,
\end{equation*}
with $c(d,\zeta) = \f{2^{2\z}\G\left(\f{d+2\z}{2}\right)}{\pi^{d/2}|\G(-\z)|}$. Notice that often in the literature on the long-range Ising model (e.g.\ \cite{Behan:2017emf,Paulos:2015jfa}) one finds the free action to be defined as above, but with $c(d,\zeta) = 1$.
}

The free propagator is
\be
C(p) = \f{1}{p^{2\zeta}}\,, \;\;\;\; C(x,y) = \f{\G\left(\Delta_{\phi}\right)}{2^{2\z}\pi^{d/2}\G(\z)} \f{1}{|x-y|^{2\D_{\phi}}}\,,
\ee
with $\Delta_{\phi}= \f{d-2\zeta}{2}$.

Perturbing the free action above by a quartic $O(\cN)$-invariant potential leads to the usual short-range ($\zeta=1$, e.g.~\cite{Moshe:2003xn}) or long-range ($\zeta<1$, e.g.~\cite{Brezin_2014,Defenu_2015}) $O(\cN)$ model.

The general type of tensor field theories we have in mind will have $\cN=N^r$, and a potential explicitly breaking the $O(\cN)$ symmetry group down to $\cG^r$, with either $\cG=O(N)$ (for real fields) or $\cG=U(N)$ (for complex fields).
For example, for $r=3$, we will write the field label as a triplet, ${\bf a}=(abc)$, and impose invariance of the action under the following transformation rule:
\be
\phi_{abc}(x) \to R^{(1)}_{aa'}\, R^{(2)}_{bb'}\,  R^{(3)}_{cc'}\,  \phi_{abc}(x)\, , \;\;\;\; R^{(i)}\in \cG \,.
\ee
Proper tensor field theories have $r>2$, otherwise we talk of vector ($r=1$) or matrix ($r=2$) field theories.
We will explicitly consider two models with sextic interactions, for $r=3$ and $r=5$. For $r=4$ we could write a model qualitatively very similar to $r=5$, but we would not learn much more, so we will not present it.

\subsection{Rank $3$}

\paragraph{Action.}

We first consider a rank-3 Bosonic tensor model in $d\leq 3$ dimensions, with $U(N)^3$ symmetry and sextic interactions.
The bare action is
\begin{align} \label{eq:action}
S[\phi,\phib] &= \int d^d x  \, \phib_{abc} (   - \p_\m\p^\m)^{\zeta} \phi_{abc} + S_{\rm int}[\phi,\phib] \, ,\\
 \label{eq:int-action}
S_{\rm int}[\phi,\phib] &= \int d^d x  \sum_{b=1}^5 \f{\l_b}{6 N^{3+\r(I_b)}} I_b \,.
\end{align}
The $U(N)^3$ invariants $I_b$ are all those that can be constructed with six fields, and their respective parameter $\r(I_b)$ will be chosen according to the optimal scaling defined in~\cite{Carrozza:2015adg}:
\be
\r(I_b)=\frac{F(I_b)-3}{2} \,,
\ee
with $F(I_b)$ counting the total number of cycles of alternating colors $i$ and $j$ with $i,j ~\in \lbrace 1,2,3\rbrace$, and the colors being introduced in the following paragraph.

It is customary to represent the tensor invariants as {\it colored graphs} \cite{Gurau-book}. To that end, we represent every tensor field as a node (black and white  for $\phi$ and $\phib$, respectively) and every contraction of two indices as an edge. Each edge is assigned a color red, blue, or green (or a label $1$, $2$, or $3$) corresponding to the positions of the indices in the tensor. We call the resulting graphs $3$-colored graphs. As a consequence of the $U(N)^3$ symmetry, such graphs are bipartite, that is, edges always go from a white to a black node.
With the aid of such representation we can write the interacting part of the action as:
\be
\begin{split} \label{eq:int-action-graph}
S_{\rm int}[\phi,\phib] = & \int d^d x  
\left( \f{\l_1}{6 N^{3}} \vcenter{\hbox{\tikzsetnextfilename{wheel2}
\input{wheel.tex}}}
+ \f{\l_2}{6 N^{4}} \vcenter{\hbox{\tikzsetnextfilename{long_pillow_rank32}
\input{long_pillow_rank3.tex}}}
+ \f{\l_3}{6 N^{4}} \vcenter{\hbox{\tikzsetnextfilename{circle2}
\input{circle.tex}}}
 \right. \\
 & \left.
 + \f{\l_4}{6 N^{5}} \vcenter{\hbox{\tikzsetnextfilename{pillow_trace_rank32}
\input{pillow_trace_rank3.tex}}}
  + \f{\l_5}{6 N^{6}} \vcenter{\hbox{\tikzsetnextfilename{triple_trace_rank32}
\input{triple_trace_rank3.tex}}}
  \right)\,,
\end{split}
\ee
where a (normalized) sum over color permutations should be understood, whenever it is non-trivial (see App.~\ref{ap:conventions} for more details on our conventions).
The graphs representing the tensor invariants are also called \emph{bubbles}. Bubbles which are composed of one, two, or three connected components are referred to as single-trace, double-trace, or triple-trace, respectively, for analogy with the matrix case, and bubbles $I_b$ for which $\rho(I_b)=0$ are called \emph{maximally single trace}  (MST), as each of their 2-colored subgraphs are single trace.
The $I_1$ invariant is the only MST bubble in our action.

\paragraph{Colored graphs and Feynman diagrams.} 

\begin{figure}[htbp]
\centering
\tikzsetnextfilename{4colored2}
\input{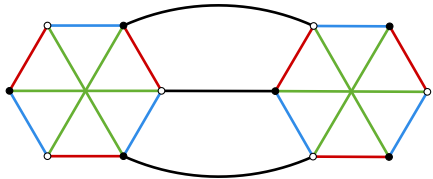}
\caption{$4$-colored graph corresponding to a two-loop Feynman diagram with external tensor contractions equivalent to $I_2$.}
\label{fig:4colored}
\end{figure}

We introduce some (mostly standard) notation for the perturbative expansion of the free energy (and the connected $n$-point functions) of the theory \cite{Bonzom:2012hw}. Each interaction invariant is represented as a $3$-colored graph as above.
Expanding around the free theory, the Gaussian average leads to the usual Wick contraction rules, for which we represent the propagators as edges of a new color, connecting a white and black node. We choose the black color for such propagators, or equivalently, the label $0$. We give an example of the resulting \emph{$4$-colored graphs} in Fig.~\ref{fig:4colored}.

\begin{figure}[htbp]
\centering
\tikzsetnextfilename{melon_dtadpole2}
\input{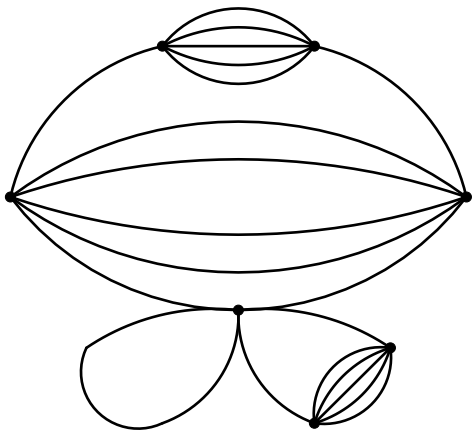}
\caption{An example of melon-tadpole Feynman diagram. Double tadpoles are based on the $I_b$ ($b \in [1,5]$) vertices and melons are based on $I_1$ vertices.}
\label{fig:melon_dtadpole}
\end{figure}

Ordinary \emph{Feynman diagrams}, the only objects that we will actually call by such name here, are obtained by shrinking each interaction bubble to a point, which we will call an  interaction vertex, or just vertex.
We give an example of such a Feynman diagram in Fig.~\ref{fig:melon_dtadpole}. While Feynman diagrams are sufficient for representing Feynman integrals, the $4$-colored graphs are necessary in order to identify the scaling in $N$. 
Indeed, in a $4$-colored graph, each propagator identifies all three indices on its two end tensors whereas each edge of color $i$ identifies only one pair of indices between its end tensors. The indices will then circulate along the cycles of color $0i$, which we call \emph{faces}, hence each face gives rise to a free sum, that is, a factor $N$. 
The amplitude of a Feynman diagram $\mathcal{G}$ thus scales as $A(\mathcal{G})\sim N^{F-3n_{1}-4n_2-4n_3-5n_4-6n_5}$, with $F$ the total number of faces in the associated $4$-colored graph and $n_i$ the number of bubbles of the interaction $i$. 
The existence of the large-$N$ limit relies on the fact that the power of $N$ is bounded from above \cite{Gurau-book,Carrozza:2015adg}.

\paragraph{Melonic graphs and melonic diagrams.} 

Melonic $k$-valent graphs are defined constructively starting from the fundamental melon, i.e.\ the unique graph built out of two $k$-valent vertices without forming self-loops (or tadpoles), and then iteratively inserting on any edge a melonic 2-point function, i.e.\ the graph obtained from the fundamental melon by cutting one edge in the middle. Notice that melonic $k$-valent graphs are always bipartite, and edge colorable with $k$ colors. 

An important result in rank-$r$ tensor models is that if one only allows for interaction bubbles which are melonic $r$-valent graphs, then in the perturbative expansion the leading order vacuum graphs at large $N$ are melonic $(r+1)$-valent graphs \cite{Bonzom:2012hw}.
However, it is important to notice that melonic $(r+1)$-valent graphs do not correspond to melonic Feynman diagrams, i.e.\  they do not remain melonic after shrinking the colors from 1 to $r$.
From the point of view of the Feynman diagrams, melonic $(r+1)$-valent graphs reduce to the same type of cactus diagrams appearing in the large-$N$ limit of vector models, and therefore field theories based on such interaction are not expected to lead to very different results than vector models.\footnote{They can nevertheless lead to new phases with patterns of spontaneous symmetry breaking which are impossible in the vector case, as we saw in the Chapter~\ref{ch:TGN}.} 

Adding non-melonic bubbles, things get more complicated, and possibly more interesting. In particular, it was found in~\cite{Carrozza:2015adg} that non-melonic interaction bubbles can be scaled in such a way that they also contribute at leading order in the $1/N$ expansion, and that for some interactions (in that specific example, the quartic tetrahedron interaction) their leading-order Feynman diagrams are melonic. The possibility of restricting the spacetime Feynman diagrams to the melonic type by means of a large-$N$ limit has been a main reason for studying tensor field theories in dimension $d\geq 1$, starting from \cite{Klebanov:2016xxf}.

\begin{figure}[htbp]
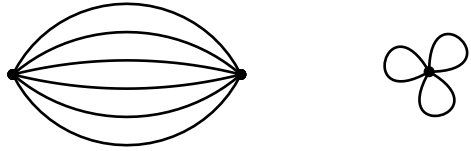

\centering
\captionsetup[subfigure]{labelformat=empty}
\subfloat[]{\tikzsetnextfilename{melon_freeenergy3}
\input{melon_freeenergy.tex}}
\hspace{1cm}
\subfloat[]{\tikzsetnextfilename{ttadpole_freeenergy2}
\input{ttadpole_freeenergy.tex}}
\caption{The two Feynman diagrams (the fundamental melon on the left, and the triple-tadpole, or trefoil, on the right) starting from which all the vacuum melon-tadpole diagrams can be built. The melon is based on the wheel vertices and the triple-tadpole is based on any of the interactions $I_i$ (for explicit examples of the corresponding colored graphs in rank 5, see Figure \ref{fig:fund_vacuum_rank5}).}
\label{fig:fund_vacuum}
\end{figure}

\begin{figure}[htbp]
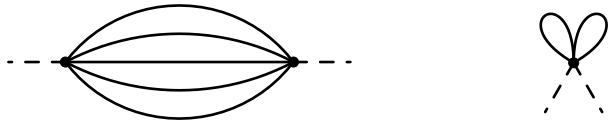

\centering
\captionsetup[subfigure]{labelformat=empty}
\subfloat[]{\tikzsetnextfilename{melonSDE2}
\input{melonSDE.tex}}
\hspace{1cm}
\subfloat[]{\tikzsetnextfilename{dtadpoleSDE2}
\input{dtadpoleSDE.tex}}
\caption{The two minimal two-point function Feynman diagrams used in the iterative construction of melon-tadpole diagrams. The melon is based on wheel vertices and the double tadpole is based on any of the interactions $I_i$.}
\label{fig:SDE}
\end{figure}

\paragraph{The large-$N$ limit.}  The $I_1$ invariant in \eqref{eq:int-action} (i.e.\ the first bubble in \eqref{eq:int-action-graph}, which we call the {\it wheel} graph, and which is also known as the complete bipartite graph $K_{3,3}$) stands out as the only non-melonic bubble in our action, and as a consequence, as the only interaction that does not lead only to tadpole corrections to the propagator at large $N$. It leads instead to Feynman diagrams which are of \textit{melon-tadpole} type \cite{Lionni:2017xvn,Lionni:2017yvi,Prakash:2019zia} (see Fig.~\ref{fig:melon_dtadpole}), i.e. diagrams obtained by repeated insertions of either melon or tadpole two-point functions (Fig.~\ref{fig:SDE}) on the propagators of either one of the two fundamental vacuum graphs in Fig.~\ref{fig:fund_vacuum}.
The 4-colored graph corresponding to the fundamental melon is built from two mirror wheel graphs (i.e.\ completing in a straightforward way Fig.~\ref{fig:4colored}), while the triple-tadpole is built on any of the interactions.\footnote{Notice that the leading 4-colored graph of the trefoil is unique for the melonic bubbles (essentially tadpoles like to be based on multilines), while there are three leading-order trefoils that can be built on the wheel. \label{foot:trefoil}}

As tadpole corrections just renormalize the mass, the effect of $I_2$ to $I_5$, and of the $I_1$ tadpoles, will be ignored in the discussion of the Schwinger-Dyson equations for  the two-point function, assuming that we are tuning the bare mass to exactly set the effective mass to zero.
Along the same line of thoughts, we have not included quartic interactions in our action, assuming that they can be tuned to zero. 
In fact, we will be using dimensional regularization, which for massless theories results in the tadpoles (and other power-divergent integrals) being regularized to zero (e.g.\  \cite{zinnjustin}); thus we will actually need no non-trivial tuning of bare parameters, and we will be able to keep mass and quartic couplings identically zero.

\subsection{Rank $5$}
\label{sec:rank5}
\paragraph{Action.}

The second sextic model we will consider is a $O(N)^5$ Bosonic tensor model in $d$ dimensions. 
We consider a real tensor field of rank $5$, $\phi_{abcde}$ transforming under $O(N)^5$ with indices distinguished by their position. The action of the model is\footnote{The optimal scaling is now defined as $\r(J_b)=\frac{F(J_b)-10}{4}$ with a straightforward generalisation of~\cite{Carrozza:2015adg}.}:
\begin{align}
S[\phi] & = \f12 \int d^d x  \, \phi_{abcde} (   - \p_\m\p^\m)^{\zeta} \phi_{abcde} + S_{\rm int}[\phi] \, ,\\
 \label{eq:int-action-rank5}
S_{\rm int}[\phi] &= \int d^d x  \sum_{b=1}^6 \f{\k_b}{6 N^{5+\r(J_b)}} J_b \,.
\end{align}

The interaction part of the action can be written with the same graphical representation as for the previous model. However, because we are now considering a rank-5 model, the graphs representing the interactions will be $5$-colored graphs, and because we have real fields with $O(N)^5$ symmetry, the graphs will not be bipartite and the nodes will have all the same color (black). An action containing all the $O(N)^5$ invariants would be rather long,\footnote{Using the code provided in~\cite{Avohou:2019qrl} (built on a generalization of the methods of~\cite{BenGeloun:2013kta,BenGeloun:2017vwn}), we can count the total number of sextic invariants, with their different coloring choices, to be 1439.} and difficult to handle. We will restrict the potential by exploiting the large-$N$ limit: we start from the interaction whose bubble is a complete graph (i.e.\ in which for every pair of nodes there is an edge connecting them), and then include only the other interactions which are generated as radiative corrections, until  we obtain a renormalizable model, at large $N$. A set of interactions of this type has been introduced in~\cite{Ferrari:2017jgw} with the name of \emph{melo-complete family}.
As we will explain further below, it turns out that besides the complete graph we need to include only the melonic bubbles (a straightforward generalization of the melonic bubbles of rank 3) and one new non-bipartite bubble:\footnote{Following the same logic for  rank-3, starting with the wheel interaction, we would have obtained the same action as in \eqref{eq:int-action-graph}. As in that case the set of interactions exhausts the sextic $U(N)^3$ invariants, we have chosen a different perspective in its presentation.}
\be
\begin{split} \label{eq:int-action-graph-rank5}
S_{\rm int}[\phi] = & \int d^d x  
\left( \f{\kappa_1}{6 N^{5}} \vcenter{\hbox{\tikzsetnextfilename{5simplex2}
\input{5simplex.tex}}}
+ \f{\kappa_2}{6 N^{8}} \vcenter{\hbox{\tikzsetnextfilename{long_pillow_rank52}
\input{long_pillow_rank5.tex}}}
+ \f{\kappa_3}{6 N^{8}} \vcenter{\hbox{\tikzsetnextfilename{wheel_pillow2}
\input{wheel_pillow.tex}}}
 \right. \\
 & \left.
 + \f{\kappa_4}{6 N^{9}} \vcenter{\hbox{\tikzsetnextfilename{pillow_trace2}
\input{pillow_trace.tex}}}
 + \f{\kappa_5}{6 N^{10}} \vcenter{\hbox{\tikzsetnextfilename{triple_trace2}
\input{triple_trace.tex}}}
  + \f{\kappa_6}{6 N^{7}} \vcenter{\hbox{\tikzsetnextfilename{triangle2}
\input{triangle.tex}}}
  \right)\,,
\end{split}
\ee
where a sum over color permutations should be understood. The conventions are detailed in App.~\ref{ap:conventions}.

\paragraph{Colored graphs and Feynman diagrams.}

\begin{figure}[htbp]
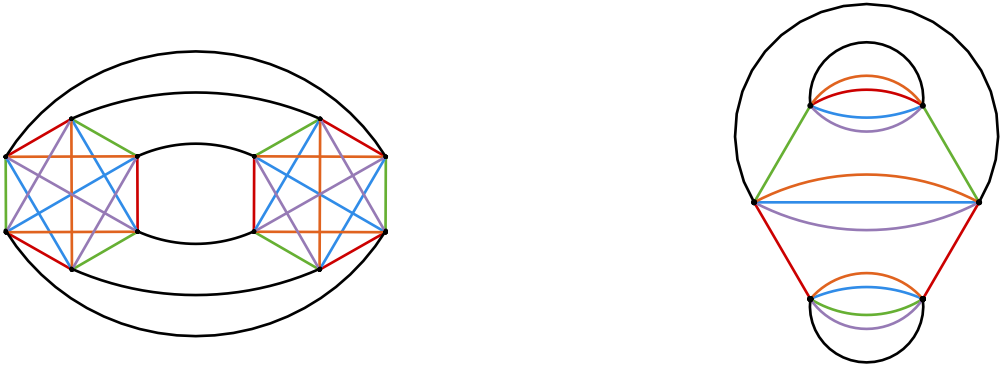

\centering
\captionsetup[subfigure]{labelformat=empty}
\vspace{-.5cm}
\begin{minipage}[c]{0.3\textwidth}
\subfloat[]{\tikzsetnextfilename{melon_freeenergy22}
\input{melon_freeenergy2.tex}}
\end{minipage}
\begin{minipage}[c]{0.5\textwidth}
\subfloat[]{\tikzsetnextfilename{tadpole_long_pillow_rank52}
\input{tadpole_long_pillow_rank5.tex}}
\end{minipage}
\vspace{-1.5cm}
\caption{Two examples of vacuum $6$-colored graphs.}
\label{fig:fund_vacuum_rank5}
\end{figure}

\begin{figure}[htbp]
\centering
\tikzsetnextfilename{rung2}
\input{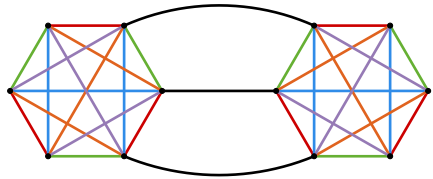}
\caption{$6$-colored graph corresponding to a two-loop correction to the six-point function, with external tensor contractions equivalent to $J_6$ (the prism).}
\label{fig:6colored}
\end{figure}

The expansion into Feynman diagrams is done similarly as for the previous model. Again, the propagators are represented by black edges. We give some examples of resulting $6$-colored graphs in Fig.~\ref{fig:fund_vacuum_rank5} and \ref{fig:6colored}.

\begin{figure}[htbp]
\centering
\tikzsetnextfilename{2rungs2}
\input{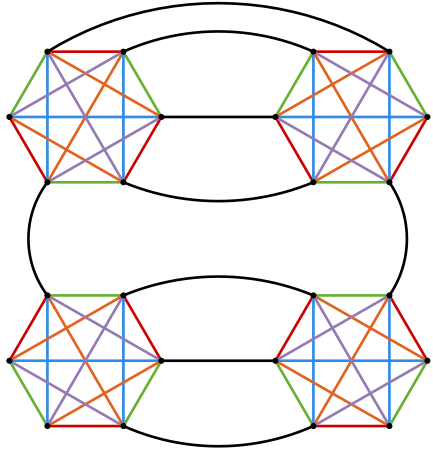}
\caption{Feynman diagram consisting in a ladder with two rungs and complete vertices. Its external tensor contractions are equivalent to $J_2$ (the long-pillow).}
\label{fig:2rungs}
\end{figure}

\paragraph{The large-$N$ expansion.} Like other tensor models, this model has also a $\frac{1}{N}$ expansion. First, we observe that every sextic interactions can be obtained as radiative corrections from the first interaction term $J_1$ (we call it the complete vertex, as its bubble is the complete graph on six vertices, also known as $K_6$). For example, the interaction $J_6$ (or the prism) is a rung with $3$ edges between two complete vertices (see Fig.~\ref{fig:6colored}), $J_2$ (or the long-pillow) and $J_4$ (or the pillow-dipole, our only double-trace interaction) are ladders made of two such rungs with different permutations of the colors between the rungs (see Fig.~\ref{fig:2rungs}). $J_5$ (or the triple-dipole, our only triple-trace interaction) is a ladder made of three rungs and $J_3$ a ladder made of four rungs. 

Then in any graph $\mathcal{G}$, we replace every interaction by their minimal representations in terms of complete vertices. This way, we obtain a new graph $\hat{\mathcal{G}}$ with only complete vertices. Since the rank of our model is a prime number, and the complete graph is the unique maximally single trace  (MST) invariant, we can use the result of~\cite{Ferrari:2017jgw} (see also \cite{Klebanov:2019jup}), where it has been proved that in this case, the leading order vacuum Feynman diagrams are the melons constructed with two mirrored complete MST interactions (see the diagram on the left in Fig.~\ref{fig:fund_vacuum_rank5}), and the usual iterative insertions of melonic two-point functions. Notice that unlike for the rank-3 wheel (which is MST, but  not a complete graph), the leading order diagrams include no tadpoles. This means that the leading order diagrams of our rank-5 model are melonic \textit{after} substituting every sextic interactions by their minimal representations in terms of the complete vertex. In terms of the original interactions, the leading order diagrams are again melon-tadpole diagrams (see Fig.~\ref{fig:melon_dtadpole}, with tadpoles now associated to  $J_b$ with $b \in [2,6]$), i.e.\ they are obtained by iterated insertions of melons and double tadpoles. The double tadpoles are based on the interactions $J_i$ vertices ($i \in [2,6]$) and the end vertices of melons are complete vertices. 
Therefore, the diagrammatics is somewhat similar to that of the quartic model \cite{Benedetti:2019eyl}, where the tetrahedron is a complete graph and it is associated to melonic diagrams, while the melonic graphs (pillow and double-trace) are associated to tadpoles.

Again, as explained for the previous model, we will ignore the effects of the tadpoles formed by $J_2$ to $J_6$, as tadpole corrections just renormalize the mass. We will also not include quartic interactions, assuming that they can be tuned to zero.

\begin{figure}[htbp]
\centering
\tikzsetnextfilename{i6non_melonic2}
\input{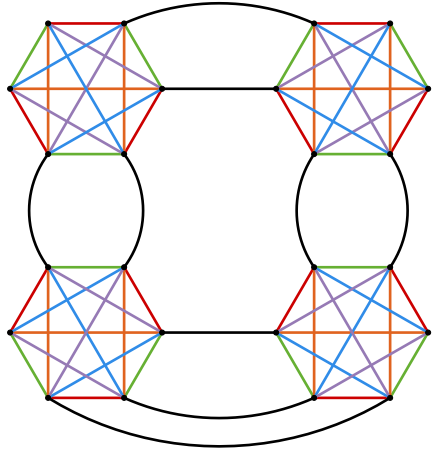}
\caption{6-colored graph corresponding to a non-melonic Feynman diagram whose exterior tensor contractions are equivalent to $J_6$.}
\label{fig:non-melonicI6}
\end{figure}

\paragraph{Radiative corrections to the prism interaction.} 
A comment is in order regarding the non-melonic interaction $J_6$. We presented in Figure~\ref{fig:6colored} a melonic contraction of two $J_1$ interactions that has $J_6$ as a boundary graph. It turns out that it is the only melonic diagram built with $J_1$ vertices that produces it. Indeed, we notice in $J_6$ the presence of two mirrored triangles (with edges red-green-blue in \eqref{eq:int-action-graph-rank5}) and each can result from 1, 2 or more complete graphs. The first case corresponds to Figure~\ref{fig:6colored}, but we see that the second case already requires non-melonic diagrams as in figure~\ref{fig:non-melonicI6}. In order to construct such a triangle from more that two $J_1$ vertices, we need at least two propagators (for the two colored edges that leave the nodes of the triangle) between each vertex, which in addition to at least two other propagators required to connect the mirror symmetric nodes of the two triangles, make the diagram non-melonic. 

\paragraph{Incomplete set of invariants.} As we said, the set of invariants we considered in the action is incomplete: there are more $O(N)^5$ invariants. However, it is closed. Indeed, we just showed that a $O(N)^5$ model with the complete interaction is dominated in the large-$N$ limit by melonic graphs. Therefore, it is enough to consider only the $O(N)^5$ invariants that can be generated from a melonic graph constructed with complete vertices. Those invariants are exactly $J_2$ to $J_6$ in the action of the model. The other $O(N)^5$ invariants will never be generated by a leading order six-point graph as they cannot be obtained from a melonic graph with complete vertices. Thus, at leading order in $N$, the set of invariants we consider is closed.

\paragraph{Rank-4 model.} 
Lastly, we notice that, as in rank 5, also in rank 4 there is a unique MST interaction \cite{Prakash:2019zia}. It turns out that the set of interactions it generates as radiative corrections are exactly of the same form as $J_2$ to $J_6$ in \eqref{eq:int-action-graph-rank5}, except that each multi-edge has one edge less than in rank 5 (for example, they can be obtained by removing the purple color in \eqref{eq:int-action-graph-rank5}).
Therefore, besides some different combinatorial factors, we do not expect important qualitative differences with respect to rank 5, and we chose to work with rank 5 as it contains a complete bubble, making the analogy to the quartic model  \cite{Benedetti:2019eyl} more evident.

\subsection{Renormalization: power counting}
\label{sec:powercount}

 We consider $\mathcal{G}$ a connected amputated Feynman diagram with $n(\mathcal{G})$ $6$-valent vertices, $E(\mathcal{G})$ edges and $r(\mathcal{G})$ external points. Computing the amplitude of the diagram $\mathcal{G}$ in momentum space, we get an independent integral $d^d p$ for every loop and a propagator $p^{-2\zeta}$ for every edge. Then, under a global rescaling of all the momenta by $t$, the amplitude is rescaled by:
\begin{align*}
t^{d(E(\mathcal{G})-n_6(\mathcal{G})-n_4(\mathcal{G})+1)-2\zeta E(\mathcal{G})}&=t^{d(n_4(\mathcal{G})+2n_6(\mathcal{G})+1-\frac{r(\mathcal{G})}{2})-\zeta\left(4n_4(\mathcal{G})+6n_6(\mathcal{G})-r(\mathcal{G})\right)}\\&=t^{d-\frac{r(\mathcal{G})}{2}(d-2\zeta)+n_6(\mathcal{G})(2d-6\zeta)+n_4(\mathcal{G})(d-4\zeta)}
\end{align*}
where we have used $2E(\mathcal{G})=6n(\mathcal{G})-r(\mathcal{G})$.

\paragraph{Short-range propagator.}

For $d=3$ and $\zeta=1$, the amplitude is rescaled as:
\begin{equation}
t^{3\left(1-\frac{r(\mathcal{G})}{6}\right)}\,.
\end{equation}

Thus, in $d=3$, the sextic interactions are marginal (the power counting does not depend on the number of internal vertices). The two-point and four-point diagrams are power divergent and the six-point diagrams are logarithmically divergent in the UV. Diagrams with more than eight external points are UV convergent. 

Therefore, in the following, we will use dimensional regularization, setting $d=3-\epsilon$. We will be interested in Wilson-Fisher type of fixed points, hence we will also consider $\epsilon$ finite, but small.

\paragraph{Long-range propagator.}

For $d<3$ and $\zeta=\frac{d}{3}$, the amplitude is rescaled as:

\begin{equation}
t^{d\left(1-\frac{r(\mathcal{G})}{6}\right)}\,.
\end{equation}

Again, the sextic interactions are marginal. The two-point and four-point diagrams are power divergent and the six-point diagrams are logarithmically divergent in the UV. Graphs with more than eight external point are UV convergent. 

We will again use dimensional regularization but in this case we will keep $d<3$ fixed and set $\zeta=\frac{d+\epsilon}{3}$.
In this case will be interested in fixed points that arise at $\epsilon=0$, as in~\cite{Benedetti:2019eyl}, by a different mechanism than in Wilson-Fisher.

\section{Two-point function}
\label{sec:SDeq}

\subsection{Rank $3$}

The standard Schwinger-Dyson equation (SDE) for the two-point function is, in momentum space:\footnote{We denote the momenta $p,q$ and so on. We define $\int_p \equiv \int \frac{d^d\; p}{(2\pi)^d}$.}
\be \label{eq:SDE}
G(p)^{-1}=C(p)^{-1}-\Sigma(p) \,,
\ee
where $G(p)$ is the Fourier transform of the full two-point function $N^{-3}\la\phib_{abc}(x)\phi_{abc}(0)\ra$, and $\Sigma(p)$ is the self-energy, i.e.\ the sum of non-trivial one-particle irreducible two-point diagrams. 

In a theory which is dominated by melon-tadpole diagrams, the self-energy at leading order in $1/N$ is obtained by summing up all the Feynman diagrams which can be obtained from those in Fig.~\ref{fig:SDE} by repeated insertions of either one of the two diagrams on internal lines. 
The resummation of all such diagrams can be represented by the same diagrams as in Fig.~\ref{fig:SDE}, but with the
edges decorated by the full two-point function. Therefore, it can be expressed in momentum space as:\footnote{See Footnote \ref{foot:trefoil} for the factor 3 in the double-tadpole contribution of the wheel.}
\be
\begin{split} \label{eq:Sigma-rank3}
\Sigma(p)&= 
\frac{\lambda_1^2}{4}\int_{q_1,q_2,q_3,q_4}G(q_1)G(q_2)G(q_3)G(q_4)G(p+q_1+q_2+q_3+q_4)\crcr
& \qquad -\frac{1}{2}(3\lambda_1+\lambda_2+\lambda_3+\lambda_4+\lambda_5)\left(\int_q G(q)\right)^2 \,.
\end{split}
\ee

\subsubsection{$\zeta=1$}

When $\zeta=1$, using a power counting argument, we see that the solution admits two regimes for  $d<3$. First, in the ultraviolet, there is a free scaling regime $G(p)^{-1}\sim p^2$: the free propagator dominates over the self energy. Second, in the infrared, there is an anomalous scaling regime $G(p)^{-1} \sim p^{2\Delta}$ with $\Delta=\frac{d}{3}$: the self energy dominates over the free propagator. Indeed, if we rescale the $q_i$ by $|p|$, the melon integral gives a global factor of $|p|^{4d-10\Delta}$ which must scale as $|p|^{2\Delta}$. This gives indeed $\Delta=\frac{d}{3}$.

We thus choose the following ansatz for the IR two-point function:\footnote{Notice that we choose a different convention for $\mathcal{Z}$ (and $Z$ below) than in~\cite{Benedetti:2019eyl}.}
\begin{equation}
G(p)=\frac{\mathcal{Z}}{p^{2d/3}}\,.
\label{eq:ansatz}
\end{equation}

Neglecting the free propagator in the IR, the SDE reduce to:
\begin{equation}
\frac{p^{2d/3}}{\mathcal{Z}}=-\frac{\lambda_1^2}{4}\mathcal{Z}^5 M_{d/3}(p)\,.
\end{equation}
The melon integral $M_{d/3}(p)$ is computed in App.~\ref{ap:melon}, giving
\begin{equation}
M_{d/3}(p)=-\frac{p^{2d/3}}{(4\pi)^{2d}}\frac{3}{d}\frac{\Gamma(1-\frac{d}{3})\Gamma(\frac{d}{6})^5}{\Gamma(\frac{d}{3})^5\Gamma(\frac{5d}{6})}~.
\end{equation}

We thus obtain:
\begin{equation}
\mathcal{Z}=\left(\frac{\lambda_1^2}{4(4\pi)^{2d}}\frac{3}{d}\frac{\Gamma(1-\frac{d}{3})\Gamma(\frac{d}{6})^5}{\Gamma(\frac{d}{3})^5\Gamma(\frac{5d}{6})}\right)^{-1/6} \,.
\end{equation}

\paragraph{Wave function renormalization.}

We introduce the wave function renormalization as $\phi=\phi_R\sqrt{Z}$ with $\phi$ the bare field and $\phi_R$ the renormalized field.
Notice that $Z$ is distinguished from $\mathcal{Z}$, as the latter is the full coefficient of the nonperturbative solution in the IR limit, while $Z$ is the usual perturbative wave function renormalization, to be fixed by a renormalization condition, as we will specify below.

After renormalization of the mass terms to zero, we have for the expansion of the renormalized two-point function at lowest order:
\begin{equation}
\Gamma^{(2)}_R(p)\equiv G_R(p)^{-1} =Zp^2-\frac{\lambda_1^2}{4} Z^{-5} M_1(p) \,.
\label{eq:gamma2}
\end{equation}
The integral $M_1(p)$ is computed in App.~\ref{ap:melon}, Eq.~\eqref{eq:M-1}. At leading order in $\epsilon$, we have:
\begin{equation}
M_1(p)=-p^{2-4\epsilon}\frac{2\pi^2}{3\epsilon(4\pi)^{6}} + \mathcal{O}(1)\,.
\end{equation}
At last, we fix $Z$ such that
\be
\lim_{\epsilon\to 0}\frac{d\Gamma^{(2)}_R(p)}{dp^2}|_{p^2=\mu^2}=1\,,
\ee
with $\mu$ the renormalization scale. 
At quadratic order in $\lambda_1$, we obtain:
\begin{equation}
Z=1+\frac{\lambda_1^2}{4}\tilde{M}_1(\mu) =1-\mu^{-4\epsilon}\frac{\lambda_1^2\pi^2}{6\epsilon(4\pi)^{6}}\,,
\label{eq:wavef3}
\end{equation}
with $\tilde{M}_1(\mu)=\frac{d}{dp^2}M_1(p)|_{p^2=\mu^2}$.

\subsubsection{$\zeta=\frac{d}{3}$}

The value of $\zeta$ in this case is chosen to match the infrared scaling of the two-point function. We now have only one regime and the full SDE is solved by the ansatz:
\begin{equation}
G(p)=\frac{\mathcal{Z}}{p^{2d/3}} \,.
\label{eq:ansatz2}
\end{equation}
For the vertex renormalization in Sec.~\ref{sec:betas} we will use analytic regularization, keeping $d<3$ fixed and setting $\zeta=\frac{d+\epsilon}{3}$, but since the two-point function is finite,  as we will now see, we can here set $\epsilon=0$.

The computations are the same as in the IR limit of the previous section, but we do not neglect the free propagator. Thus, we obtain:
\begin{equation}
\frac{1}{\mathcal{Z}^6}-\frac{1}{\mathcal{Z}^5}=\frac{\lambda_1^2}{4(4\pi)^{2d}}\frac{3\Gamma(1-\frac{d}{3})\Gamma(\frac{d}{6})^5}{d\Gamma(\frac{d}{3})^5\Gamma(\frac{5d}{6})} \,.
\label{Z-norm-d/3}
\end{equation}
At the first non-trivial order in the coupling constant, this gives:
\begin{equation}  \label{eq:wavef3-LR}
\mathcal{Z}=1-\frac{\lambda_1^2}{4(4\pi)^{2d}}\frac{3\Gamma(1-\frac{d}{3})\Gamma(\frac{d}{6})^5}{d\Gamma(\frac{d}{3})^5\Gamma(\frac{5d}{6})}+\mathcal{O}(\lambda_1^4).
\end{equation}
This expression is finite for $d<3$. Moreover, as we did not neglect the free propagator, $\mathcal{Z}$ has an expansion in $\lambda_1$, as the perturbative wave function renormalization, with which it can be identified in our non-minimal subtraction scheme. Therefore, in the case $\zeta=d/3$, the wave function renormalization is finite.

\subsection{Rank $5$}

For the $O(N)^5$ model, the Schwinger-Dyson equation in the large-$N$ limit is:
\be
G(p)^{-1}=C(p)^{-1}-\Sigma(p) \,,
\ee
with
\be
\begin{split} \label{eq:Sigma-rank5}
\Sigma(p)&= 
\frac{\kappa_1^2}{6}\int_{q_1,q_2,q_3,q_4}G(q_1)G(q_2)G(q_3)G(q_4)G(p+q_1+q_2+q_3+q_4)\\
& \qquad -(\kappa_2+\kappa_3+\kappa_4+\kappa_5+\kappa_6)\left(\int_q G(q)\right)^2  \,.
\end{split}
\ee
The only differences with the rank-3 model are the combinatorial factors in front of the melon and tadpole integrals. Thus, we can use the results of the previous section.

\subsubsection{$\zeta=1$}

In the IR limit, the SDE is solved again by $G(p)=\frac{\mathcal{Z}}{p^{2d/3}}$ with:
\begin{equation}
\mathcal{Z}=\left(\frac{\kappa_1^2}{6(4\pi)^{2d}}\frac{3}{d}\frac{\Gamma(1-\frac{d}{3})\Gamma(\frac{d}{6})^5}{\Gamma(\frac{d}{3})^5\Gamma(\frac{5d}{6})}\right)^{-1/6} \,.
\end{equation}
The wave function renormalization is given by:
\begin{equation} \label{eq:wavef5}
Z=1+\frac{\kappa_1^2}{6}\tilde{M}_1(\mu) =1-\mu^{-4\epsilon}\frac{\lambda_1^2\pi^2}{9\epsilon (4\pi)^{6}}\,.
\end{equation}

\subsubsection{$\zeta=\frac{d}{3}$}

The full SDE is solved again by $G(p)=\frac{\mathcal{Z}}{p^{2\zeta}}$ with:
\begin{equation} \label{eq:wavef5-LR}
\mathcal{Z}=1-\frac{\kappa_1^2}{2(4\pi)^{2d}}\frac{\Gamma(1-\frac{d}{3})\Gamma(\frac{d}{6})^5}{d\Gamma(\frac{d}{3})^5\Gamma(\frac{5d}{6})}+ \mathcal{O}(\kappa_1^3) \,.
\end{equation}
This is directly the wave function renormalization which is thus finite.

\section{2PI effective action and four-point kernels}
\label{sec:kernels}

In this section, we compute the four-point kernels of both models using the 2PI formalism (see \cite{Benedetti:2018goh}).
We will make use of them first for showing that indeed there is no need of counterterms with quartic interactions, and then, in the next section, for the discussion of the all-orders beta functions for the sextic couplings.

\subsection{Rank 3}

In rank 3 and at leading order in $1/N$, the 2PI effective action is given by: 
\begin{equation}
\label{eq:2PIrank3}
\begin{split}
-\G^{2PI}[\mbG] = &-\f{1}{6}\left(\f{3 \l_1}{N^3}\d^{(1)}_{\mba\mbd\mbb\mbc\mbe\mbf}+\sum_{i=2}^5 \f{\l_i}{N^{3+\rho(I_i)}}\d^{(i)}_{\mba\mbb;\mbc\mbd;\mbe\mbf}\right) \int\dd x ~\mbG_{(\mba,x)(\mbb,x)}\mbG_{(\mbc,x)(\mbd,x)}\mbG_{(\mbe,x)(\mbf,x)} \\
&+\f{1}{2}\left(\f{\l_1}{6N^3}\right)^2 3\, \d^{(1)}_{\mba\mbb\mbc\mbd\mbe\mbf} \d^{(1)}_{\mbg\mbh\mbj\mbk\mbm\mbn} \times\\
&\quad \int\dd x\dd y ~\mbG_{(\mba,x)(\mbg,y)}\mbG_{(\mbb,x)(\mbh,y)} \mbG_{(\mbc,x)(\mbj,y)}\mbG_{(\mbd,x)(\mbk,y)}\mbG_{(\mbe,x)(\mbm,y)}\mbG_{(\mbf,x)(\mbn,y)} \,.
\end{split}
\end{equation}
This is obtained by summing the contributions of the leading-order vacuum diagrams which are also two-particle irreducible (2PI), i.e.\ cannot be disconnected by cutting two lines, and with arbitrary propagator $\mbG$ on each line. As we already know, all the leading-order vacuum diagrams are obtained from the diagrams in Fig.~\ref{fig:fund_vacuum} by repeated insertions of the two-point diagrams in Fig.~\ref{fig:SDE}, but since all such insertions lead to two-particle reducible diagrams, we are left with just the two fundamental diagrams of Fig.~\ref{fig:fund_vacuum}, whose evaluation leads to Eq.~\eqref{eq:2PIrank3}.

One recovers the self-energy from (using the further condensed notation $A=(\mba,x)$):
\begin{equation} \label{eq:Sigma2PI-complex}
\Sigma[\mbG]_{AB} = -\f{\d\G^{2PI}[\mbG]}{\d\mbG_{AB}} \,,
\end{equation}
which can be seen to reproduce Eq.~\eqref{eq:Sigma-rank3} in momentum space.

\begin{figure}[htbp]
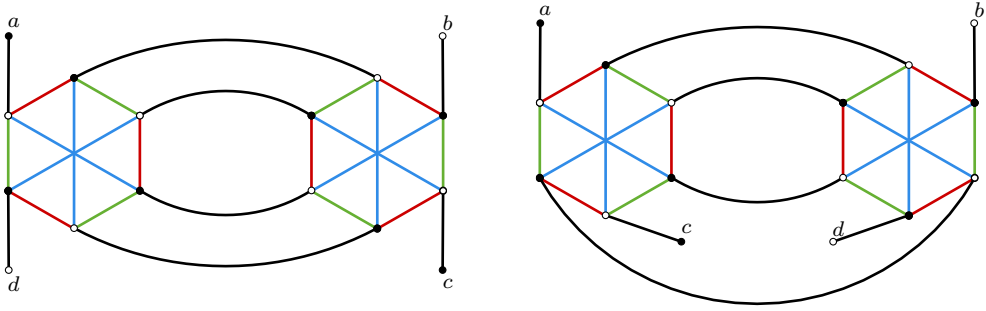

\captionsetup[subfigure]{labelformat=empty}
\vspace{-.5cm}
\begin{minipage}[c]{0.4\textwidth}
\subfloat[]{\tikzsetnextfilename{kernel_wheel2}
\input{kernel_wheel.tex}}
\end{minipage}
\hspace{.5cm}
\begin{minipage}[c]{0.4\textwidth}
\subfloat[]{\tikzsetnextfilename{kernel_wheel_dt2}
\input{kernel_wheel_dt.tex}}
\end{minipage}
\vspace{-.5cm}
\caption{The contractions corresponding to a 4-point rung in the ladder making the 4-point function (a pillow on the left and a double trace on the right).}
\label{fig:kernel_wheel}
\end{figure}

The right-amputated four-point kernel on-shell is obtained by taking two derivatives of $\G^{2PI}[\mbG]$ with respect to $\mbG$  and then multiplying by two full propagators on the left:
\be \label{eq:defK}
K[\mbG]_{AB,CD} = \mbG_{AA'}\mbG_{BB'}\f{\d\Sigma[\mbG]_{CD}}{\d\mbG_{A'B'}}\,.
\ee
Applying such definition to Eq.~\eqref{eq:2PIrank3} we obtain:
\be
\begin{split}
K_{(\mba,x)(\mbb,y)(\mbc,z)(\mbd,w)}=&\int dx'dy' \, G_{xx'}G_{yy'}\left[ 
 -\f{1}{3}(9\l_1+ 2\l_2 + 3\l_3 + \l_4)\delta_{x'y'}\delta_{x'z}\delta_{x'w}G_{x'x'}\hat{\delta}^p_{\mba \mbb;\mbc\mbd}\right.\crcr
&\quad -\f{1}{3}(\l_2 + 2 \l_4+3\l_5)\delta_{x'y'}\delta_{x'z}\delta_{x'w}G_{x'x'}\hat{\delta}^d_{\mba \mbb;\mbc\mbd} \crcr
&\quad \left.+\frac{\l_1^2}{4}G_{x'y'}^4(3\hat{\delta}^p_{\mba \mbb;\mbc\mbd} \delta_{x'w}\delta_{y'z}+ 2\hat{\delta}^d_{\mba \mbb;\mbc\mbd}\delta_{x'z}\delta_{y'w})\right] \,,
\end{split}
\ee
where we defined the rescaled pillow and double-trace contraction operators $\hat{\delta}^p_{\mba \mbb;\mbc\mbd}=\frac{1}{N^2}\delta^p_{\mba \mbb;\mbc\mbd}$ and $\hat{\delta}^d_{\mba \mbb;\mbc\mbd}=\frac{1}{N^3}\delta^d_{\mba \mbb;\mbc\mbd}$ (see App.~\ref{ap:conventions}).
The colored graphs corresponding to the last line are depicted in Fig.\ref{fig:kernel_wheel}.

In momentum space this four-point kernel becomes:
\be
\begin{split}
K_{(\mba,p_1)(\mbb,p_2)(\mbc,p_3)(\mbd,p_4)}=&(2\pi)^d\delta(p_1+p_2+p_3+p_4)G(p_1)G(p_2)\left[ -\f{1}{3}(9\l_1+ 2\l_2 + 3\l_3 + \l_4)\int_q G(q)\hat{\delta}^p_{\mba \mbb;\mbc\mbd} \right.\crcr
&\quad -\f{1}{3}(\l_2 + 2 \l_4+3\l_5)\int_q G(q)\hat{\delta}^d_{\mba \mbb;\mbc\mbd} \crcr
&\quad +\frac{\l_1^2}{4}\left(3\hat{\delta}^p_{\mba \mbb;\mbc\mbd} \int_{q_1,q_2,q_3}G(q_1)G(q_2)G(q_3)G(-p_1-p_4-q_1-q_2-q_3)\right.\\
&\qquad \left.\left.  + 2\hat{\delta}^d_{\mba \mbb;\mbc\mbd}\int_{q_1,q_2,q_3}G(q_1)G(q_2)G(q_3)G(-p_1-p_3-q_1-q_2-q_3)\right) \right] \,.
\end{split}
\ee
For convenience, we introduce also a reduced kernel, with the tadpoles set to zero, i.e.:
\begin{equation}
\hat{K}_{(\mba,x)(\mbb,y)(\mbc,z)(\mbd,w)}=\frac{\l_1^2}{4}G_{zw}^4(3\hat{\delta}^p_{\mba \mbb;\mbc\mbd} G_{xw}G_{yz} + 2\hat{\delta}^d_{\mba \mbb;\mbc\mbd} G_{xz}G_{yw}) \,.
\label{kernel-rank3}
\end{equation}
In fact, since the propagator is massless, the tadpoles are zero in dimensional regularization, hence the reduced kernel is all we need.

\begin{figure}[htbp]
\centering
\vspace{.4cm}
\tikzsetnextfilename{ladders2}
\input{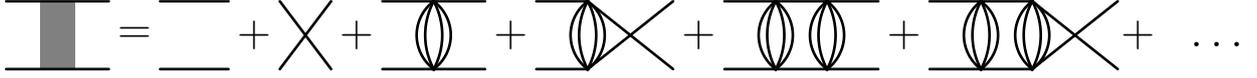}
\caption{The full forward four-point function as a series of ladders. The crossings should be omitted for our rank-3 model, because it is built on complex fields, with bipartite graphs. However, they contribute for the rank-5 model, which has real fields.
}
\label{fig:ladders}
\end{figure}

The full four-point function  at leading order in $1/N$ is obtained by summing ladders of arbitrary lenghts with the (reduced) four-point kernel acting as rung (see \cite{Benedetti:2019eyl,Gurau:2019qag}). 
More precisely, defining the forward four-point function as
\be \label{eq:fw4pt}
\cF_{(\mba,x)(\mbb,y)(\mbc,z)(\mbd,w)} \equiv  \langle{\phi_{\mba}(x) \bar{\phi}_{\mbb}(y) \phi_{\mbc}(z) \bar{\phi}_{\mbd}(w)} \rangle - G(x-y) G(z-w) \d_{\mba\mbb} \d_{\mbc\mbd} \,,
\ee
one finds that at leading order in $1/N$ it is given by a geometric series on the (reduced) kernel:
\be \label{eq:fw4pt-ladders}
\cF_{(\mba,x)(\mbb,y)(\mbc,z)(\mbd,w)} = \int dz'dw' \,  (\mathbf{1} - \hat{K})^{-1}_{(\mba,x)(\mbb,y)(\mbc,z')(\mbd,w')} 
 \, G_{w'w}G_{z'z}\,.
\ee
We represent the series of ladder diagrams in Fig.~\ref{fig:ladders}, where the crossings do not contribute here because we consider a bipartite model with $U(N)^3$ symmetry.

For dimensional reasons, the propagators being massless and by the use of dimensional regularization, we do not expect the four-point function to require a renormalization of the quartic couplings, which are dimensionful. We verify this explicitly at lowest order in perturbation theory, that is by considering the fully amputated four-point kernel, with $G(q)$ replaced by the bare propagator $C(q)$.
Therefore, the reduced kernel writes:
\begin{equation}
\frac{\lambda_1^2}{4}\mathcal{Z}^4 (3\hat{\delta}^p_{\mba \mbb;\mbc\mbd}U_{\zeta}(p_1+p_4) + 2\hat{\delta}^d_{\mba \mbb;\mbc\mbd}U_{\zeta}(p_1+p_3))\,,
\end{equation}
with 
\begin{equation}
U_{\zeta}(p_1+p_4)=\int_{q_1,q_2,q_3}\frac{1}{q_1^{2\zeta}q_2^{2\zeta}q_3^{2\zeta}(p_1+p_4+q_1+q_2+q_3)^{2\zeta}}\,.
\end{equation}

Using Eq.~\eqref{eq:intG}, we find:
\begin{equation}
U_{\zeta}(p_1+p_4)=\frac{|p_1+p_4|^{3d-8\zeta}}{(4\pi)^{3d/2}}\frac{\Gamma(d/2-\zeta)^4\Gamma(4\zeta-3d/2)}{\Gamma(\zeta)^4\Gamma(2d-4\zeta)}\,.
\end{equation}

\paragraph{Standard propagator.} For $\zeta=1$ and $d=3-\epsilon$, this is finite (no poles in $\epsilon$):
\begin{equation}
U_1(p_1+p_4)=-\frac{|p_1+p_4| \pi}{4} \,.
\end{equation}

\paragraph{Long-range propagator.} For $\zeta=d/3$ and $d<3$, this is also finite:
\begin{equation}
U_{d/3}(p_1+p_4)=\frac{|p_1+p_4|^{d/3}}{(4\pi)^{3d/2}}\frac{\Gamma(d/6)^4\Gamma(-d/6)}{\Gamma(d/3)^4\Gamma(2d/3)} \,.
\end{equation}

In both cases, there are no divergences in the four-point kernel. We thus do not need to renormalize the four-point couplings and we can take them to be zero from the beginning.

\subsection{Rank 5}

In rank $5$ and at leading order in $1/N$, the 2PI effective action is given by: 
\begin{equation}
\label{eq:2PIrank5}
\begin{split}
-\G^{2PI}[\mbG] = &-\f{1}{6}\left(\sum_{i=2}^6 \f{\kappa_i}{N^{5+\rho(J_i)}}\d^{(i)}_{\mba\mbb;\mbc\mbd;\mbe\mbf}\right) \int\dd x ~\mbG_{(\mba,x)(\mbb,x)}\mbG_{(\mbc,x)(\mbd,x)}\mbG_{(\mbe,x)(\mbf,x)} \\
&+\f{1}{2}\left(\f{\kappa_1}{6N^5}\right)^2 \d^{(1)}_{\mba\mbb\mbc\mbd\mbe\mbf} \d^{(1)}_{\mbg\mbh\mbj\mbk\mbm\mbn}\times\\
&\qquad \int\dd x\dd y ~\mbG_{(\mba,x)(\mbg,y)}\mbG_{(\mbb,x)(\mbh,y)} \mbG_{(\mbc,x)(\mbj,y)}\mbG_{(\mbd,x)(\mbk,y)}\mbG_{(\mbe,x)(\mbm,y)}\mbG_{(\mbf,x)(\mbn,y)} \,.
\end{split}
\end{equation}

One recovers the self-energy from:
\begin{equation}
\Sigma[\mbG] = -2\f{\d\G^{2PI}[\mbG]}{\d\mbG} \,,
\end{equation}
where the extra factor  2 with respect to Eq.~\eqref{eq:Sigma2PI-complex} is due to the difference between real and complex fields.
The amputated four-point kernel is still obtained by derivating the self energy with respect to $\mbG$.

The right-amputated four-point kernel on-shell is then:
\be
\begin{split}
K_{(\mba,x)(\mbb,y)(\mbc,z)(\mbd,w)}=&G_{xx'}G_{yy'}\left[ -2\left(\kappa_{6}+\kappa_{3}+\frac{2\kappa_{2}}{3}+\frac{\kappa_{4}}{3}\right)\delta_{x'y'}\delta_{x'z}\delta_{x'w}G_{x'x'}\hat{\delta}^p_{\mba\mbb;\mbc\mbd} \right.\crcr
&-2\left(\frac{\kappa_{2}}{3}+\frac{2\kappa_{4}}{3}+\kappa_{5}\right)\delta_{x'y'}\delta_{x'z}\delta_{x'w}G_{x'x'}\hat{\delta}^d_{\mba\mbb;\mbc\mbd} \\
&\left.+\frac{5\kappa_1^2}{6}\delta_{x'w}\delta_{y'z}G_{x'y'}^4\hat{\delta}^p_{\mba\mbb;\mbc\mbd}\right] \,. 
\end{split}
\ee

The structure is the same as for the rank-$3$ model: the only difference comes from the combinatorial factors. Then, the Feynman amplitudes are the same as before and there are still no divergences. We can thus again take the four-point couplings to be zero from the beginning. Eliminating also the tadpoles, the four-point kernel is reduced to:
\begin{equation}
\hat{K}_{(\mba,x)(\mbb,y)(\mbc,z)(\mbd,w)}=G_{xw}G_{yz}\frac{5\kappa_1^2}{6}G_{zw}^4\hat{\delta}^p_{\mba\mbb;\mbc\mbd} \,.
\end{equation}

\section{Beta functions}
\label{sec:betas}

We have seen in Sec.~\ref{sec:SDeq} that the Schwinger-Dyson equations for the two-point functions admit a conformal IR limit for $\zeta=1$, and a conformal solution valid at all scales for $\zeta=d/3$. The argument is by now quite standard in theories with a melonic large-$N$ limit, and in one dimension, for the SYK model or its tensor generalizations, it is sufficient for concluding that the theory is conformal (in the IR limit or at all scales). However, for field theories in higher dimensions we should also consider the renormalization of the couplings. In particular, it is not possible to restrict the model to having only one interaction (the one leading to melonic diagrams), as we have seen that other interactions are generated by radiative corrections, and these lead to a renormalization group flow of the other couplings, which hence cannot be set to zero. In fact, in order to claim that we found a non-trivial conformal field theory, we should identify an interacting fixed point of the renormalization group.\footnote{In principle  a fixed point provides us only with a scale invariant theory, full conformal invariance having to be proved case by case or on the basis of the available theorems in dimensions two and four. See for example \cite{Nakayama:2013is} for a review.}
Therefore, in this section we will study the beta functions for the full actions \eqref{eq:int-action-graph} and \eqref{eq:int-action-graph-rank5}, and their relative fixed points.

We will explain the general structure of the beta functions in the rank-3 case. As we will see, the rank-5 case is very similar, except for the presence of an additional type of interaction, $J_6$ (the prism), a difference which however turns out to be crucial.

\subsection{Rank $3$}

The amputated connected six-point function can be decomposed into the different interaction terms:
\begin{equation}
\Gamma^{(6)}(p_1,\ldots , p_6)=\sum_{i=1}^5\Gamma^{(6,i)}(p_1,\ldots , p_6)\hat{\delta}^{i}\,.
\end{equation}

The renormalized sextic couplings $g_i$ are defined in terms of the  bare expansion of the six-point functions by the renormalization condition:
\begin{equation}
\label{eq:renorcouplings}
g_i=\mu^{-2\epsilon}Z^3\Gamma^{(6,i)}(p_1,\ldots , p_6)
\end{equation}
where the power of the renormalization scale $\m$ is  fixed by demanding that $g_i$ are dimensionless, and it is the same both for $\zeta=1$ in $d=3-\epsilon$ dimensions  and for $\zeta=\frac{d+\epsilon}{3}$ in general $d<3$.
For the external momenta we choose a symmetric subtraction point (generalizing the quartic case, see \cite{Brezin:1974eb,Kleinert:2001ax}):
\be
p_i \cdot p_j =\frac{\mu^2}{9}\left(6\delta_{ij}-1\right) \,.
\ee
This choice of external momenta satisfies the momentum conservation, $\sum_{i=1}^6 p_i=0$, and it is non-exceptional, in the sense that $\sum_{i\in I} p_i\neq 0$ for any subset $I$ of the set of indices $\{1,\ldots,6\}$, therefore avoiding IR divergences in all diagrams.

The beta functions are defined by $\b_i = \m\p_\m g_i$. We will obtain them by differentiating the bare expansion with respect to $\mu$ (using the fact that the bare couplings are independent of the renormalization scale $\mu$) and then replacing the bare couplings by their expressions in terms of the renormalized one. 

At leading order in $1/N$ the wheel vertex does not receive any radiative corrections, i.e.:
\be
g_1=\mu^{-2\epsilon}Z^3\l_1\,.
\ee
Since $Z=1+\cO(\l_1^2)$, the inverse $\l_1(g_1)$ is guaranteed to exist in the perturbative expansion, at least.\footnote{For the long-range model, it is actually easier to write the inverse relation, because at $\epsilon=0$ we can solve the exact equation \eqref{Z-norm-d/3} by multiplying it by $\mathcal{Z}^6$ and using $\mathcal{Z}^6\l_1^2=g_1^2$:
\begin{equation*}
\mathcal{Z} = 1- \f{g_1^2}{g_c^2}\,, \;\;\;\; g_c^{-2} = \frac{1}{4(4\pi)^{2d}}\frac{3\Gamma(1-\frac{d}{3})\Gamma(\frac{d}{6})^5}{d\Gamma(\frac{d}{3})^5\Gamma(\frac{5d}{6})} \,.
\end{equation*}
Therefore, $\l_1=g_1/\mathcal{Z}^3$ exists for $g_1<g_c$.
\label{foot:g_c}
}
Its beta function will then be
\be
\beta_1 \equiv \m \partial_\m g_1 = (-2\epsilon+3 \eta)  g_1\,,
\ee
where we defined the anomalous dimension $\eta= (\m\partial_\m \ln Z)|_{\l_1(g_1)}$. Clearly, if $\epsilon=0$ and $Z$ is finite, as in the long-range case $\z=d/3$ with $d<3$, then the beta function is identically zero, and we have a chance of finding a one-parameter family of fixed points, as in~\cite{Benedetti:2019eyl}. On the other hand, for $\epsilon>0$, in order to find a non-trivial fixed point we have to rely on a Wilson-Fisher type of cancellation between the tree level term and the quantum corrections, hence we need $\eta\neq 0$, that is, we need a short-range propagator.

\begin{figure}[htbp]
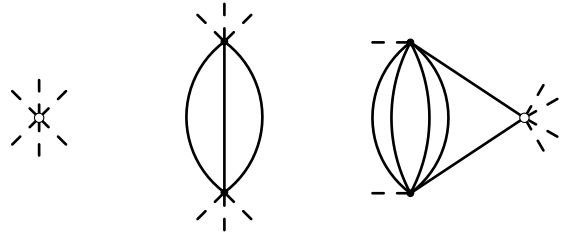

\centering
\captionsetup[subfigure]{labelformat=empty}
\subfloat[]{\tikzsetnextfilename{bare_vertex2}
\input{bare_vertex.tex}}
\hspace{1cm}
\subfloat[]{\tikzsetnextfilename{rung32}
\input{rung3.tex}}
\hspace{1cm}
\subfloat[]{\tikzsetnextfilename{order32}
\input{order3.tex}}
\caption{The three diagrams that contribute to the bare expansion of the six-point couplings up to order three (bare coupling, $D$ and $S$, see Sec.~\ref{sec:betas1}). The black circles represent wheel vertices and the white circles can be any of the $I_i$ interactions (including the wheel itself).}
\label{fig:bare3}
\end{figure}

We now compute the bare expansion of the other couplings. The expansion starts of course at tree level, with a bare vertex with any $I_i$ interaction.
At order two in the coupling constants, there is only one diagram which contributes: two wheel vertices connected by three internal edges (we call this Feynman diagram the \textit{candy}). At order three, we have one more diagram: two wheel vertices connected to each other by four internal edges and each of them connected by another internal edge to a vertex with any $I_i$ interaction (including the wheel itself). These diagrams are the only tadpole-free six-point diagrams that can be obtained by cutting edges of vacuum melon-tadpole diagrams, at this order in the couplings, and they are pictured in Fig.~\ref{fig:bare3}.

We can actually construct the leading order $6$-point graphs at all orders using the forward four-point function introduced in Eq.~\eqref{eq:fw4pt-ladders}. 
Indeed, the amputated connected six-point functions can be obtained by deleting  three different lines in the vacuum diagrams, without disconnecting the diagrams. On the other hand, vacuum diagrams are given in Fig.~\ref{fig:fund_vacuum}, with lines decorated by melonic and tadpole insertions, but we should not leave any closed tadpoles otherwise the diagram will evaluate to zero in dimensional regularization. Therefore, we can have at most one tadpole vertex; this fact does not limit the number of wheel vertices, as they can appear in melonic insertions as well, but it has the important consequence that the couplings $\l_2$ to $\l_5$ appear at most linearly in $\Gamma^{(6)}$.
Equivalently, we can just consider the two diagrams in Fig.~\ref{fig:fund_vacuum} with only melonic insertions. Furthermore, for the trefoil on the right of Fig.~\ref{fig:fund_vacuum}, we should cut an internal line on each of the three (decorated) leaves. At last, we should notice that each time we delete a line in a melonic two-point function, we generate a ladder diagram. In fact, starting from the SDE equation $G=(C^{-1}-\Sigma[G])^{-1}$, and using \eqref{eq:defK}, we obtain
\be
\f{\d G_{AB}}{\d C_{EF}} = (1-K)^{-1}_{AB,A'B'} G_{A'E'} C^{-1}_{E'E} G_{B'F'} C^{-1}_{F'F} + (E\leftrightarrow F)\,.
\ee
When using this formula on vacuum diagrams, we should then strip off the factors $G\cdot C^{-1}$ in order to obtain amputated $n$-point functions. We thus obtain the right-amputated version of Eq.~\eqref{eq:fw4pt-ladders}.

\begin{figure}[htbp]
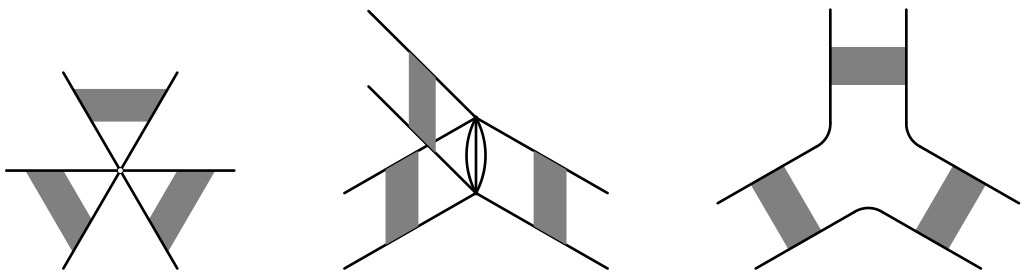

\centering
\captionsetup[subfigure]{labelformat=empty}
\subfloat[]{\tikzsetnextfilename{bare_vertex_kernel2}
\input{bare_vertex_kernel.tex}}
\hspace{1cm}
\subfloat[]{\tikzsetnextfilename{candy_kernel2}
\input{candy_kernel.tex}}
\hspace{1cm}
\subfloat[]{\tikzsetnextfilename{3kernels2}
\input{3kernels.tex}}
\caption{The three types of diagrams contributing to the bare expansion of the six-point couplings in the large-$N$ limit at all order in the coupling constants. The black circles represent wheel vertices and the white circles can be any of the $I_i$ interactions (including the wheel). The grey squares represent the full forward four-point function.}
\label{fig:allorders}
\end{figure}

In conclusion, we then have three different types of leading-order $6$-point graphs. First, we can connect three full forward four-point functions on every pairs of external legs of the bare vertex of Fig.~\ref{fig:bare3}, thus obtaining the graph on the left of Fig.~\ref{fig:allorders}. We can also do the same with the candy and obtain the graph in the middle of Fig.~\ref{fig:allorders}.  Finally, we can also connect three full forward four-point functions and obtain the graph on the right of Fig.~\ref{fig:allorders}. The last two have been encountered for example in  \cite{Gross:2017hcz,Gross:2017aos}, where they have been called \emph{contact} and \emph{planar} diagrams, respectively.

This implies that renormalized couplings $g_i$, with $i>1$, have a bare expansion of the form:
\be
g_i = \m^{-2\epsilon} Z^3 \left( \l_i +  A_i(\l_1^2 )+ \sum_{j=1\ldots 5}  B_{ij}(\l_1^2) \l_j \right)\,,
\ee
with $A_i(x)$ and $B_{ij}(x)$ starting at linear order in $x$. The term $ \l_i + \sum_{j=1\ldots 5}  B_{ij}(\l_1^2) \l_j$ is a resummation of contribution from the graphs on the left of Fig.~\ref{fig:allorders}, while $A_i(\l_1^2 )$ resums the other two. Although we could at least write the relative Feynman integral expressions in terms of forward four-point functions and six-point kernels, as we will not need them, and the combinatorics is different for different bubbles, we will not be more precise than that.

For $i>1$ the relation between bare and renormalized couplings is linear and thus it can be easily inverted:
\be
\l_i = (\mathbf{1}+B)^{-1}_{ij} \left( \f{\m^{2\epsilon} g_j}{Z^3} - A_j \right)\big|_{\l_1=\l_1(g_1)}\,,
\ee
where $\mathbf{1}_{ij}=\delta_{ij}$.

Using the fact that the flow of $g_1$ is independent of the others, then one arrives at the conclusion that the beta functions of the other couplings are linear combinations of the couplings, with coefficients that are functions of $g_1^2$:
\be
\beta_i = (-2\epsilon+3 \eta)  g_i + \tilde{A}_i(g_1^2) + \sum_j  \tilde{B}_{ij}(g_1^2)  g_j \,,
\ee
where 
\begin{align}
\tilde{A}_i(g_1^2)  &=  \m^{-2\epsilon} Z^3 \left(  \m\p_\m A_i - \sum_{j,k}  (\m\p_\m B_{ij}) (\mathbf{1}+B)^{-1}_{jk} A_k \right)\big|_{\l_1=\l_1(g_1)} \,,\\
\tilde{B}_{ij}(g_1^2) &=  \sum_{j}  (\m\p_\m B_{ij}) (\mathbf{1}+B)^{-1}_{jk} \,.
\end{align}

As we saw above, the combination $-2\epsilon+3 \eta$ is either identically zero, or it is zero at the fixed point of $g_1$.
In order to find the fixed points we are left with a linear problem.

In the following we will compute explicitly the beta functions at lowest order in the perturbative expansion, i.e.\ we will only include the contribution of the diagrams in Fig.~\ref{fig:bare3}.

\subsubsection{The $\zeta=1$ case}
\label{sec:betas1}

We will now compute the beta function only up to order $3$ in the coupling constants for $\zeta=1$. 

Expanding the six-point functions of Eq. \eqref{eq:renorcouplings} to order three, we have the bare expansions:
\begin{align}
g_{2}&=\mu^{-2\epsilon}Z^3\left(\l_{2}-\frac{9}{2}D_1(\mu)\l_1^2+S_1(\mu)\l_1^2\left(9\l_1 + \frac{1}{2}\l_2\right)\right)\,,\crcr
g_{3}&=\mu^{-2\epsilon}Z^3\left(\l_{3}+ S_1(\mu)\frac{3}{4}\l_1^2\l_{3}\right)\,,\crcr
g_{4}&=\mu^{-2\epsilon}Z^3\left(\l_{4}+ S_1(\mu)\l_1^2\left(\f{27}{2}\l_1 +3 \l_2+3\l_3+\f{11}{4}\l_4\right)\right)\,,\crcr
g_{5}&=\mu^{-2\epsilon}Z^3\left(\l_{5}-D_1(\mu)\f{1}{6}\l_1^2+ S_1(\mu)\l_1^2\left(\f{1}{4}\l_2+\l_4+\f{15}{4}\l_5\right)\right)\,,
\end{align}
where $Z$ is given in Eq.~\eqref{eq:wavef3}, $D_1(\mu)$ is the candy integral, and $S_1(\mu)$ the integral corresponding to the Feynman diagram on the right of Fig.~\ref{fig:bare3}. The integrals are both computed in App.~\ref{ap:betafun4}.
It is convenient to rescale the couplings as $\tilde{g}_i=g_i/(4\pi)^d$ (and we immediately forget the $\sim$). To compute the beta functions we need:
\begin{gather}
\mu\partial_{\mu} D_1=-\frac{4\pi}{(4\pi)^3}\mu^{-2\epsilon}+\mathcal{O}(\epsilon)\,,\\
\mu\partial_{\mu} S_1 =\frac{8\pi^{2}}{(4\pi)^{6}}\mu^{-4\epsilon}+\mathcal{O}(\epsilon)\,, \\
\mu\partial_{\mu} Z=\mu^{-4\epsilon}\frac{2g_1^2\pi^2}{3}\,.
\end{gather}
Then, using $\mu\partial_{\mu} \l_i=0$, the beta functions $\b_i\equiv \mu\partial_{\mu} g_i$ come as:
\begin{align}
\b_1 &= -2 g_1\left(\eps-g_1^2\pi^2\right)\,, \\
\b_2 &= -2 g_2\left(\eps-g_1^2\pi^2\right) + 4g_1^2\pi^2\left(\f{9}{2\pi}+18g_1+g_2\right)\,,\\
\b_3 &= -2 g_3\left(\eps-4\, g_1^2\pi^2\right)\,,\\
\b_4 &= -2 g_4\left(\eps-g_1^2\pi^2\right) + g_1^2\pi^2\left(108g_1 +24g_2 + 24g_3 + 22g_4\right)\,,\\
\b_5 &= -2 g_5\left(\eps-g_1^2\pi^2\right)+g_1^2\pi^2\left(\f{2}{3\pi} +2g_2+8g_4+30g_5\right)\,.
\end{align}
First, we notice that if $g_1=0$, then all the other coupling have a trivial running, dictated by their canonical dimension ($2\epsilon$).
In such case, for $\epsilon>0$ we have only the trivial fixed point,  $g^*_i=0~\forall i$.
For $\epsilon=0$ instead, we have a 4-dimensional manifold of fixed points. This is a generalization of the vector model case, where the $(\phi_i \phi_i)^3$ interaction is exactly marginal at large $N$, and which in fact corresponds to the case in which we retain only the triple-trace term $I_5$ in our action. In that case it is known that at some critical coupling non-perturbative effects lead to vacuum instability with a consequent breaking of conformal invariance  \cite{Bardeen:1983rv,Amit:1984ri}. It would be interesting to study the vacuum stability of our model with $g_1=0$ in order to explore the possibility of a similar phenomenon, but we leave this for future work.

We are here interested in melonic fixed points, with $g_1\neq 0$, for which we need  $\epsilon>0$.
In that case, we obtain two interacting fixed points:
\begin{gather}
g^*_1=\pm\f{\sqrt{\eps}}{\pi}; \qquad g^*_2=\f{9}{2\pi}\left(-1\mp 4\sqrt{\eps}\right); \qquad g^*_3=0;
\\
g^*_4=\f{54}{11\pi}\left(1 \pm 3\sqrt{\eps}\right);\qquad g^*_5 = \f{-1021\mp 2700\sqrt{\eps}}{990\pi}.
\end{gather}
The standard linear stability analysis of the system of beta functions consists in diagonalizing the stability matrix $\cB_{ij} \equiv \p\b_i/\p g_j|_{g*}$,
thus identifying the scaling operators and their scaling dimensions, from its right-eigenvectors and eigenvalues, respectively. 
In the present case, we find the slightly unusual situation of having a non-diagonalizable stability matrix.
In fact, we find that both melonic fixed points have the same eigenvalues (critical exponents):
\begin{equation}
(4 \eps;\; 4 \eps;\; 6 \eps;\; 22 \eps;\; 30 \eps)\,,
\end{equation}
with the $4 \eps$ eigenvalue having algebraic multiplicty two, but geometric multiplicity one; hence the stability matrix is not diagonalizable.
In terms of the couplings $\{g_1,g_2,g_3,g_4,g_5\}$, the associated eigendirections are, respectively:
\begin{gather}
 \{0,3,0,-4,1\};\quad \{\mp\f{1}{6\sqrt{\eps}}, \f{2(- 236 \pm 5265\sqrt{\eps})}{6435\eps}, 0, \f{1108\mp 12285\sqrt{\eps}}{19305 \eps} ,0 \}; \\
 \{0,0,2,-3,1\} \quad
 \{0,0,0,-1,1\};\quad \{0,0,0,0,1\}\,,
\end{gather} 
with the first two forming a Jordan chain of length two.
Each (generalized) eigendirection, by its scalar product with the vector of renormalized operators arranged in the same order as the corresponding couplings, defines a scaling operator $\cO_i$ of dimension $\Delta_i=d+\theta_i$, with the $\theta_i$ being the critical exponent associated to that eigendirection.
As our critical exponents are all positive, all the scaling operators are irrelevant at the fixed points, and therefore the latter are infrared stable.
The fact that the stability matrix is not diagonalizable implies that the fixed point theory is a logarithmic conformal field theory (see for example \cite{Hogervorst:2016itc}). Therefore, although we find real exponents, as opposed to the complex ones of the quartic model \cite{Giombi:2017dtl}, the fixed-point theory is still non-unitary.

\subsubsection{The $\zeta=\frac{d}{3}$ case}
Using the results of App.~\ref{ap:betafun4}, along with the fact that there is no diverging wave-function renormalization in this case, the bare expansion gives:
\begin{align}
g_{1}&=\mu^{-2\epsilon}\cZ^3\l_{1}\,,\crcr
g_{2}&=\mu^{-2\epsilon}\cZ^3\left(\l_{2}-\frac{9}{2}\cZ^3D_{d/3}(\mu)\l_1^2+\cZ^6S_{d/3}(\mu)\l_1^2\left(9\l_1 + \frac{1}{2}\l_2\right)\right)\,,\crcr
g_{3}&=\mu^{-2\epsilon}\cZ^3\left(\l_{3}+ \cZ^6S_{d/3}(\mu)\frac{3}{4}\l_1^2\l_{3}\right)\,,\crcr
g_{4}&=\mu^{-2\epsilon}\cZ^3\left(\l_{4}+\cZ^6S_{d/3}(\mu)\l_1^2\left(\f{27}{2}\l_1 + 3\l_2+3\l_3+\f{11}{4}\l_4\right)\right)\,,\crcr
g_{5}&=\mu^{-2\epsilon}\cZ^3\left(\l_{5}-\cZ^3D_{d/3}(\mu)\f{1}{6}\l_1^2+ \cZ^6S_{d/3}(\mu)\l_1^2\left(\f{1}{4}\l_2+\l_4+\f{15}{4}\l_5\right)\right)\,,
\end{align}
with $\cZ$ given in Eq.~\eqref{eq:wavef3-LR}.
After rescaling of the coupling constants by $(4\pi)^d$, the beta functions  at $\epsilon=0$ read:
\begin{align}
\b_1 &= 0\,, \\
\b_2 &= g_1^2\f{\Gamma(d/6)^3}{\Gamma(d/3)^3\Gamma(d/2)}\left(-\f{2\Gamma(-d/6)\Gamma(d/6)}{\Gamma(d/3)\Gamma(2d/3)}\left(9g_1+\frac{1}{2}g_2\right)+9\right)\,,\\
\b_3 &= - 3g_1^2 g_3\frac{\Gamma(-d/6) \Gamma(d/6)^4}{2 \Gamma(d/3)^4 \Gamma(d/2)\Gamma(2d/3)}\,,\\
\b_4 &= -g_1^2 \frac{\Gamma(-d/6) \Gamma(d/6)^4}{\Gamma(d/3)^4 \Gamma(d/2)\Gamma(2d/3)}\left(27g_1+6g_2+6g_3+\frac{11}{2}g_4\right)\,,\\
\b_5 &= g_1^2\f{\Gamma(d/6)^3}{\Gamma(d/3)^3\Gamma(d/2)}\left(-\frac{2\Gamma(-d/6)\Gamma(d/6)}{\Gamma(d/3)\Gamma(2d/3)}\left(\frac{1}{4} g_2+g_4+\frac{15}{4}g_5\right)+\frac{1}{3}\right)\,.
\end{align}
This time, in addition to a 4-dimensional manifold of fixed points (set by $g_1^*=0$, and thus analogue to what we discussed for the case $\zeta=1$ at  $\epsilon=0$), we also find a line of fixed points parametrized by the exactly marginal coupling $g_1$: 
\begin{gather}
g_2^*=-18g_1+\f{9\Gamma(d/3)\Gamma(2d/3)}{\Gamma(-d/6)\Gamma(d/6)};\qquad g_3^*=0;\\
g_4^*=\f{54}{11}\left(3g_1 - \f{2\Gamma(d/3)\Gamma(2d/3)}{\Gamma(-d/6)\Gamma(d/6)}\right);\\
g_5^*=-\f{30}{11}g_1 + \f{1021\Gamma(d/3)\Gamma(2d/3)}{495\Gamma(-d/6)\Gamma(d/6)}.
\end{gather}
The critical exponents are: 
\begin{equation}
 \left( \f{55 g_1^2\a}{ 2};\;
   \f{11 g_1^2\a}{ 2 };\; 
   \f{3 g_1^2\a}{2 };\;
    g_1^2\a\right)\,,
\end{equation}
with
\be
\a= - \f{\Gamma(-d/6) \Gamma(d/6)^4}{
  \Gamma(d/3)^4 \Gamma(d/2) \Gamma(2 d/3)} >0\,, \;\;\; \text{for } d<3\,.
\ee
The respective eigendirections in terms of $\{g_2,g_3,g_4,g_5\}$ are: 
\begin{equation}
\{0,0,0,1\};\quad \{0,0,-1,1\};\quad\{0,2,-3,1\};\quad\{3,0,-4,1\}.
\end{equation}
Since the critical exponents are again positive, the eigendirections correspond again to irrelevant perturbations.
In this case, the stability matrix is diagonalizable, with real exponents, hence we have so far no signal of non-unitarity.

\subsection{Rank $5$}

The diagrams contributing to the six-point function at large $N$ are again the ones of Fig.~\ref{fig:bare3} (or Fig.~\ref{fig:allorders} at all orders). However, now the black vertices represent the complete interaction and the white vertices represent only the other interactions $J_i$ for $i>1$. This will slightly change the bare expansion of the couplings and their beta functions. 

\subsubsection{The $\zeta=1$ case}

There is no radiative corrections for the coupling of the complete interaction, the renormalized coupling is just rescaled by the wave function renormalization of Eq.~\eqref{eq:wavef5}:
\begin{equation}
g_1=\mu^{-2\epsilon}Z^3\kappa_1 \,.
\end{equation}

Then, we obtain the following bare expansions up to order three in the coupling constants:
\begin{align}
g_{2}&=\mu^{-2\epsilon}Z^3\left(\kappa_{2}+\kappa_1^2S_1(\mu)\left(2\kappa_{6}+\frac{2}{3}\kappa_{2}\right)\right)\,,\crcr
g_{3}&=\mu^{-2\epsilon}Z^3\left(\kappa_{3}+\kappa_1^2\kappa_{3}S_1(\mu)\right)\,,\crcr
g_{4}&=\mu^{-2\epsilon}Z^3\left(\kappa_{4}+\kappa_1^2S_1(\mu)\left(3\kappa_{6}+4\kappa_{3}+\frac{10}{3}\kappa_{2}+\frac{7}{3}\kappa_{4}\right)\right)\,,\crcr
g_{5}&=\mu^{-2\epsilon}Z^3\left(\kappa_{5}+\kappa_1^2S_1(\mu)\left(\kappa_{2}+\frac{8}{3}\kappa_{4}+5\kappa_{5}\right)\right)\,,\crcr
g_{6}&=\mu^{-2\epsilon}Z^3\left(\kappa_{6}-\frac{10}{3}\kappa_1^2 D_1(\mu)\right) \,,
\end{align}
with $D_{1}(\mu)$ and $S_{1}(\mu)$ defined in the previous section.

Let us rescale all the coupling constants as $\tilde{\kappa_1}=\frac{\kappa_1}{(4\pi)^d}$ and forget about the tilde.
Then the beta functions are:
\begin{align}
\beta_{g_1}&=-2\epsilon g_1+\frac{4}{3}\pi^2g_1^3 \,,\crcr  
\beta_{g_{2}}&=-2\epsilon g_{2}+\frac{4}{3}\pi^2g_1^2g_{2}+\frac{8\pi^2}{3}g_1^2\left(6g_{6}+2g_{2}\right)\,, \crcr
\beta_{g_{3}}&=-2\epsilon g_{3}+\frac{4}{3}\pi^2g_1^2g_{3}+8\pi^2g_1^2g_{3} \,, \crcr
\beta_{g_{4}}&=-2\epsilon g_{4}+\frac{4}{3}\pi^2g_1^2g_{4}+\frac{8\pi^2}{3}g_1^2\left(9g_{6}+12g_{3}+10g_{2}+7g_{4}\right)\,,\crcr
\beta_{g_{5}}&=-2\epsilon g_{5}+\frac{4}{3}\pi^2g_1^2g_{5}+\frac{8\pi^2}{3}g_1^2\left(3g_{2}+8g_{4}+15g_{5}\right)\,, \crcr
\beta_{g_{6}}&=-2\epsilon g_{6}+\frac{4}{3}\pi^2g_1^2g_{6}+\frac{40\pi}{3}g_1^2 \,.
\end{align}

The only fixed point when $\epsilon \neq 0$ is the trivial one: $g_i^*=0,~\forall i$. We do not find any Wilson-Fisher like fixed point. 
This is due to the beta function of the prism. The non-zero fixed point of $\beta_{g_1}$ is $g_1^*=\frac{\sqrt{\epsilon}}{2\pi}$. If we put it in the beta function of the prism, we obtain an expression independent of $g_6$ and proportional to $g_1^2$. This would imply $g_1=0$ which is incompatible with $g_1^*=\frac{\sqrt{\epsilon}}{2\pi}$ when $\epsilon \neq 0$. This solution is not a fixed point of the whole system.  

\subsubsection{The $\zeta=\frac{d}{3}$ case}

When $\zeta=d/3$, the wave function renormalization is finite and equal to $\mathcal{Z}$, given in Eq.~\eqref{eq:wavef5-LR}. In this case the bare expansion is:
\begin{align}
g_1&=\mu^{-2\epsilon}\mathcal{Z}^3\kappa_1  \,,\crcr
g_{2}&=\mu^{-2\epsilon}\left(\mathcal{Z}^3\kappa_{2}+\kappa_1^2\mathcal{Z}^9S_{d/3}(\mu)\left(2\kappa_{6}+\frac{2}{3}\kappa_{2}\right)\right)\,,\crcr
g_{3}&=\mu^{-2\epsilon}\left(\mathcal{Z}^3\kappa_{3}+\kappa_1^2\kappa_{3}\mathcal{Z}^9S_{d/3}(\mu)\right)\,,\crcr
g_{4}&=\mu^{-2\epsilon}\left(\mathcal{Z}^3\kappa_{4}+\kappa_1^2\mathcal{Z}^9S_{d/3}(\mu)\left(3\kappa_{6}+4\kappa_{3}+\frac{10}{3}\kappa_{2}+\frac{7}{3}\kappa_{4}\right)\right)\,,\crcr
g_{5}&=\mu^{-2\epsilon}\left(\mathcal{Z}^3\kappa_{5}+\kappa_1^2\mathcal{Z}^9S_{d/3}(\mu)\left(\kappa_{2}+\frac{8}{3}\kappa_{4}+5\kappa_{5}\right)\right)\,,\crcr
g_{6}&=\mu^{-2\epsilon}\left(\mathcal{Z}^3\kappa_{6}-\frac{10}{3}\mathcal{Z}^6\kappa_1^2 D_{d/3}(\mu)\right)\,.
\end{align}

Then, the beta function of the complete interaction is again exactly zero. The other beta functions are, after rescaling of the coupling constants by $(4\pi)^d$:
\begin{align}
\beta_{g_2}&=-2g_1^2\frac{\Gamma(d/6)^4 \Gamma(-d/6)}{\Gamma(d/3)^4\Gamma(d/2)\Gamma(2d/3)}\left(2g_6+\frac{2}{3}g_2\right) \,,\crcr
\beta_{g_3}&=-2g_1^2g_3\frac{\Gamma(d/6)^4 \Gamma(-d/6)}{\Gamma(d/3)^4\Gamma(d/2)\Gamma(2d/3)} \,,\crcr
\beta_{g_4}&=-2g_1^2\frac{\Gamma(d/6)^4 \Gamma(-d/6)}{\Gamma(d/3)^4\Gamma(d/2)\Gamma(2d/3)}\left(3g_6+4g_3+\frac{10}{3}g_2+\frac{7}{3}g_4 \right)\,,\crcr
\beta_{g_5}&=-2g_1^2\frac{\Gamma(d/6)^4 \Gamma(-d/6)}{\Gamma(d/3)^4\Gamma(d/2)\Gamma(2d/3)}\left(g_2+\frac{8}{3}g_4+5g_5 \right) \,,\crcr
\beta_{g_6}&=\frac{20}{3}\frac{\Gamma(d/6)^3}{\Gamma(d/3)^3\Gamma(d/2)}g_1^2 \,.
\end{align}

The beta function for $g_6$ admits a unique fixed point with $g_1^*=0$. The other beta functions are then exactly zero. 
Starting from nonzero couplings, we find that the flow is driven by $g_6$ flowing to minus infinity in the IR, and the other couplings flow towards: 
\begin{equation}
g_2^*=-3g_6; \quad  g_3^* = 0; \quad g_4^*=3g_6; \quad g_5^* = -3g_6.
\end{equation}

\section{Spectrum of operators}
\label{sec:BSeq}

For the rank-3 case we found IR fixed points with non-zero wheel coupling, both in the short-range and long-range versions of the model.
In order to better understand the conformal field theory at such IR fixed points,\footnote{We assume here that our fixed points correspond to conformal field theories.} we wish to compute the spectrum of operators that appear in the operator-product expansion (OPE) of $\phi_{abc}(x)\bar{\phi}_{abc}(0)$. Schematically, these are expected to be the bilinear operators $\phi_{abc}(\partial^2)^n\bar{\phi}^{abc}$, and their spectrum can  be obtained using the conformal Bethe-Salpeter (BS) equation \cite{Klebanov:2016xxf,Giombi:2017dtl}, or equivalently, the spectral decomposition of the four-point function \cite{Liu:2018jhs,Gurau:2019qag,Benedetti:2019ikb}.

The four-point function of our CFT can be written in a standard representation-theoretic form as \cite{Simmons-Duffin:2017nub,Liu:2018jhs,Gurau:2019qag}:
\begin{equation} \label{eq:4pt}
\begin{split}
  \frac{1}{N^6} \langle{\phi_{abc}(x_1) \bar{\phi}_{abc}(x_2) \phi_{a'b'c'}(x_3) \bar{\phi}_{a'b'c'}(x_4)} \rangle
 = & G(x_1-x_2) G(x_3-x_4)
 + \crcr
    +  \frac{1}{N^3}\sum_J 
  \int_{\frac{d}{2}-\imath \infty}^{\frac{d}{2}+\imath\infty} \frac{dh}{2\pi \imath} 
  &
  \;\frac{1}{1-k_{\zeta}(h,J)} \; \mu_{\Delta_{\phi}}^d(h,J)
     \cG^{\Delta_{\phi}}_{h,J}(x_i) + (\text{non-norm.})\,,
\end{split}
\end{equation}
with $\cG^{\Delta_{\phi}}_{h,J}(x_i) $ the conformal block, $\mu_{\Delta_{\phi}}^d(h,J)$ the measure, and $k_{\zeta}(h,J)$ the eigenvalues of the two particle irreducible four-point kernel. The non-normalizable contributions are due to operators with dimension $h<d/2$, and they should be treated separately \cite{Simmons-Duffin:2017nub}.\footnote{As we will see below, we will actually encounter an operator with dimension $h_0<d/2$. We will be cavalier in its treatment.} The subleading term is  the most interesting part, and it is related to the forward  four-point function that we introduced in Eq.\eqref{eq:fw4pt}. The appearance of $k_{\zeta}(h,J)$ should be clear from Eq.\eqref{eq:fw4pt-ladders}.
Closing the contour to the right, we pick poles at $k_{\zeta}(h,J)=1$ (other poles are spurious and they cancel out \cite{Simmons-Duffin:2017nub}), and we recover an operator-product expansion in the $t$-channel ($12\to34$):
\begin{equation} 
  \frac{1}{N^6} \langle{\phi_{abc}(x_1) \bar{\phi}_{abc}(x_2) \phi_{a'b'c'}(x_3) \bar{\phi}_{a'b'c'}(x_4)} \rangle
 =  G(x_1-x_2) G(x_3-x_4)
 +  \frac{1}{N^3} \sum_{m,J} c_{m,J}^2  \; \cG^{\Delta_{\phi}}_{h_{m,J},J}(x_i) \,,
\end{equation}
where $h_{m,J}$ are the poles of $(1-k_{\zeta}(h,J))^{-1}$, and the squares of the OPE coefficients are the residues at the poles \cite{Liu:2018jhs,Gurau:2019qag,Benedetti:2019ikb}.
We will limit ourselves to just studying the location of the poles, i.e.\ the spectrum of operators in the OPE.

Eigenfunctions of the kernel are known to take the form of three-point functions of two fundamental scalars with an operator. For example, in the case of spin zero we have:
\begin{equation}
    v_0(x_0,x_1,x_2) = \la\cO_h(x_0)\phi_{abc}(x_1)\bar{\phi}_{abc}(x_2)\ra = \f{C_{\cO\phi\bar{\phi}}}{(x_{01}^2 x_{02}^2)^{h/2}(x_{12}^2)^{\f12(\f{d}3-h)}} \,.
    \label{3-pt}
\end{equation}
Therefore, we need to find the eigenvalues $k(h,J)$ of the kernel from the equation:
\begin{equation}
    k_{\zeta}(h,J) v_J(x_0,x_1,x_2) = \int\,d^dx_3\,d^dx_4\, K(x_1,x_2;x_3,x_4)v_J(x_0,x_3,x_4),
    \label{BS}
\end{equation}
where the form of the kernel is obtained from \eqref{kernel-rank3} to be
\begin{equation}
    K(x_1,x_2;x_3,x_4) = \f{\l_1^2}{4} \left[3G(x_{14})G(x_{23})+2G(x_{13})G(x_{24})\right]G(x_{34})^4\,,
    \label{kernel}
\end{equation}
and since we integrate over $x_3$ and $x_4$, both terms can be combined into one.

\subsection{$\z=1$}

Since the corresponding integrals are simpler to solve in position space, we wish to set up the eigenvalue equation in position space. For that, we need the two-point function in position space, which for the case $\zeta=1$ is as follows:
\begin{align}
 G(x) &= \int\,\f{d^d p }{(2\pi)^d} e^{-i p\cdot x}\,G(p)=\cZ\int\,\f{d^d p }{(2\pi)^d} \f{e^{-i p\cdot x}}{p^{2d/3}}\nn\\
 &= \cZ\f{2^{d/3}}{(4\pi)^{d/2}}\f{\Gamma(\f{d}6)}{\Gamma(\f{d}3)}\f1{(x^2)^{d/6}}= F_1 \f1{(x^2)^{d/6}} \,,
 \label{2-pt-x}
\end{align}
where $F_1 =\cZ\f{2^{d/3}}{(4\pi)^{d/2}}\f{\Gamma(\f{d}6)}{\Gamma(\f{d}3)} $.
To perform the integrals at $J=0$, we shall use the following identity \cite{Giombi:2017dtl},
\begin{equation}
    \int\,d^d x_0 \f1{(x_{01}^2)^{\a_1}(x_{02}^2)^{\a_2}(x_{03}^2)^{\a_3}} = \f{L_d(\a_1,\a_2)}{(x_{12}^2)^{d/2-\a_3}(x_{13}^2)^{d/2-\a_2}(x_{23}^2)^{d/2-\a_1}},
    \label{identity}
\end{equation}
with $\a_1 + \a_2 +\a_3 =d$, and $L_d(\a_1,\a_2)=\pi^{d/2}\f{\Gamma(\f{d}2-\a_1)\Gamma(\f{d}2-\a_2)\Gamma(\f{d}2-\a_3)}{\Gamma(\a_1)\Gamma(\a_2)\Gamma(\a_3)}$.

To solve for the eigenvalues, let us first perform the integral over $x_3$ using \eqref{identity},
\begin{equation}
     \int\,d^d x_3 \f1{(x_{03}^2)^{h/2}(x_{23}^2)^{d/6}(x_{34}^2)^{5d/6-h/2}}=\f{L_d\big(\f{h}2,\f{d}6\big)}{(x_{02}^2)^{-d/3+h/2}(x_{04}^2)^{d/3}(x_{24}^2)^{d/2-h/2}}
     \label{int-1}
\end{equation}
with,
\begin{equation}
   L_d\bigg(\f{h}2,\f{d}6\bigg) =\pi^{d/2}\f{\Gamma(\f{d}3)\Gamma(-\f{d}3+\f{h}2)\Gamma(\f{d}2-\f{h}2)}{\Gamma(\f{d}6)\Gamma(\f{5d}6-\f{h}2)\Gamma(\f{h}2)}\,.
   \label{part-1}
\end{equation}
Now, performing the remaining integral over $x_4$, we get
\begin{equation}
     \int\,d^d x_4 \f1{(x_{04}^2)^{d/3+h/2}(x_{24}^2)^{d/2-h/2}(x_{14}^2)^{d/6}}=\f{L_d(\f{d}3+\f{h}2,\f{d}2-\f{h}2)}{(x_{02}^2)^{d/3}(x_{01}^2)^{h/2}(x_{12}^2)^{d/6-h/2}},
     \label{int-2}
\end{equation}
with
\begin{equation}
     L_d\bigg(\f{d}3+\f{h}2,\f{d}2-\f{h}2\bigg) =\pi^{d/2}\f{\Gamma(\f{d}3)\Gamma(\f{h}2)\Gamma(\f{d}6-\f{h}2)}{\Gamma(\f{d}3+\f{h}2)\Gamma(\f{d}2-\f{h}2)\Gamma(\f{d}6)}.
   \label{part-2}
\end{equation}
Collecting the terms from the first and second integrals, and combining their coefficients from Eq.~\eqref{2-pt-x}, Eq.~\eqref{part-1} and Eq.~\eqref{part-2}, we get the $J=0$ eigenvalues of the kernel to be
\begin{align}
    k_1(h,0) &= \f54\, \l_1^2\,F_1^6\, \pi^d\,\f{\Gamma(\f{d}3)^2\Gamma(-\f{d}3+\f{h}2)\Gamma(\f{d}6-\f{h}2)}{\Gamma(\f{d}6)^2\Gamma(\f{5d}6-\f{h}2)\Gamma(\f{d}3 + \f{h}2)}\nn\\
    &=\f54 \l_1^2\bigg(\f1{\pi^{d}}\f{4}{\lambda_1^2}\f{d}{3}\f{\Gamma(\frac{d}{6})\Gamma(\frac{5d}{6})}{\Gamma(1-\frac{d}{3})\Gamma(\frac{d}{3})}\bigg)\bigg(\pi^d\f{\Gamma(\f{d}3)^2\Gamma(-\f{d}3+\f{h}2)\Gamma(\f{d}6-\f{h}2)}{\Gamma(\f{d}6)^2\Gamma(\f{5d}6-\f{h}2)\Gamma(\f{d}3 + \f{h}2)}\bigg)
    \nn\\
    &= -5\times \f{\Gamma(\f{5d}6)\Gamma(\f{d}3)\Gamma(-\f{d}3+\f{h}2)\Gamma(\f{d}6-\f{h}2)}{\Gamma(-\f{d}3)\Gamma(\f{d}6)\Gamma(\f{5d}6-\f{h}2)\Gamma(\f{d}3 + \f{h}2)}.
    \label{eq:eigenvalue}
\end{align}
To find the spectrum of the bilinears, we must solve the above equation for $k_1(h,0)=1$, with $d=3-\epsilon$. We use the method of~\cite{Benedetti:2019ikb},  setting $h=1+2n+2z$, and treating $z$ as a perturbation of the classical dimension, which is justified for small $\epsilon$. 

For $n=0$ and $n=1$, we find the following solutions:
\begin{align}
h_{0}&=1+\frac{29}{3}\epsilon + \mathcal{O}(\epsilon^2), \crcr
h_1&=3+3\epsilon + \mathcal{O}(\epsilon^2).
\end{align}
We also find a solution looking like a quartic operator (as we deduce from the dimension at $\epsilon=0$):
\begin{equation}
h_{q}=2-\frac{32}{3}\epsilon + \mathcal{O}(\epsilon^2) = d-h_0.
\end{equation}
However, at the time of the paper wrongly interpreted, this pole should not be included in the spectrum of operators. Indeed it corresponds to the \emph{shadow} of the quadratic operator and the contour should have been deformed appropriately to exclude it. We comment more on this question in the concluding remarks. Lastly, for $n>1$ we find:
\begin{equation}
h_n=1+2n-\frac{\epsilon}{3}+\frac{20}{3n(n-1)(4 n^2-1)}\epsilon^2+\mathcal{O}(\epsilon^3)~~, ~~ n>1.
\end{equation}
The solutions we just found are exactly the ones found in~\cite{Giombi:2017dtl}, which is not surprising, as their equation 4.6 for $q=6$, giving the eigenvalues of the kernel, is the same as Eq.~\eqref{eq:eigenvalue}. However, it was assumed to hold for rank $5$, but as we have seen, it turns out that in that case there is no Wilson-Fisher fixed point, hence no interacting CFT to which these equations might apply. On the other hand, we have shown here that we still recover the same spectrum for the model in rank $3$, which admits a melonic Wilson-Fisher fixed point.

\begin{figure}[htbp]
\centering
\captionsetup[subfigure]{labelformat=empty}
\subfloat[]{\includegraphics[scale=0.45]{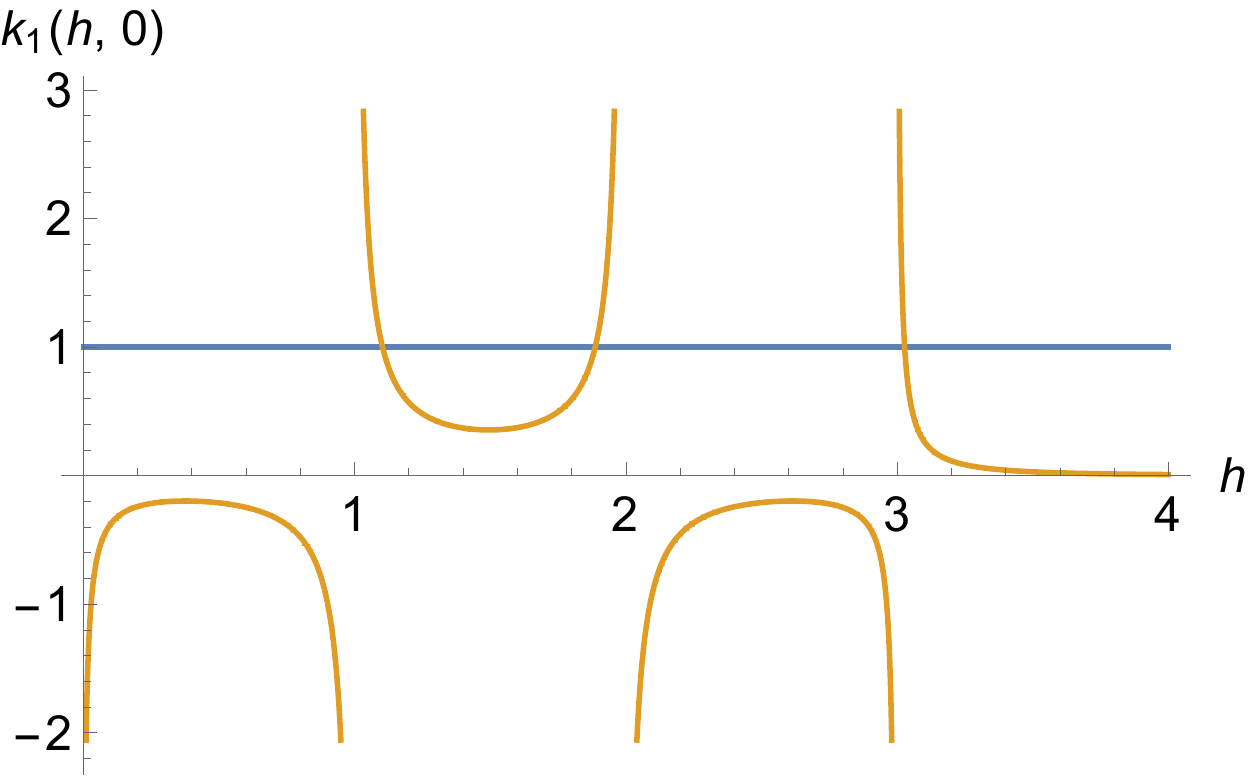}}
\subfloat[]{\includegraphics[scale=0.45]{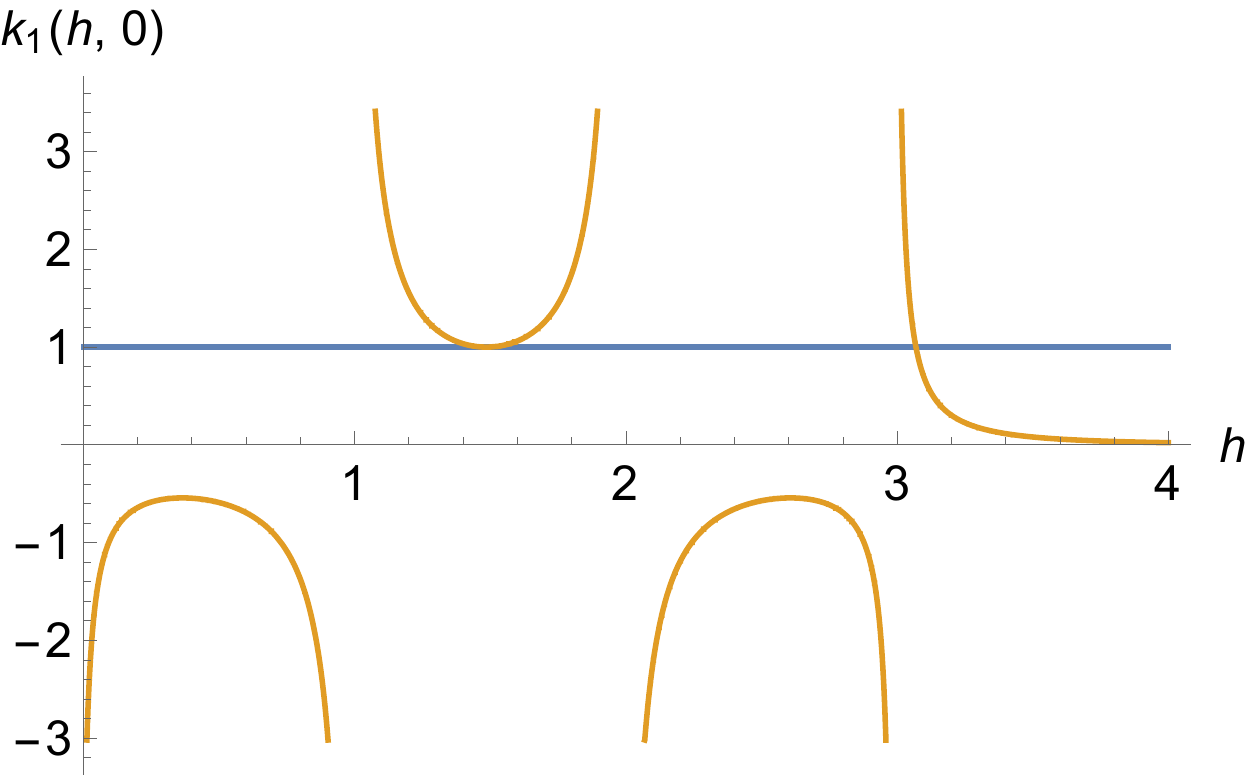}}
\subfloat[]{\includegraphics[scale=0.45]{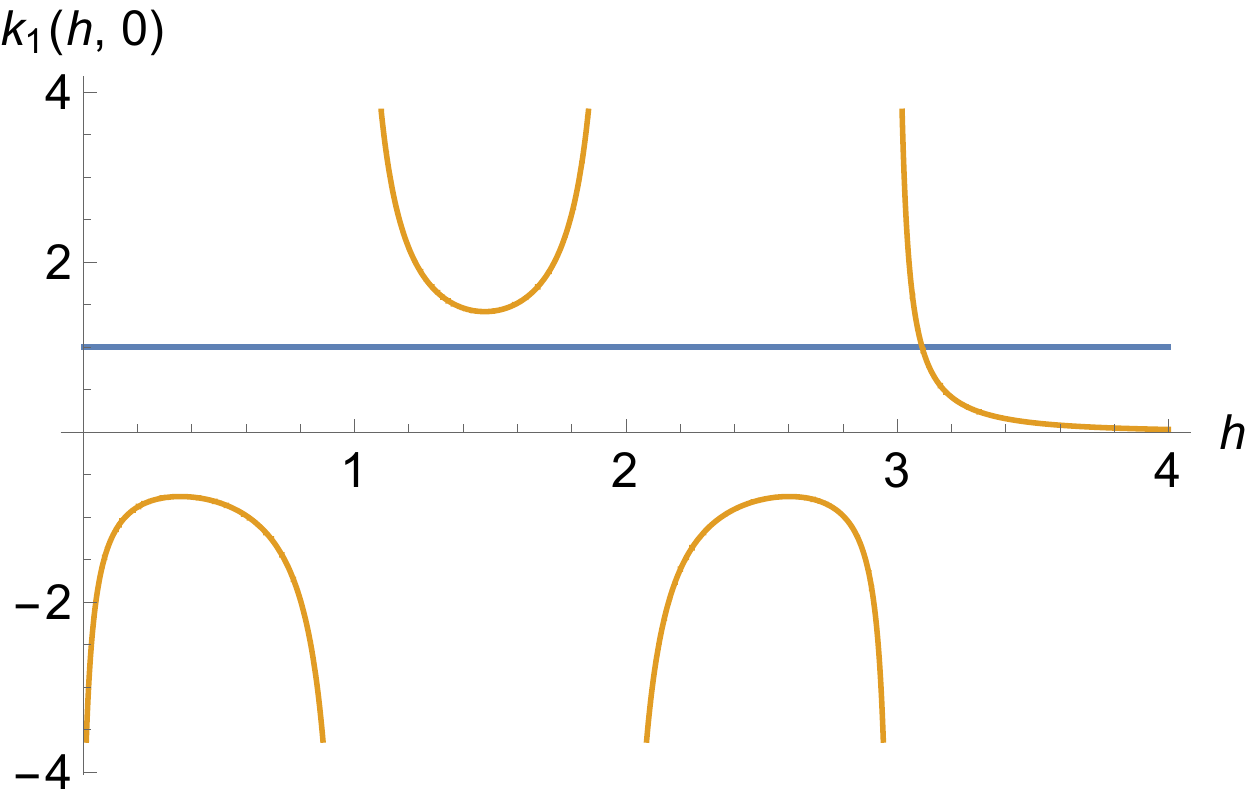}}
\caption{Plots of $k_1(h,0)$ at $d=3-\epsilon$ for, from left to right, $\epsilon=0.01$, $\epsilon=0.02819$, and $\epsilon=0.04$. On the left panel, the intersections with the blue line  correspond to $h_0$, $h_q$ and $h_1$. On the middle panel, $h_0$ and $h_q$ have merged, and on the right panel only $h_1$ remains.}
\label{fig:complexsol}
\end{figure}

As $\epsilon>0$, all the solutions we found are real. If we send $\epsilon$ to zero, we recover the classical dimensions $h^{classical}_n=1+2n$ of the bilinear operators $\phi_{abc}(\partial^2)^n\phi^{abc}$, except for $h_{q}$ corresponding to a quartic operator. 
However, this is only true for $\epsilon$ small enough. As $\epsilon$ increases, the two solutions $h_{0}$ and $h_q$ merge and become complex, see Fig.~\ref{fig:complexsol}. This happens around $\epsilon=0.02819$. Again, the same phenomenon appeared in~\cite{Giombi:2017dtl}.

\paragraph{Higher spins.}
We can also compute the spectrum of bilinears at higher spin. Using \cite{Gurau:2019qag}, $k_1$ becomes:
\begin{equation} \label{eq:f1J-zeta1}
k_1(h,J)=-5\times \f{\Gamma(\f{5d}6)\Gamma(\f{d}3)\Gamma(-\f{d}3+\f{h}2+J/2)\Gamma(\f{d}6-\f{h}2+J/2)}{\Gamma(-\f{d}3)\Gamma(\f{d}6)\Gamma(\f{5d}6-\f{h}2+J/2)\Gamma(\f{d}3 + \f{h}2+J/2)}.
\end{equation}
We find the following solutions for $k_1(h,J)=1$:
\begin{align}
\label{eq:h0J-zeta1}
h_{0,J}&=1+J-\frac{4J^2+29}{3(4J^2-1)}\epsilon + \mathcal{O}(\epsilon^2), \\
h_{1,J}&=3+J+\frac{-4J^2-8J+27}{3(2J+3)(2J+1)}\epsilon + \mathcal{O}(\epsilon^2), \\
h_{n,J}&=1+2n+J-\frac{\epsilon}{3}+\frac{5\epsilon^2}{3n(n-1)(n+1/2+J)(n-1/2+J)} + \mathcal{O}(\epsilon^3)~~, \quad n>1.
\end{align}
Notice that these can all be written in the form $h_{n,J}=d-2+2n+J+2z_{n,J}$, with $d=3-\epsilon$.

For $J=0$, we recover the solutions we found in the beginning of this section, except for $h_q$. This is due to the fact that the factor $\Gamma(-\f{d}3+\f{h}2+J/2)$ in Eq.~\eqref{eq:f1J-zeta1} only leads to a singularity for $h>0$ if $J=0$. Therefore, for $J> 0$, we only have dimensions corresponding to bilinear operators and no longer have a dimension corresponding to a quartic operator. 

One can check at leading order in $\epsilon$ from Eq.~\eqref{eq:h0J-zeta1}, or to all orders directly from Eq.~\eqref{eq:f1J-zeta1}, that the spin-2 operator with $n=0$ has the classical dimension $h_{0,2}=3-\epsilon=d$, as expected from a conserved energy-momentum tensor.

\subsection{$\z=\frac{d}{3}$}

The computation of the spectrum of bilinears of the long range model with the modified propagator goes exactly along the same lines as the one with the normal propagator. 
The only difference lies in the structure of the two-point function. The position space expression for the renormalized propagator (or two-point function) is:
\begin{equation}
	G(x) = \f{F_{d/3}}{(x^2)^{d/6}},\hspace{.2cm}F_{d/3} =\cZ\f{2^{d/3}}{(4\pi)^{d/2}}\f{\Gamma(\f{d}6)}{\Gamma(\f{d}3)},
\end{equation}
where $\cZ$ is the solution of \eqref{Z-norm-d/3}. \\
Once again we solve the same eigenvalue Eq.~\eqref{BS} using the same kernel \eqref{kernel}. The resulting eigenvalue, for $J=0$, is:
\be
\begin{split}
	k_{d/3}(h,0) &= \f54\, \l_1^2\,F_{d/3}^6\, \pi^d\,\f{\Gamma(\f{d}3)^2\Gamma(-\f{d}3+\f{h}2)\Gamma(\f{d}6-\f{h}2)}{\Gamma(\f{d}6)^2\Gamma(\f{5d}6-\f{h}2)\Gamma(\f{d}3 + \f{h}2)}\\
	&=\f54\, \l_1^2\,\cZ^6\f1{(4\pi)^{2d}}\bigg(\f{\Gamma(\f{d}6)}{\Gamma(\f{d}3)}\bigg)^4 \,\f{\Gamma(-\f{d}3+\f{h}2)\Gamma(\f{d}6-\f{h}2)}{\Gamma(\f{5d}6-\f{h}2)\Gamma(\f{d}3 + \f{h}2)}\\
	&= \f54 g_1^2\bigg(\f{\Gamma(\f{d}6)}{\Gamma(\f{d}3)}\bigg)^4 \,\f{\Gamma(-\f{d}3+\f{h}2)\Gamma(\f{d}6-\f{h}2)}{\Gamma(\f{5d}6-\f{h}2)\Gamma(\f{d}3 + \f{h}2)} \,,
	\label{eigenvalue-2}
\end{split}
\ee
where in the last line we used the renormalized coupling defined in \ref{sec:betas1}, namely  $g_1 = \f1{(4\pi)^d}\l_1\,\cZ^3$.

In order to find the OPE spectrum we have to solve for $k_{d/3}(h,0)=1$. The main difference with respect to the previous case is that the spectrum will now depend on the value of the exactly marginal coupling, which will replace $\epsilon$ in the role of small parameter. 

Again we use the method of~\cite{Benedetti:2019ikb} to solve $k_{d/3}(h,0)=1$, and we find the following solutions:
\begin{align}
h_0&=\frac{d}{3}+\frac{15\Gamma(1-d/6)}{d\Gamma(2d/3)\Gamma(d/2)}\left(\frac{\Gamma(d/6)}{\Gamma(d/3)}\right)^4g_1^2 + \mathcal{O}(g_1^4)\,, \crcr
h_n&=\frac{d}{3}+2n+\frac{(-1)^{n+1}}{n!}\frac{5\Gamma(n-d/6)}{2\Gamma(2d/3-n)\Gamma(d/2+n)}\left(\frac{\Gamma(d/6)}{\Gamma(d/3)}\right)^4g_1^2+ \mathcal{O}(g_1^4)\,.
\end{align}
Notice that at $g_1=0$, we recover the classical dimensions $h^{classical}_n=d/3+2n$. At $g_1 \neq 0$, all dimensions are real, and they are greater than $d/3$ for $g_1^2>0$ and small. 

As above, the solution that looks like a quartic operator:
\begin{equation}
h_{q}=\frac{2d}{3}-\frac{15\Gamma(1-d/6)}{d\Gamma(2d/3)\Gamma(d/2)}\left(\frac{\Gamma(d/6)}{\Gamma(d/3)}\right)^4g_1^2 + \mathcal{O}(g_1^4)\,,
\end{equation}
has to be discarded as a contribution from the shadow of the quadratic operator $h_q=d-h_0$. 

\begin{figure}[htbp]
\centering
\includegraphics[scale = 0.5]{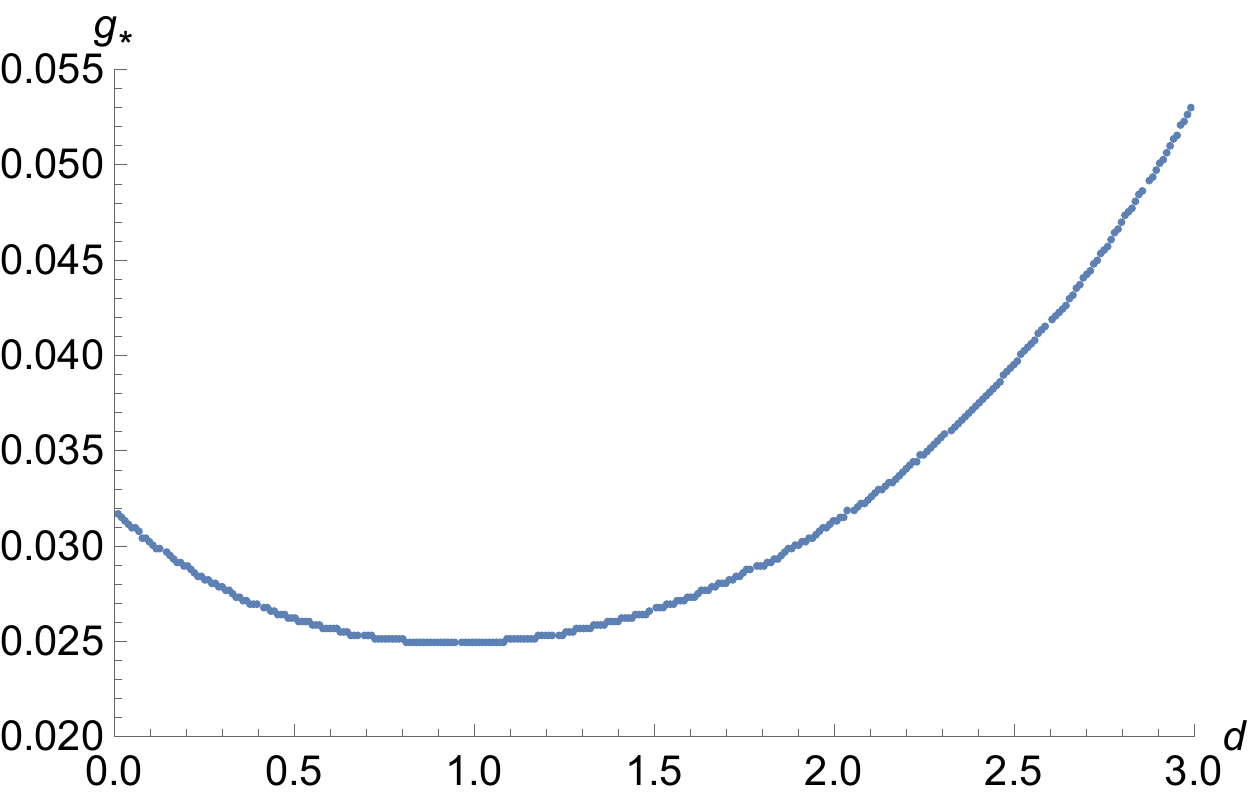}
\hspace{1cm}
\includegraphics[scale = 0.5]{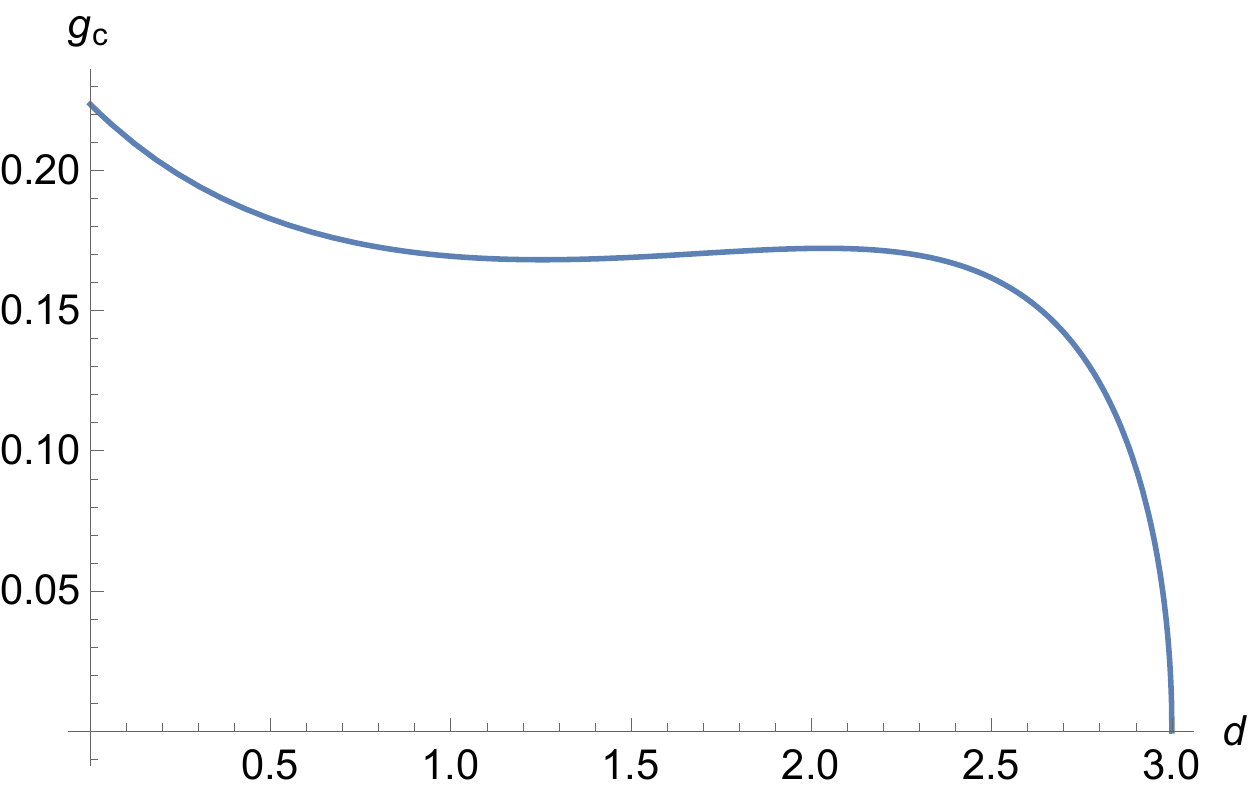}
\caption{Plots of $g_\star$ and $g_c$ as functions of $d$. The two curves cross at $d\simeq 2.97$.}
\label{fig:g_star}
\end{figure}

The plots of $k_{d/3}(h,0)$ are qualitatively similar to those in Fig.~\ref{fig:complexsol}, and we find the appearance of a pair of complex solutions for $g_1> g_\star >0$. For $d=2$, we have $g_\star \simeq 0.0313$, which is smaller than the value $g_c$ defined in Footnote \ref{foot:g_c}, at which the relation between bare $\l_1$ and renormalized $g_1$ becomes non-invertible, and which for $d=2$ is $g_c\simeq 0.1722$. A similar situation is found for any $d\lesssim 2.97$, while for $d\gtrsim 2.97$ we find $g_c<g_\star$. Comparative plots of $g_\star$ and $g_c$ as functions of $d$ are shown in Fig.~\ref{fig:g_star}. Therefore, for $d\lesssim 2.97$, the scenario differs from the one encountered in~\cite{Benedetti:2019eyl}, as in the present case the complex transition lies within the regime of validity of the fixed point solution. Furthermore, at the transition where the first two solutions of $k_{d/3}(h,0)=1$ merge (and then become complex) their value is $d/2$, within numerical precision. Such transition thus seems to be compatible with the scenario advanced in~\cite{Kim:2019upg}, where the appearance of complex dimensions of the form $d/2+\im f$ for a given operator has been conjectured to be a signal that such operator acquires a non-zero vacuum expectation value.

\paragraph{Higher spins.}

Again we can compute the spectrum of bilinears for spin $J>0$. The eigenvalue becomes:
\begin{equation}
k_{d/3}(h,J) = \f54 g_1^2\bigg(\f{\Gamma(\f{d}6)}{\Gamma(\f{d}3)}\bigg)^4 \,\f{\Gamma(-\f{d}3+\f{h}2+J/2)\Gamma(\f{d}6-\f{h}2+J/2)}{\Gamma(\f{5d}6-\f{h}2+J/2)\Gamma(\f{d}3 + \f{h}2+J/2)}\,.
\end{equation}
We find the following solutions for $k_{d/3}(h,J)=1$:
\begin{align}
h_{0,J}&=\frac{d}{3}+J-\frac{5}{2}\frac{\Gamma(-d/6+J)}{\Gamma(2d/3)\Gamma(d/2+J)}\left(\frac{\Gamma(d/6)}{\Gamma(d/3)}\right)^4g_1^2 + \mathcal{O}(g_1^4)\,, \crcr
h_{n,J}&=\frac{d}{3}+2n+J+\frac{(-1)^{n+1}}{n!}\frac{5\Gamma(n-d/6+J)}{2\Gamma(2d/3-n)\Gamma(d/2+n+J)}\left(\frac{\Gamma(d/6)}{\Gamma(d/3)}\right)^4g_1^2+ \mathcal{O}(g_1^4)\,.
\end{align}

Again, when $J=0$ we recover the dimensions we computed in the beginning of this section, except for the one corresponding to a quartic operator.

However, differently from the $\zeta=1$ case, and as in  \cite{Benedetti:2019ikb}, we find no spin-two operator of dimension $d$. This is due to the fact that the energy momentum tensor is not a local operator.

\section{Conclusions}
\label{sec:concl}

In this chapter, we presented an analysis of the melonic large-$N$ limit in various versions of Bosonic tensor models with sextic interactions.
We considered explicitly tensors of rank 3 and 5, but we expect rank 4 to behave similarly to rank 5. And we chose as free propagator either the standard short-range propagator, or a critical long-range propagator. We discussed in detail some standard properties of melonic theories, as the closed Schwinger-Dyson equation for the two-point function, and the Bethe-Salpeter equation for the spectrum of bilinear operators. However, as we emphasized, the conformal solution of these equations are only justified if the quantum field theory actually admits a fixed point of the renormalization group. In this respect, we found a striking difference between the rank-3 and rank-5 models, as only the former (both in the short-range and long-range versions) admits a non-trivial (and real) fixed point for $d<3$, with an interaction leading to melonic dominance. 
The rank-5 model instead has only one trivial (i.e.\ non-interacting) fixed point. It would be interesting to check whether such conclusion would remain valid after including in the action \eqref{eq:int-action-graph-rank5} the other possible sextic interactions that we have omitted by restricting to a melo-complete family.

Comparing our findings for the short range model with those of the sextic model in~\cite{Giombi:2017dtl}, we observe similar results for two-point function and spectrum of operators. However, we do so for the rank-3 model, where such analysis is justified by the existence of a melonic fixed point, whereas their analysis was formally based on a rank-5 model, which we showed is inconsistent. The fact that we find the same result is not a coincidence: our kernel eigenvalue \eqref{eq:eigenvalue} coincides with the $q=6$ case of the eigenvalue computed in~\cite{Giombi:2017dtl} for a general $q$-valent melonic theory. Such eigenvalue depends only on the assumption that a $q$-valent interaction leads to melonic dominance. The latter can for example be obtained with a rank-$(q-1)$ model with a complete interaction, as assumed in~\cite{Giombi:2017dtl}. However, as argued in~\cite{Prakash:2019zia}, and as we saw also here, rank $q-1$ is not necessary: a $q$-valent interaction can lead to a melonic limit even in a tensor model of rank $r<q-1$ (in which case the model was called \emph{subchromatic} in~\cite{Prakash:2019zia}); this is the case of our rank-3 model with wheel interaction.

Comparing instead our long-range model to the quartic long-range model of~\cite{Benedetti:2019eyl}, we see some similarity but also an important difference: on one hand, both models admit a line of fixed points, parametrized by the interaction that leads to melonic dominance; on the other, in the quartic case, the fixed point and conformal dimensions are real only for purely imaginary tetrahedral coupling \cite{Benedetti:2019eyl}, while in our sextic model, we have a real fixed point and real spectrum for a real wheel coupling.
Furthermore, unlike in~\cite{Benedetti:2019eyl}, in the present case the appearance of complex dimensions at some critical value of the marginal coupling seems to be compatible with the scenario conjectured in~\cite{Kim:2019upg}, according to which it is a signal of an instability of the vacuum.

We have also encountered some of the recurring aspects of melonic theories (for rank 3, at least): for the short-range version, reality of the CFT constrains $\epsilon$ to stay very small; in the long-range version, we have instead the freedom to reach an integer dimension ($d=2$ in this case), by keeping the marginal coupling small, but at the price of loosing the energy-momentum tensor (as usual in long-range models \cite{Paulos:2015jfa}).
It would be interesting to get a better understanding of how general these features are.

One new feature that we found is that the fixed point of the short-range model has a non-diagonalizable stability matrix, even in the range of $\epsilon$ for which the exponents are real. This is an indication that the fixed-point theory is a logarithmic CFT, and thus it is non-unitary. We hope to explore this aspect more thoroughly in the near future.

Concerning the question of gauging, since unfortunately in rank 3, the short-range tensorial fixed points (with non-zero $\l_1$) occur
in $d < 3$ dimensions, gauging with Chern-Simons would not make sense. Then strictly speaking, writing a covariant derivative in the long-range model would introduce an infinite number of interactions between the gauge field and the tensors.

We briefly noticed that certain eigenvalues of the ladder-kernel, earlier wrongly identified as indicating mixing with higher order operators, had to be discarded since corresponding to shadow operators. We are currently investigating with D. Benedetti, a non-disordered cubic interacting vector model, that also shows melonic dominance. After the quartic and sextic models, it provides us with a third category of melonic field theories for drawing a clearer picture of how is to be deformed the contour using the partial wave decomposition of the four-point function, in order to pick the physical poles. 

Lastly, it would be important to understand the fate of our line of fixed points (in the long-range model) at higher orders in the $1/N$ expansion. 
At some order in the expansion we expect to find vertex corrections also to the wheel interaction, and therefore a non-zero beta function $\beta_1$.
A similar situation occurs in the vector $\phi^6$ model, where the leading-order beta function vanishes identically, but already at next-to-leading order in $1/N$ one finds a non-zero beta function \cite{Pisarski:1982vz}, thus reducing the leading-order line of fixed points to an isolated fixed point.

%

\newpage

\begin{subappendices}

\section{Conventions for the interaction terms}
\label{ap:conventions}

We write here in an explicit form the interactions appearing in Eq.~\eqref{eq:int-action} and \eqref{eq:int-action-rank5}, as well as the quartic invariants, in terms of contraction operators built as linear combinations of products of Kronecker delta functions.

\subsection{Rank 3}
Using the compact notation $\mathbf{a}=(a_1a_2a_3)$,
the $U(N)^3$ quartic invariants, also known as pillow and double-trace invariants, respectively, are:
\begin{align}
I_p &=  \delta^p_{\mba\mbb; \mbc\mbd}\phi_{\mba}(x) \phib_{\mbb}(x)  \phi_{\mbc}(x) \phib_{\mbd }(x)\,,\\
I_d &= \delta^d_{\mba\mbb; \mbc\mbd }  \phi_{\mba}(x) \phib_{\mbb}(x)  \phi_{\mbc}(x) \phib_{\mbd }(x)\,,
\end{align}
with: 
\be
    \delta^p_{\mba\mbb; \mbc\mbd }=\frac{1}{3} \sum_{i=1}^3  \delta_{a_id_i} \delta_{b_ic_i} \prod_{j\neq i}  \delta_{a_jb_j}\delta_{c_j d_j} \; , \quad\quad  \delta^d_{\mba\mbb; \mbc\mbd }  = \delta_{\mba \mbb}  \delta_{\mbc \mbd}\,,
\ee
and $\delta_{\mba \mbb}  = \prod_{i=1}^3 \delta_{a_i b_i} $.

The sextic invariants depicted in Eq.~\eqref{eq:int-action} are instead:
\begin{align}
I_1 &=  \delta^{(1)}_{\mba\mbb \mbc\mbd\mbe\mbf} \phi_{\mba}(x) \phib_{\mbb}(x)  \phi_{\mbc}(x) \phib_{\mbd }(x)\phi_{\mbe }(x)\phib_{\mbf }(x)\,,\\
I_b &=\delta^{(b)}_{\mba\mbb; \mbc\mbd; \mbe\mbf }  \phi_{\mba}(x) \phib_{\mbb}(x)  \phi_{\mbc}(x) \phib_{\mbd }(x)\phi_{\mbe }(x)\phib_{\mbf }(x)\,,   \qquad b=2,\ldots,5\,,
\end{align}
with
%
\begin{align}
   & \delta^{(1)}_{\mba\mbb\mbc\mbd \mbe\mbf }= \d_{a_1b_1}\d_{a_2f_2}\d_{a_3d_3}\d_{c_1d_1}\d_{c_2b_2}\d_{c_3f_3}\d_{e_1f_1}\d_{e_2d_2}\d_{e_3b_3}\, ,\crcr
  & \delta^{(2)}_{\mba\mbb; \mbc\mbd; \mbe\mbf }= \frac{1}{9}\left( \sum_{i=1}^3 \sum_{j \neq i}  \delta_{a_if_i} \delta_{b_ic_i} \delta_{c_j d_j}\delta_{e_jf_j}\left(\prod_{k\neq i}  \delta_{a_kb_k} \right)\left(\prod_{l\neq j}  \delta_{e_l d_l} \right) \left( \prod_{m \neq i,j} \delta_{c_m f_m}\right)+ \mbc\mbd \leftrightarrow \mbe\mbf + \mbc\mbd \leftrightarrow \mba\mbb \right) \, ,\crcr
  & \delta^{(3)}_{\mba\mbb; \mbc\mbd; \mbe\mbf }= \frac{1}{3} \sum_{i=1}^3 \delta_{a_if_i} \delta_{b_ic_i} \delta_{d_i e_i}\prod_{j\neq i}  \delta_{a_j b_j}\delta_{c_j d_j}\delta_{e_j f_j} \, ,\crcr
  & \delta^{(4)}_{\mba\mbb; \mbc\mbd; \mbe\mbf }=\frac{1}{3}\left(\delta_{\mba\mbb}\delta^p_{\mbc\mbd;\mbe\mbf} + \delta_{\mbc\mbd}\delta^p_{\mba\mbb;\mbe\mbf}+ \delta_{\mbe\mbf}\delta^p_{\mba\mbb;\mbc\mbd}\right)\, ,\crcr
  & \delta^{(5)}_{\mba\mbb; \mbc\mbd; \mbe\mbf }=\delta_{\mba\mbb}\delta_{\mbc\mbd}\delta_{\mbe\mbf}\, .
 \end{align}
Besides the color symmetrization, to simplify the computation of the beta-functions, we have included a symmetrization with respect to exchange of pairs of black and white vertices.

\subsection{Rank 5}

Using the compact notation $\mathbf{a}=(a_1a_2a_3a_4a_5)$, the $O(N)^3$ melonic quartic invariants are:
\begin{align}
I_p &=  \delta^p_{\mba\mbb; \mbc\mbd}\phi_{\mba}(x) \phi_{\mbb}(x)  \phi_{\mbc}(x) \phi_{\mbd }(x)\,,\\
I_d &= \delta^d_{\mba\mbb; \mbc\mbd }  \phi_{\mba}(x) \phi_{\mbb}(x)  \phi_{\mbc}(x) \phi_{\mbd }(x)\,,
\end{align}
with: 
\be
   \delta^p_{\mba\mbb; \mbc\mbd }=\frac{1}{5} \sum_{i=1}^5  \delta_{a_id_i} \delta_{b_ic_i} \prod_{j\neq i}  \delta_{a_jb_j}\delta_{c_j d_j} \; , \quad\quad  \delta^d_{\mba\mbb; \mbc\mbd }  = \delta_{\mba \mbb}  \delta_{\mbc \mbd} \,,
\ee
and $\delta_{\mba \mbb}  = \prod_{i=1}^5 \delta_{a_i b_i} $.

The sextic invariants depicted in Eq.~\eqref{eq:int-action-rank5} are instead:

\begin{align}
J_1 &=  \delta^{(1)}_{\mba\mbb \mbc\mbd\mbe\mbf} \phi_{\mba}(x) \phi_{\mbb}(x)  \phi_{\mbc}(x) \phi_{\mbd }(x)\phi_{\mbe }(x)\phi_{\mbf }(x)\,,\\
J_b &=\delta^{(b)}_{\mba\mbb; \mbc\mbd; \mbe\mbf } \phi_{\mba}(x) \phi_{\mbb}(x)  \phi_{\mbc}(x) \phi_{\mbd }(x)\phi_{\mbe }(x)\phi_{\mbf }(x)\,,   \qquad b=2,\ldots,6\,,
\end{align}
with
\begin{align}
   & \delta^{(1)}_{\mba\mbb\mbc\mbd\mbe\mbf}  = \delta_{a_1 b_1}  \delta_{a_2f_2} \delta_{a_3 e_3}  \delta_{a_4 d_4 } \delta_{a_5 c_5}   \delta_{b_2 c_2}\delta_{b_3 d_3}\delta_{b_4 f_4} \delta_{b_5 e_5} \delta_{c_3 f_3}\delta_{c_4 e_4}\delta_{c_1 d_1}\delta_{e_1 f_1}\delta_{e_2 d_2}\delta_{d_5 f_5} \,  , \crcr
   & \delta^{(2)}_{\mba\mbb; \mbc\mbd; \mbe\mbf }= \frac{1}{60}\left( \sum_{i=1}^5 \sum_{j \neq i}  \delta_{a_ic_i} \delta_{b_id_i} \delta_{c_j e_j}\delta_{d_jf_j}\left(\prod_{k\neq i}  \delta_{a_kb_k} \right)\left(\prod_{l\neq j}  \delta_{e_l f_l} \right) \left( \prod_{m \neq i,j} \delta_{c_m d_m}\right)+ \mbc\mbd \leftrightarrow \mbe\mbf + \mbc\mbd \leftrightarrow \mba\mbb \right) \, ,\crcr
   & \delta^{(3)}_{\mba\mbb; \mbc\mbd; \mbe\mbf }= \frac{1}{5} \sum_{i=1}^5 \delta_{a_if_i} \delta_{b_ic_i} \delta_{d_i e_i}\prod_{j\neq i}  \delta_{a_j b_j}\delta_{c_j d_j}\delta_{e_j f_j} \, ,\crcr
  & \delta^{(4)}_{\mba\mbb; \mbc\mbd; \mbe\mbf }=\frac{1}{3}\left(\delta_{\mba\mbb}\delta^p_{\mbc\mbd;\mbe\mbf} + \delta_{\mbc\mbd}\delta^p_{\mba\mbb;\mbe\mbf}+ \delta_{\mbe\mbf}\delta^p_{\mba\mbb;\mbc\mbd}\right)\, ,\crcr
  & \delta^{(5)}_{\mba\mbb; \mbc\mbd; \mbe\mbf }=\delta_{\mba\mbb}\delta_{\mbc\mbd}\delta_{\mbe\mbf}\, , \\
  & \delta^{(6)}_{\mba\mbb; \mbc\mbd; \mbe\mbf }= \frac{1}{60} \sum_{i=1}^5 \sum_{j\neq i}\sum_{k\neq i,j}  \delta_{a_ic_i} \delta_{b_id_i} \delta_{c_j e_j}\delta_{d_jf_j}\delta_{a_k e_k}\delta_{b_k f_k}\left(\prod_{l\neq i,k}  \delta_{a_l b_l} \right)\left(\prod_{m\neq j,k}  \delta_{e_m f_m} \right) \left( \prod_{n \neq i,j} \delta_{c_n d_n}\right) \, . \nn
 \end{align}

\section{The melon integral}
\label{ap:melon}

In this section we compute the melon integral contributing to the wave function renormalization. 

We want to compute:
\begin{equation}
M_{\Delta}(p)=\int_{q_1,q_2,q_3,q_4}G_0(q_1)G_0(q_2)G_0(q_3)G_0(q_4)G_0(p+q_1+q_2+q_3+q_4) \,,
\end{equation}
with $G_0(p)=\frac{1}{p^{2\Delta}}$.

We will use the following formula to compute $M(p)$:
\begin{equation}
\int\f{\dd[d] k}{(2\pi)^d} \f{1}{k^{2\a}(k+p)^{2\b}}  = \f{1}{(4\pi)^{d/2}}\f{\G(d/2 - \a)\G(d/2 - \b)\G(\a +\b -d/2)}{\G(\a)\G(\b) \G(d - \a - \b)}\f{1}{|p|^{2(\a+\b - d/2)}} \,.
\label{eq:intG}
\end{equation}
We obtain:
\begin{equation}
M_{\Delta}(p)=\frac{p^{4d-10\Delta}}{(4\pi)^{2d}}\frac{\Gamma(d/2-\Delta)^5\Gamma(5\Delta-2d)}{\Gamma(\Delta)^5\Gamma(5d/2-5\Delta)} \,.
\end{equation}
For $\Delta=\frac{d}{3}$, this simplifies to:
\begin{equation} \label{eq:M-d/3}
M_{d/3}(p)=-\frac{p^{2d/3}}{(4\pi)^{2d}}\frac{3}{d}\frac{\Gamma(1-\frac{d}{3})\Gamma(\frac{d}{6})^5}{\Gamma(\frac{d}{3})^5\Gamma(\frac{5d}{6})} \,.
\end{equation}
We will also need the melon integral for $d=3-\epsilon$ and $\Delta=1$:
\begin{equation}  \label{eq:M-1}
M_1(p)=\frac{p^{2-4\epsilon}}{(4\pi)^{6-2\epsilon}}\frac{\Gamma(2\epsilon-1)\Gamma(\frac{1-\epsilon}{2})^5}{\Gamma(\frac{5}{2}(1-\epsilon))}\,.
\end{equation}
At first order in $\epsilon$, this gives:
\begin{equation}
M_1(p)=- \frac{p^{2-4\epsilon}}{(4\pi)^{6}}\frac{2\pi^2}{3\epsilon} + \mathcal{O}(1)\,.
\end{equation}

\section{Beta functions details}
\label{ap:betafun4}

\subsection{2-loop amplitude}

We want to compute the two-loop amputated Feynman integral (the candy) represented in the middle of Fig.~\ref{fig:bare3}.
We use the subtraction point defined in Sec.~\ref{sec:betas1}. Then, respecting the conservation of momenta, we can write the candy integral as:
\begin{equation}
D_{\Delta}(\mu)=\int_{q_1,q_2} G_0(q_1)G_0(q_2)G_0(-p_1-p_2-p_3-q_1-q_2).
\end{equation}
This gives with $G(q)=\frac{1}{q^{2\Delta}}$:
\begin{equation}
D_{\Delta}(\mu)=\int_{q_1,q_2} \frac{1}{q_1^{2\Delta}q_2^{2\Delta}(p_1+p_2+p_3+q_1+q_2)^{2\Delta}}\,.
\end{equation}
We use twice Eq.~\eqref{eq:intG} and obtain (using $|p_1+p_2+p_3|=\mu$):
\begin{equation}
D_{\Delta}(\mu)=\f{1}{(4\pi)^d}\f{\Gamma(d/2 - \Delta)^3\Gamma(3\Delta- d)}{\Gamma(\Delta)^3\Gamma(3d/2 - 3\Delta)}\f{1}{\mu^{2(3\Delta - d)}}\,.
\end{equation}
For $\zeta=1$, we set $\Delta=1$ and $d=3-\epsilon$. We obtain at first order in $\epsilon$:
\begin{equation}
D_1(\mu)=\mu^{-2\epsilon}\frac{1}{(4\pi)^3}\frac{2\pi}{\epsilon}+ \mathcal{O}(1)\,.
\end{equation}
For the modified propagator case, we set $\Delta=\frac{d+\epsilon}{3}$ and $d<3$. We obtain at first order in $\epsilon$:
\begin{equation}
D_{d/3}(\mu)=\mu^{-2\epsilon}\frac{1}{(4\pi)^d}\frac{\Gamma(\frac{d}{6})^3}{\Gamma(\frac{d}{3})^3\Gamma(\frac{d}{2})\epsilon}+ \mathcal{O}(1)\,.
\end{equation}

\subsection{4-loop amplitude}

We compute the following four-loop amputated Feynman integral: 
\begin{equation*}
\int \dd[d]x\dd[d]y ~ G(x-y)^4G(x-z)G(y-z)\,. 
\end{equation*}
Again we use the symmetric subtraction point and we can write the integral in momentum space as:
\begin{equation}
S_{\Delta}(\mu)=\int_{q_1,q_2,q_3,q_4} G_0(q_1)G_0(q_2)G_0(q_3)G_0(q_4)G_0(-p_1-p_2-q_4)G_0(-p_1-q_1-q_2-q_3-q_4) \,.
\end{equation}
With $G_0(q)=\frac{1}{q^{2\Delta}}$, this gives: 
\be
S_{\Delta}(\mu)=\int_{q_1,q_2,q_3,q_4} \f{1}{(p_1+q_1+q_2+q_3+q_4)^{2\Delta}}\f{1}{(q_4 + p_1 +p_2)^{2\Delta}}\f{1}{(q_1q_2q_3q_4)^{2\Delta}}\,.
\ee
We integrate loop by loop using Eq.~\eqref{eq:intG}, until we are left with a triangle-type one-loop integral:
\be
S_{\Delta}(\mu)=\f{1}{(4\pi)^{3d/2}}\left(\f{\G(d/2 - \Delta)}{\G(\Delta)}\right)^4 \f{\G(4\Delta - 3d/2)}{\G(2d - 4\Delta)}\int_{q_4} \f{1}{(q_4 + p_1 +p_2)^{2\Delta}} \f{1}{q_4^{2\Delta}}\f{1}{(p_1 +q_4)^{2(4\Delta - 3d/2)}}\,.
\ee
We use a Mellin-Barnes representation \cite{Davydychev:1995mq,ODwyer:2007brp} to rewrite the remaining integral as: 
\be
\begin{split}
\int_{q_4} \f{1}{(q_4 + p_1 +p_2)^{2\Delta}} \f{1}{q_4^{2\Delta}} \f{1}{(p_1 +q_4)^{2(4\Delta - 3d/2)}} & = \\
\f{\pi^{d/2} ((p_1+p_2)^2)^{d/2 - \sum_i \n_i}}{\Gamma(d - \sum_i \n_i) \prod_i \Gamma(\n_i)(2\pi i)^2}\int_{-i\infty}^{i\infty}&\frac{\dd s\dd t }{(2\pi)^d} x^s y^t\Gamma(-s)\Gamma(-t) \Gamma(d/2 - \n_2 - \n_3 - s)\\
&\times \Gamma(d/2 - \n_1 - \n_3 - t)\Gamma(\n_3 + s+t)\Gamma(\sum_i\n_i -d/2 + s+t)\,,
\end{split}
\ee
with $\n_1=\n_2=\Delta,\n_3=4\Delta - 3d/2$ and $x=p_1^2/(p_1+p_2)^2,y=p_2^2/(p_1+p_2)^2$. 

In the case $\z=1$, we set $\Delta=1$ and deforming the contour on the right and picking the residue at $s=t=0$,\footnote{The other contributions from poles at $(s=n\geq 1,t=m\geq 1)$, $(s=n-2\eps, t=m-2\eps)$ and $(s=n-2\eps,t=m)$, $(s=n,t=m-2\eps)$ (assuming $n,m \in \mathbb{N}$) cancel, as well as those at $(s=0,t=m\geq1)$ with $(s=0, t=m-2\eps)$ or $(s=n\geq1,t=0)$ with $(s=n-2\eps, t=0)$.} we find in the last $\Gamma$ function the only contribution to the pole in $1/\eps$ from a $d = 3-\eps$ expansion. Putting everything together, we find, at first order in $\epsilon$: 
\begin{equation}
S_1(\mu)=\f{\mu^{-4\epsilon}}{(4\pi)^{6}}\f{\Gamma(1/2)^4}{\Gamma(3/2)}\f{\Gamma(-1/2)}{2\eps} = \f{\mu^{-4\epsilon}}{(4\pi)^6}\f{-2\pi^2}{\eps}+ \mathcal{O}(1)\,.
\end{equation}

In the case $\z=(d+\epsilon)/3$, we set $\Delta=\z$ and again only the residue at $s=t=0$ gives a contribution to the pole in $1/\epsilon$. We find:

\begin{equation}
S_{d/3}(\mu)=\frac{\mu^{-4\epsilon}}{(4\pi)^{2d}}\frac{\Gamma(d/6)^4\Gamma(-d/6)}{2\epsilon\Gamma(d/2)\Gamma(d/3)^4\Gamma(2d/3)}+ \mathcal{O}(1) \,.
\end{equation}

\end{subappendices}

%% file: RenGW.tex
\newcommand{\R}{\mathbb{R}} 
\newcommand{\C}{\mathbb{C}} 
\newcommand{\Z}{\mathbb{Z}} 
\newcommand{\N}{\mathbb{N}} 
\newcommand{\mP}{{\mathbb{P}}}
\newcommand{\mE}{{\mathbb{E}}}

\newcommand{\tah}{{\, {\rm th}\, }}

\newcommand{\veps}{\varepsilon}
\newcommand{\bbbone}{{\mathds{1}}}

The chapter is structured as follows. In Section \ref{section:QFTonGraph}, we introduce the ensemble of random trees that will be of concern as well as the random walk approach to the propagator of the theory. We also recall the multiscale point of view for renormalization towards an infrared fixed point and motivate the rescaling of the Laplacian appropriate for just renormalizable models. After presenting briefly in Section \ref{section:proba} the needed results of \cite{BarlowKumagai}, we prove upper and lower bounds on completely convergent graphs. In Section \ref{section:localization}, we obtain upper bounds on differences on amplitudes when transporting external legs, important in order to assure local counterterms. We discuss in Section \ref{section:unicycle} the setting that we think would stand for an analog of finite temperature field theory in this framework and the description of a model that would naturally serve as a concrete playground for the methods exposed below. Finally, the last Appendix~\ref{app:proba} details the probabilistic results that we relied upon, first focusing on branching processes, then on heat kernels over fixed and random graphs.

\section{Quantum Field Theory on a Graph}
\label{section:QFTonGraph}

\subsection{$\phi^{q}$ QFT on a graph}

\begin{figure}
\centerline{\includegraphics[width=13cm]{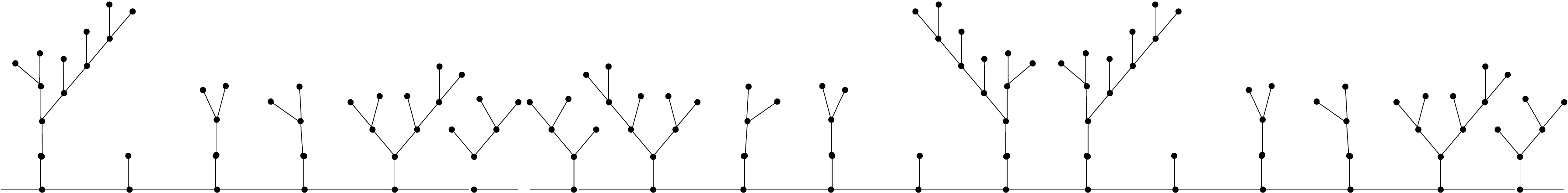}  }
\medskip
\centerline{\Large $\Downarrow$}
\vskip .5cm
\centerline{\includegraphics[width=13cm]{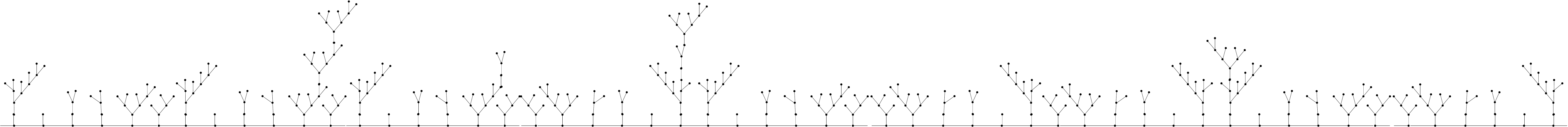}  }
\caption{Zooming towards the continuum random tree}
\label{treezooming}
\end{figure}

For this introductory section we follow \cite{Gurau:2014vwa} (in particular its section 3.3.2).
Let us consider a space-time which is a proper \emph{connected} graph $\Gamma$, 
with vertex set $V_\Gamma $ and edge set
$E_\Gamma$. It can be taken finite or infinite. The word ``proper" means that the graph has neither multiedges nor self-loops (often called tadpoles in physics). In the finite case we often omit to write cardinal symbols such as
$\vert V_\Gamma \vert $, $\vert E_\Gamma \vert $ when there is no ambiguity.
In practice we shall here consider mostly trees, more precisely 
either finite trees $\Gamma$ for which $V_\Gamma = E_\Gamma +1$, or infinite trees
in the sense of \cite{DJW}
which can be also interpreted as conditioned percolation clusters or Galton-Watson trees conditioned on non-extinction
in the sense of \cite{BarlowKumagai}. The main characteristic of such infinite trees is to have a single 
\emph{infinite spine} $\cS(\Gamma) \subset V(\Gamma)$. This spine is decorated all along by lateral independent Galton-Watson finite critical trees, which we call the branches. An artistic view of a cut on the spine is depicted on Fig.~\ref{treezooming}, and the approach towards a continuum limit. 

On any such graph $\Gamma$, there is a natural notion of the Laplace operator $\cL_\Gamma$.
We recall that on a directed graph $\Gamma$
the \emph{incidence matrix} is the rectangular $V$ by $E$ matrix 
with indices running over vertices and
edges respectively, such that 
\begin{itemize}
\item
${\epsilon_\Gamma(v,e)}$ is +1 
if $e$ ends at $v$, 
\item
${\epsilon_\Gamma(v,e)}$ is -1 if $e$ starts at $v$,
\item
${\epsilon_\Gamma(v,e)}$ is 0 otherwise.
\end{itemize}

The $V $ by $V$ square matrix 
with entries $d_v$ on the diagonal is called the \emph{degree} or coordination matrix
$D_\Gamma$. 
The {\emph{adjacency matrix}} is the symmetric $V \times V$ matrix $A_\Gamma$ made of zeroes on the diagonal: $A_\Gamma(v,v) = 0 \;\; \forall v\in V$, and such that 
if $v \ne w$ then $A_\Gamma(v,w)$ is the number of edges of $G$ which have vertices $v$ and $w$ as their ends. 
Finally the {\emph{Laplacian matrix}} of $\Gamma$ is defined to be $\cL_\Gamma = D_\Gamma - A_\Gamma$.
Its positivity properties stem from the important fact that it is a kind of square of the incidence matrix,
namely 
\be
L_\Gamma = \epsilon_\Gamma \cdot \epsilon_\Gamma^t\;, \label{laplacian}
\ee
where $\eps^t$ stands for the transpose of $\eps$.
Remark that this Laplacian is a positive rather than a negative operator (the sign convention being opposite to the one of differential geometry). Its kernel (the constant functions) has dimension 1 since $\Gamma$ is connected.

The kernel $C_\Gamma (x, y) $ of the inverse of this operator is formally given by the sum over random paths $\omega$ from $x$
to $y$
\be \cL_\Gamma^{-1} = C_\Gamma (x, y) =\left[\sum_n \left(\frac{1}{D_\Gamma}A_\Gamma\right)^n \frac{1}{D_\Gamma} \right](x,y)= \sum_{\omega: x \to y} \; \prod_{v\in \Gamma} \;\biggl[\frac{1}{d_v}\biggr]^{n_v (\omega)}
\label{pathrep}
\ee
where $d_v= D_\Gamma(v,v) = \cL_\Gamma (v,v)$ is the coordination at $v$ and $n_v (\omega)$ is the number of visits of $\omega$ at $v$. We sometimes omit the index $\Gamma$ when there is no ambiguity.

As we know this series is not convergent without an infrared regulator (this is related to the Laplacian having
a constant zero mode). 
For a finite $\Gamma$ we can take out this zero mode by fixing a root vertex in the graph and deleting 
the corresponding line and column in $\cL_\Gamma$.
But it is more symmetric to use the mass regularization. It 
adds $m^2 \bbbone$ to the Laplacian, where $\bbbone$ is the identity operator on $\Gamma $, with kernel $\delta (x,y)$. Defining $C_\Gamma^m (x, y)$ as the kernel of $(\cL_\Gamma + m^2 \bbbone )^{-1}$
we have the \emph{convergent} path representation
\be C_\Gamma^m (x, y) = \sum_{\omega: x \to y} \; \prod_{v\in \Gamma} \;\biggl[\frac{1}{d_v + m^2}\biggr]^{n_v (\omega)}
\label{masslap}
\ee
and the infrared limit corresponds to $m \to 0$.

A scalar Bosonic free field theory $\phi$ on $\Gamma$ 
is a function $\phi : V_\Gamma \to \R$ defined on the vertices of the graph and
measured with the Gaussian measure 
\be d\mu_{C_\Gamma}(\phi) = \frac{1}{Z_0} e^{-\frac{1}{2} \phi (\cL_\Gamma + m^2 \bbbone ) \phi 
} \prod_{x\in V_\Gamma} d\phi(x) , \label{eq:mesuremu}
\ee
where $Z_0$ a normalization constant.
It is obviously well-defined as a finite dimensional probability measure for $m >0$ and $\Gamma$ finite. 
We meet associated infrared divergences in the limit of $m=0$ and they are 
governing the large distance behavior of the QFT in the limit of infinite graphs $\Gamma$. The systematic
way to study QFT divergences is through a multiscale expansion 
in the spirit of \cite{FMRS,Rivabook,books,ChandraHairer}. No matter whether an ultraviolet or an infrared limit is considered, the renormalization group always flows from ultraviolet to infrared and the same techniques apply in both cases.

The $\phi^{q}$ interacting theory is then defined by the formal functional integral \cite{Gurau:2014vwa}:
\be
d\nu_\Gamma (\phi ) = \frac{1}{Z(\Gamma, \lambda)} e^{- \lambda \sum_{x \in V_\Gamma} \phi^{q} (x) } d \mu_{C_\Gamma} (\phi) \;,
\ee
where the new normalization is
\be Z(\Gamma, \lambda)= \int e^{- \lambda \sum_{x \in V_\Gamma} \phi^{q} (x) } d \mu_{C_\Gamma} (\phi) = \int d\nu_\Gamma (\phi ) .
\label{mesurenu}\ee

The correlations (Schwinger functions) of the $\phi^{q}$ model on $\Gamma$ are
the normalized moments of this measure:
\be S_{N} (z_1,...,z_N ) = \int   
\phi(z_1)...\phi(z_N) \; d\nu_\Gamma (\phi ) , 
\ee
where the $z_i$ are external positions hence fixed vertices of $\Gamma$.
The case of fixed flat $d$-dimensional lattice corresponds to $\Gamma ={\mathbb Z}^d$.
As well known the Schwinger functions expand in the formal series of Feynman graphs
\be S_{N} (z_1,...,z_N ) = \sum_{V=0}^\infty \frac{(- \lambda)^V}{V!} \sum_G A_G (z_1,...,z_N ), 
\ee
where the sum over $G$ runs over Feynman graphs with $V$ internal vertices of valence $q$ and $N$ external leaves of valence 1.
Beware not to confuse these Feynman graphs with the ``space-time" 
graph $\Gamma$ on which the QFT lives.
More precisely $E_G$ is the disjoint union of a set $I_G$ of internal edges and of a set 
$N_G$ of external edges, and for the interaction $\phi^{q}$ these Feynman graphs have $V_G =V$ internal 
vertices which are regular with total degree $q$ and $N_G=N$ external leaves of degree 1. Hence $qV_G= 2E_G+N_G$. If $q$ is even this as usual implies parity
rules, namely $N_G$ has also to be even. We often write simply $V$, $E$, $N$ instead of $V_G$, $E_G$ and $N_G$
when there is no ambiguity.
Besides, for our purpose, we will neither focus on exact combinatoric factors nor about convergence of this series
although these are of course important issues treated elsewhere \cite{Rivabook}. We also shall
consider only connected Feynman graphs $G$, which occur in the expansion of the connected Schwinger functions.

As usual the treatment of external edges is attached to a choice for the external arguments of the graph.
Our typical choice here is to use external edges which all link a $q$-regular internal vertex to a 1-regular leaf with 
fixed external positions $z_1$, \dots $z_N$ in $\Gamma$. The (unamputated) graph amplitude is then a function of the external arguments obtained 
by integrating all positions $x_v$ of internal vertices $v$ of $G$ over our space time, which is $V(\Gamma)$. Hence
\begin{align}  A_{G}(z_1, \cdots , z_N) = \sum_{\substack{x_v \in V_\Gamma \\ v\in V_G}} \prod_{\ell \in
E_G } C_\Gamma^m (x_{\ell},y_{\ell}) 
\end{align}
where $x_\ell$ and $y_\ell$ is our (sloppy, but compact!) notation for the vertex-positions at the two ends of edge $\ell$. 

We consider now perturbative QFT on \emph{random trees}, which instead of $\Gamma$ we note from now as $T$. 
The universality class of random trees \cite{aldous}
 is the Gromov-Hausdorff limit of any critical Galton-Watson tree process with fixed branching rate \cite{harris},
and conditioned on non-extinction.
 It has a unique \emph{infinite spine}, decorated with a product of independent Galton-Watson
measures for the branches along the spine \cite{DJW}. We briefly recall the corresponding probability measure,
following closely \cite{DJW}, but instead of half-infinite rooted trees with spine labeled by ${\mathbb N}$ we consider
trees with a spine infinite in both directions, hence labeled by ${\mathbb Z}$.

The order $\vert T \vert$ of a rooted tree is defined as its number of edges.
To a set of non-negative {\it branching weights} $w_i,\,i\in\mathbb
N^\star$ is associated the weights generating function $g(z):= \sum_{i \ge 1} w_i z^{i-1} $ and the
{\it finite volume partition function} 
$Z_n$ on the set $\cT_n$ of all rooted trees $T$ with root $r$ of order $\vert T \vert =n$
\be
Z_n = \sum_{T \in \cT_n}\prod_{u\in T \setminus r}w_{d_u}\,,
\ee
where $d_u$ denotes the degree of the vertex $u$. The generating function for all $Z_n$'s
is 
\be Z(\zeta ) = \sum_{n=1}^\infty Z_n \zeta^n .
\ee It satisfies the equation (cf. App.~\ref{app:proba})
\be\label{fixZ}
Z(\zeta ) =\zeta g(Z (\zeta)).
\ee
Assuming a finite radius of convergence $\zeta_0$ for $Z$ one defines 
\be
Z_0 = \lim_{\zeta\uparrow \zeta_0}Z(\zeta ).
\ee
The critical Galton-Watson
probabilities $p_i := \zeta _0 w_{i+1}Z_0^{i-1}$ for $i\in\mathbb N$ are then normalized: $\sum_{i=0}^\infty p_i = 1$. 
We then consider the class of infinite random trees defined by an \emph{infinite} spine of vertices $s_k$, $k\in \mathbb Z$,
plus a collection of $d_k -2$ \emph{finite} branches $T^{(1)}_k, \dots , T^{(d_k -2)}_k$, at each vertex $s_k$
of the spine (recall the degree of $k$ is indeed $d_k$). The set of such infinite trees is called $\cT_\infty$. It is equipped with a probability measure $\nu$ that we now describe.
This measure is obtained as a limit of measures $\nu_n$ on finite trees of order $n$. These measures $\nu_n$
are defined by \emph{identically and independently} distributing branches around a spine with measures 
\be \mu ( T) = Z_0^{-1}\zeta_0^{|T|}\prod_{u \in T \setminus r}w_{d_u} = \prod_{u\in T\setminus r}p_{d_u -1} \,.
\ee

\begin{theorem}
Viewing $\nu_n(T) = Z_n^{-1}\prod_{u\in T \setminus r}w_{d_u}\,,\quad T\in\cT_n\,,$ as a probability measure on $\cT$ we have
\be
\nu_n \to \nu\quad \text{as}\quad n\to\infty\,,
\ee
where $\nu$ is the probability measure on $\cT$ concentrated on the
subset of infinite trees $\cT_\infty$  \cite{DJW}. 
Moreover the spectral dimension of generic infinite tree ensembles is
$ d_{spec} = 4/3\,$.
In fact the trees in $\cT_\infty$ conditioned to have a single infinite spine can be constructed by redefining the branching weights $w_i$, let us say of mean $m$, through
\be 
w_i^*= \f{iw_i}{m}
\ee 
for one vertex per generation,
such that the probability to have no child vanishes \cite{Kesten, Abraham:2015}.
\end{theorem}

From now on we write $\mE (f)$ for the average according to the measure
$d \nu$ of a function $f$ depending on the tree $T$, and $\mP$ for the probability of an event $A$ according to $d\nu$. Hence $\mP (A) = \mE ( \chi_A)$ where $\chi_A$ is the characteristic function for the event $A$
to occur.
For simplicity and in order not to loose the reader's attention into unessential details we shall also
restrict ourselves from now on to the case of critical \emph{binary} Galton-Watson trees. 
It corresponds to weights $w_1 = w_3 = 1$, and $w_i = 0$ for all other values of $i$. 
In this case the above formulas simplify. 
The critical Galton-Watson process corresponds to offspring probabilities 
$p_0 = p_2 = \frac{1}{2} $, $p_i = 0$ for $i \ne 0, 2$.
The generating function for the 
branching weights is simply $g(z) = 1 + z^2$ and the generating function for the
finite volume trees $Z(\zeta ) = \sum_{n=1}^\infty Z_n \zeta^n $
obeys the simple equation $
Z(\zeta ) =\zeta (1 + Z^2 (\zeta))$,
which solves to the Catalan function $Z= \frac{1 - \sqrt{1 - 4 \zeta^2}}{2\zeta}$. In the above notations 
the radius of convergence of this function is $\zeta_0 = \frac{1}{2}$. Moreover $Z_0 = \lim_{\zeta\uparrow \zeta_0}Z(\zeta ) =1$ and the independent measure on each branch of our random trees is simply
\be
\mu (T) = 2^{- \vert T \vert }\,.
\ee

\subsection{Fractional laplacians}

Since the most interesting QFTs (including, in dimension 1, the tensorial theories \`a la Gurau-Witten) are the ones with just renormalizable power counting, 
we want to state our result in that case. A time-honored method for that is to raise the ordinary Laplacian to a suitable fractional power $\alpha$ 
in the QFT propagator \cite{Abd,Gross:2017vhb}. We assume from now on that this fractional power obeys $0 < \alpha <1$ and call $C^\alpha$
the corresponding propagator, i.e. the kernel of $\cL^{-\alpha}$. It is most conveniently computed using the identity
\be \cL^{-\alpha} = \frac{\sin\pi\alpha}{\pi} \int_0^\infty \frac{2m^{1-2\alpha }}{\cL + m^2} dm \label{kallen}
\ee
since this ``K\"allen-Lehmann" representation, with density $m^{1-2\a}$ respects the positivity
properties of the random path representation of the ordinary Laplacian inverse.

In the continuum $\mathbb{R}^d$ case, we have the ordinary heat-kernel integral representation
\begin{align}  
C^\alpha_{{\mathbb R}^d}(x,y) = \frac{\sin\pi\alpha}{\pi} \int_0^\infty 2m^{1-2\alpha }dm \int_0^\infty e^{-m^{2}t - 
\frac{\vert x-y \vert^{2} }{4t }} \frac{d t}{t^{d/2}} ,
\end{align}
On ${\mathbb Z}^d$ the rescaled kernel of the Laplacian between points
$x$ and $y$ is similarly obtained from eq. \eqref{kallen}, 
using the random walk representation:
\be C_{{\mathbb Z}^d}^{\alpha}(x,y) = \frac{\sin\pi\alpha}{\pi} \int_0^\infty 2m^{1-2\alpha } dm\sum_{\omega: x \to y}\prod_v \;\biggl[\frac{1}{2d + m^2}\biggr]^{n_v (\omega)} 
\ee
where $n_v (\omega)$ is the number of visits of $\omega$ at $v$. Notice that each vertex on ${\mathbb Z}^d$ has degree $2d$.

As remarked above in the case of a general graph $\Gamma$ we no longer have translation invariance of Fourier integrals but still the 
random path expansion, so that
\be C_{\Gamma}^{\alpha}(x,y) = \frac{\sin\pi\alpha}{\pi} \int_0^\infty 2m^{1-2\alpha }dm\sum_{\omega: x \to y}\prod_v \;\biggl[\frac{1}{d_v + m^2}\biggr]^{n_v (\omega)} 
\ee
where the walks $\omega$ now live on $\Gamma$ and $d_v$ is the degree at vertex $v$. 

\subsection{The random tree critical power $\alpha = \frac{2}{3} - \frac{4}{3q}$}
\label{subsection: powercounting}

In integer dimension $d$, standard QFT \emph{power counting} with propagator $C^\alpha$ relies on 
the standard notion of degree of divergence. For a regular Feynman graph of degree $q$ with $N$ external legs, this degree is defined as
\be \omega(G) = (d-2 \alpha )E - d (V-1) = (d-2 \alpha )(qV-N)/2 - d (V-1) . \label{divdegree}
\ee

This power counting is neutral (hence does not depend on $V)$ in the critical or just-renormalizable case
\be \alpha =\frac{ (q-2)d }{2q } 
\ee
in which case we have
\be \omega(G) =d \left(1- \frac{N}{q}\right). \label{divdegree2}
\ee

For instance if $q=d=4$ we recover that the $\phi^4_4$ theory with propagator $p^{-2}$ is critical,
and if $d=1$ we recover the critical index 
$\alpha= \frac{1}{2} - \frac{1}{q}$ of the infrared SYK theory with $q$ interacting Fermions \cite{Sachdev:1992fk,Kitaev2015,Maldacena:2016hyu,Polchinski:2016xgd}. 

As we will show in section \ref{section:localization}, a just-renormalizable $\phi^q$ theory is obtained by substituting in the above formulas the \emph{spectral dimension} $d=4/3$
of random trees, namely 
\be\alpha = \frac{2}{3} - \frac{4}{3q}\;, \quad \omega(G) =\frac{4-N}{3} \;. \label{crtiticalpha}
\ee 
This is not surprising since this spectral dimension is precisely related to the short-distance, long-time behavior of the inverse Laplacian
averaged on the random tree. We shall fix from now the fractional power $\alpha$ to its critical value.
Nevertheless this simple rule requires justification, which is precisely provided by the next sections.

\subsection{Slicing into scales}

The multiscale decomposition of Feynman amplitudes is a systematic tool
to establish power counting and study perturbative and constructive renormalization in quantum field theory
\cite{Gurau:2014vwa,FMRS,Rivabook}. It relies on a sharp slicing into a geometrically growing sequence of scales of the
Feynman parameter for the propagator of the theory. This parameter is nothing but the \emph{time}
in the random path representation of the Laplacian. The short time behavior of the propagator is unimportant 
since the graph $\Gamma$ is an ultraviolet regulator in itself. We are therefore interested in infrared problems, namely the long distance behavior of the theory
(in terms of the graph distance). In the usual discrete
random walk expansion of the inverse Laplacian, the total time is the length of the path hence an 
integer. This integer when non trivial cannot be smaller than 1. However the results of \cite{BarlowKumagai}
are formulated in terms of a continuous-time random walk
which should have equivalent infrared properties. In what follows we shall use 
both points of view.

\begin{definition}[Time-of-the-Path Slicing]
We introduce the infrared parametric slicing of the propagator $1/(\cL+m^2)$:
\begin{align} C =& \sum_{j=0}^{\infty} C^j \;; \quad C^0= \bbbone\;, \nonumber\\
C^j =& \sum_{\omega: x \to y \atop M^{2(j-1)} \le n(\omega) < M^{2j} } 
\; \prod_{v\in \Gamma} \;\biggl[\frac{1}{d_v + m^2}\biggr]^{n_v (\omega)} \quad \forall j \ge 1\;. \label{decoa}
\end{align}
\end{definition}
$M$ is a fixed constant which parametrizes the thickness of a renormalization group slice 
(the craftsman trademark of \cite{Rivabook}).
Each propagator $C^{j}$ indeed corresponds to a theory with
both an ultraviolet and an infrared cutoff, which differ by the fixed 
multiplicative constant $M^2$. 
An infrared cutoff on the theory is then obtained by setting a maximal value $\rho= j_{max}$ for the index $j$.
The covariance with this cutoff is therefore 
\be C_{\rho} = \sum_{j=0}^{\rho} C^{j} \;. \label{decocut1} 
\ee
In the continuum $\R^d$ case we have the ordinary heat-kernel representation
hence the explicit integral representation
\begin{align}  
C^{\alpha,j}_{{\mathbb R}^d}(x,y) &= \frac{\sin\pi\alpha}{\pi} \int_0^\infty 2m^{1-2\alpha }dm \int_{M^{2j}}^{M^{2(j+1)}} e^{-m^{2}t - 
\frac{\vert x-y \vert^{2} }{4t }} \frac{dt}{t^{d/2}} \;,
\end{align}
from which it is standard to deduce scaling bounds such as 
\begin{align} C_{{\mathbb R}^d}^{\alpha,j}(x,y) &\le K M^{(2 \alpha -d)j} e^{- c M^{-2j}\vert x -y \vert^2} \;, \label{flatbound}
\end{align}
for some constants $K$ and $c$. From now on, we use most of the time $c$ or $K$ as generic names for
any inessential constant (therefore they are the same as the $O(1)$ notation 
in the constructive field theory literature). 
We shall also omit from now on to keep
inessential constant factors such as $ \frac{\sin\pi\alpha}{\pi} $. 
In ${\mathbb Z}^d$ the sliced propagator then writes
\be C_{{\mathbb Z}^d}^{\alpha,j}(x,y) = \int_0^\infty 2m^{1-2\alpha } dm \sum_{\omega: x \to y \atop M^{2(j-1)} \le n(\omega) < M^{2j} } \prod_v \;\biggl[\frac{1}{2d + m^2}\biggr]^{n_v (\omega)} .
\ee
It still can be shown easily to obey the same bound \eqref{flatbound}. For
a general tree $T$ the sliced decomposition of the propagator then writes
\begin{align} C_{T}^{\alpha}(x,y) &= \sum_{j=0}^{\infty} C_{T}^{\alpha,j}(x,y) ; \quad C_{T}^{\alpha,0}= \bbbone, \quad {\rm and\;\; for }\;\; j \ge 1\;, \\
C_{T}^{\alpha,j}(x,y) &=  \int_0^\infty 2m^{1-2\alpha } dm \sum_{\omega: x \to y \atop M^{2(j-1)} \le n(\omega) < M^{2j} } 
\; \prod_{v\in T} \;\biggl[\frac{1}{d_v + m^2}\biggr]^{n_v (\omega)}\;.
\label{slicedprop2}
\end{align}
Remark that after $n$ steps a path cannot reach farther than distance $n$ (for the discrete time random walk). In particular we can safely include the
function $\chi_j (x,y)$ in any estimate on $C_{T}^{\alpha,j}$, where $\chi_j (x,y)$ is the characteristic function for
$d(x,y) \le M^{2j}$. \footnote{The graph distance $d(x,y)$ denotes the smallest number of steps on the tree needed to connect $x$ to $y$.}
A generic tree $T$ in $\cT$ has spectral dimension $4/3$ so that we should expect for such a tree 
\begin{align} C_{T}^{\alpha,j}(x,y) &\le K M^{\left(2 \alpha -\frac{4}{3}\right)j} \chi_j (x,y) \;. \label{flatbound1}
\end{align}
A fixed tree can nevertheless be non-generic, hence has no a priori well defined dimension $d$. At the same time, since it always contains an infinite spine which has dimension 1, the propagator on any tree $T$ in $\cT$ should obey
the following bound:
\begin{align} C_{T}^{\alpha,j}(x,y) &\le K M^{(2 \alpha -1)j}\chi_j (x,y) \;. \label{flatbound2}
\end{align}
However we do not need a very precise bound for exceptional trees since as we will see in the next section, they will 
be wiped by small probabilistic factor. 
In fact a very rough ``dimension zero" bound can be obtained for all points $x$, $y$ on $T$:
\begin{align} C_{T}^{\alpha,j}(x,y) &\le K M^{2 \alpha j} \chi_j (x,y)\;. \label{flatbound3}
\end{align}
Indeed, overcounting the number of paths from $x$ to $y$ in time $t$ as the total number of paths from $x$ in time $t$ leads to this inequality. 
In the binary tree case each vertex degree is bounded by 3. At a visited vertex $v$ we have $d_v$ choices for the next random path step so that
\begin{align}
\sum_{\omega: x \to y \atop M^{2(j-1)} \le n(\omega) < M^{2j} } \; \prod_{v\in T} \;\biggl[\frac{1}{d_v + m^2}\biggr]^{n_v (\omega)} &\leq
\sum_{M^{2(j-1)} \leq n< M^{2j} } \; \left[\frac{3}{3+m^2}\right]^{n}\\
&\leq K \int_{M^{2(j-1)}}^{M^{2j}} dt e^{-ctm^2}
\end{align}
where $K$ and $c$ are some inessential constants. \footnote{For the upper inequality, we used that for any $m>0$, ~$ 3/(3+m^2) > 1/(1+m^2)$. The lower one is obtained by comparing the Taylor expansions of both members around $m=0$. $K$ is chosen such that the inequality between the rational function and the exponential holds. Whereas $c$ is independent of $m$, if $m<1$, $K>5$ is enough. } Then the naive inequality
\be K \int_0^\infty m^{1-2 \alpha}dm 
\int_{M^{2(j-1)}}^{M^{2j}} dt e^{-ctm^2} \le K' M^{2j\alpha},
\ee
allows to conclude.

Yet none of the bounds \eqref{flatbound1}-\eqref{flatbound3} are sufficient to establish the correct
power counting of Feynman amplitudes averaged on $T \in \cT$. We need to combine the multiscale decomposition
(best tool to estimate general Feynman amplitudes on a fixed space) with probabilistic estimates 
to show that the prefactor $ M^{\left(2 \alpha -\frac{4}{3}\right)j} $ in \eqref{flatbound1}
is indeed the typical one and that the typical volume factors for the integrals on vertex positions
correspond also to those of a space of dimension 4/3.

\subsection{Multiscale analysis}

From this subsection onward, we will drop the $\alpha$ of the propagator $C^{\a,j}$ to lighten the notation. Consider a fixed connected Feynman graph $G$ with $n$ internal 
 vertices, all with degrees $q=4$, $N$ external edges and $L =2n -N/2$ internal edges. 
There are in fact several possible prescriptions to treat external arguments in a Feynman amplitude \cite{FMRS,Rivabook},
but they are essentially equivalent from the point of view of integrating over inner vertices the product of propagators. 
A convenient and simple choice is to put all external legs in the most infrared scale, namely the infrared cutoff scale $\rho$
(similar to a zero external momenta prescription in a massive theory),
and to work with amputated amplitudes which no longer depend
on the external positions $z_1, \dots, z_N$ but only on the position 
$x_0$ of a fixed inner \emph{root vertex} $v_0$. It means we 
forget the $N(G)$ external propagators $C_{T}(x_{v(k)},z_k) $ factors in $A_G$ and shall integrate only the $n-1$ positions 
$x_v, v \in \{1, \dots, n-1 \}$. In this way we get an amplitude
$A_G^{amp} (x_0)$ which is solely a function of $x_0$. \footnote{In a usual theory there is no $x_0$ dependence because of translation invariance, but for a particular tree $T$ there is no such invariance.}
However we should remember that fields and 
propagators at the external cutoff scale have a canonical dimension
which in our case for a field of scale $j$ is $M^{-j/3}$. To compensate 
for the missing factors after amputation
we shall multiply this 
amputated amplitude by $M^{-\rho N /3}$, and for the fixing of 
position $x_0$, we shall add another global factor $M^{4\rho/3}$. 
Hence we define
\be \tilde A_G^{amp} (x_0) := M^{\rho (4-N) /3}
\sum_{\substack{x_v \in V(T)\\1\leq v\leq n-1}} \prod_{\ell \in I(G) } C_{T} (x_\ell, y_\ell )\;. \label{agmu0}
\ee
For simplicity, we write now $A_G$ again, instead of $\tilde A^{amp}_G$.
The decomposition \eqref{decoa1} leads to the multiscale representation for a Feynman 
graph $G$, which is:
\begin{align}  A_{G}(x_0) &= M^{\rho (4-N) /3} \sum_{\mu} A_{G,\mu}(x_0)\; \; ,\label{agmu1}\\
A_{G,\mu }(x_0) & = \sum_{\substack{x_v \in V(T)\\1\leq v\leq n-1}} 
\prod_{\ell \in I(G) }  C_{T}^{j_\ell} (x_\ell, y_\ell ) \;. \label{agmu2}
\end{align}
$\mu$ is called a ``scale assignment" (or simply ``assignment").
It is a list of integers $\{ j_\ell \}$, one for each internal edge of $G$, which provides for each internal edge $\ell$ of $G$ the scale $j_\ell$ of that edge. $A_{G,\mu}$ is the amplitude associated to the pair $(G,\mu)$, and \eqref{agmu1}-\eqref{agmu2} is the multiscale representation of the Feynman amplitude.

We recall that the key notion in the multiscale analysis of a Feynman amplitude is that of ``high" subgraphs. In our infrared setting, this means the connected components of $G_{j}$, the subgraph of $G$ made of all edges $\ell$ with index $j_\ell \le j$. These connected components are labeled as $G_{j,k}$, $k=1,...,k(G_j)$, where $k(G_j)$ denotes the number of connected components of the graph $G_j$.

A subgraph $g \subset G$ then has in 
the assignment $\mu$ internal and external indices defined as
\be  i_g(\mu) = \sup\limits_{l {\rm \ internal \ edge \ of \ } g} \mu(l)\;, \label{II.1.10}
\ee
\be  e_g(\mu) = \inf\limits_{l {\rm \ external \ edge \ of \ } g} 
\mu(l) \;. \label{II.1.11}
\ee
Connected subgraphs verifying the condition 
\be  e_g(\mu) > i_g(\mu) \quad \quad {\rm (high \ condition)} 
\label{II.1.12}
\ee
are exactly the \emph{high} ones. This definition depends on the assignment $\mu$. For
a high subgraph $g$ and any value of $j$
such that $i_g (\mu ) < j \le e_g(\mu ) $ there exists exactly one value of
$k$ such that $g$ is equal to a $G_{j,k}$. 
High subgraphs are partially ordered by inclusion and form a forest in the sense of 
inclusion relations \cite{FMRS,Rivabook}.

The key estimates then keep only the spatial decay of a $\mu$-optimal spanning tree $\tau(\mu)$ of $G$,
which minimizes $\sum_{\ell \in \tau(\mu)} j_\ell (\mu)$
(we use the notation $\tau$ for spanning trees of $G$
in order not to confuse them with the random tree $T$). The important 
property of $\tau (\mu)$ is that it is a spanning tree
within each high component $G_{j,k}$ \cite{FMRS,Rivabook}.
It always exists and can be chosen according to Kruskal greedy algorithm \cite{kruskal}.
It is unique if every edge is in a different slice; otherwise there may be several such trees in which case one simply picks one of them.

Suppose we could assume bounds similar to the $\mathbb{R}^d$ case. It would mean that 
a sliced propagator in the slice $j_\ell$ would be bounded as
\be C_{T}^{j_\ell} (x_\ell, y_\ell ) \simeq K M^{-2j_\ell/3} e^{-M^{-j_\ell} d(x_\ell,y_\ell) }
\ee
and that spatial integrals over each $x_v$ would be really 4/3 dimensional, i.e cost $M^{4j_v/3}$ if performed with the decay of a scale $j_v$ propagator.
Picking a Kruskal tree $\tau (\mu) $ with a fixed root vertex $v_*$, and forgetting the spatial decay of all the edges not in $\tau$, one can then recursively organize
integration over the position $x_v$ of each internal vertex $v$ from the leaves towards the root. This can be indeed done using for each $v$ 
the spatial decay of the propagator joining $v$ to its unique towards-the-root-ancestor 
$a(v)$ in the Kruskal tree. In this way calling $j_v$ the scale of that propagator 
we would get as in \cite{FMRS,Rivabook} an estimate
\begin{align}
|A_{G,\mu}| &\le K^{V(G)} M^{-N \rho/3}  \prod_{\ell \in I(G)} M^{-2j_\ell/3} \prod_{v \in V(G)\setminus v_*} M^{4j_v/3} \\
&= K^{V(G)} \prod_{j =1}^\rho \prod_{k=1}^{k(G_j)}  M^{\omega(G_{j,k})} \label{sketch}
\end{align}
where the divergence degree 
of a subgraph $S \subset G$ is defined as
\be \omega(S) = \frac{2}{3}E(S) - \frac{4}{3} (V(S)-1) = \frac{4-N(S)}{3} \;.
\ee

Standard consequences of such bounds are 
\begin{itemize}
\item uniform exponential bounds for completely convergent graphs \cite{Rivabook}.
\item renormalization analysis: when high subgraphs have positive divergent degree 
we can efficiently replace them by local counterterms, which create a flow for marginal and relevant operators.
The differences are remainder terms which become
convergent and obey the same bounds as for convergent graphs, provided we use
an effective expansion which renormalizes only high subgraphs \cite{FMRS,Rivabook}.
\end{itemize}

In fact these bounds cannot be true for all particular trees $T$ since they depend on the Galton-Watson branches being typical. In more exceptional cases, for instance for a
tree reduced to the spine plus small lateral branches the effective spatial dimension is 1 rather than 4/3.
Such exceptional cases become more and more unlikely when we consider larger and larger sections of the spine. Our probabilistic analysis below proves that for the \emph{averaged} Feynman amplitudes everything happens as in equation \eqref{sketch}. To give a meaning to these averaged amplitudes, we fix the position of the root vertex $x_0$ to lie on the spine of $T$. Averaging over $T$
restores translation invariance along the spine, so that we have finally to evaluate averaged amplitudes $\mE (A_G)$ which are simply numbers. It is for these amplitudes that 
we shall prove in the next sections our main results
Theorem \ref{theoconv} and \ref{theoconv1}.
But we need to introduce first our essential probabilistic tool, namely the $\lambda$-good conditions on trees of \cite{BarlowKumagai}.

\section{Probabilistic Estimates}
\def\lam {\lambda}
\label{section:proba}
We have first to recall the probabilistic estimates on random trees from
\cite{BarlowKumagai} that we are going to use, simplifying slightly some aspects inessential for our discussion. More details on those techniques are collected in the Appendix~\ref{app:proba}, where we provide a framework applicable to a larger class of graphs. 
As mentioned above, \cite{BarlowKumagai} mostly considers random paths 
which are Markovian processes with \emph{continuous} times, but 
those are statistically equivalent to above discrete processes in the interesting long-time infrared limit, as is discussed in the remark 5.3 of \cite{BarlowKumagai}.

For $x \in T$, we note $B(x,r)$ the ball of $T$ centered on $x$ and containing points at most at distance $r$ from $x$,
and $M(x,r)$ the number of points of $T$ at distance $1+[r/4]$ of $x$, where $[.]$
means the integer part. For a subgraph $A\subset T$, we define the volume $V(A) = \sum_{v\in A}d_v$ and more concisely $V(x,r) = V(B(x,r))$. For $(x,y) \in T^2$, we also write $q_t (x,y)$ (or sometimes $q_{t,x}(y)$ to emphasize the starting point $x$) for the sum over random paths in time $t$. More precisely given a continuous time random walk $Y$ on $T$, starting at $x$ at $t=0$ and jumping from a vertex $v$ to its neighbours with probability $1/d_v$, waiting at $v$ for a time sampled from a Poisson distribution of mean 1,  the heat-kernel writes
\be 
q_t(x,y) = \f{\mP^x(Y_t = y)}{d_y}\;,
\ee
where $\mP^x(Y_t=y)$ denotes the probability that the random walk $Y$ sits at $y$ at the time $t$.

For  $\lam\ge 64$,
the ball $B(x,r)$ is said {\sl $\lam$--good} (Definition 2.11 of \cite{BarlowKumagai}) if:
\begin{gather}
r^2\lam^{-2} \le V(x,r) \le r^2\lam \;, \label{goodvol} \\
M(x,r) \le \frac{1}{64} \lam\;, \quad
V(x, r/\lam) \ge r^2 \lam^{-4}\;, \quad V(x, r/\lam^2) \ge r^2 \lam^{-6}\;.
\end{gather}
See the Appendix~\ref{app:proba} for details.
Remark that if $B(x,r)$ is {\sl $\lam$--good} for some $\lambda$,
it is {\sl $\lam'$--good} for all $\lambda' > \lambda$. We will also say $\lam$-bad for a ball $B(x,r)$ that is not $\lam$-good.

Corollary 2.12 of \cite{BarlowKumagai} proves that
\be
\mP (B(x,r) \hbox{ is not $\lam$--good}) \le c_1 e^{-c_2 \lam}\;. 
\label{smallproba1}
\ee
This inequality together with the Borel-Cantelli lemma (cf. App.~\ref{app:proba}) imply that given $r$ and a real monotonic sequence $\{\lam_{l}\}_{l\geq 0}$ with $\lim_{l \to \infty} \lam_{l}= +\infty $, there is, with probability one, a finite $l_0$ such that $B(x,r)$ is $\lam_{l_0}$-good. In particular
\begin{lemma}\label{smallproba3}
Defining the random variable $L=\min \{l : B(x,r) \text{ is } \lam_{l}\text{-good}\}$
we have
\be \mP [ L=l]
\le c_1 e^{-c_2 \lam_{l-1}}. \label{smallproba2}
\ee
\end{lemma}
\proof This is because the ball $B(x,r)$
must then be $\lam_{l-1}\text{-bad}$.
\qed

\medskip
Besides, the conditions of $\lambda$-goodness allow to bound with the right scaling the random path factor $q_t (x,y)$ for $y$ not too far from $x$. More precisely the main part of Theorem 4.6 of \cite{BarlowKumagai} reads
\begin{theorem}
\label{theoBK1}
Suppose that $B=B(x,r)$ is $\lam$--good for 
$\lambda\ge 64$, and let $I(\lam,r)=[r^3 \lam^{-6}, r^3 \lam^{-5}]$. Then 

\begin{itemize}
\item 
for any $K \ge 0$ and any $y\in T$ with $d(x,y) \le Kt^{1/3}$
\be q_{2t}(x,y) \le c\left(1+{\sqrt K}\right) t^{-2/3} \lam^{3}
\quad \hbox{\rm for } t\in I(\lambda, r) \;, \label{gooddecay1}
\ee

\item
for any $y\in T$ with $d(x,y) \leq c_2 r \lambda^{-19}$
\be q_{2t}(x,y) \ge c t^{-2/3} \lam^{-17}
\quad \hbox{\rm for } t\in I(\lambda, r)\;. \label{gooddecay2}
\ee

\end{itemize}

\end{theorem}
Notice that these bounds are given for $q_{2t}(x,y)$ 
but the factor $2$ is inessential (it can be gained below
by using slightly different values for $K$) and we omit it from now on for simplicity.

\subsection{Preliminaries: two-point function}

To translate these theorems into our multiscale setting, we introduce
the notation $I_j = [ M^{2(j-1)}, M^{2j}] $ and we have the infrared equivalent continuous time representation\footnote{We refer to Ch. 2 of \cite{ambjorn} for details on going from the discrete to continuous time propagators, the exponential factor stemming from the mass regulator.}
\be C^j_T (x,y) = \int_0^\infty u^{-\alpha } du \int_{I_j} q_t (x,y) e^{-ut } dt = \Gamma(1- \alpha)\int_{I_j} q_t (x,y)t^{\alpha -1} dt \;.
\label{trans}
\ee
This relates our sliced propagator 
\eqref{slicedprop2}
to the kernel $q_t$ of \cite{BarlowKumagai}. We forget from now on 
the inessential $\Gamma(1- \alpha)$
factor.
In our particular case $q=4,\alpha= 1/3$, \eqref{trans} means that 
we should simply multiply the estimates on $q_t$ established in \cite{BarlowKumagai} 
by $c M^{2j/3}$ to obtain similar estimates for $ C^j_T$. However we have also to perform
spatial integrations not considered in \cite{BarlowKumagai}, which complicate the probabilistic analysis.
As a warm up, let us therefore begin with a few very simple examples. Recall
that we do not carefully track inessential constant
factors in what follows, and that we can use the generic letter $c$ for any such constant
when it does not lead to confusion.

\begin{lemma} [Single Integral Upper Bound]\label{oneline}
There exists some constant $c$ such that
\be 
\mE \left[ \sum_y C_T^j (x,y)f(x,y) \right] \le c M^{2j/3} \;,
\label{eq:lemma-singleline}
\ee
for any $L^1$ function $f$ with $0\leq f(x,y)\leq 1,~ \forall x,y\in T$. 
\end{lemma}
\proof
We introduce two indices $k \in \N$, and $l \in \N$ with the condition $l\ge l_0 := \sup \{M^{2}, 64\}$ and parameters
$\lambda_{k,l}:= k+ l $. We also define radii
\begin{align} r_{j,k} &:= M^{2j/3} k^{5/3} \;,\\ 
r_{j,k,l} &:= M^{2j/3}(k+ l)^{5/3}\;,
\end{align}
and the balls $B^T_{j,k}$ and $B^T_{j,k,l}$ centered on $x$
with radius $r_{j,k}$ and $r_{j,k,l}$ (we put an upper index $T$
to remind the reader that these sets depend on our random space, 
namely the tree $T$). We also define the annuli
\be
A^T_{j,k} := \{y : d(x,y) \in  [r_{j,k},r_{j,k+1}[
\},
\ee
so that the full tree is the union of the annuli $A^T_{j,k}$
for $k \in \N$:
\be 
T = \cup_{k \in \N} \; A^T_{j,k}\;. 
\label{sumT}
\ee
Remark that $A^T_{j,k} \subset B^T_{j,k+1} \subset B^T_{j,k,l}$ 
for any $l \in \N^\star$.
Remark also that with these definitions
\be 
I_j = [M^{2j-2},M^{2j}] \subset I(\lambda_{k,l},r_{j,k,l})=
[r_{j,k,l}^3\lambda_{k,l}^{-6}, r_{j,k,l}^3\lambda_{k,l}^{-5}]\;,
\ee
where $I(\lambda,r)$ is as in Theorem \ref{theoBK1},
since our condition $l \ge l_0 \ge M^{2}$ ensures that $r_{j,k,l}^3\lambda_{k,l}^{-6} \le M^{2j-2}$. 
Finally defining $K_k := M^{2/3} (k+1)$ we have
\be 
d(x,y ) \le K_k t^{1/3},\quad \forall t \in I_j,\;\forall y \in A^T_{j,k}\;. \label{distcond}
\ee
Since the propagator is pointwise positive we can commute any sum or integral as desired. Taking \eqref{sumT} into account we can
organize the sum over $y$ according to the annuli $A^T_{j,k}$. Commuting the 
sum $\mE$ and the sum over $k$, 
according to the Borel-Cantelli argument in the section above, there exists (almost surely in $T$) a smallest finite $l$ such that the $B^T_{j,k,l}$ ball is $\lam_{k,l}$-good. Defining the random variable $L=\min \{l \ge l_0 : B^T_{j,k,l} \text{ is } \lam_{k,l}\text{-good}\}$, we can
partition our $\mE$ sum according to
the different events $L=l$. We now fix this $l$ so as to evaluate, according to \eqref{trans}
\be 
\mE \left[\sum_y C_T^j (x,y) f(x,y) \right] = \sum_{k=0}^\infty \sum_{l=l_0}^\infty \mP[L=l] \mE\vert_{L=l} \Bigl[\sum_{y \in A^T_{jk}} \int_{I_j} dt t^{\alpha-1} q_t (x,y) f(x,y)
\Bigr]\;,
\ee
where $\mE\vert_{A}$ means conditional expectation with respect to the event $A$.
We are in position to apply Theorem \ref{theoBK1} since all hypotheses and conditions are fulfilled (including 
$\lambda_{k,l} \ge 64$ since $l_0 \ge 64$). We have for some inessential constant $c$, under condition $L=l$
\be 
q_t (x,y) \le c (1 + \sqrt{K_k} )
M^{-4j/3} \lam_{k,l}^3 , \quad \forall t \in I_j,\;\forall y \in A^T_{j,k} \;.
\ee
Hence integrating over $t \in I_j$ 
\be 
C_T^j (x,y) \le c (k+l)^{7/2} M^{-2j/3},\quad \forall y \in A^T_{j,k}\;, 
\label{cjesti1}
\ee
for some other inessential constant $c$.
We can now sum over $y \in A^T_{j,k}$, overestimating the volume of the annulus $A^T_{j,k}$ by the volume of the $B^T_{j,k,l}$ ball (the number of vertices it contains), to obtain
\be 
\sum_{y \in A^T_{j,k}} C_T^j (x,y) f(x,y)
\le c (k+l)^{7/2} M^{-2j/3} vol(B^T_{j,k,l}) \;, 
\label{cjesti12}
\ee
since $f$ is bounded by one.
The condition $L=l$
allows to control the volume $vol(B^T_{j,k,l}) $ by the $\lam_{k,l}\text{-good}$ condition.
More precisely \eqref{goodvol} implies
\be
\mE\vert_{L=l} \, [ vol(B^T_{j,k,l}) ] \le r_{j,k,l}^2 \lambda_{k,l} \;.
\ee
Using Lemma \ref{smallproba3} we conclude that
\begin{align}
\mE \left[\sum_y C_T^j (x,y)f(x,y) \right] &\le c\sum_{k=0}^\infty \sum_{l=l_0}^\infty \mP[L=l]
(k+l)^{7/2}M^{-2j/3} r_{j,k,l}^2 \lam_{k,l}\nonumber
 \\
&\leq c M^{2 j/3}\sum_{k=0}^\infty \sum_{l=l_0}^\infty e^{-c'(k+l)} (k+l)^{47/6} \le c M^{2 j/3} \;. 
\end{align}
\qed 

\begin{cor}[Tadpole]\label{tadpolecor}
There exists some constant $c$ such that
\be 
\mE \left[ C_T^j (x,x) \right] \le c M^{-2j/3} \;.
\label{cortadpole}
\ee
\end{cor}
\proof
Taking $f(x,y)= \delta_{xy}$ in Lemma~\ref{oneline} gives the bound.
\qed

\medskip
A lower bound of the same type is somewhat easier, as we do not need to exhaust the full spatial integral but can restrict to a subset, in fact a particular $\lambda$-good ball.
\begin{lemma}[Single Integral Lower Bound] \label{lowerbou}
\be \mE \left[ \sum_y C_T^j (x,y) \right] \ge c M^{2j/3} .\label{lowerbou1}
\ee
\end{lemma}

\proof 
We follow the same strategy than for the upper bound but we do not need
the index $k$ and the annuli $A_{j,k}$, since most of the volume is typically in the first annulus - 
namely the $k=0$ ball $B_j$. Restricting the sum over $y$ this ball is typically enough 
for a lower bound of the \eqref{lowerbou1} type. So we work at $k=0$ but we need again probabilistic estimates to tackle
the case of untypical volume of the ball $B_j$.
Therefore we define 
for $l\ge l_0 := \sup \{M^{2}, 64\}$, the parameter 
$\lambda_l = l$ and the two balls $B^T_{j,l} = B(x,r_{j,l})$ and 
$\tilde B^T_{j,l} = B(x,\tilde r_{j,l})\subset B^T_{j,l}$ of radii
respectively $r_{j,l}:=M^{2j/3}\lambda_l^{5/3}$ and $\tilde r_{j,l}:= c_2r_{j,l}\lambda_l^{-19}$ (in order for \eqref{gooddecay2} to apply below). We introduce the random variable 
\be L=\min \{l \ge l_0 : B^T_{j,l} \text{ and } \tilde B^T_{j,l} \text{ are both } \lam_{l}\text{-good}\}\;.
\ee

Again, our choice of $r_{j,l}$ ensures that
\be I_j = [M^{2j-2},M^{2j}] \subset I(\lambda_{l},r_{j,l})=
[r_{j,l}^3\lambda_{l}^{-6}, r_{j,l}^3\lambda_{l}^{-5}]\;,
\ee
and the summands being positive, we will restrict the sum over $y$ to the smaller ball $\tilde B^T_{j,l} \subset B^T_{j,l}$, in order for \eqref{gooddecay2} to apply. We get
\begin{align} 
\mE \left[\sum_y C_T^j (x,y) \right] 
&\ge 
\mP[L\le l] \mE\vert_{L \le l} \
\Bigl[\sum_{y \in \tilde B^T_{j,l} } \int_{I_j} dt t^{\alpha-1} q_t (x,y)
\Bigr], \quad \forall l,\ \\
&\ge c M^{-2j/3} l^{-17} \mP[L\le l] \;\mE\vert_{L=l} 
[ vol( \tilde B^T_{j,l} ) ],\quad\quad \forall l\;,\ \\
&\ge cM^{2j/3}\mP[L\le l] l^{-161/3}, \quad\quad\quad\quad\quad\quad\quad\quad\ \forall l\;, \ \\
&\ge cM^{2j/3} \;.
\end{align}
Indeed for the last inequality we remark that 
$\lim_{l \to \infty}\mP[L\le l] = 1$ (by Lemma \ref{smallproba3})
hence $\sup_{l \ge l_0}\mP[L\le l]l^{-161/3} $ is a \emph{strictly positive} constant that we absorb in $c$.
\qed 

\subsection{Bounds for convergent graphs}\label{fullgraphs}

In this section we prove our first main result, namely 
the convergence of Feynman amplitudes 
of the type \eqref{agmu0}-\eqref{agmu2} as the infrared cutoff $\rho$ is lifted.
Therefore we consider a fixed completely convergent graph $G$ with $n$ inner vertices and $N$ external lines, hence for which $N(S) \ge 6\; \forall S \subset G$. In this graph we mark a root vertex $v_0$ with fixed position $x_0$, 
lying on the spine, i.e. common to all trees $T$. By translation invariance of the infinite spine, the resulting amplitude $A_G (x_0)$
is in fact independent of $x_0$ and we have

\begin{theorem}\label{theoconv} For a completely superficially convergent graph (i.e. with no 2- or 4-point subgraphs)
$G$ of order $V(G) = n$, the limit as 
$\lim_{\rho \to \infty} \mE ( A_G)$
of the averaged amplitude exists 
and obeys the uniform bound
\be \mE ( A_G) \le K^n (n!)^\beta \label{uniconv}
\ee
where $\beta =\frac{52}{3}$. \footnote{We do not try to make $\beta$ optimal. We expect that a tighter probabilistic analysis could prove
subfactorial growth in $n$ for $\mE ( A_G)$.}
\end{theorem}

\proof
From the linear decomposition $A_{G} = \sum_\mu A_{G,\mu}$
follows that $\mE (A_G) = \sum_\mu \mE (A_{G,\mu} )$.
As mentioned above we use
only the decay of the propagators of an optimal Kruskal 
tree $\tau (\mu)$ to perform the spatial integrals over the position of the inner vertices. It means that we first 
apply Cauchy-Schwarz inequalities to the $n+1 - N/2$
edges $\ell\not\in \tau (\mu)$. 

To be exact, the first Cauchy-Schwarz inequality applies to the Markovian random walk with heat-kernel $q_{2t}(x,y)$ which rewrites as an inner-product by the Chapman-Kolmogorov property
\begin{align}
q_{2t}(x,y) &= \sum_{z\in V(T)}q_t(x,z)q_t(z,y)= \expval{q_{t,x} ,q_{t,y}}_2 \nonumber \\
&\leq  \sqrt{\expval{q_{t,x}^2}_2\expval{q_{t,y}^2}_2} =\sqrt{ q_{2t}(x,x)q_{2t}(y,y)}.
\end{align}
We refer to \cite{Kumagai} for more details and will again use this inner product in Section \ref{section:localization}. 
A second Cauchy-Schwarz inequality is then used for the scalar product $(f,g)= \int_0^\infty dt t^{\alpha - 1}f(t) g(t)$ with $f$ standing for $\sqrt{q_{2t}(x,x)}$ and analogously for $g$.

Labeling all the corresponding half-edges (not in $\tau(\mu)$) as fields $f= 1, \cdots 2n+2 - N$ and their positions and scale as $x_f$ and $j_f$ we have
\begin{align} \prod_{\ell \not \in \tau (\mu)} 
C_T^{j_\ell} (x_\ell,y_\ell) &\le c^n \prod_{\ell \not \in \tau (\mu)} 
\sqrt{C_T^{j_\ell}(x_\ell, x_\ell) C_T^{j_\ell}(y_\ell, y_\ell) }\nonumber
\\
&=c^n\prod_{f=1}^{2n+2 - N} [C_T^{j_f}(x_f,x_f)]^{1/2},
\label{loopbou}
\end{align}
making use of eq. \eqref{trans}.

Each inner vertex $v\in \{1, \cdots n-1\}$ to integrate over is linked to the root by a single path in $\tau (\mu)$. The first line, $\ell_v$, in this path relates $v$ to a single ancestor $a(v)$ by an edge $\ell_v \in \tau (\mu)$. This defines a scale $j_v := j_{\ell_v}(\mu)$ for the sum over the position $x_v$. 

Taking \eqref{loopbou} into account, we write therefore 
\be \mE [A_{G,\mu }] \le \mE \Big[ 
c^n \sum_{\{x_{v} \}} 
\prod_{v=1}^{n-1} C_{T}^{j_v} (x_v, x_{a(v)})
\prod_{f =1}^{2n+2 - N} [C_T^{j_f}(x_f,x_f)]^{1/2} \Big] \; . \label{agmu3}
\ee

We apply now to the $n-1$ spatial integrals exactly the same analysis
than for the single integral of Lemma \ref{oneline}.
The main new aspect is that the events of the previous section do not provide \emph{independent} small factors for each spatial integral. 
For instance if two positions $x_v$ and $x_{v'}$ 
happen to coincide and the smallest-$l$ $\lambda_{l}$-good 
event occur for a ball centered at $x_v$, it \emph{automatically implies}
the $\lambda_{l-1}$-bad event for the ball centered at $x_{v}$
and at $x_{v'}$,
because it is the \emph{same event}. Therefore in this case we do not get twice the same small associated probabilistic factor
of Lemma \ref{smallproba3}. This is why we loose a (presumably spurious)
factorial $[n!]^\beta$ in \eqref{uniconv}.

More precisely we introduce for each $v \in [1, n-1]$ two
integers $k_v$ and $l_v \ge l_0$, 
the radii $r_{j_v, k_v}$, $r_{j_v, k_v, l_v}$
and the parameters $\lambda_{k_v,l_v}$ exactly as before.
 We introduce also
all these variables for every field $f \in [ 1, \cdots 2n+2 - N ]$ not in $\tau( \mu)$. We define again the random variable $L_v$ for $v \in [1, n-1]$ as the first integer $\ge l_0$ such that the ball 
$B^T_{j_v,k_v,l_v}$ is $\lam_{k_v,l_v}\text{-good}$ and
$L_f$ for $f \in [1, 2n+2-N]$ as the first integer $\ge l_0$ such that the ball 
$B^T_{j_f,k_f,l_f}$ is $\lam_{k_f,l_f}\text{-good}$.
The integrand is then bounded according to Theorem \ref{theoBK1},
leading to
\begin{align} \mE [A_{G,\mu }] \le& c^n
\sum_{\{k_v\},\{l_v\}\atop\{k_f\},\{l_f\}}\mP (L_v= l_v, L_f = l_f) 
\Big[ \prod_{v=1}^{n-1} M^{2j_v/3}
[k_v+l_v]^{47/6} \prod_{f =1}^{2n+2 - N} M^{-j_f/3}[k_f+l_f]^{7/4} \Big].
\end{align}
Now as mentioned already the $3n +1 -N$ events 
$L_v=l_v$ or $L_f=l_f$ are not independent so we use only the single \emph{best} probabilistic factor for one of them. It means we define
$m= \sup_{v,f}\{k_v+l_v, k_f+l_f \}$ and use that $\mP [L_v= l_v, L_f = l_f] \le c' e^{-cm}$
to perform all the sums
with the single probabilistic factor $e^{-cm}$
from \eqref{smallproba2}. Since each index is bounded by $m$, the big
sum
\be \sum_{\{k_v \le m\},\{l_v \le m\}\atop\{k_f \le m\},\{l_f \le m\}} \prod_{v=1}^{n-1}
[k_v+l_v]^{47/6} 
\prod_{f =1}^{2n+2 - N} [k_f+l_f]^{7/4} 
\ee
is bounded by 
$c^n m^{\frac{59}{6} (n-1) + \frac{15}{4} (2n+2 - N) }$ hence by 
$c^n m^{\frac{52n}{3}}$. Finally since
\be\sum_m e^{-cm} m^{\frac{52n}{3}}
\le c^n [n!]^\beta, \quad \beta =\frac{52}{3} ,
\ee
we obtain the usual power counting estimate 
up to this additional factorial factor:
\be \mE [A_{G,\mu }] \le c^n [n!]^{\beta}
\sum_{\mu}\prod_{v=1}^{n-1} M^{2j_v/3}\prod_{f =1}^{2n+2 - N} M^{-j_f/3}.
\ee
From now on we can proceed to the standard infra-red
analysis of a just renormalizable theory exactly similar to the usual $\phi^4_4$ analysis of \cite{FMRS,Rivabook,Gurau:2014vwa}. 
Organizing the bound according to the inclusion forest of the high subgraphs $G_{j,k}$ we rewrite 
\be \prod_{v=1}^{n-1} M^{2j_v/3}\prod_{f =1}^{2n+2 - N} M^{-j_f/3} = \prod_{j,k} M^{\omega(G_{j,k})}
\ee
with $\omega(S) = \frac{2}{3}E(S) - \frac{4}{3} (V(S)-1) = \frac{4-N(S)}{3}$ and get 
therefore the bound
\be
\mE [A_{G,\mu }] \le c^n [n!]^{\beta}
\sum_{\mu}\prod_{j,k} M^{[4-N(G_{j,k})]/3}.
\ee
The sum over $\mu$ is then performed with the usual strategy of \cite{FMRS,Rivabook,Gurau:2014vwa}.
We extract from the factor $\prod_{j,k} M^{[4-N(G_{j,k})]/3} $ an independent exponentially decaying factor (in our case at least 
$M^{-\vert j_f -j_{f'}\vert /54}$ for \emph{each vertex} $v$ and \emph{each pair of fields $(f,f')$ 
hooked to $v$} of their scale difference $\vert j_f -j_{f'}\vert$
\footnote{The attentive reader wondering about
the factor 54 will find that it comes from the fact that $(N-4)/3\ge N/9$ for $N\ge 6$ and that 
there are 6 different pairs at a $\phi^4$ vertex.}). 
We can then organize and perform easily the sum over all scales assigned to all fields, hence over $\mu$, and it results only in still another $c^n$ factor.
This completes the proof of the theorem.
\qed
\medskip

A lower bound 
\be
\mE \left[\sum_y [C_T^j (x,y) ]^2 \right] 
\ge c
\ee
can be proved exactly like Lemma \ref{lowerbou} and implies that the elementary one loop 4-point function is truly logarithmically divergent when $\rho \to \infty$.

Taken all together the results of this section prove that for the $\phi^q$ interaction at $q=4$ the value $\alpha= \frac{1}{3}$
is the only one for which the theory can be just renormalizable. 
Extending to any $q$ can also be done 
following exactly the same lines and proves that $\alpha= \frac{2}{3} - \frac{4}{3q}$, as in \eqref{crtiticalpha},
is the only exponent for which the theory is just renormalizable in the infrared regime.

\section{Localization of High Subgraphs}
\label{section:localization}
When the graph contains $N=2$ or $N=4$ subgraphs, we need to renormalize. According to the Wilsonian strategy, renormalization has to be performed only on \emph{high} divergent subgraphs,
and perturbation theory is then organized into a multi-series in effective constants, one for each scale, all related through a flow equation.
This is standard and remains true either for an ultraviolet 
or for an infrared analysis \cite{Rivabook}. 

Two key facts power the renormalization machinery and their combination allows to compare efficiently the contribution of a high divergent subgraph to its Taylor expansion around local 
operator \cite{Rivabook,Gurau:2014vwa}:

\begin{itemize}
\item the quasi-locality (relative to the internal 
scale $i_S (\mu)$) between external vertices of any high subgraph $S=G_{j,k}$ provided by the Kruskal tree
(because it remains a spanning tree when restricted to any high subgraph);

\item the small change in an external propagator 
of scale $e_S (\mu) =j_M$ when one of its arguments is moved by a distance typical of the much smaller internal ultraviolet scale $i_S (\mu)=j_m\ll j_M$. 
\end{itemize}
Taken together these two facts explain why the contribution
of a high subgraph is quasi-local from the point of view of its external scales, hence explain why renormalization by \emph{local} counterterms works.

However usual tools of ordinary quantum field theory such as translation invariance and
momentum space analysis are no longer available on random trees, and
we have to find the probabilistic equivalent of the two above facts in our random-tree setting:

\begin{itemize}
\item in our case, the proper time of the path of a propagator at scale $j$ is $t_j \simeq M^{2j}$ and the ordinary associated distance scale is $r_j \simeq t_j^{1/3} \simeq M^{2j/3}$. We expect the associated scaled decay between external vertices of any high subgraph $G_{j,k}$ 
provided by the Kruskal tree to be true only for typical trees. However we prove below that the techniques used in Lemma \ref{oneline} to sum over $y$ validate this picture;

\item in our case, the small change in an external propagator
of scale $j_M$ should occur when one of its arguments is moved by a distance of order $r_{j_m} \simeq M^{2j_m/3}$. We shall prove that in this case we gain a small factor $M^{-(j_M-j_m)/3}$ compared to the ordinary estimate in $M^{-2j_M/3}$ of \eqref{cortadpole} for $C_T^{j_M}$. This requires comparing propagators with different arguments hence some additional work.
\end{itemize}

Hence, the following analysis justifies the heuristic power counting argument given in Subsection \ref{subsection: powercounting} and that the subtraction of local counterterms allows indeed to control the diverging amplitude in this context of random trees (with some additional subtleties in the 2-point function case).

\subsection{Preliminaries: subtractions}
We explain first on a simplified example how to implement these ideas, then give a general result.
Our first elementary example
consists in studying the effect of a small move of one of the arguments of a sliced propagator $C^{j}_T (x, y)$. We need to check that it leads, after averaging on $T$, to a relatively smaller and smaller effect on the sliced propagator when $j \to \infty$. 

Consider three sites $x$, $y$ and $z$ on the tree and
the difference 
\be \Delta^j_T (x;y,z) := \vert C^{j}_T (x, y) - 
C^{j}_T (x, z)\vert .
\ee
We want to show that when $d(y,z) \ll r_j = M^{2j/3}$, we gain
in the average $\mE [\Delta (x,y,z) ]$ a small factor 
compared to the ordinary estimate in $M^{-2j/3}$ for a single propagator without any difference.

This is expressed by the following Lemma.
\begin{lemma}\label{egainlemma}
There exists some constant $c$ such that for any $T$ and 
any $t \in I_j$
\be\vert q_t (x,y) - q_t (x,z) \vert \le 
c M^{-j}\sqrt{d(y,z) q_t (x,x)}. \label{egain0}
\ee
Moreover
\be \mE [\Delta^j_T (x;y,z) ]\le c M^{-2j/3} M^{-j/3 } \sqrt{d(y,z)} . \label{egain}
\ee
This bound is uniform in $x \in \cS$ and the factor
$ M^{-j/3 } \sqrt{d(y,z)}$ is the gain, provided 
$d(y,z) \ll r_j = M^{2j/3}$.
\end{lemma}
\proof
We use again results of \cite{BarlowKumagai}.
With their notations, it is proved in their Lemma 3.1 that
\be \vert f(y) - f(z) \vert^2 \le R_{eff} (y,z) \cE (f,f) 
\ee
where the effective graph resistance $R_{eff} (y,z)$ in the case of a tree $T$ is nothing but the natural distance $d(y,z)$
on the tree, and noting as earlier $\langle f,g\rangle_2 $
the $L_2(T)$ scalar product $\sum_{y\in T} f(y)g(y) $,
\be\cE (f,f) := \langle f, \cL f \rangle_2
\ee
is the natural positive quadratic form associated to the Laplacian. Applying this estimate to the function
$f_{t,x}$ defined by $f_{t,x}(y) = q_t(x, y)$ exactly as in the 
proof of Lemma 4.3 of \cite{BarlowKumagai} leads to 
\be\vert f_{t,x}(y) - f_{t,x}(z) \vert^2 \le 
d(y,z)\frac{q_t (x,x)}{t}
\ee
hence to 
\be\vert f_{t,x}(y) - f_{t,x}(z) \vert \le 
c M^{-j}\sqrt{d(y,z) q_t (x,x)}
\ee
for any $t \in I_j$.
From there on \eqref{egain} follows easily by an analysis similar to 
Corollary \ref{tadpolecor}.
\qed
\medskip

The next Lemma describes a simplified renormalization situation:
a single propagator $C^{j_M}_T (x, y)$ mimicks a single external propagator at an ``infrared" scale $j_M$ and another propagator 
$C^{j_m}_T (y, z)$ mimicks a high subgraph at an ``ultraviolet" scale $j_m \ll j_M$.
The important point is to gain a factor $M^{-(j_M - j_m)/3}$
when comparing the ``bare" amplitude 
\be
A^{b}_T (x,z) := \sum_{y\in T} C^{j_M}_T (x, y) C^{j_m}_T (y, z) 
\ee
to the ``localized" amplitude at $z$
\be A^l_T (x,z):= 
C^{j_M}_T (x, z) \sum_{y\in T} C^{j_m}_T (y, z)
\ee
in which the argument $y$ has been moved to $z$ in the external 
propagator $C^{j_M}_T$. Introducing the averaged ``renormalized" amplitude
\be \bar A^{ren}_T(x,z):= \mE [A^{b}_T (x,z) - A^l_T (x,z)],
\ee we have
\begin{lemma}
\be \vert\bar A^{ren}_T(x,z)\vert \le
c M^{- (j_M - j_m)} .
\ee
\end{lemma}
This Lemma shows a net gain $M^{- (j_M - j_m)/3}$
compared with the ordinary estimate $M^{- 2(j_M - j_m)/3}$
which we would get for $A^{b}_T $ or $A^{l}_T $ separately.
\proof
We replace the difference $ C^{j_M}_T (x, y) - C^{j_M}_T (x, z)$ by
the bound of Lemma \ref{egainlemma}. Taking out of $\mE$ the trivial scaling factors
\be \vert\bar A_{ren}(x,z)\vert \le
c M^{-j_M/3+2j_m/3}\mE\Big[ \sum_{y \in T} \sqrt{d(y,z)} 
\sup_{t \in I_{j_M} \atop 
t' \in I_{j_m}}[\sqrt{q_t (x,x)}
 q_{t'}(y,z) ] \Big].
\ee
We apply the same strategy that in the previous sections, hence we introduce the radii $r_{j_m,k_m}$ and 
$r_{j_m,k_m,l_m}$ and the corresponding balls and annuli as in the proof of Lemma \ref{oneline} to perform the sum over $y$ using the 
$q_{t'}(y,z)$ factor. 
We also introduce the radii $r_{j_M,k_M,l_M}$ to tackle the 
$\sqrt{q_t (x,x)}$ which up to trivial scaling is exactly similar to a field factor in $[ C_T^{j_f}(x_f,x_f)]^{1/2}$ in \eqref{loopbou},
hence leads to a $M^{-2j_M/3}$ factor. The $\sum_{y\in T} $ then costs an 
$M^{4j_m/3}$ factor, the $\sqrt{d(y,z)}$ factor costs an $M^{j_m/3}$ factor
and the $q_{t'}(y,z)$ brings an $M^{-4j_m/3}$. 
Gathering these factors leads to the result.
\qed

\medskip

\subsection{Renormalization of four-point subgraphs}

The 4-point subgraphs $N(S) =4$ in this theory have $\omega (S) = \frac{4-N(S)}{3}$ hence 
are logarithmically divergent. Consider now a graph $G$ which has no 2-point subgraphs, hence with $N(S) \ge 4$ for any subgraph $S$. Recall the previous evaluation
\begin{align}
|A_{G,\mu}| &\le K^{V(G)} M^{-N \rho/3}  \prod_{\ell \in I(G)} M^{-2j_\ell/3} \prod_{v \in V(G)} M^{4j_v/3} \\
&= K^{V(G)} \prod_{j =1}^\rho \prod_{k=1}^{k(G_j)}  M^{\omega(G_{j,k})} \label{sketch1}
\end{align}
of its bare amplitude. When there are 4-point subgraphs this amplitude, which is finite 
at finite $\rho$, diverges when $\rho \to \infty$ since there is no decay factor between the internal scale $i_\mu (S) $ and the external scale...

In the effective series point of view we fix a scale attribution $\mu$ 
and renormalization is only performed for the
high subgraphs $G_{j,k}$ with $N(G_{j,k})=4$. 
They form a \emph{single forest} $\cF_\mu$ for the 
inclusion relation. Therefore 
in this setting the famous ``overlapping divergences" problem is completely solved from the beginning. Such divergences are simply an artefact of the BPHZ theorem and completely disappear in the effective series organized according to the Wilsonian point of view \cite{Rivabook}.

In other words, for every 4-point subgraph $S$ we choose a root vertex $v_S$, with a position noted $x^S_1$,
to which at least one external propagator, 
$C(z_1, x^S_1)$ of $S$ hooks, and we 
introduce the localization operator $\tau_S$
which acts on the three of the four external propagators $C$ attached to $S$ through the formula
\be \tau_S C(z_2, x^S_2)C(z_3, x^S_3)C(z_4, x^S_4) := C(z_2, x^S_1)C(z_3, x^S_1)
C(z_4, x^S_1).
\ee
The effectively renormalized amplitude 
with global infrared cutoff $\rho$ is then defined as
\begin{align}  A^{eff}_{G,\rho}(x_0) &:= M^{\rho (4-N) /3} \sum_{\mu} A^{eff}_{G,\rho, \mu}(x_0)\; \; ,\label{agmu6}\\
A^{eff}_{G,\rho, \mu }(x_0) &:= \prod_{S \in \cF_\mu } (1 - \tau_S) \prod_{v=1}^{n-1} \sum_{x_v \in V(T)} 
\prod_{\ell \in I(G) }  C_{T}^{j_\ell} (x_\ell, y_\ell ) . \label{agmu7}
\end{align}
The result on a given tree still depends on the choice of the root vertex (because there is no longer translation invariance on a fixed given tree). 
Nevertheless translation invariance is recovered 
along the spine for the averaged amplitudes
and our second main result is:

\begin{theorem}\label{theoconv1} For a graph $G$ with $N (G) \ge 4$ and no 2-point subgraph
$G$ of order $V(G) = n$, the averaged effective-renormalized amplitude 
$\mE[ A_G^{eff}] =\lim_{\rho \to \infty} 
\mE[ A_{G,\rho}^{eff}]$ is convergent 
as $\rho \to \infty$ and obeys the same uniform bound than in the completely convergent case, namely
\be \mE ( A_G^{eff}) \le K^n (n!)^\beta .
\ee
\end{theorem}
\proof
Since the renormalization operators $1-\tau_S $ 
are introduced only for the high subgraphs, they always bring by estimates \eqref{egain0}-\eqref{egain} a factor $M^{- (e_g(\mu) - i_g (\mu))/3}$.

Exactly like in the previous section, we obtain
therefore a bound
\begin{align}
|A^{eff}_{G,\mu}| &\le K^{V(G)} M^{-N \rho/3}  \prod_{\ell \in I(G)} M^{-2j_\ell/3} \prod_{v \in V(G)} M^{4j_v/3} \\
&= K^{V(G)} \prod_{j =1}^\rho \prod_{k=1}^{k(G_j)}  M^{\omega^{ren}(G_{j,k})} \label{sketch2}
\end{align}
with $\omega^{ren}(G_{j,k}) = \omega(G_{j,k}) =\frac{4-N(G_{j,k})}{3} $ if $N(G_{j,k}) > 4$
and $\omega^{ren}(G_{j,k}) = - \f{1}{3}$ if $N(G_{j,k}) = 4$.
Therefore 
$ A^{eff}_{G} = \sum_\mu A^{eff}_{G,\mu } $ can be bounded exactly like $A_G$, using the same single 
$\lambda$-good condition as for the proof of Theorem
\ref{theoconv}. It therefore obeys the same estimate.
\qed
\medskip

The perturbative theory can be organized in terms of these effective amplitudes provided the bare coupling constant at a vertex $v$ with highest scale $j^h(v)$
is replaced by an effective constant $\lambda_{j^h(v)}$.

Remember that in the usual BPHZ renormalized amplitude
we must introduce the Zimmermann's forest sum, that is introduce $\tau_S $ counterterms also for subgraphs that are not high. Such counterterms cannot be combined efficiently with anything so have to be bounded independently, using the cutoff provided by the condition that they are not high. This unavoidably leads to additional factorials
which this time are not spurious, as they correspond to the so-called renormalons.
These renormalons disappear in the effective series \cite{Rivabook}, and the problem is exchanged for another question, namely whether the flow of the effective constants remains bounded or not.

\subsection{Multiple subtractions}
\label{sub:multiple}
Finally in the general perturbative series there occurs also 2-point subgraphs. For them we need to perform multiple subtractions. In the $\phi^q$ theory with $q=4$ the 2-point function has divergence degree $\omega = 2/3$ so it is not cured by a single difference as above. We need a kind of systematic analog of an operator product expansion
around local or quasi-local operators.
In our model the Laplacian is the main actor
which replaces ordinary gradients in fixed space models. It is also the one that can be transported easily from one point to another, gaining each time small factors. Therefore if our problem requires renormalization beyond strictly local terms (such as wave function renormalization) we shall describe now a possibly general method
to apply.

For any function $f$ we can write the expansion
\be f(u) = \overline{ f} (u) + \mathscr{L} f (u) 
\ee
where $\overline f$ is the local average 
$\frac{1}{d_u}\sum_{v \sim u} f (v)= \frac{1}{D}Af$ over the neighbors of $u$, and $\mathscr{L} := \frac{1}{D}\cL = \bbbone - \frac{1}{D}A$
is the normalized operator that appears in the 
discretized heat equation on $T$. 
Remark indeed that from \eqref{pathrep} we deduce
\be [C_{n+1} -C_{n}] (x,y) = \left[\left(\frac{1}{D}A-\bbbone\right)
C_{n} \right](x,y) = - [\mathscr{L} C_{n}](x,y) \label{heat0}
\ee
where $C_n (x,y)$ is the sum over discrete random walks
from $x$ to $y$ in exactly $n$ steps.

Iterating we can define 
for any fixed $p \in \N$ (where we simply put $d$ for $d_u$ when there is no ambiguity) an expansion:
\be f = \bar f + \overline{\mathscr{L} f}
+ \overline{\mathscr{L}^2 f} + \cdots + \overline{\mathscr{L}^p f}
+ \mathscr{L}^{p+1} f.
\ee
From now on we forget the discretized notations and return to the infrared
continuous time notation in which the heat equation
reads
\be \frac{d}{dt} q_{t} = - \cL q_t .\label{heat1}
\ee
\begin{lemma}
\label{lem:HKdecreasing}
Consider the function $\psi_x(t)= \langle q_{t,x}^2\rangle_2 = q_{2t}(x,x)$.
The $r$-th time derivatives $\phi_r = (-1)^r\psi^{(r)}$ are all positive monotone decreasing. 
\end{lemma}
\proof
The heat equation \eqref{heat1} means by induction that 
\be\phi_r =2^r \langle q_{t,x}, \cL^r q_{t,x}\rangle_2 \ge 0.
\ee
\qed\medskip
\begin{cor}
\be \langle q_{t,x}, \cL^r q_{t,x}\rangle_2 \le c_r q_{c'_rt}(x,x) t^{-r}. \label{corpsi}
\ee
\end{cor}
\proof
For any $r$ since $\phi_r$ is positive monotone decreasing, we have 
\be\phi_r (t) \le \frac{2}{t}\int_{\frac{t}{2}}^t \phi_r (s) ds
= \frac{2}{t} [\phi_{r-1} \left(\frac{t}{2}\right) - \phi_{r-1} (t) ]\le\frac{2}{t} \phi_{r-1} \left(\frac{t}{2}\right)
\ee
so that \eqref{corpsi} follows by induction with 
$c_r = 2^{r (r+1)/2}$ and $c'_r = 2^{1-r}$.
\qed
\medskip

Local transport up to $p$-th order of the function $f$ from 
point $z$ to $y$ is then defined as 
\begin{align} f(z) =& \Big[ \bar f + \overline{\cL f}
+\overline{\cL^2 f} + \cdots + \overline{\cL^p f}\Big](y)\\
&+ \Delta_{yz}\Big[\bar f + \overline{\cL f}
+ \overline{\cL^2 f} 
+ \cdots + \overline{\cL^p f}
\Big] + \cL^{p+1}f(z)
\end{align}
where $\Delta_{yz} g := g(z) - g(y)$.
Each difference term is then evaluated in the case
$f=q_{t,x}$ as
\begin{align} \vert \Delta_{yz}\overline{\cL^r q_{t,x}}\vert 
&\le \sum_{u\sim y \atop v \sim z}
\vert \cL^r q_{t,x}(u) -\cL^r q_{t,x} (v) \vert \\
&\le c_r \sqrt{d(y,z) \cE(\cL^r q_{t,x}, \cL^r q_{t,x}) }\\
&\le c_r\sqrt{d(y,z) q_{c'_rt}(x,x)} t^{-r-1/2}
\end{align}
and the last term $ \cL^{p+1}f(z)$ is a finite sum of differences of the type $\cL^{p}_{\cdot}q_{t,x}(z)- \cL^{p}_{\cdot}q_{t,x}(u)$
for $u$ close to $z$. It does not need to be transported, since again
\be \vert \cL^{p}q_{t,x}(z)- \cL^{p}q_{t,x}(u) \vert
\le c_p \sqrt{d(z,u)q_{c'_pt}(x,x)} t^{-p-1/2} .
\ee
The constants in these equation may 
grow very fast with $p$, but renormalization shall
require such bounds only up to a very small order $p$, typically two.

Applying now the usual probabilistic estimates in the manner of the previous section means that the $\sqrt{q_{c't}(x,x)}$ averages to a $c M^{-2j/3}$ factor
uniformly for $t_j \in I_{j}$. Therefore we have the following analogs of Lemma \ref{egainlemma}:
\begin{cor}\label{egainlemma2}
There exists some constant $c_r$ such that uniformly for $t_j \in I_{j}$
\begin{align} \mE [\vert \Delta_{yz}\overline{\cL^r q_{t,x}}
\vert ]&\le c_r M^{-2j/3} M^{-(2r+1)j } \sqrt{d(y,z)}, \label{egain3}
\\ \mE [\vert \Delta_{yz}\overline{\cL^r C_j^T (x,z)}\vert ]
&\le  c_r M^{-(2r+1)j } \sqrt{d(y,z)}, \label{egain4}
\\
 \mE [\vert \cL^{p+1} C_j^T (x,z) \vert]
&\le c_p M^{-(2p+1)j } .
\label{egain5}
\end{align}
\end{cor}
These bounds coincide with those of Lemma \ref{egainlemma} for $r=0$ but improve rapidly with $r$. They should be useful for further renormalization, such as the one of the more divergent 2-point function. 
In the $\phi^4$ model above, since our propagator is a fractional power of the Laplacian, 
the corresponding ``wave function renormalization" is not the standard one of the Laplacian. 
Moreover, physics is not directly associated to perturbative renormalization 
but rather to renormalization group flows, which require the
computation of beta functions that are model dependent. 
For all these reasons we shall not push further the study of the scalar $\phi^4$ model here.

\section{Field Theory on Unicycles}
\label{section:unicycle}
We now begin the investigation of a quantum field on unicycles dressed by Galton-Watson trees, the unicycle being the compactified spine encountered above. We first define our unicycle ensemble, then precise how should be set an interacting Fermionic model on it. Nevertheless, we would need extra imputs to write a random walk expansion, which would differ from our preceding discussion. Hence, we turn to a Bosonic version of it and look to solve for a melonic Schwinger-Dyson equation. 
\subsection{Unicycles}
\label{unicycles}

The cycle $\cC_\ell$ of length $\ell$ is the connected graph with $\ell$ vertices and $\ell$ edges forming a single circuit.
Unicyclic graphs $\Gamma$ are very mild modifications of trees. Instead of having \emph{no} cycle  they have a \emph{single cycle} $\cC(\Gamma) $. 
They can therefore be embedded on the sphere as planar graphs with \emph{two faces} 
(recall that trees have a \emph{single} ``external" face). 
The order $n= \vert \Gamma \vert$ of a unicycle $\Gamma$ 
is still defined as its total number of edges 
which is also its total number of vertices. 
Another important integer for a unicycle $\Gamma$ is its length $\ell  \ge 1$ which is defined as the
length of  $\cC(\Gamma) $. Hence $\ell \le n$.

\begin{figure}
\centerline{\includegraphics[width=8cm]{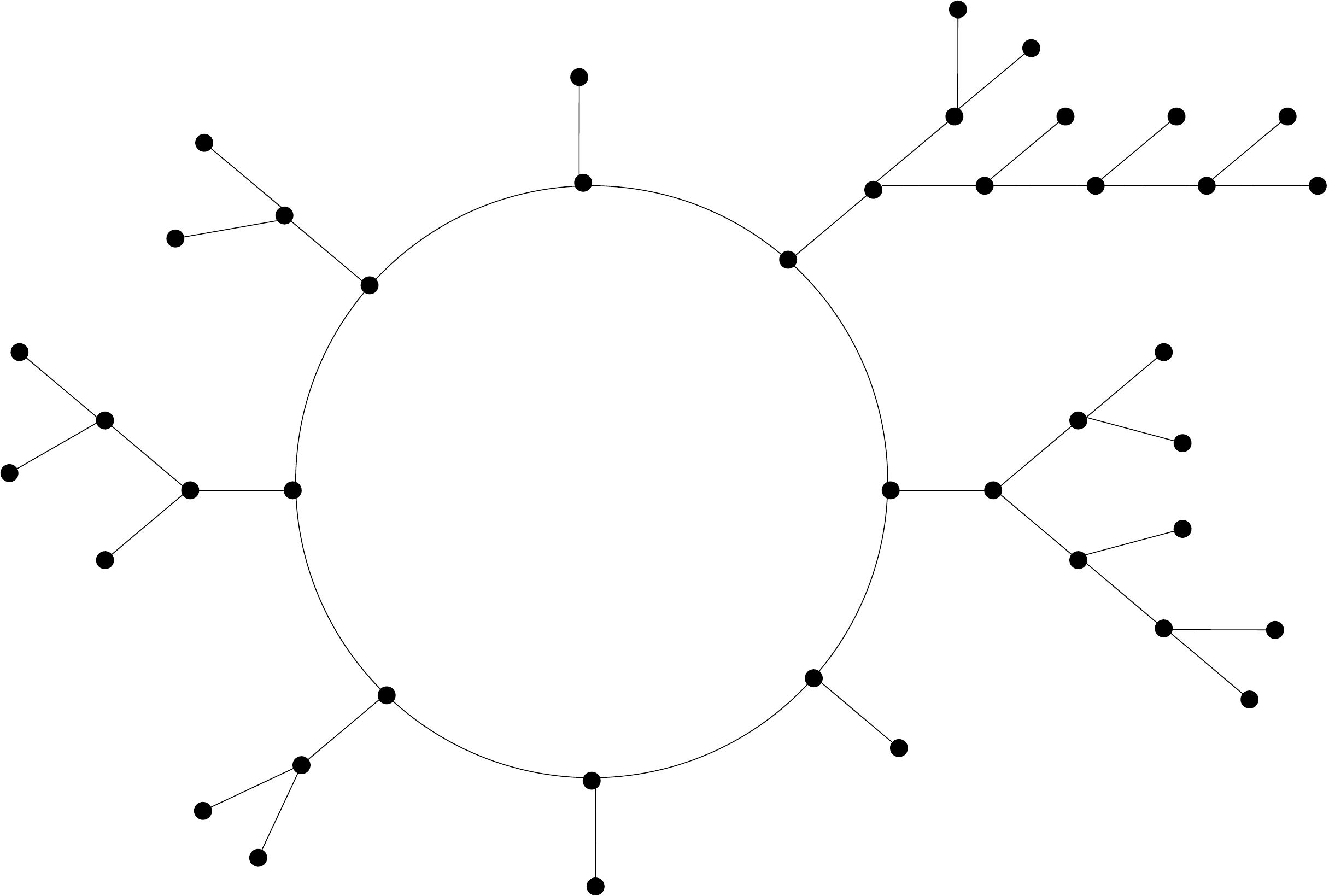}}
\caption{This unicycle of length $\ell=8$ and order $n=42$ is binary, every vertex has degree either 3 or 1.}
\label{unicyc}
\end{figure}
 
For simplicity we choose an orientation of the cycle and we orient every decorating tree from leaves to root, hence we 
can consider $\Gamma$ as an oriented graph or digraph.
Again for simplicity we shall restrict to the \emph{binary case}. It means 
that we shall consider unicycles whose vertices have degree either 1 or 3. 
All vertices of a cycle have degree $\ge 2$, hence in the binary case they must have degree three.
A binary unicycle of length $\ell$ is therefore 
characterized by the set of $\ell$ cyclically ordered rooted trees $T_0, \cdots , T_{\ell -1}$
attached to it.  It really means that the set $\cU_\ell$ of
unicyclic graphs of length $\ell$ can be identified to the set $[ \prod_{k=0}^{\ell -1} \cT_k]/{\rm Cyc}(\ell)$ of 
$\ell$ identical copies $\cT_k$ of the set of rooted binary trees $\cT$,
quotiented by the group ${\rm Cyc}(\ell)$ of cyclic permutations of $\{0, \cdots , \ell-1\}$.

Calling $C= \{t_0, \cdots t_{\ell-1} \}$ the set of vertices of the cycle,
since each $t_k$ is of degree 3, there is a set $\{t'_0, \cdots t'_{\ell-1}\}$ 
of tree vertices, each $t'_k$ being joined to $t_k$ by a single
edge not belonging to the cycle; therefore the order of 
a binary unicycle of length $\ell$ is at least $2\ell$ (see Figure \ref{unicyc}).

This characterization of binary unicycles suggests to define a probability measure for finite unicycles whose 
infinite order limit is closely related to the previous class $\cT_\infty$ of infinite binary random trees. \footnote{Of course we expect that this limit is universal i.e. would be the same for $p$-ary random unicycles, but our emphasis here is not on this point.}
The  cycle of the unicycle should be thought of as a finite analog of the \emph{spine} of an infinite tree in $\cT_\infty$.
Each $s_k$ is indeed the root of a binary Galton Watson tree
$T_k$, and we can equip these trees with independent probability measures $2^{- \vert T_k \vert}$.

We can then in the same vein as earlier define a measure $d\nu_\ell$ over the set $\cU_\ell$ of
unicyclic graphs of length $\ell$. It is just the product of independent 
Galton-Watson critical measures over the  attached trees $T_k$:
\be  d \nu_\ell = \prod_{k=0}^{\ell -1}  \mu (T_k ).
\ee
We shall not consider directly the limit $ \lim_{\ell \to \infty} d \nu_\ell$ since it is delicate to define
an analog of the \emph{infinite} spine \emph{with a periodic boundary condition}. Instead we can work 
with asymptotics of expectation values for $ d \nu_\ell$ as $\ell \to \infty$, just like
a thermodynamic or infrared limit can be defined as the large size limit of \emph{finite size} partition functions 
and correlation functions.

In the next section we shall therefore define the SYK model on unicycles in $\cU_\ell$ with finite $\ell$, 
define their correlation functions averaged over $d \nu_\ell$
and we shall study the infrared limit of these correlations when $\ell \to \infty$.
The cycle in the unicycle is the analog of
the lattice-regularized Euclidean time at a certain non-zero temperature. The trees of the unicycle
introduce the new ``random gravity" aspects of this Euclidean time.

\subsection{Lattice-regularized SYK}
\label{Fermions}

Consider a graph $\Gamma_0$ of length $\ell$ and order $n=\ell$, that is without any decoration by trees.
We want to define the {\it Fermionic} SYK model on $\Gamma_0$. 
For the moment consider a one component Majorana Fermion $\psi$, the generalization to $N$
components being straightforward.

The $U(1)$ Euclidean circle of length $\beta$
is replaced by an oriented finite cycle $C_\ell = \{t_0 , \cdots t_{\ell -1} \}$ with  $t_k = ak$
and $\ell = \beta/a$. The ultraviolet limit $a \to 0$
and infrared limit $\beta \to \infty$ then both imply $\ell \to \infty$, by keeping constant the perimeter of the circle.

We have first to implement the antiperiodic boundary conditions. Antiperiodicity on the lattice means that 
$\psi$ is in fact periodic but with period $2 \beta$ instead of $\beta$ hence should be analyzed in terms of 
$2 \ell$ frequencies $ \frac{\pi q}{\beta}  =  \frac{\pi q }{a \ell}  $ with $q = 1 , \cdots 2 \ell$, but that 
only odd Matsubara frequencies $\omega_p =(2p+1) \frac{\pi }{\beta} $ for $p = 0, \cdots \ell -1$ contribute.
The discrete Fourier transforms are defined as
\be \hat \psi (\omega_p) = 
\frac{1}{\sqrt \ell} \sum_{k=0}^{\ell -1} e^{- i \omega_p \cdot t_k} \psi (t_k) = \frac{1}{\sqrt \ell} \sum_{k=0}^{\ell -1}
e^{- \frac{(2p+1)i \pi  k}{\ell} }  \psi (ak) .
\ee
Remark that 
\be \hat \psi (\omega_{\ell -p -1}) = \overline{\hat \psi (\omega_p)} \label{compconj}
\ee
and that the inverse Fourier law gives antiperiodic fields of antiperiod $\beta$
\be
\psi (t_k) =  \frac{1}{\sqrt \ell} \sum_{p=0}^{\ell -1}  
e^{i\omega_p \cdot t_k} \hat \psi ( \omega_p) = 
\frac{1}{\sqrt \ell} \sum_{p=0}^{\ell -1} e^{\frac{(2p+1) i \pi  k}{\ell} } \hat \psi \left( \frac{(2p+1)\pi }{\beta}  \right) = - \psi (t_k + \beta) .
\ee

The action should be discretized in the usual way, that is turning each derivative at $t_k$ into
a lattice difference 
\be \frac{1}{a} [ \psi (ka +a) -\psi (ka) ] =  \frac{1}{a}  [ \psi (t_{k+1}) -\psi (t_k) ] 
\ee and integrals such as $\int dt f(t)$
into Riemann lattice sums $ a \sum_{k=0}^{\ell -1} f(t_k)$.
To discretize the quadratic part $ \frac{i}{2}   \int dt \sum_{i=1}^N \psi_i (t) \frac{d}{dt}\psi_i (t)   $ of the SYK action one
should take  into account the anticommutation of Fermions.
Factoring out $\frac{i}{2}$ leads to consider the quadratic form  
\be Q_{lat} (\psi) = \,\left[ \,  
\sum_{k=0}^{\ell -2} \psi (t_k)  \psi (t_{k+1})   - \psi (t_{\ell-1}) \psi ( t_0) \right]  , \label{SYKlataction}
\ee 
where the last term is subtle: because of  antiperiodicity, $\psi(t_\ell)$ should be identified with $ - \psi (t_0)$.
The total interacting SYK lattice action is therefore
\be I_{lat} =\sum_{i=1}^N  \frac{i}{2}  Q_{lat} (\psi_i) -a i^{q/2}  \sum_{k=0}^{\ell -1}
\sum_{1 \le i_1 <  \cdots < i_q \le N}	J_{i_1, \cdots , i_q}	\psi_{i_1}(t_k) \cdots \psi_{i_q} (t_k) .
\ee
Remark that in order to have a non zero normalization (for $q$ even) the total number $\ell$
should be even, a condition which we assume from now on.

We can rewrite the quadratic (free) action in Fourier space. Forgetting the trivial $i$ index, it means
\begin{align} Q_{lat} (\psi) &= \Bigl[ \,  
\sum_{k=0}^{\ell -2} \psi_i (t_k)  \psi (t_{k+1})   - \psi (t_{\ell-1}) \psi( t_0) \Bigr]  \\
&= \frac{1}{\ell} \sum_{p=0}^{\ell -1}  \sum_{q=0}^{\ell -1} 
\biggl[\sum_{k=0}^{\ell -2}  e^{\frac{i \pi }{\ell} [(2p+1)k +(2q +1) (k+1)]} 
 -  e^{\frac{i \pi }{\ell} [(2p+1) (\ell -1) }  \Biggr]
 \hat \psi (\omega_p ) \hat \psi (\omega_q ) \\
 &= \frac{1}{\ell} \sum_{p=0}^{\ell -1}  \sum_{q=0}^{\ell -1} 
\biggl[ e^{\frac{i \pi (2q+1)}{\ell}}  \sum_{k=0}^{\ell -1}  e^{\frac{2i \pi }{\ell} (p+q +1) k} \Biggr]
 \hat \psi (\omega_p ) \hat \psi (\omega_q )\\
 &=  \sum_{p=0}^{\ell -1} 
\biggl[ e^{- \frac{i \pi ( 2p +1)}{\ell}} -1   \Biggr]
 \hat \psi (\omega_p )  \overline{\hat \psi (\omega_p )} \;.
\end{align} 
In the last line, we took advantage of the fact that the sum over $k$ gives zero
unless $e^{\frac{2i \pi }{\ell} (p+q +1)} =1$ hence $p+q +1 = \ell$.
We also added the $-1$ because it is the Fourier transform of a mass term 
hence is zero and we used \eqref{compconj} .

Remark  that the factor $e^{\frac{-i \pi ( 2p +1)}{\ell}}  - 1$ is never zero and behaves as $-i \pi (2p+1)$ for small $p$. Hence
the  free lattice propagator is invertible and approximates at small $p$ 
the inverse of the continuous free propagator, hence the 
inverse of the Matsubara frequency.

Consider now any unicycle $\Gamma$ with decorating trees
oriented from leaves to the cycle. We want to define the Fermionic SYK model on $\Gamma$. 

We impose two conditions. First we want the free action of the $N$-component Majorana field $\psi$ to be a quadratic form 
$\frac{i}{2}  Q_\Gamma (\psi)$
with a good non-zero normalization. Second we want to impose
the anti-periodic boundary conditions along the spine to
coincide with the ones of the ordinary free SYK model on the trivial unicycle $\Gamma_0$ (not decorated by the trees).

A naive quadratic form would couple a Fermion on each vertex to its nearest neighbours. However it does not work, since as soon 
as a single branch of $\Gamma$ is non trivial, the corresponding free theory has zero normalization. 
Indeed in this case the tree 
has a non trivial terminal branch with two leaves $s_1$ and $s_2$ related to a node $s_3$, and the Grassmann integral contains a term such as  $\int d\psi (s_1 ) d \psi(s_2) 
d\psi(s_3)   e^{\psi(s_1) \psi(s_3)  + \psi(s_2) \psi(s_3)}$, which is zero.

In fact our two conditions lead to the same conclusion, namely that we need some kind of Fermion doubling. The SYK model on $\Gamma$ requires  a single $N$-component Fermion variable not only for the $n$ vertices of $\Gamma$
but also for the $n$ edges of $\Gamma$. 
With this convention we can define
\be Q_{\Gamma} (\psi) = \,\Bigl[ \,  
\sum_{e,v} \epsilon_{ev} \psi (e)  \psi (v)  \Bigr]  .\label{SYKlataction}
\ee 

To implement the anti-periodicity we fix a particular  root vertex $v_0$ on the spine  and
we reverse the last half-arrow $e_\ell $ into that vertex. 

\begin{lemma}
The normalization  of the free theory at any unicycle $\Gamma$, $Z_{\Gamma}$, is $2^{N\ell}$, so that 
\begin{align}
Z_{\Gamma}^{-1} \int e^{Q_{\Gamma} (\psi)} \prod d \psi = 1.
\end{align}
\end{lemma}
\proof First treat the spine, then continue by induction, adding a leaf. \qed

There are really $2\ell$ Fermions on the spine, $\ell$ ones for the vertices and $\ell$ ones for the edges. 
This allows to automatically implement the evenness condition, hence we can use the normalized quadratic form $Q_C$
in the form
\be Q_{lat} (\psi) = \,\Bigl[ \,  
\sum_{k=0}^{\ell -2} \psi (t_k)  \psi (t_{k+1})   - \psi (t_{\ell-1}) \psi ( t_0) \Bigr]  . \label{SYKlataction}
\ee 

Let us note that this is the simplest way to introduce Fermions on a graph, the general case taking the Dirac operator as acting on the cliques of the graph \cite{Knill(2013)}. \footnote{An $i$-clique being a subset of $i$ vertices of the graph, such that any two vertices in the set are adjacent in the graph. See also \cite{Gubser:2018cha} for a similar definition of Fermions on a tree graph in the context of p-adic AdS/CFT.}

\subsection{Bosons}
\label{Bosons}

For the moment let us focus on the two-point function for Bosonic version of SYK. The bare SYK model on 
$\Gamma$ can be defined by the discretized action
\be I_\Gamma = \sum_{u \in V(\Gamma)} \biggl[ \frac{1}{2}\phi (u)  ( L_\Gamma + m^2 \bbbone) \phi (u) 
-\frac{i^{q/2}}{q!} \sum_{1 \le i_1 <  \cdots < i_q \le N}	J_{i_1 \cdots  i_q}	\phi_{i_1}(u) \cdots \phi_{i_q} (u) \biggr]\;, 
\ee
where $L_\Gamma$ is the lattice Laplacian on $\Gamma$ and the disordered coupling has second moment:
\be
\expval{J_{i_1 \cdots  i_q}^2} = \frac{(q-1)!J^2}{N^{q-1}}. 
\ee 
In the $N \to \infty$ limit we get a self-consistent melonic equation for the two-point function $G^{mel} (x,y)$
\be 
[G^{mel} (x,y)]^{-1}  = - [G_0^{mel} (x,y) ]^{-1} - \int \dd \tau^\prime J^2 [G^{mel}(x,y)]^{q-1} \;, 
\label{eq:convolution1}
\ee
which simplifies in the infrared limit \cite{Gurau:2017qna}
into the convolution equation:
\be 
\sum_{z \in \Gamma}  J^2 G_{ir}^{mel}(x,z)  [G_{ir}^{mel}(z,y)]^{q-1}= -\delta (x,y).
\label{eq:convolution2}
\ee
We shall now \emph{assume} that the effective infrared two point function $G^{mel}_{ir} (x,y)$  is 
asymptotic to  an $\alpha$-regularized propagator on $\Gamma$, namely
$ G_\Gamma^{\alpha} (x, y) = 2\int_0^\infty  m^{-2\alpha + 1}  G_\Gamma^m (x, y) dm $. We shall average
over unicycles $\Gamma$ of given length $\ell $ and search for the right value of $\alpha$
to fulfill \eqref{eq:convolution2}. Since we average over different $\Gamma$'s but which share all the same 
cycle of length $\ell$ it makes sense to consider $x$ and $y$ in that cycle, but the intermediate point $z$ 
can be anywhere on $\Gamma$, including in the decorating trees.

So we are searching for the value of $\alpha $ such that for $x, y \in \cC_\ell$ the equation
\be 
\expval{\sum_{z \in \Gamma}  J^2 G_\Gamma^{\alpha} (x, z)  [G_\Gamma^{\alpha} (z, y) ]^{q-1}}_{\ell} = -\delta (x,y) 
\label{eq:convolution3}
\ee
holds. In this discrete setting, the right hand side is a Kronecker delta. Consequently we take $x \simeq y$ on the left and the Kruskal tree is made of a single propagator connecting $x$ to $z$.
The average $\expval{.}_\ell$ means  averaging with $d\nu_\ell$, hence over all unicycles of length $\ell$ with independent 
critical binary Galton-Watson 
measure on the trees decorating the cycle.  
 The $d_{sp}$-dimensional $\delta$ function has a representation as:
 \begin{equation}
 \label{eq:deltarep}
     \delta (x,y) = \lim_{\epsilon \to 0} \frac{\Gamma(\frac{d_{sp}}{2} - \epsilon)}{2^{2\epsilon}\pi^{d/2}\Gamma(\epsilon)}\frac{1}{\abs{x-y}^{d_{sp}-2\epsilon}}.
 \end{equation}
 For the scope of these notes, as we remain cavalier with the constants, we will for now only track down the space scaling read from the above equation.  

On a fixed graph $\Gamma$ it is convenient to express the regulated two-point function as 
\begin{equation}
G^{\alpha}(x,y)= \int_0^\infty dt t^{\alpha-1} q_t(x,y)\;.
\label{eq:integrated-kernel}
\end{equation}
Then the multiscale analysis is especially interesting. 
According to the earlier Lemma~\ref{oneline}, it results into
\be   \mE \left[ \sum_z G_T^{\alpha , j} (x,z) \right] \simeq M^{\frac{4j}{3}} M^{-\frac{4j}{3}+2\alpha j} ,
\label{singleline1}
\ee
and the tadpoles count each for 
\be   \mE \left[  G_T^{\alpha , j} (x,x) \right] \simeq  M^{-\frac{4j}{3}+2\alpha j} .
\label{singleline2}
\ee
Remembering eq.~\eqref{agmu3} in the proof of Theorem~\ref{theoconv}, which can be framed as
\be \mE  \left[ \left(\sum_z G_T^{\alpha , j} (x,z) \right) (  G_T^{\alpha , j} (x,z) )^{q-1} \right]  \simeq \mE \left[  G_T^{\alpha , j} (x,z) \right]\mE \left[ \sum_z G_T^{\alpha , j} (x,x) \right] ^{q-1} \;,
\label{eq:factorization}
\ee
and collecting all factors,
we have
\be   \mE \left[ \left(\sum_z G_T^{\alpha , j} (x,z) \right) (  G_T^{\alpha , j} (x,x) )^{q-1} \right] \simeq  M^{-\frac{4(q-1)j}{3} +2q\alpha j} .
\label{singleline3}
\ee
Returning to eq.~\eqref{eq:convolution3}, we find
\be   M^{-\frac{4(q-1)j}{3} +2q\alpha j} \simeq   M^{-\frac{4j}{3}}  \implies \alpha = \frac{2}{3}  - \frac{4}{3q}.
\label{singleline4}
\ee
Let us note that the scaling associated to the Kronecker delta that selects the critical value of $\alpha$ making the $q$-th order interaction marginal, must correspond to the spectral dimension (here $4/3$). 

The effective infrared propagator, Fourier transformed on the spine, then behaves on the unicycle as
\be
\expval{G^{\alpha}(p) }_\ell \simeq p^{-2 \alpha} = p^{-\frac{4}{3} +  \frac{8}{3q}} ,
\ee
hence very differently from the flat deterministic case $p^{-1+2/q}$. An interesting exercise might be, in order to better control eq.~\eqref{eq:factorization} to reproduce this scaling from the convolution of propagators \eqref{eq:integrated-kernel}, using the quenched estimates for $q_t$ and decomposing the walks until the point $z$ inside the branches as walks inside the branches, plus walks on the unicycle, similarly to the ideas of \cite{Kesten}, except that a single random walk was considered, whereas we are looking at the intersection of $q$ distinct ones.

\begin{subappendices}
\section{An Excursion inside Probability Theory}
\label{app:proba}
This appendix compiles many useful results, well-known to probabilists, exploited in this chapter. We emphasize that none of them are ours, but we think that presenting them can be helpful to understand how such derivations are obtained. 
First we will discuss generic estimates of number of vertices at and below given heights for the Galton-Watson processes that constituted our underlying geometry preceedingly, determining the typical volumes that occur. We will then switch to on-diagonal and off-diagonal heat-kernel bounds on fixed graphs. Aiming at describing random graphs, we will subsequently specify conditions under which those estimates, quenched or annealed, will hold with sufficiently high probability, and comment on their application to the Galton-Watson trees. The main reference for this part is \cite{BarlowKumagai} and for the second, we relied on \cite{Kumagai,BarlowKumagai}. \footnote{We also refer to \cite{KumProceedings}for a short although very comprehensive picture for other inequalities equivalent to heat-kernel bounds -- such as Parabolic Harnack, and the important questions in the field of propagation on random media.} Since all detailed proofs can be found in the references, we will stay casual about them.

\vspace{5mm}
Before we start, let us pull out the Borel-Cantelli lemma, in the toolkit of every probabilist \cite{Feller}. 
\begin{lemma}
[Borel-Cantelli] For a sequence of events $\{E_n\}_{n\geq 0}$ whose probability converges 
\be \sum_{n\geq 0}\mP(E_n)<\infty\;, \ee the probability that an infinite subset of events occurs is zero
\be
\mP\left(\lim_{n\to \infty}\bigcap_{n\geq 0}\bigcup_{k\geq n} E_k\right) =0\;.
\ee
\end{lemma} We also recall the use of denoting $c_i$ for appropriate $\cO(1)$ constants, and if they change when varying some parameter, say $\l$, they will be denoted by $c_i(\l)$. 

\subsection{Aspects of random trees and branching processes}
We define the branching process $\{X_{n\geq 0}\}$ with root $X_0=1$, and critical offspring distribution $Bin(n_0,1/n_0)$, $n_0\geq 2$, such that each vertex has mean number of children $m=1$. We also take the random variable $Y_n=\sum_{k=0}^nX_k$, counting all vertices until the height $n$. Two generating functions of the momenta are useful to consider: 
\be
f_n(s)=\mE[s^{X_n}]\;, \quad g_n(s) = \mE[s^{Y_n}]\;.
\ee
They both obey recursive relations
\be
\label{eq:recursion_genfun}
f_{n+1}(s) = f_n(f(s))\;, \quad g_{n+1}(s) = sf(g_n(s))\;, 
\ee
denoting $f(s)$ for $f_1(s)$.
The first equation comes from conditioning on the preceding generation and using that descendants are identically distributing: 
\begin{align}
f_{n+1}(s) &= \mE[\mE[s^{X_{n+1}}|X_n]] \\
&= \mE\left[\prod_{i=1}^{X_n}\mE[s^{X_1}]\right] = \mE[f(s)^{X_n}]\;.
\end{align}
The second comes by decomposing on all possible combinations leading to the total population, using again that descendants are i.i.d. :
\begin{align}
g_{n+1}(s) &= \mE [s^{X_0+\dots X_n}\mE[s^{X_{n+1}}|X_n]]\\
 &= \mE [s^{X_0+\dots X_n}f(s)^{X_n}]\;,
\end{align}
and proceeding by recursion.
Kesten, Ney, Spitzer \cite{KestenNeySpitzer} proved that asymptotically the non-extinction probability is driven by the variance (if finite) of the critical branching process
\be
\label{eq:kesten}
\lim_{n\to\infty}n\mP[X_n>0] =2/f''(1)\;.
\ee
Denoting, for an integer $k$, $X[k]=\sum_{i=1}^kX_i$, where the variables $X_i$ are identically distributed with the random variable $X$, we can estimate the probability that the number of vertices at and below a certain height $n$ doesn't exceed respectively the height and the height square:
\begin{lemma} For all $\l>0$, there are constants $c_1, c_2>0$ such that
\begin{gather}
\mP[X_n[n]\geq \l n]\leq c_1 e^{-\l/6}\;,\\
\mP[Y_n[n]\geq \l n^2]\leq c_2 e^{-\l/5}\;.
\label{eq:lowbound_yn}
\end{gather} 
\end{lemma}
\vspace{2mm}
Those are consequences of two inequalities
\begin{align}
&\log f_n(e^\theta)\leq \theta + 2n\theta^2\;, \quad \theta \in (0,1/6n]\;,\\
&\log g_n(e^\theta)\leq (1+\a n)\theta\;, \quad  \a\in (1,2], \theta \in \left[0,\f{\a-1}{(1+\a n)^2}\right]\;,
\end{align}
coming from the recusive relations \eqref{eq:recursion_genfun} and of the Chernoff bound saying that for a random variable $X$ and all $t>0$, $P[X\geq a]\leq E[e^{tX}]e^{-ta}$. More precisely, decomposing the branching process into those that go extinct before height $n/2$ and those that survive after, relying on the estimate~\eqref{eq:kesten} and on the backwards Chebyshev inequality $P(\x\geq E[\x]/2)\geq E[\x]^2/4E[\x^2]$, Barlow and Kumagai obtain a finer control on the number of vertices of the tree up to height $n$:
\begin{lemma}
\label{lemma:bounds_yn}
\begin{itemize}
\item[(1)] There are constants $c_0,p_0>0$ such that
\be
\mP[Y_n>c_0n^2] \geq p_0/n\;.
\ee
\item[(2)] For random variables $\eta_n$ with distributions $Bin(n,p_0/n)$, then for all $\l>0$
\be
\mP[Y_n[n]\geq \l n^2] \geq \mP[c_0\eta_n\geq \l]\;.
\ee
\item[(3)] For $0<\l<1$ and $n\geq c_1/\l$ 
\be
\label{eq:upbound_yn}
\exp(-c_2/\l)\leq \mP[Y_n[n]\leq \l n^2]\leq \exp(-c_3/\l^{1/2}) \;.
\ee
\end{itemize}
\end{lemma}
\vspace{2mm}
The last upper bound follows from the dominance (2) on the binomial variable $\eta_n$, \footnote{Notice how the lower bound metamorphoses into an upper bound.} while to show the lower bound, it is enough to condition on the trees that have stoped before a height $m=n/k$, for $k\geq 1$.
Finally in order to estimate the average volume of a ball $B(x,r)$ around any point $x$ on the graph, one takes the size-baised processes of the off-the-spine branches $\tilde{X}_{n\geq 0}$ and analogous $\tilde{Y}_n=\sum_{k=1}^n\tilde{X}_k$, such that $\tilde{X}_0=1$ on the spine, $\tilde{X}_1$ distributed as $Bin(n_0-1, 1/n_0)$, (since the spine is growing in another direction) and $\tilde{X}_n$ distributed with $Bin(n_0, 1/n_0)$ as before. Entirely similar bounds as those of Lemma~\ref{lemma:bounds_yn} were obtained for the sized-baised $\tilde{Y}_n$. The point of this was to be able to decompose any ball $B(x,r)$ as three types of processes $\tilde{X}$ growing on the geodesic from $x$ to the root of the graph, off the geodesic but still on the spine and descendants of $x$ after the geodesic. For the three cases, the largest volume is obtained from at most $r$ independent processes. This drives to the following: 
\begin{theorem} For $\l>0, r\geq 1$, there are constants $c_0, c_1, c_2, c_3>0$
\begin{gather} 
\mP(V(x,r)>\l r^2) \leq c_0 \exp (-c_1\l)\;,\\
\mP(V(x,r)<\l r^2) \leq c_2 \exp (-c_3/\sqrt{\l})\;.
\end{gather}
\end{theorem}
To be complete, we will also have to estimate the minimal number of ``gates" at distance $r$ from which to escape a ball $B(x,r)$, or more precisely, $M(x,r)$ being the smallest integer $m$ such that for $A=\{z_1, \dots z_m\}$ with the points $z_i$ at distance $d(x,z_i)\in [r/4,3r/4]$, any geodesic from $x$ to $B(x,r)^c$ passes through $A$. 
\begin{theorem}
There are constants $c_1,c_2>0$ such that for each $r\geq 1$ and $x\in X$
\be
\mP(M(x,r)\geq m) \leq c_1\exp (-c_2 m)\;.
\ee
\end{theorem}
\noindent In order to show this, the idea remains similar to the preceding theorem. That is to count descendants and ancestors of $x$ outwards and towards the spine at the distance $r/4$, themselves still having descendants at distance $r/4$. From the estimate~\eqref{eq:kesten}, such vertices $z_i$ have distributions $Ber(p_r)$ with $p_r\leq c_0/r$. By rewriting the distribution of the total number of vertices $z_i$ as a martingale (plus a controlled correction), \cite{BarlowKumagai} could conclude.

Since all conditions for the $\l$-goodness of the ball $B(x,r)$, eq.~\eqref{goodvol}, are assured with probabilities exponential in $\l$, leading to the inequality~\eqref{smallproba1}.

\subsection{Heat-kernel bounds on random graphs}
Here, we want to share a broader view on determination of heat-kernel estimates on random graphs, that hopefully can be useful for pursuing the application of quantum field theoretic techniques on more generic random graphs than trees. We recall that the heat-kernel $q_t(x,y)$ gives the probability that a random walker starting at $x$ reaches $y$ in a time $t$. The goal is to give conditions under which one can say that the heat-kernel behaves as
\be
\label{eq:HK}
q_t(x,y) \sim c^\prime t^{-d_H/d_w} \exp\left[-c\left(\f{d(x,y)^{d_w}}{t}\right)^{1/(d_w-1)} \right]\;, 
\ee
in the sense of providing upper and lower bounds for a certain range of time $t$ and for points $x,y$ separated by distance $d(x,y)$, where $c,c^\prime$ are inessential constants. 
There are two dimensions entering the game. The \emph{Hausdorff} dimension $d_H$ which tells about the number of points in a ball of given radius that $x$ can reach and the \emph{walking} dimension $d_w$ that tells the time it takes to escape the ball. Compared to a Brownian random walk on $\mathbb{R}^d$, if $d_w>2$, then the walk is said subdiffusive. 
We see that the \emph{spectral} dimension, setting $y=x$, corresponds to $d_s = 2d_H/d_w$. When the probability to return at the starting point in the limit of infinite time doesn't vanish, the random walk is called \emph{recurrent}. Otherwise it is said \emph{transient}. A classic result of P\'{o}lya \cite{Polya:1921} states that:
\begin{theorem}[P\'{o}lya]
A random walk on $\mathbb{Z}^d$ is recurrent for $d=1,2$ and transient for $d\geq 3$. 
\end{theorem}

\subsubsection{Heat-kernels on graphs}
We first set a few definitions. We start with a weighted graph $(X,\mu)$, where weights $\mu_{xy}$ are assigned to all edges $xy$. Two adjacent vertices $x, y$ will be written $x\sim y$. 
The graph will be taken infinite, \footnote{Otherwise its spectral dimension vanishes.} locally finite, connected and having a marked vertex, the root. Also, we assume a metric function $d$, such that for a ball $B(x,r)=\{y\in X: d(x,y)\leq r\}$, we will write $V(x,r)= V(B(x,r))= \sum_{y,z\in B(x,r)}\mu_{yz}$ for the analog of the volume of the ball. $B^c$ will mean the complement of $B$. And as in propositional logic, we will write $a\wedge b$ or $a\lor b$ for the mininum or maximum between the values $a$ and $b$. 

We introduce the energy function
\be
\cE(f,g)= \f{1}{2}\sum_{x,y\in X:x\sim y} (f(x)-f(y))(g(x)-g(y))\mu_{xy}, 
\ee
as well as the space $H^2:=\{f:\cE(f,f)<\infty\}$.
We define the effective resistance between two points 
\be 
R_{eff}(x,y)^{-1}:= \underset{f\in H^2}{\inf}\{\cE(f,f):f(x)=1 \text{ and } f(y)=0\}\;.
\ee
The effective resistance presents some nice properties.
\begin{lemma}
\label{lem:resistance}
\begin{itemize}
\item[(1)] If $c_1:= \inf_{x,y\in X}\mu_{xy}>0$, then $R_{eff}(x,y)\leq d(x,y)/c_1$, for all $x,y\in X$.
\item[(2)] If $(X,\mu)$ is a tree and $c_2 := sup_{x,y\in X}\mu_{xy}<\infty$, then $R_{eff}(x,y)\geq d(x,y)/c_2$, for all $x,y\in X$.
\item[(3)] $\abs{f(x)-f(y)}^2\leq R_{eff}(x,y) \cE(f,f)$, for all $x,y\in X$ and $f\in H^2$.
\item[(4)] $R_{eff}(\cdot,\cdot)$ is a metric on $X$. 
\end{itemize}
\end{lemma}
\vspace{2mm}
\noindent Hence, from (1) and (2), the effective resistance between two points of a tree is, uniformely on the graph, proportional to their distance.
For the trees considered above, the effective resistance is linked to the earlier number of ``gates" $M(x,r)$ around $x$ through
\be
R_{eff}(x,B(x,r)^c)\geq c\f{r}{M(x,r)}\;,
\ee
for a constant $c$, which follows straightforwardly from Ohm's laws on electric circuits ($r$ being the length of the wire and $M(x,r)$ giving the minimal number of parallel components). 

\vspace{5mm}
Let us consider a continuous-time random walk\footnote{There are subtleties on going from discrete to continuous time random walks, the first requiring care with respect to the parity of the number of steps and the latter requiring a proper definition of the distribution from which the time of each jump is taken. Holding to \cite{BarlowKumagai}, we take the second one, sampling the jump time from an exponential law of mean one. For more details on the first case, see \cite{Kumagai}.} $Y_{\{t\geq 0\}}$ on $(X,\mu)$, launched at $Y_0=x$, that is \be q_t(x,y)=\mathbb{P}^x(Y_t=y)/\mu_{xy}\;. \ee $\mP^x(A)$ and $\mE^x(A)$ will stand for the probability and expectation of event $A$ conditioned under a random walk starting at $x\in X$. 
We will also need the ``escape time" from the ball $B$,\footnote{The name assumes that $Y_t$ starts inside the ball, but it is not necessary.} $\tau_B= \inf\{t\geq 0 : Y_t\notin B\}$, its counterpart $\sigma_B= \inf\{t\geq 0 : Y_t \in B\}$ and the stricter $\sigma^+_B= \inf\{t >0 : Y_t \in B\}$. 
The green density on a ball $B$ is defined as 
\be
g_B(x,y)= \int_{t} \mP^x(Y_t=y : t<\tau_B)/\mu_y\;,
\ee
restricting over walks from $x$ to $y$ that stay inside $B$ for all times. It obeys the property: 
\be
g_B(x,y)=\mP^x(\sigma_y<\t_b)g_B(y,y)\;.
\label{eq:green_property}
\ee
One can relate the effective resistance with conductance and escape time to the following results, that we do not prove:
\begin{gather}
R_{eff}(B,A)^{-1}= \sum_{x\in B} \mu_x \mP^x(\sigma_A<\sigma_B^+)\;,\quad \text{if $A^c$ is finite,}\\
\mP^x(\sigma_A<\sigma_B)\leq \f{R_{eff}(x,A\cup B)}{R_{eff}(x,A)}\;,\quad \text{if $A^c$ and $B^c$ are finite,}\\
\mE^x[\t_B]\leq R_{eff}(x,B^c)V(B)\;, \quad \text{if $B$ is finite,}
\label{eq:resistance_property}
\end{gather}
giving to the effective resistance its interpretation in terms of ``current" (of probability).

We now have the next crudest possible bounds on the on-diagonal heat-kernel.
\begin{theorem}
(1) If $V(x,r)\geq c_1 r^{d_H}$ for all $x\in X$ and $r\geq 1$, with $d_H\geq 1$, then 
\be
\underset{x\in X}{\sup}~q_t(x,x)\leq c_2 \f{1}{t^{d_H/(d_H+1)}}\;, \quad \forall t\geq 1\;.
\ee
(2) For any ball $B(x,r)$, one also has the lower bound
\be 
q_{2t}(x,x) \geq \f{\mathbb{P}^x(\tau_B>t)^2}{V(x,r)}\;, \quad \forall t>0 \;.
\label{eq:up-bound}
\ee
\end{theorem}
\vspace{2mm}
\noindent The first bound can be seen arising from a Faber-Krahn inequality on the smallest eigenvalue of the Laplace operator outside some set $\Omega\subset X$, cf.~\cite{Kumagai}. The second bound arises from 
\begin{align}
\mP(\tau_B>t)^2&\leq \mP(Y_t \in B)^2 = \left(\int_{y\in B} q_t(x,y)\right)^2\;, \\
&\leq V(B)\int_{y\in B} q_t(x,y)^2\leq V(B)q_{2t}(x,x)\;,
\end{align}
first unconditioning the walk, then Cauchy-Schwarz, then Chapman-Kolmogorov (see also the earlier proof of Th.~\ref{theoconv}).

\vspace{5mm}
However, if one knows more about the connectivity of the graph, a finer upper bound can be constructed, constituting the plinth of what will come. 
\begin{theorem}
Let us consider graphs for which is assumed a bound on the resistance
\be
R_{eff}(x,y) \leq c_* d(x,y)^\a\;, \quad \forall y, r\in \mathbb{N}\;, 
\ee
then the heat-kernel obeys
\be
q_t(x,x)\leq c^\prime \f{1}{t^{d_H/(d_H+\a)}} \left(c_*\vee \f{r^{d_H}}{V(x,r)}\right)\;.
\label{eq:upper-bound-res}
\ee
\end{theorem}
\vspace{2mm}
\noindent The inequality results from defining the functions $f_t(y):=q_{t}(x,y)$ and $\psi(t):=f_{2t}(x)$, which obeys the heat equation (see also below, the remark at eq.~\eqref{eq:off-diagonal}). Then, one can write for any radius $r$ and ball $B(x,r)$
\be 
\psi(t) \leq \f{2}{V(r)}\lor \f{2r^\a}{t}\;, 
\ee
the left term coming from the normalizability of $f_t(y)$ in the ball and the right one from the monotonicity of $\psi(t)$ and eq.~\eqref{eq:off-diagonal}. Taking the radius $r$ such that $t=r^{d_H+\a}$, one concludes.

\vspace{5mm}
Precise definitions vary slightly but this behaviour of the resistance, combined with a lower bound with the same power $\a$, essentially characterizes \emph{strongly-recurrent} graphs \cite{KumagaiMisumi(2008)}. 
We combine the bounds on the volume and the resistance to form the $F_{R,\l}$ condition 
\be
F_{R,\l} = \left\{
\begin{aligned}
V(x,R)\leq \l R^{d_H}~&; \quad R_{eff}(x,z)\leq c_*d(x,z)^\a\;, \hspace{5mm}\forall z\in \overbar{B_R}\\
V(x,\eps_\l R)\geq \f{\left(\eps_\l R\right)^{d_H}}{\l}~&; \quad R_{eff}(x,B_R^c)\geq \f{R^\a}{\l}
\end{aligned}
\right\}
\ee
where $\eps_\l = 1/(3c_*\l)^{1/\a}$.
These assumptions allow to sharpen the lower bound on the heat-kernel. 
\begin{theorem}
\label{th:low-boundFRL}
If $F_{R,\l}$ holds, then there are constants $c_1(c_*), q_0,q_1>0$, such that $\forall y\in~B(x,\eps_\l R)$: 
\begin{gather}
q_{2t}(y,y)\geq \f{c_1}{\l^{q_1}}\f{1}{t^{d_H/(d_H+\a)}}\;, \quad \text{for } t\in \left[\f{1}{4\l^{q_0}}R^{d_H+\a},\f{1}{2\l^{q_0}} R^{d_H+\a}\right]\;.
\label{eq:lower-bound-res}
\end{gather}
\end{theorem}
\vspace{2mm}
\noindent One relies on the upper bound \eqref{eq:up-bound} and on the following lemma, that controls the probability that the escape time is not too large:
\begin{lemma} Assuming $F_{R,\l}$, there are constants $c_1(c_*), q_0>0$ such that 
\begin{gather}
\mE^x(\t_{B(0,R)})\geq c_1\l^{-q_0}R^{d_H+\a}\;,\\
\mP^x(\t_{B(0,R)}>t)\geq \f{c_1\l^{-q_0}R^{d_H+\a}-t}{2c_*\l R^{d_H+\a}} \;, \quad \forall t\geq 0, x\in B(0,\eps_\l R)\;.
\end{gather}
\end{lemma}
\noindent Let us only say that those result from the inequalities introduced in eq.~\eqref{eq:green_property} and \eqref{eq:resistance_property} as well as the assumptions in $F_{R,\l}$.

\vspace{5mm}
From the linear behaviour of the resistance with the distance, we find the spectral dimension $4/3$ on trees. 
It is possible to generalize those bounds on the heat-kernel from less strict upper and lower bounds on volumes and resistances, by only delimiting upper and lower polynomial bounds \cite{KumagaiMisumi(2008)}.

\paragraph{Remark.} How does one go from on-diagonal to off-diagonal bounds? As in Lemma~\ref{egainlemma} of the main text, this derives from (3) of Lemma~\ref{lem:resistance}, by defining the function $f_t(y):=q_t(x_0,y)$ and having
\be
\label{eq:off-diagonal}
\abs{f_t(y)-f_t(x_0)}^2\leq R_{eff}(x_0,y) f_t(x_0)/t\;.
\ee
As in Lemma~\ref{lem:HKdecreasing}, one uses the heat-kernel $\psi(t):=f_{2t}(x_0)$ that obeys the heat equation 
\be 
\psi(t)^\prime= -2\cE(f_t,f_t)\;. 
\ee 
In order to conclude, this last equation and the monotonicity it implies for $\psi(t)$ aid to show
\be
t\cE(f_t,f_t)\leq 2 \int_{t/2}^t ds \cE(f_s,f_s) \leq \psi(t/2) \leq f_t(x_0)\;.
\ee

\subsubsection{Bounds for random weighted graphs}
Let us now take an ensemble of weighted graphs $\{(X(\omega),\mu^\omega):\om \in \Om\}$ satisfying the same conditions as previously and determined by a probability space $(\Omega, \cF, \mathbb{P})$. Then, given a sampled graph $(X(\om),\mu^\om)$ we are interested into \emph{quenched} estimates on graphs for which we can characterize typical properties such as the typical volume of balls or the scaling of the resistance. This generalizes the earlier encountered $\l$-good balls. 
\begin{theorem}
For constants $R_0,\l_0\geq 1$ and assuming that there exists a function $p(\l)$, with some constants $c_1, q_0>0$, such that $0\leq p(\l)\leq c_1\l^{-q_0}$. If additionally 
\be
\label{eq:probaFRL}
\mathbb{P}\left[\{\om : (X(\om,\mu^\om) \text{ satisfies } F_{R,\l}\}\right]\geq 1-p(\l) \;,
\ee
then there are constants $\a_1,\a_2$ and a dense subset $\Om_0\subset \Om$, with $\mP(\Om_0)=1$ such that: 
$\forall\om~\in~\Om_0$ and $x\in X(\om)$, there are $T_x(\om)\in \mathbb{R_{+}}$, for which 
\be
(\log t)^{-\a_1}t^{-d_H/(d_H+\a)}\leq q_{2t}^\om(x,x)\leq (\log t)^{\a_1}t^{-d_H/(d_H+\a)}\;, \quad \forall t\geq T_x(\om)\;.
\label{eq:encadrement}
\ee
Besides, if one has $p(\l)\leq \exp(-c_2\l^{q_0})$ then the $\log$ are replaced by $\log\log$ oscillations. 
\end{theorem}
\vspace{2mm}
\noindent To simplify, we focus on the case where $x$ is the root, $x=0$. By the two bounds~\eqref{eq:upper-bound-res} and~\eqref{eq:lower-bound-res}, taking $t=c_1(\l)R^{d_H+\a}$, the probability that the heat kernel stays confined in the corresponding range is high
\be
\mP\left(c_1\l^{q_1}\leq t^{d_H/(d_H+\a)}q^\om_{2t} (0,0)\leq c_1\l^{q_1}\right)\geq 1 - 2p(\l)\;.
\ee
From Borel-Cantelli on $p(\l)$, one can find a constant $K_0(\om)$ and sequences $t_k = e^k$, $\l_k = k^{2/q_0}$, such that for any $k\geq K_0(\om)$, the bounds~\eqref{eq:encadrement} are obeyed for the appropriate $T_0(\om)$, absorbing the constant $c_1$ inside the logarithms. 

\vspace{5mm}
As we saw in~\eqref{eq:lowbound_yn} and \eqref{eq:upbound_yn}, for Galton-Watson trees, it is actually this last exponential bound that occurs. 

At last, the \emph{annealed} bounds recover the exponential form \eqref{eq:HK} without logarithmic factors.
\begin{theorem}
Taken the $F_{R,\l}$ condition holding with a function $p(\l)\geq 0$ in eq.~\eqref{eq:probaFRL}, such that $\lim_{\l\to\infty}p(\l)=0$ 
then 
\begin{gather}
c_4t^{-d_H/(d_H+\a)}\leq \mE[q_{2t}^\om(x,x)]\;, \quad \forall t>0\;.
\end{gather}
If added to the condition $F_{R,\l}$~\eqref{eq:probaFRL}, there are constants $c_5>0,\l_0>1$ and $q_0^\prime>2$ controlling the lower bounds 
\be
\mP\left[\f{R^d_H}{\l}\leq V(B(x,R)), R_{eff}(x,y)\leq \l d(x,y)^\a, ~\forall y \in B(x,R)\right]\geq 1 -\f{c_5}{\l^{q_0^\prime}}\;, 
\label{eq:additiontoFRL}
\ee
for all $R\geq 1, \l\geq \l_0$. Then with a constant $c_6$
\be
\mE[q^\om_{2t}(x,x)] \leq c_6 t^{-d_H/(d_H+\a)}\;, \quad \forall t>0\;.
\ee
\end{theorem}
\vspace{2mm}
\noindent The technique for obtaining lower bounds on an average, is to restrict to events that behave nicely, namely that already satisfy the bound. This happens, as we have seen in Th.~\ref{th:low-boundFRL}, if we take events say $F$ obeying $F_{R,\l}$, for $\l$ and $R$ chosen sufficiently large, such that their probability to occur is large enough (bounded is enough). To extend the result for any $t>0$, one has to change appropriately the radius $R$, that will change the probability $\mP(F)$ of $F$ to occur.

Finally, for the last upper bound, taking a ball $B(x,R)$ such that $t=R^{d_H+\a}$, we consider all its possible volumes and resistances controlling them with a sequence $\l= k$ and call $H_k$ the corresponding event in eq.~\eqref{eq:additiontoFRL}, event which ensures the validity of the upper bound~\eqref{eq:upper-bound-res}. This means
\begin{align}
\mE[q^\om_{2t}(x,x)]&\leq \sum_k c k t^{d_H/(d_H+\a)}\mP(H_k \backslash H_{k-1})\\
&\leq \sum_k c k t^{d_H/(d_H+\a)}\mP(H_k^c)\;, 
\end{align}
that is summable from the assumption~\eqref{eq:additiontoFRL}.

\end{subappendices}

%% file: conclusion.tex
\begin{flushright}
\hfill\begin{minipage}{8cm}
{\footnotesize {\it     The Road goes ever on and on\\
    Down from the door where it began.\\
    Now far ahead the Road has gone,\\
    And I must follow, if I can,\\
    Pursuing it with eager feet,\\
    Until it joins some larger way\\
    Where many paths and errands meet.\\
    And whither then? I cannot say.}
\vspace{0.2cm}
\newline 
{\footnotesize \textbf{J. R. R. Tolkien}, \it{The Fellowship of the Ring.}}}
\end{minipage}
\end{flushright}

\section{Melonic Phase Transitions, CFTs and Holography}

Tensor models unveil a new large-$N$ limit that presents evidently conformal interacting fixed points, melonic CFTs. With a short-range kinetic term, they seem generically non-unitary, whereas when long-ranged and the interaction tuned marginal, unitarity is not ruled out. To the well-studied quartic models, we have added sextic ones, allowing real fixed points with real couplings. We understand well the spectrum of bilinears \footnote{We are studying with D. Benedetti how, in the language of conformal partial wave decomposition of the four-point function, the contour needs to be deformed, when varies the order of the interaction, to pick the physical poles.} and it remains to obtain OPE coefficients and to compute correlation functions of primaries to substantiate conformal symmetry. Particularly regarding the sign of logarithmic CFT of the short-range case, it would be neat to find the logarithmic multiplet that displays logarithmic factors in its correlation functions. Leading order four-point functions of the logarithmic fishnet theories resum also chains of ladders \cite{Gromov:2018hut}. Are there more connections to draw?

We have also seen that the perturbative parameter is bound to a small window in order to preserve reality of the fixed points and outside the window, the fixed points merge to form complex CFTs. It would be interesting to frame better this behaviour. 

Moreover, besides the bilinear operators, the spectrum contains many others, some not reducible using the equations of motion. Can we show that the latter are irrelevant at leading order in the CFT data? It is important to understand how the spectrum of all invariants organises inside the CFT. 
This question is also relevant to the holographic content of tensor models. Rephrasing the theory in terms of its vast number of invariants requires more ingenuity that in its lower rank relatives. Perhaps, inspiration may come from considering analogs of matrix eigenvalues for tensors. \footnote{First shots on this program, applied to Gaussian tensors, studied the distribution of the largest eigenvalue \cite{Evnin:2020ddw} or generalized the matrix resolvent \cite{Gurau:2020ehg}.} This problem pends on determining from first principles the low-energy effective action of tensor models. Also what are the implications of such growth in the bulk? 

Regarding the symmetries we have explored, we went through large-$N$ fixed points preserving the groups $U(N)\times U(N^2)$, $U(N)^3$ and $O(N)^3$. The interactions we have picked are specific instances of multifield potentials of the form $\l_{ijkl}\phi_i\phi_j\phi_k\phi_l$ or their sextic analog, see e.g. \cite{Osborn:2017ucf}. There have been recent progress for finding the CFT data of $O(m)\times O(n)$ fixed points using analytical and numerical bootstrap \cite{Henriksson:2020fqi}. \footnote{There $m$ is taken fixed and depending on the scheme, when $n$ is taken finite or large.} Since they present a method that could be applied to fixed points with any global symmetry, it might be  an interesting exercise to revisit the large-$N$ conformal data of the quartic Bosonic tensor fields. This technique may perhaps simplify the tackling of $1/N$ corrections to the conformal spectrum.

In the Fermionic color-symmetric $U(N)^3$ model, we found the phase diagram of the vacuum, with two phases spontaneously generating mass: one $U(N)^3$ symmetric and the second with residual symmetry $U(N/2)^2\times U(N)^2$. In order to probe how general such symmetry breaking might be, the phases of the sextic Bosonic theory could be analysed with a multi-matrix intermediate field \cite{Bonzom:2019kxi}.
As discussed in \cite{Benedetti:2017fmp}, the Goldstone Bosons of the phase with broken $U(N)$ subgroup are governed by a complex Grassmannian non-linear sigma model, more precisely with Grassmannian $\mathrm{Gr}(N,N/2)\equiv U(N)/(U(N/2)\times U(N/2))$. Although we have not made further use of this fact, it is interesting to notice the appearance of such models in the context of tensor models, something that might deserve further study.\footnote{A different pattern of spontaneous symmetry breaking has been discussed in \cite{Diaz:2018eik}, in which a $U(N)^2$ subgroup of the symmetry group gets broken down to its diagonal subgroup $U(N)$.}

Lastly, it would be interesting to continue working on models with different symmetry groups allowing MST interactions \cite{Carrozza:2018psc}, as is the tetrahedral quartic interaction. Finding healthy nontrivial fixed points for Fermionic systems in the presence of such interaction has so far remained challenging (see \cite{Benedetti:2017fmp,Prakash:2017hwq}), but it is worth persevering, as such models would have higher chance of displaying genuinely new physical behavior \cite{Ferrari:2019ogc}.

\section{QFT on Random Geometry}

We have only started the renormalization analysis of a field theory on a random geometry through this random walk formulation, and immediate questions concern actual determination of correlation functions and beta functions, providing therefore a concrete realization of Wilson's QFT in non-integer dimension. \footnote{Constructive issues, such as the analyticity domain of the annealed partition function, are also intriguing, but these would require some non-perturbative analysis such as the Loop-Vertex-Expansion, described in \cite{Rivasseau:2017hpg}, and we don't have any idea yet on how to use it on trees.} We are wondering if other types of expansion, such as the current expansion used in spin systems (e.g. \cite{Aizenman:2019yuo}), could be of use and how generic inequalities, as Griffiths-Hurst-Sherman, would be modified once taken quenched or annealed on random graphs. Naturally for this more general purpose, one should return also to short-range propagators.

Extensions to more interesting models would be next on the list. To name one, melonic models (for which long range propagators were also studied as we saw) would then lead to trees on trees. Nevertheless, tackling Fermionic matter will need special care, since the Fermionic random walk expansion differs from the Bosonic one, the first having Hausdorff dimension $d_H=1$  \cite{Jaroszewicz:1992js} as opposed to $d_H=2$ for the latter. Thus the preceding bounds on propagators require adequate reconsideration. Perhaps, since after all a Fermionic propagator can be obtained by differentiating a scalar one with respect to the momentum in the proper time formalism, a starting point may be to study those derivatives.

Different ensembles of trees can be looked at. For instance, we know that under specific conditioning of the branching process, it is possible to force the Galton-Watson tree to grow a finite number $p>1$ of infinite spines \cite{abraham2018}. It would be interesting to characterize heat-kernel bounds relying on the techniques of \cite{BarlowKumagai}, their scaling limit (since Aldous' CRT has a single spine) and determine the renormalization group properties of field theories on such trees. Are they related to the supercritical processes or those generated by stable laws with index $\b \in (1,2)$ for which the spectral dimension are $2\b/(2\b-1)$ \cite{CroydonKumbeta}?

We have argued that similar techniques could generalize directly to the strongly-recurrent graphs, among which are the random planar maps of spectral dimension $d_s=2$ \cite{Gwynne and Miller(2017)}. Computation of correlation functions in this formalism may present another approach to the KPZ exponents. 

Additionally, we are now quite familiar that single and double scaling limits of large-$N$ tensor models roam between the Aldous tree and the Brownian map. 
The analog of the genus-expansion of correlation functions of matrix models for tensors is given by a half-integer (so-called degree) that combines geometric with topological information. 
We hope, that decomposing a single tensor model into its large $N$ limit and subleading contributions seen as a fluctuating scalar field on top of this background, this point of view may help explore different scaling limits and indicate a way to probe a larger set of families within the degree expansion. 

Ultimately, returning to the unicycles, we think that albeit modest, this work may serve the subject of a ``random holography". Up to now, having in mind the SYK model, the disorder has been introduced through the interaction, whereas here it is the underlying geometry that is being averaged on. In the former case, an effective geometry, fleshed by wormholes connecting the boundaries where the quantum systems live, seems to emerge out of the disorder average \cite{Penington:2019kki}. Could a similar property occur in our case? In particular, in melonic models, is there an analogous reparameterization invariance on the spine? The next step will consist in computing the ladder kernel on which rest dynamical properties such as, perhaps, chaos. In the original papers, the technique relies on an analytic continuation from Euclidean to Lorentzian propagators as described in \cite{Murugan:2017eto}. Whether a similar procedure makes sense for the effective propagator on the spine requires more work. Within the same framework, could we recover the periodic expression of the propagator in our random walk setting? Still, it is not yet clear what type of decorating trees one should use on the spine, if non-critical are enough or if one needs to resort to critical trees as well. To have good control on the computation, it would be important to set up a proper continuum limit of those objects.

A closely related perspective considers the partition function of a looping random walk on a two dimensional hyperbolic surface as the quantization of JT gravity (\cite{Stanford:2020qhm} and references therein) from what they derived the density of states in three distinct regimes, depending on the pressure and length of the loop. Allowing branches to grow on the spine or loop changed for us the resulting boundary propagator. How would the corresponding partition function change?

\vspace{5mm}
As a closing remark, we find quite amusing that models that were designed to generate higher-dimensional geometries as emerging from a continuum limit, reappear from a holographic point of view at the boundary... For the moment, it looks only coincidental. The Tensor Track may reserve many surprises yet to come. 

%% file: Resume.tex
Les phénomènes naturels aux échelles mesurables par l'être humain peuvent être décrits à partir de quatre forces fondamentales qui, classées dans ordre de l'intensité relative de leur couplage à l'échelle de $10^{-15}$m, sont les forces: forte ($10^{38}$), électromagnétique ($10^{36}$), faible ($10^{32}$) et gravitationnelle (1). Cet incroyable écart semble leur suggérer une nature très différente. En effet, les trois premières sont aujourd'hui très bien décrites par le Modèle Standard, une théorie quantique des champs, fondamentalement probabiliste, pour laquelle les particules observées sont des excitations de champs quantiques. À partir de l'échelle de l'ordre du mètre (et jusqu'aux confins de l'univers observable), la force gravitationnelle prend le dessus sur les trois autres. La meilleure théorie que nous en ayons est la relativité générale, dont les équations relient la géométrie de l'espace-temps à la distribution de matière qui s'y propage. Parmi ses postulats, s'inscrivent la localité des interactions et le respect de la causalité. Il semble ainsi que notre conception de la Nature distingue l'espace-temps de la matière qu'il contient. Le domaine de la ``gravité quantique" vise à unifier ces deux points de vue. Suivant le postulat mis en doute, différentes théories sont développées, à l'instar de la théorie des cordes (dont les particules fondamentales sont des cordes) ou de la gravitation quantique à boucles (qui part d'une description discrète de la géométrie). 

\vspace{5mm}
Les modèles de tenseurs sur lesquels nous nous sommes penchés sont curieusement reliées à ces deux théories. Ce sont des théories de champs quantiques tensoriels $T_{i_1i_2\dots i_q}$ de rang $q\geq 3$, se transformant sous $G^{\otimes q}$, à l'aide d'un groupe classique $G$ de taille $N$ (nous avons regardé les groupes $U(N)$ et $O(N)$). Leur point de départ est une fonction de partition euclidienne: \footnote{Les indices répétés sont supposés sommés.}
\be
Z = \int DT e^{-S[T]}\;, \quad S[T] = T_{i_1i_2\dots i_q}C^{-1}T_{i_1i_2\dots i_q} + S_{int}[T]\;, 
\ee
de laquelle découlent toutes les fonctions de corrélation du système, caractérisant la dynamique de la théorie. L'interaction $S_{int}[T]$, typiquement un polynôme en $T$, est prise invariante sous $G^q$. Les fonctions de corrélations sont alors obtenues par une expansion de Taylor du terme d'interaction et par l'intégration de l'exponentielle gaussienne, conduisant aux diagrammes de Feynman. Certains modèles de tenseurs de rang $q$ dits ``colorés", avec une interaction d'ordre $q+1$ correspondant à un $q$-simplexe, voient leurs diagrammes de Feynman discrétiser des variétés linéaires par morceaux de dimension topologique $q$. On peut alors chercher à construire une limite continue telle que la dynamique de la géométrie résultante serait décrite par les équations d'Einstein. Attaqué de front, c'est un problème difficile. Une approche consiste à remarquer que les modèles colorés sont solubles dans la limite de $N$ grand. Une telle limite simplifie dans de nombreux cas des équations de théorie des champs, une fois que le couplage de l'interaction redimensionné avec $N$ de façon appropriée. Il faut ensuite regarder les corrections en $1/N$. Par exemple, pour des modèles de matrices ($q=2$), une telle expansion conduit à une série où le terme d'ordre \footnote{À un facteur global de $N$ près.} $N^{2-2g}$ resomme toutes les surfaces de genre $g$, le terme dominant donné par les surfaces planaires \cite{DiFrancesco:1993cyw}. Un développement similaire, indicé par le degré de Gurau, fut construit pour les modèles colorés \cite{Gurau:2010ba, Gurau:2011aq, Gurau:2011xq}, puis non-colorés \cite{Bonzom:2012hw}, et dont les fonctions de corrélation dominantes possèdent une structure récursive, formant la classe des ``melons" (voir par exemple la figure \ref{fig:melons-intro} dans le texte). La structure diagrammatique des fonctions de corrélation d'ordres supérieurs fut décomposée en termes de schémas réduits, en nombre fini à chaque ordre en $1/N$ \cite{GurauSchaeffer}.

\begin{figure}[htbp]
\centering
\begin{minipage}{0.3\textwidth}
           \centering 
            \includegraphics[width=0.5\textwidth]{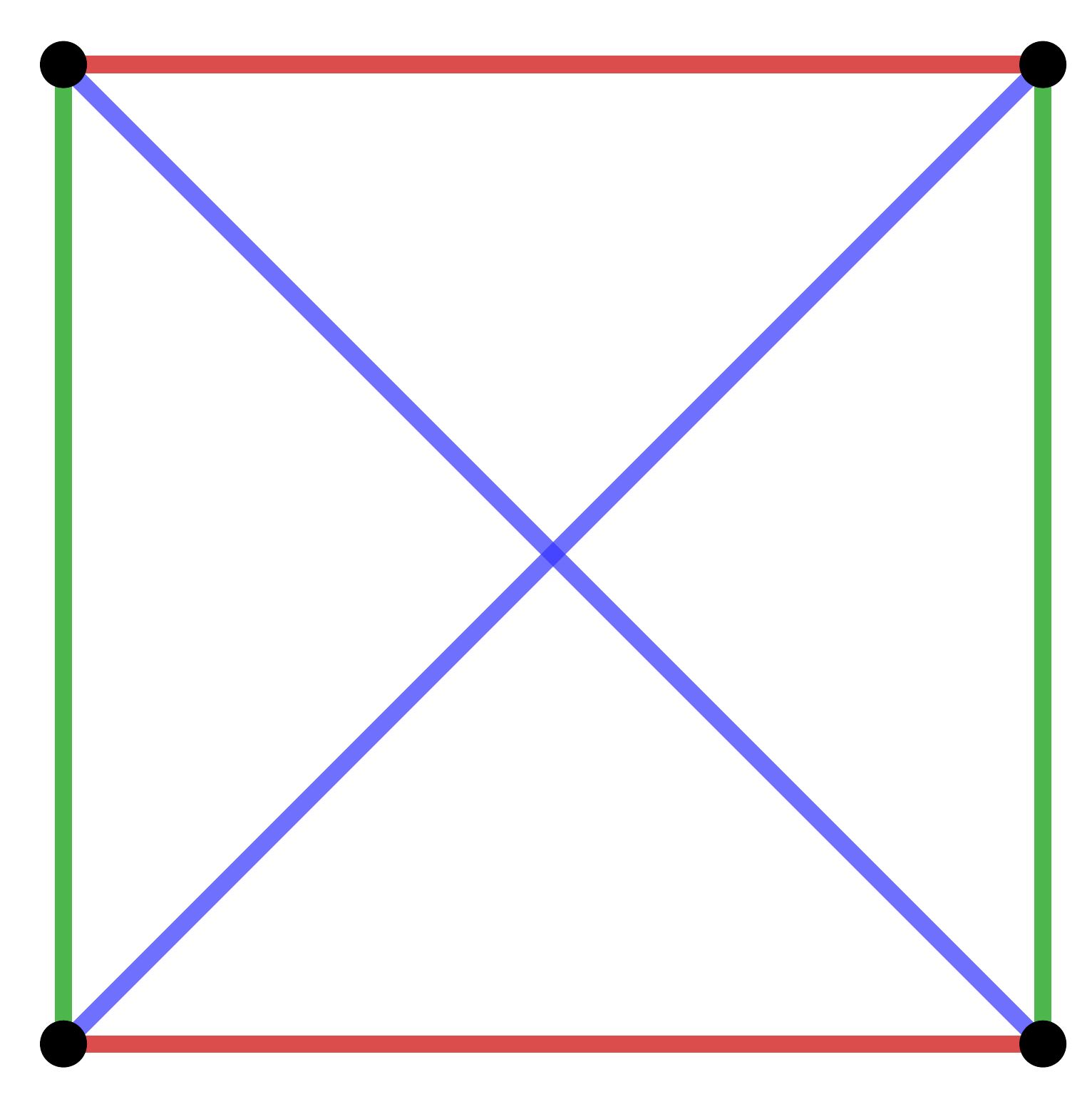}
        \end{minipage}
\begin{minipage}{0.3\textwidth}
            \centering
            \includegraphics[width=0.5\textwidth]{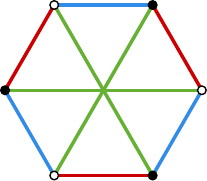}
        \end{minipage}
\begin{minipage}{0.3\textwidth}
            \centering
            \includegraphics[width=0.5\textwidth]{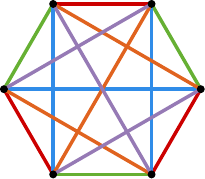}
        \end{minipage}
        \caption{Exemples d'interactions MST (respectivement en rangs 3, 3 et 5 ; seule celle du milieu est bipartite).}
        \label{fig:exempleMST}
\end{figure}

Une question importante dans ce domaine est la détermination de la puissance de $N$ associée à chaque couplage qui rend la limite de grand $N$ non-triviale. La puissance originale (ici pour des tenseurs $(T, \bar{T})$, pris complexes de rang $r$)
\be
S (T,\bar T)  = N^{r-1}\left[\sum  T_{b^1\dots b^r} \bar T_{q^1\dots q^r} \prod_{c=1}^r \delta_{b^cq^c} 
      + \sum_{\text{graphes $r$-colorés } \cB}    t_{\cB} \Tr_{\cB}( T, \bar T)\right] \;, 
\ee
où les couplages $t_\cB$ sont indépendants de $N$ et les interactions sont des graphes $r$-colorés, 
favorisait les interactions meloniques, les autres étant sous-dominantes. Pour des interactions dites ``Maximally Single Trace", il fut prouvé \cite{Carrozza:2015adg, Ferrari:2017jgw} que la puissance optimale était
\be 
S(T,\bar T) = N^{r/2}\left[T \bar{T} + \sum_{\text{graphes $r$-colorés } \cB} t_\cB N^{-\rho(\cB)} \Tr_{\cB}( T, \bar T)\right], \quad \rho(\cB) = \f{F_{\cB}}{r-1} - \f{r}{2},
\ee
agrandissant la classe de graphes dominants à des melons généralisés (cette fois, les melons se dessinent lorsque l'interaction est réduite à un vertex du graphe). Ici, $F_{\cB}$ compte les cycles de deux couleurs alternées dans les interactions, aussi appelés \emph{faces}. Ces interactions possèdent la particularité de n'avoir qu'une seule face par paire de couleur. Des exemples de ces interactions sont présentés en Figure \ref{fig:exempleMST}. Chaque tenseur est représenté par un vertex (noir pour $T$ et blanc pour $\bar{T}$) et chaque contraction d'indice est associée à une arête dont la couleur suit la position de l'indice. En présence d'un groupe de symétrie complexe, les interactions sont bipartites (les vertex blancs ne sont reliés qu'aux vertex noirs et réciproquement).

\vspace{5mm}
Depuis les années 1930, s'étoffe l'idée que les constantes d'interactions des théories de champs changent en fonction de l'échelle d'énergie considérée. Cette dépendance se traduit par un flot dans l'espace des couplages possibles décrit par le groupe de renormalisation, culminant avec les travaux de Wilson et al. \cite{Wilson:1973jj, WilsonNobel}. Pour les théories renormalisables, on a un contrôle du flot à toute énergie. De fait, on cherche les points fixes de ce flot, fonction entre autres des symétries de la théorie, qui souvent possèdent en plus une symétrie conforme fortement contraignante sur la forme des fonctions de corrélation, auquel cas on parle de théorie conforme (ou CFT). L'intérêt physique de ces points fixes est qu'ils décrivent des propriétés critiques de systèmes avec les symétries correspondantes,\footnote{Et un certain nombre fini d'autres paramètres tel que le nombre de degrés de liberté.} définissant de la sorte des classes d'universalité. On peut ainsi se demander s'il existe des points fixes propres aux théories tensorielles, quelles sont leurs caractéristiques critiques et s'ils correspondent à des théories conformes et unitaires. 

Les premiers travaux abordant cette question ont considéré un modèle quartique de rang $3$ sous la dimension critique $4$, pour lequel l'action générale renormalisable est donnée par 
\begin{equation*}
S = \int \dd[d]x \left(\f{1}{2}\phi(-\D)^\zeta\phi + \f{g_d}{N^3}\vcenter{\hbox{\includegraphics[height=.7cm]{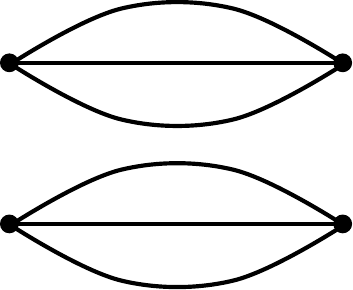}}}
+ \f{g_p}{N^2}\vcenter{\hbox{\includegraphics[height=1cm]{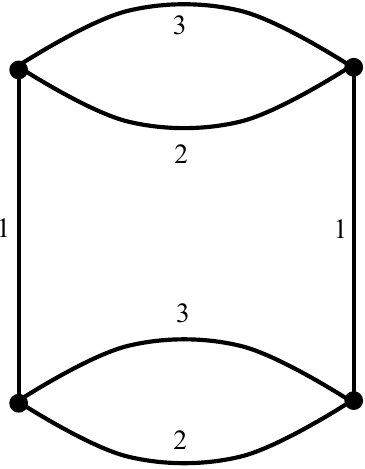}}} +\f{g_t}{N^{3/2}}\vcenter{\hbox{\includegraphics[height=1cm]{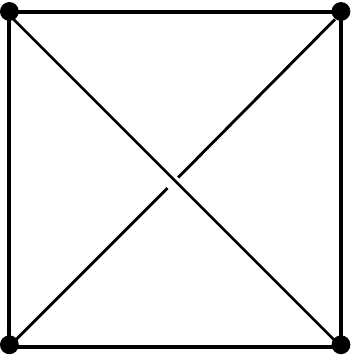}}} \;\right)
\end{equation*}
Pour $\z=1$, le modèle est dit de courte portée, alors que s'il est pris entre $0<\z<1$, on parle de modèle à longue portée. D'abord regardant le cas de courte portée \cite{Giombi:2017dtl}, des points fixes avec des valeurs imaginaires de $g_d$ et $g_p$ furent obtenus, ennuyeux pour la stabilité du point fixe car les opérateurs associés sont positifs. Par ailleurs, le spectre des dimensions conformes des opérateurs bilinéaires en les champs $\phi$ possédait une valeur complexe. Le cas de longue portée avec $\z=d/4$ fut également considéré \cite{Benedetti:2019eyl}, où autorisant une valeur imaginaire au couplage $g_t$, quatre lignes de points fixes réels furent trouvés pour $g_d$ et $g_p$ (et exposants critiques réels). Plus tard, les coefficients OPE furent démontrés réels \cite{Benedetti:2019ikb} et l'invariance conforme du modèle fut établie \cite{Benedetti:2020yvb}, indiquant que à grand $N$, cette CFT melonique est unitaire.

\begin{figure}[htbp]
\centering 
\includegraphics[width=.8\textwidth]{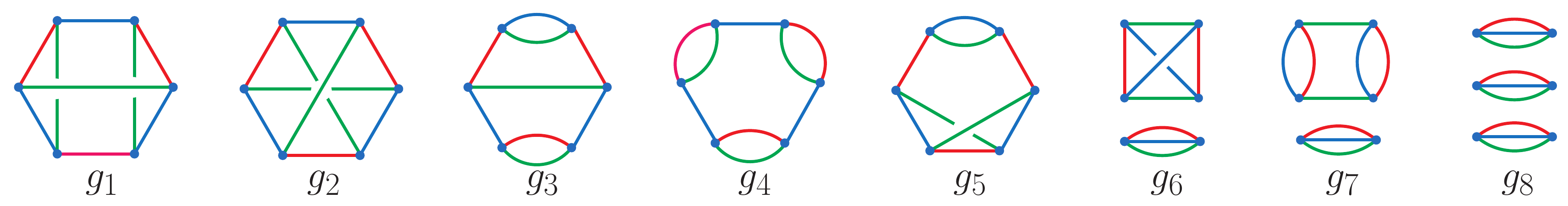}
\caption{Toutes les interactions d'ordre 6 pour une théorie invariante sous $O(N)^3$. Seules les interactions $g_2, g_3, g_4, g_7, g_8$ forment des graphes bipartites.}
\label{fig:phi6-vertices}
\end{figure}

Sur les modèles sextiques de rang $3$ sous la dimension critique $3$, seul le cas de courte portée fut étudié. D'abord fut prise une limite dite ``prismatique" \cite{Giombi:2018qgp} qui favorisait une interaction dont l'opérateur était positif (avec la forme d'un prisme). Des points infrarouges non-triviaux fixes incluant les 8 couplages autorisés (voir la figure \ref{fig:phi6-vertices}) furent obtenus par une expansion en $\eps = d-3$ et pour $N$ grand. Cela dit, la condition d'un spectre de dimensions conformes réelles ne fut satisfaite que pour des dimensions $d<1.68$ et $2.81<d<3$.
Toutefois, cela laissait ouverte la construction explicite de points fixes meloniques pour des modèles sextiques. 

La première partie de ma thèse traite, à trois dimensions, de théories renormalisables tensorielles fermionique (en rang 3) et bosoniques (en rangs 3 et 5) avec interactions respectivement d'ordre 4 et 6, en nous basant sur l'expansion melonique. Pour la théorie fermionique, de symétrie $U(N)^3$, on trouve deux points fixes (dont un stable à haute énergie) qui préservent des interactions purement tensorielles, en plus des points fixes obtenus par le modèle tridimensionnel de Gross-Neveu \cite{ROSENSTEINreview}, vectoriel. Par contre, la dimension conforme du champ intermédiaire bilinéaire semble inchangée aux quatre points fixes. Par l'introduction de champs intermédiaires matriciels, réécrivant les interactions  quartiques comme résultant d'un couplage à des champs ``intermédiaires" gaussiens, une fois ceux-ci diagonalisés, on a pu étudier le potentiel effectif du système à toute valeur des couplages et en déduire un diagramme des phases de l'état du vide, minimum du potentiel. Celui-ci présentait une phase sans masse symétrique sous $U(N)^3$ et deux phases brisant spontanément la symétrie chirale, générant une masse. L'une conserve la symétrie globale initiale tandis que l'autre la brise en $U(N/2)^2\times U(N)^2$ (brisant aussi la symétrie sous permutation des couleurs). Cela suggère l'étude plus générale des brisures des symétries globales continues et discrètes des modèles de tenseurs. La question de jauger la symétrie globale est aussi discutée, d'où on conclut qu'une telle jauge ne change pas le comportement du modèle à grand $N$. 

Dans les modèles sextiques, de groupes de symétrie $U(N)^3$ et $O(N)^5$, un couplage non-melonique adjoint d'une puissance de $N$ optimale nous conduit à une expansion melonique généralisée, suivant les travaux de \cite{Carrozza:2015adg,Ferrari:2017jgw}.
Les termes cinétiques sont pris de courte ou longue portée et on étudie, à grand $N$, perturbativement les différents groupes de renormalisation des couplages d'ordre 6, jusqu'à quatre boucles. Une des difficultés techniques réside dans la détermination des coefficients combinatoires devant les différentes contributions de chaque couplage. Tandis que le modèle de rang 5 ne présente pas de point fixe non-trivial, celui de rang 3 possède deux points fixes non-triviaux réels de type Wilson-Fisher dans le cas à courte portée (que l'on relie à des théories conformes logarithmiques, non-unitaires) et une ligne de points fixes dans l'autre. On obtient enfin les dimensions conformes réelles des opérateurs primaires bilinéaires en le champ fondamental. On remarque toutefois que le paramètre d'expansion doit se restreindre à une étroite fenêtre pour préserver la réalité du point fixe. 

Une autre raison qui a motivé l'étude des théories tensorielles définies sur $\mathbb{R}^d$ fut l'apparition de diagrammes meloniques dans la limite de grand $N$ d'un modèle quantique de $N$ fermions en interaction sujet à du désordre, le modèle de Sachdev-Ye-Kitaev \cite{Kitaev2015}. Celui-ci possède dans un régime de basse température des propriétés analogues à celles de régions proches de l'horizon de trous noirs presqu'extrêmes, reliées à la notion de chaos quantique \cite{Maldacena:2016hyu, Maldacena:2016upp}. Ainsi, celui-ci fournit un exemple ``élémentaire", i.e. à basse dimension, de correspondance holographique où une théorie quantique est duale à une théorie gravitationnelle de dimension supérieure. Dans le cas où la théorie gravitationnelle est asymptotiquement Anti-de Sitter, celle quantique possède une invariance conforme \cite{AdS}. Cette année, ce modèle a suscité d'importants progrès concernant le paradoxe de l'information impliqué par l'existence de trous noirs \footnote{À savoir, que devient l'information qui a traversé l'horizon, une fois le trou noir évaporé.}, cf. la revue \cite{Almheiri:2020cfm}. Toutefois, la présence de désordre dans le modèle le rend exotique par rapport aux cas standards de la dualité. De ce fait, les modèles de tenseurs, sans désordre et dont la limite melonique implique le même comportement de basse énergie que SYK, ont suscité l'intérêt de la communauté de physique de haute énergie (par exemple \cite{Witten:2016iux,Klebanov:2016xxf}) et l'étude de leurs propriétés conformes ou autour dudit point fixe. Notre travail s'inscrit dans ce registre.

Parmi les questions encore ouvertes, nous comptons l'effet des corrections en $1/N$ sur les propriétés décrites plus haut (retrouvons-nous des points fixes meloniques non-triviaux?) ou encore l'étude d'opérateurs d'ordre supérieur à deux en les champs fondamentaux, qu'il est nécessaire de comprendre pour une description complète de la CFT tensorielle. 

\vspace{5mm}
La deuxième partie de la thèse aborde le groupe de renormalisation à l'aide de la technique constructive de l'analyse multi-échelle \cite{Rivabook}, qui offre un contrôle essentiel pour se débarrasser de divergences de type ``renormalons" ou étudier des propriétés d'analyticité de la fonction de partition. Par là, nous revenons à la question de la gravité quantique en formalisant la notion de théorie des champs sur une géométrie aléatoire à partir de techniques probabilistes \cite{BarlowKumagai} pour borner des propagateurs, exprimés sous forme de marches aléatoires, une idée de Symanzik \cite{Symanzik:1969}. Notons que des expansions similaires sont utilisées pour étudier de façon rigoureuse des propriétés de fonctions de corrélations. Cela permit l'obtention de bornes sur des fonctions beta, des relations entre des exposants critiques sous la dimension critique des modèles ou encore la preuve de la trivialité de la classe d'universalité de $\phi^4$ en dimensions $d\geq 4$ (voir par exemple \cite{Fernandez:1992jh, Aizenman:2019yuo}). En règle générale, il s'agit de déterminer le volume, la probabilité d'intersections de marcheurs aléatoires sur le graphe ou d'autres propriétés de structures typiques étant donné un ensemble aléatoire de graphes. Nous nous concentrons sur un modèle scalaire quartique avec un terme cinétique à longue portée rendant l'interaction marginale, sur des arbres de Galton-Watson critiques. Au point critique, l'émergence d'une spine infinie fournit un espace sur lequel calculer des fonctions de corrélations moyennées. Nos bornes sur les amplitudes renormalisées et notre procédure de soustraction des divergences confirment l'attribution d'une dimension effective $4/3$ à la spine. 
On esquisse aussi l'extension du formalisme à des fermions et à une spine compactifiée afin de généraliser les résultats de renormalisation à des modèles plus riches, comme les tenseurs discutés plus haut. 

Détaillant l'obtention des bornes probabilistes sur le noyau de la chaleur dans un graphe aléatoire, nous espérons faciliter l'usage de ces techniques en théorie des champs, sur des ensembles aléatoires de graphes plus généraux. Au long terme, les techniques développées pourraient servir à étendre les relations de KPZ \cite{Knizhnik:1988ak} obtenues à deux dimensions, à un ensemble de géométries plus large, précisant l'effet d'une géométrie fluctuante sur la matière quantique qui s'y propage.